%% file: top.tex
\documentclass[acmtog,nonacm]{acmart}
\AtBeginDocument{%
  \providecommand\BibTeX{{%
    \normalfont B\kern-0.5em{\scshape i\kern-0.25em b}\kern-0.8em\TeX}}}

\setcopyright{rightsretained} 
\acmJournal{TOG}
\acmYear{2023} \acmVolume{42} \acmNumber{4} \acmMonth{8} \acmPrice{15.00} \acmArticle{1} \acmDOI{10.1145/3592430}

\usepackage{pifont}
\usepackage{tikz}
\usepackage{enumitem}
\usepackage{tabularx}
\usepackage{multirow}

\usepackage{mathtools}
\mathtoolsset{centercolon}

\newcommand*{\HQIMAGES}{}

\newcommand{\TODO}[1]{{\color{red}{[TODO: #1]}}}

\newcommand{\updates}[1]{{{#1}}}

\newcommand{\ours}{\textsc{FlexiCubes}}
\newcommand{\DMTet}{\textsc{DMTet}}
\newcommand{\nvdiffrec}{\textsc{nvdiffrec}}
\newcommand{\nvdiffrecmc}{\textsc{nvdiffrecmc}}

\newcommand{\eg}{\emph{e.g.}} %
\newcommand{\vs}{\emph{vs.}} %
\newcommand{\etc}{\emph{etc.}} %
\newcommand{\etal}{\emph{et al.}} %

\input{sources/figures}

\input{sources_suppl/supp_figs.tex}

\input{sources/tables}
\acmSubmissionID{662}

\begin{document}

\title{Flexible Isosurface Extraction for Gradient-Based Mesh Optimization}

\author{Tianchang Shen}
\email{frshen@nvidia.com}
\orcid{0000-0002-7133-2761}
\affiliation{%
  \institution{NVIDIA},
  \institution{University of Toronto},
  \institution{Vector Institute}
  \country{Canada}
}

\author{Jacob Munkberg}
\orcid{0009-0004-0451-7442}
\affiliation{%
  \institution{NVIDIA}
  \country{Sweden}
}

\author{Jon Hasselgren}
\orcid{0009-0002-3423-190X}
\affiliation{%
  \institution{NVIDIA}
  \country{Sweden}
}

\author{Kangxue Yin}
\orcid{0009-0007-0668-2769}
\affiliation{%
  \institution{NVIDIA}
  \country{Canada}
}

\author{Zian Wang}
\orcid{0000-0003-4166-3807}
\affiliation{%
  \institution{NVIDIA},
  \institution{University of Toronto},
  \institution{Vector Institute}
  \country{Canada}
}

\author{Wenzheng Chen}
\orcid{0009-0008-5623-1963}
\affiliation{%
  \institution{NVIDIA},
  \institution{University of Toronto},
  \institution{Vector Institute}
  \country{Canada}
}

\author{Zan Gojcic}
\orcid{0000-0001-6392-2158}
\affiliation{%
  \institution{NVIDIA}
  \country{Switzerland}
}

\author{Sanja Fidler}
\orcid{0000-0003-1040-3260}
\affiliation{%
  \institution{NVIDIA},
  \institution{University of Toronto},
  \institution{Vector Institute}
  \country{Canada}
}

\author{Nicholas Sharp}
\orcid{0000-0002-2130-3735}
\authornote{Authors contributed equally.}
\affiliation{%
  \institution{NVIDIA}
  \country{USA}
}

\author{Jun Gao}
\orcid{0000-0002-3521-0417}
\authornotemark[1]
\affiliation{%
  \institution{NVIDIA},
  \institution{University of Toronto},
  \institution{Vector Institute}
  \country{Canada}
}

\renewcommand\shortauthors{Shen \etal{}}

\begin{abstract}

This work considers gradient-based mesh optimization, where we iteratively optimize for a 3D surface mesh by representing it as the isosurface of a scalar field, an increasingly common paradigm in applications including photogrammetry, generative modeling, and inverse physics.
Existing implementations adapt classic isosurface extraction algorithms like Marching Cubes or Dual Contouring; these techniques were designed to extract meshes from fixed, known fields, and in the optimization setting they lack the degrees of freedom to represent high-quality feature-preserving meshes, or suffer from numerical instabilities.
We introduce \ours{}, an isosurface representation specifically designed for optimizing an unknown mesh with respect to geometric, visual, or even physical objectives.
Our main insight is to introduce additional carefully-chosen parameters into the representation, which allow local \emph{flexible} adjustments to the extracted mesh geometry and connectivity.
These parameters are updated along with the underlying scalar field via automatic differentiation when optimizing for a downstream task.
We base our extraction scheme on Dual Marching Cubes for improved topological properties, and present extensions to optionally generate tetrahedral and hierarchically-adaptive meshes.
Extensive experiments validate \ours{} on both synthetic benchmarks and real-world applications, showing that it offers significant improvements in mesh quality and geometric fidelity.

\end{abstract}


\figTeaser

\maketitle
\input{sources/01_intro}

\input{sources/02_related}

\input{sources/03_method}

\input{sources/04_experiments}

\input{sources/05_applications}

\input{sources/06_conclusion.tex}

\begin{acks}
The authors are grateful to Aaron Lefohn and David I.W. Levin for their support and discussions during this research, as well as the anonymous reviewers for their valuable comments and feedback.
\end{acks}

{\small
\bibliographystyle{plainnat}

\input{output.bbl}
}

\newpage
\appendix
\Huge
\section*{Supplemental Material}
\normalsize
\setcounter{section}{0}
\renewcommand{\thesection}{\Alph{section}}
\input{sources_suppl/overview.tex}
\input{sources_suppl/method_details.tex}
\input{sources_suppl/analysis_details.tex}
\input{sources_suppl/main_experiments_details.tex}
\input{sources_suppl/application_details.tex}

\end{document}

%% file: sources/figures.tex
\newcommand{\figTeaser}{
    \begin{teaserfigure}
    \centering
    \includegraphics[width=0.98\textwidth,trim=0 0 0 30,clip]{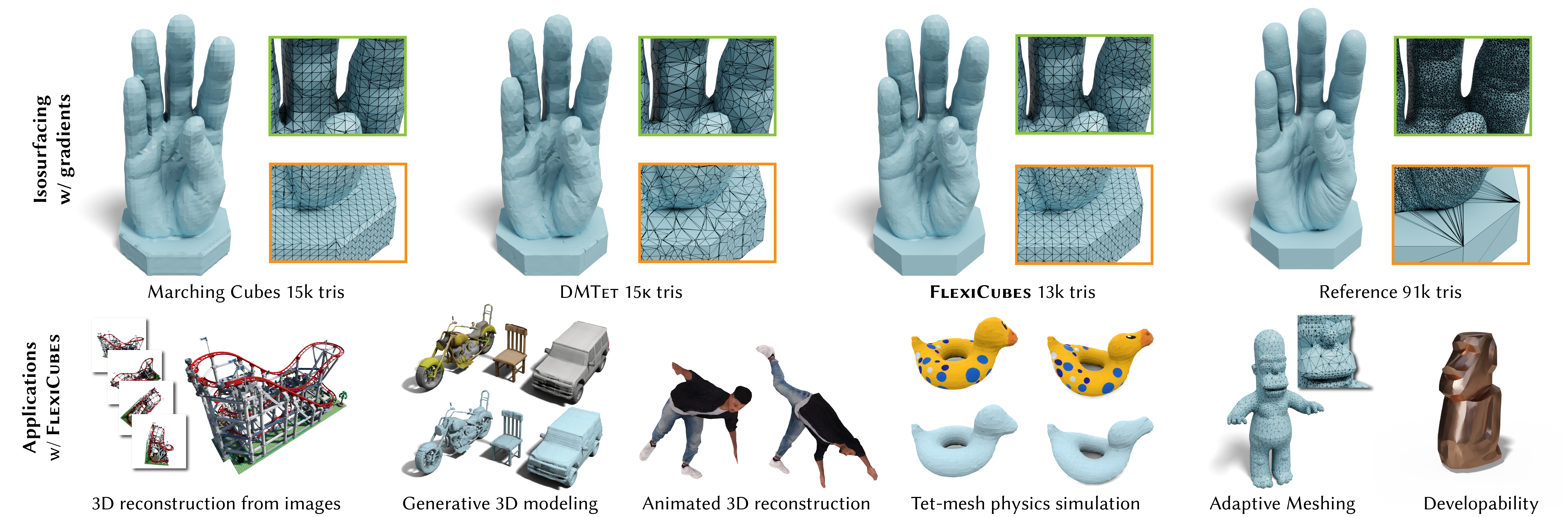}
    \caption{We introduce \ours{}, a high quality isosurface representation specifically designed for gradient-based mesh optimization with respect to geometric, visual, or even physical objectives. 
    We present a detailed quality evaluation and demonstrate that \ours{} improves the results in a range of applications.}
    \Description{\ours{}}
    \label{fig:teaser}
    \end{teaserfigure}
}

\newcommand{\figDMC}{
    \begin{figure}
    \centering
    \includegraphics[width=0.45\textwidth,trim=0 0 0 0,clip]{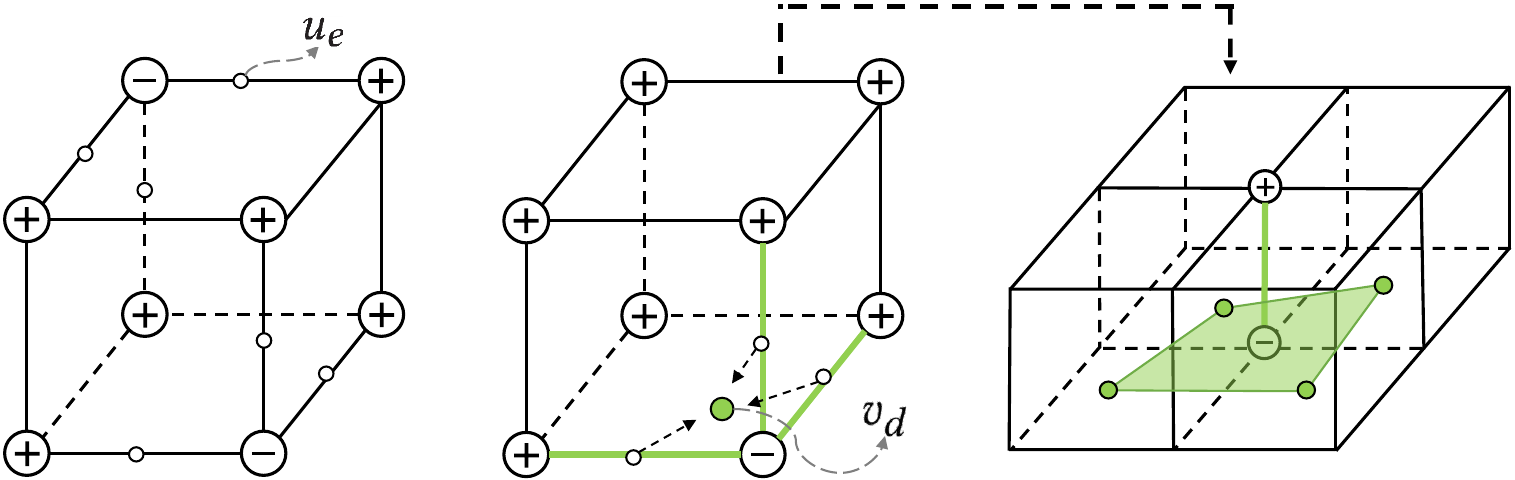}
    \caption{Dual Marching Cubes first interpolate vertex along the edge to obtain $u_e$. The dual vertex $v_d$ is computed via Equation~\ref{eq:v_d}. We connect four neighboring dual vertices to obtain a quadrilateral.}
    \label{fig:dmc}
    \end{figure}
}

\newcommand{\figDMCConfigs}{
    \begin{figure}
    \centering
    \includegraphics[width=0.45\textwidth,trim=0 0 0 0,clip]{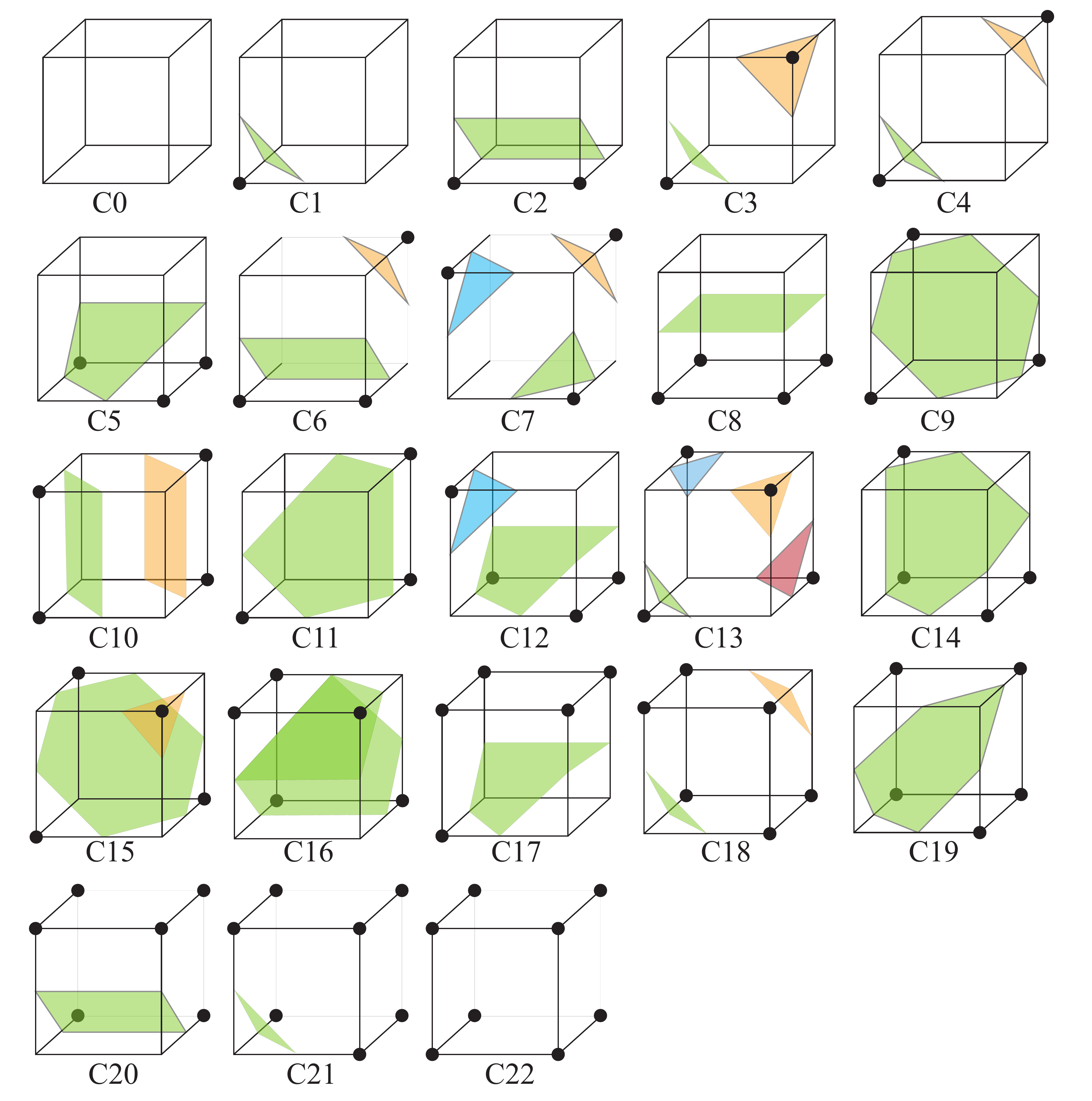}
    \caption{All configurations for the Dual Marching Cubes surface, with rotation symmetric cases removed. 
    Each colored polygon is a primal face, in which a single dual vertex is extracted as output.
    Marked vertices indicate negative signed distance values, $s(x) < 0$, and unmarked  vertices indicate positive values. 
    Figure adapted from ~\citet{nielson2004dual}. 
    }
    \label{fig:dmc_configs}
    \end{figure}
}

\newcommand{\figDualVertex}{
    \begin{figure}
    \centering
    \includegraphics[width=0.45\textwidth,trim=0 0 0 0,clip]{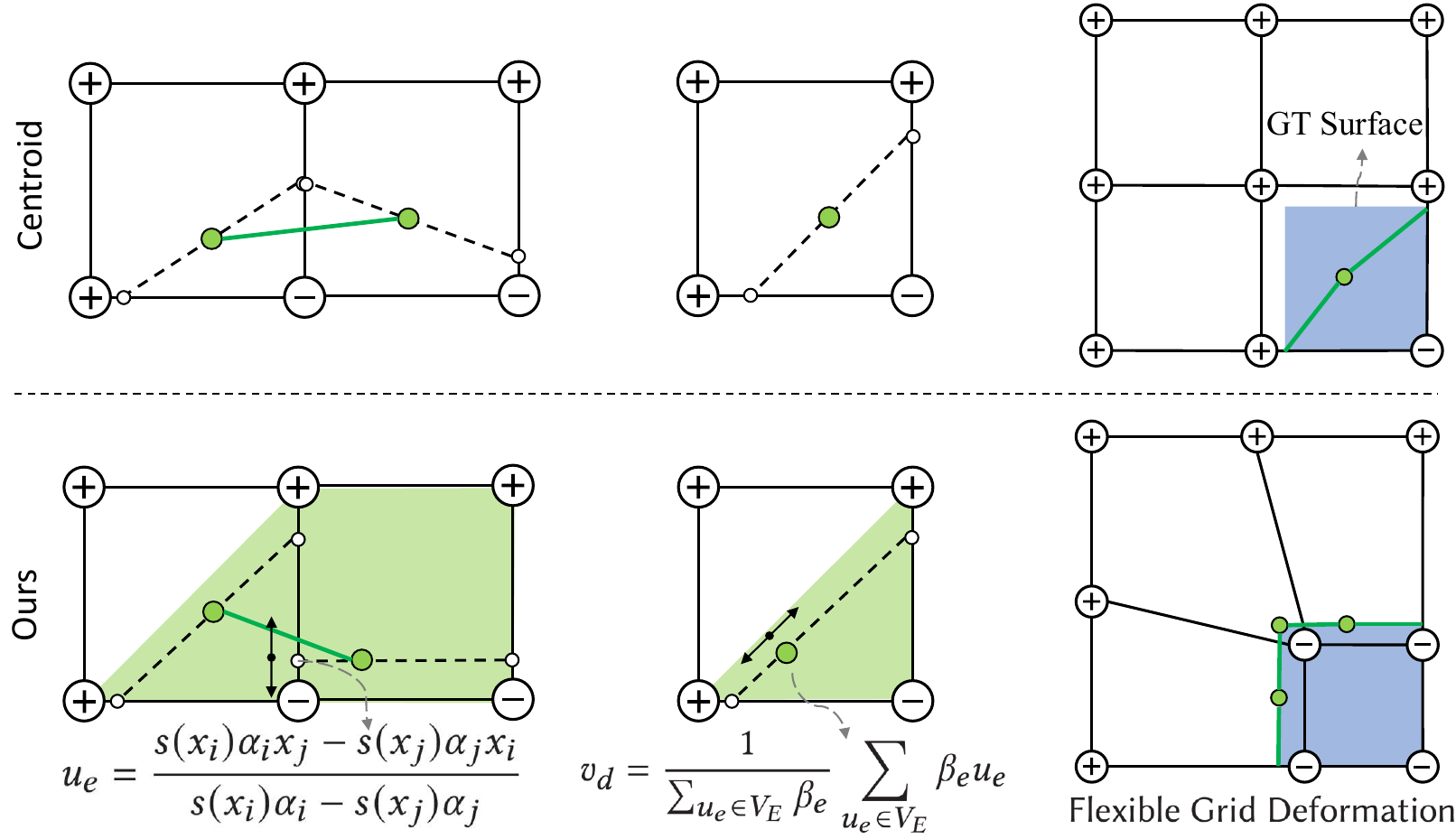}
    \caption{Formulation of determining the position of dual vertex. The dual vertex in our formulation can be placed anywhere within the green region.    }
    \label{fig:dual_vertex}
    \end{figure}
}

\newcommand{\figSpliting}{
    \begin{figure}
    \centering
    \includegraphics[width=0.45\textwidth,trim=0 0 0 0,clip]{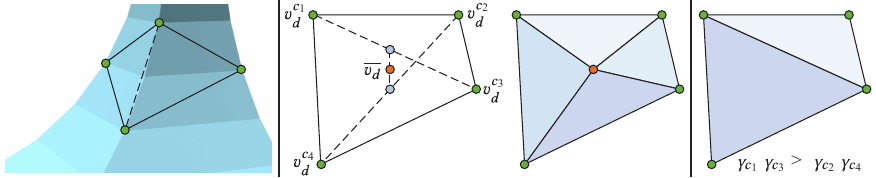}
    \caption{\textbf{Left}: \ours{} generates non-planar quadrangles, which may need to be triangulated e.g. for rendering. Note that in this example, there is a clearly favourable triangulation, naturally aligning with the curvature of the reference surface. 
    \textbf{Middle}: During optimization we use a differentiable strategy to select triangulation by tessellating each quad into four triangles with 
    an interpolated midpoint. 
    \textbf{Right}: During inference we tessellate each quad into two triangles along the dominant diagonal. 
    } 
    \label{fig:spliting}
    \end{figure}
}
\newcommand{\figTetmesh}{
    \begin{figure}
    \centering
    \includegraphics[width=0.45\textwidth,trim=0 0 0 0,clip]{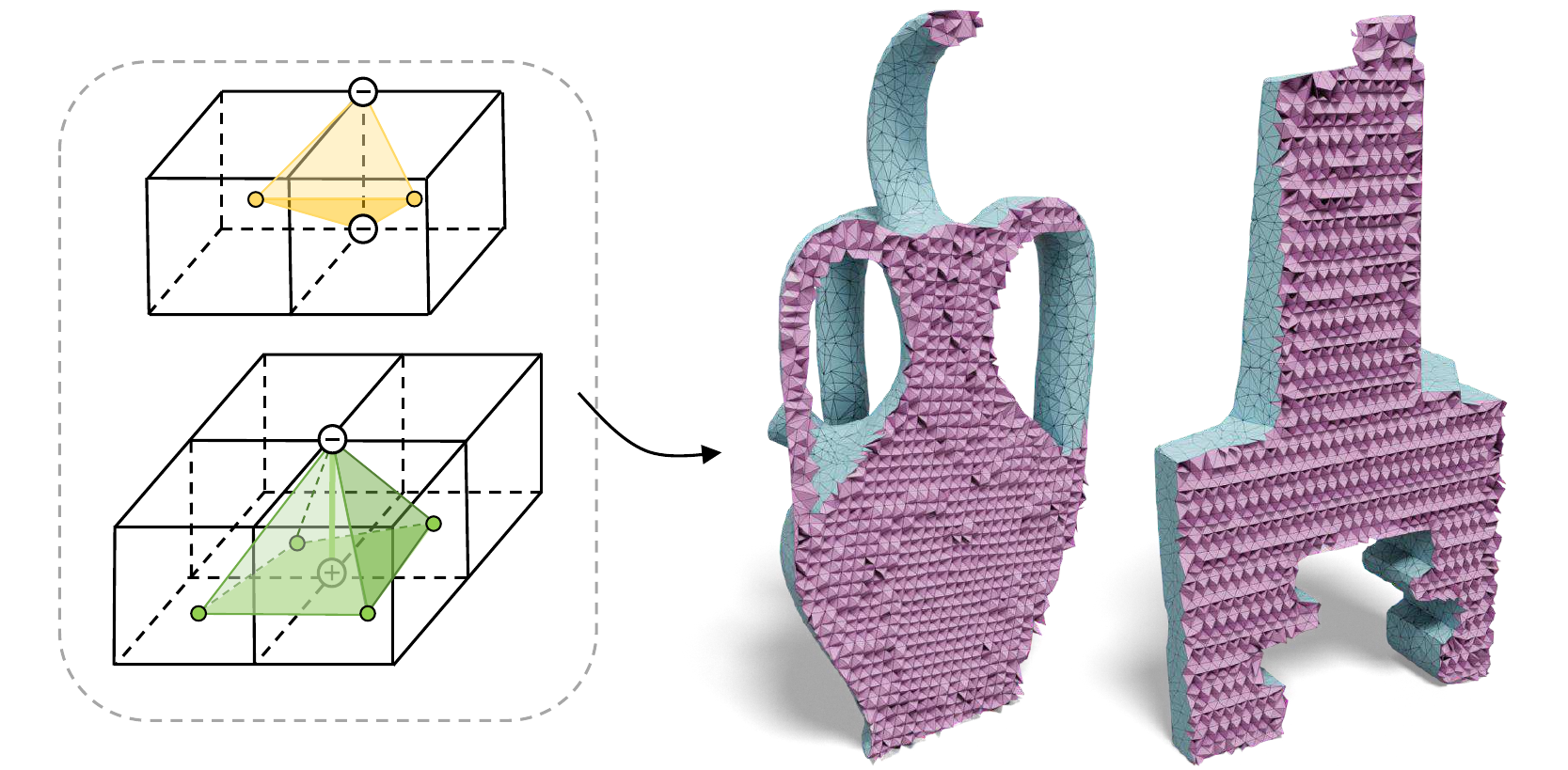}
    \caption{We extend \ours{} to generate a tetrahedral mesh by dividing the interior volume illustrated \emph{left}. Example outputs are shown \emph{right}.}
    \label{fig:tetmesh}
    \end{figure}
}

\newcommand{\figAblQual}{
    \begin{figure}
    \centering
    \includegraphics[width=0.45\textwidth,trim=0 0 0 0,clip]{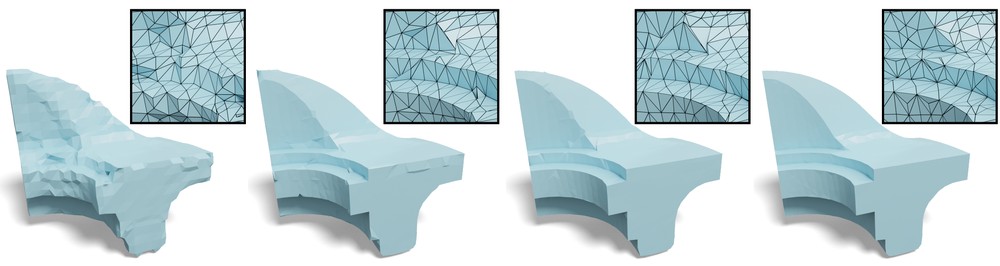}
    \begin{tabularx}{\columnwidth}{XXXX}
    \small{DMC Centroid} & \small{+flex vertex} & \small{+grid deform} & \small{+flex quad split}
    \end{tabularx}
    \caption{Ablating the effect of parameters in \ours{}. Adding flexible dual vertex positioning, the mesh edges align much better with the sharp geometric features, which are further improved with grid deformation. With flexible quad splitting, \ours{} split the quadrangles along the diagonals that align with the features. }
    \label{fig:ablqual}
    \end{figure}
}

\ifdefined\HQIMAGES
    \newcommand{\figDifferentiability}{
        \begin{figure}[tb]
            \centering
            \setlength{\tabcolsep}{1pt}
            \begin{tabular}{cccccc}
            & 0 & 50 & 100 & 200 & 1000 \\
            \rotatebox[origin=c]{90}{\small{MC}}  &
            \raisebox{-0.5\height}{ \includegraphics[width=0.17\columnwidth, trim=0.5cm 2cm 0.5cm 2cm,clip]{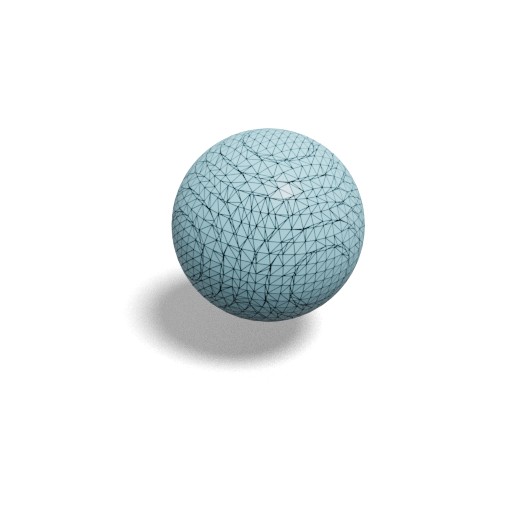}} &
            \raisebox{-0.5\height}{ \includegraphics[width=0.17\columnwidth, trim=0.5cm 2cm 0.5cm 2cm,clip]{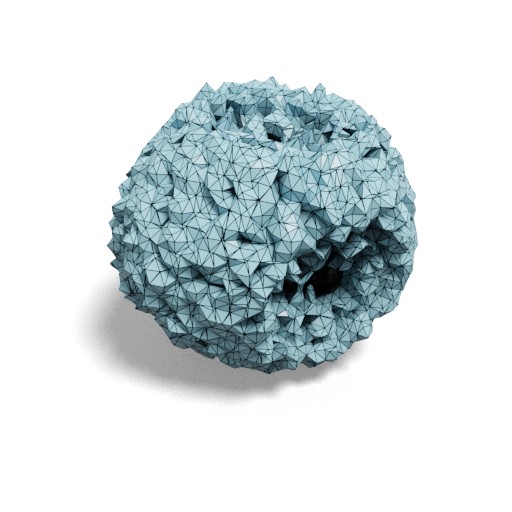}} &
            \raisebox{-0.5\height}{ \includegraphics[width=0.17\columnwidth, trim=0.5cm 2cm 0.5cm 2cm,clip]{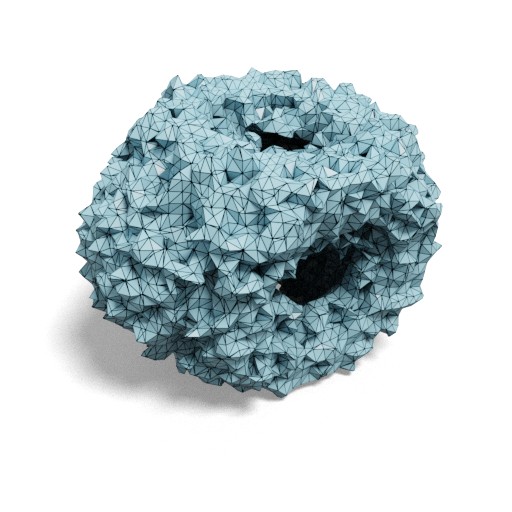}} &
            \raisebox{-0.5\height}{ \includegraphics[width=0.17\columnwidth, trim=0.5cm 2cm 0.5cm 2cm,clip]{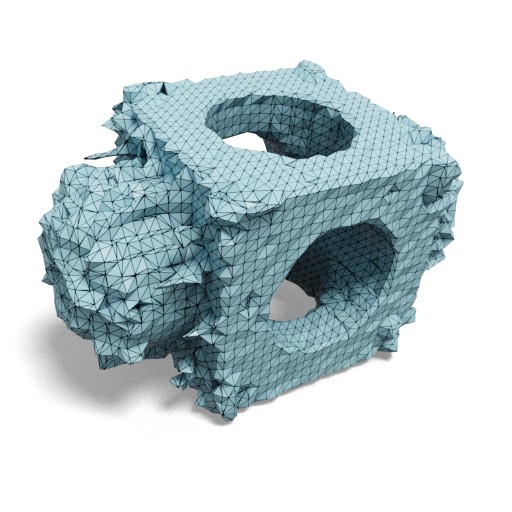}} &
            \raisebox{-0.5\height}{ \includegraphics[width=0.17\columnwidth, trim=0.5cm 2cm 0.5cm 2cm,clip]{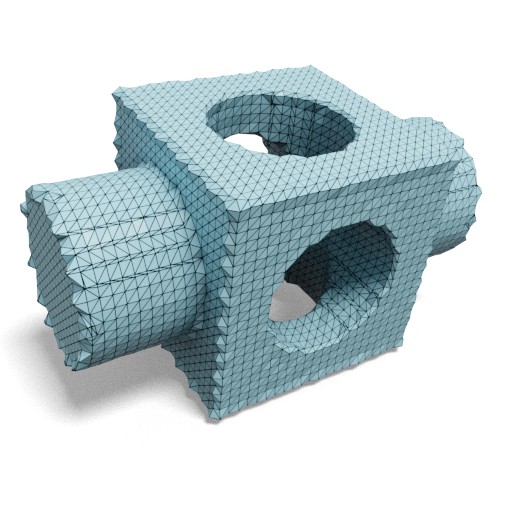}} \\
            \rotatebox[origin=c]{90}{\small{DC$_\mathrm{reg001}$}}  &
            \raisebox{-0.5\height}{ \includegraphics[width=0.17\columnwidth, trim=0.5cm 2cm 0.5cm 2cm,clip]{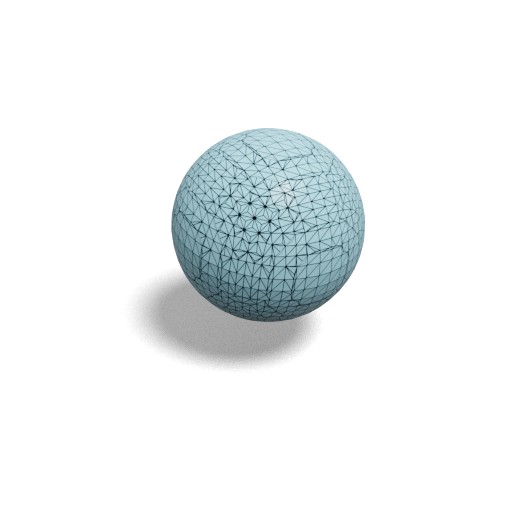}} &
            \raisebox{-0.5\height}{ \includegraphics[width=0.17\columnwidth, trim=0.5cm 2cm 0.5cm 2cm,clip]{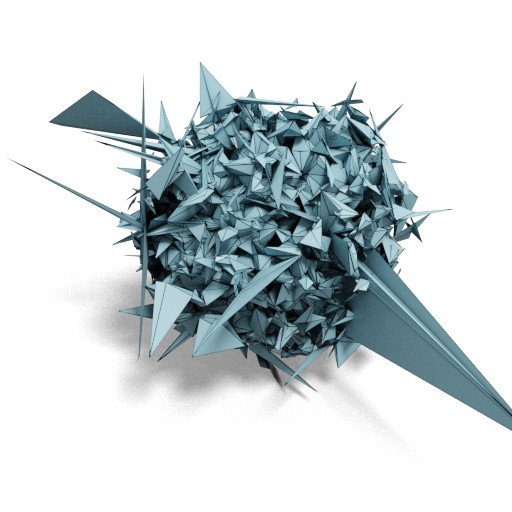}} &
            \raisebox{-0.5\height}{ \includegraphics[width=0.17\columnwidth, trim=0.5cm 2cm 0.5cm 2cm,clip]{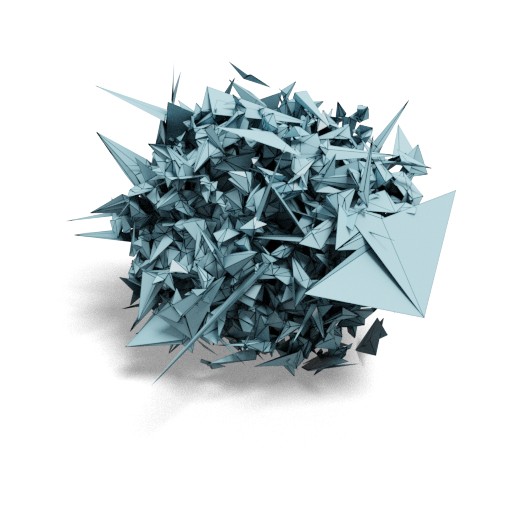}} &
            \raisebox{-0.5\height}{ \includegraphics[width=0.17\columnwidth, trim=0.5cm 2cm 0.5cm 2cm,clip]{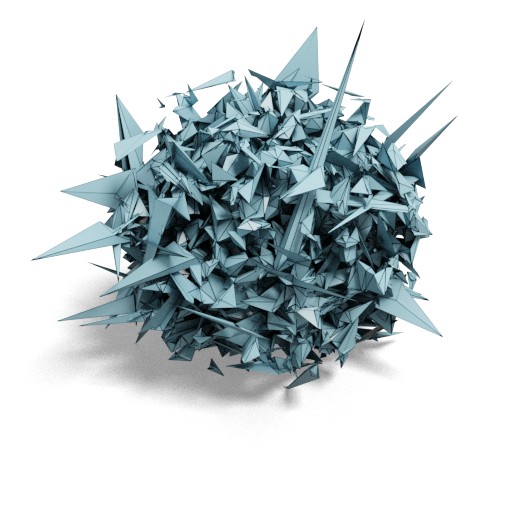}} &
            \raisebox{-0.5\height}{ \includegraphics[width=0.17\columnwidth, trim=0.5cm 2cm 0.5cm 2cm,clip]{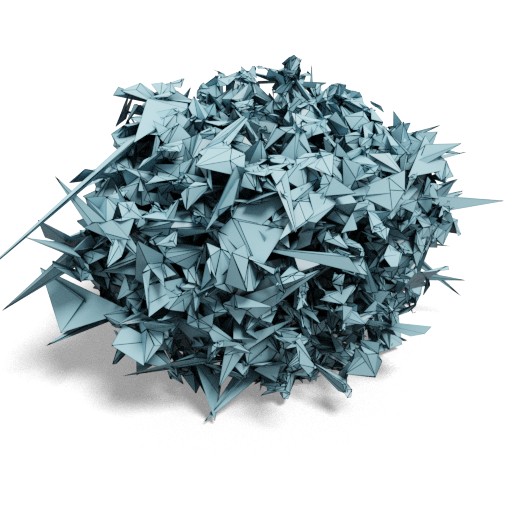}} \\
            \rotatebox[origin=c]{90}{\small{DC$_\mathrm{reg1}$}}  &
            \raisebox{-0.5\height}{ \includegraphics[width=0.17\columnwidth, trim=0.5cm 2cm 0.5cm 2cm,clip]{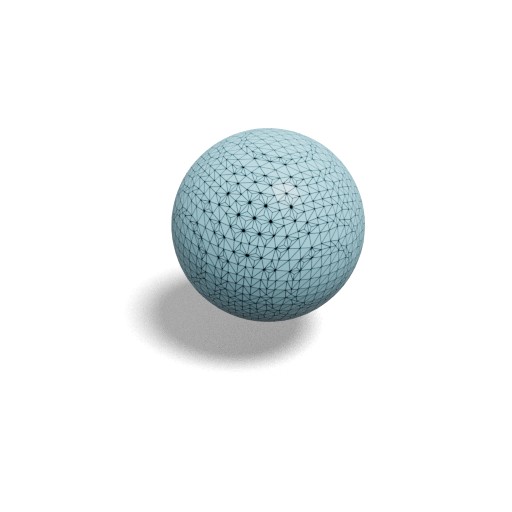}} &
            \raisebox{-0.5\height}{ \includegraphics[width=0.17\columnwidth, trim=0.5cm 2cm 0.5cm 2cm,clip]{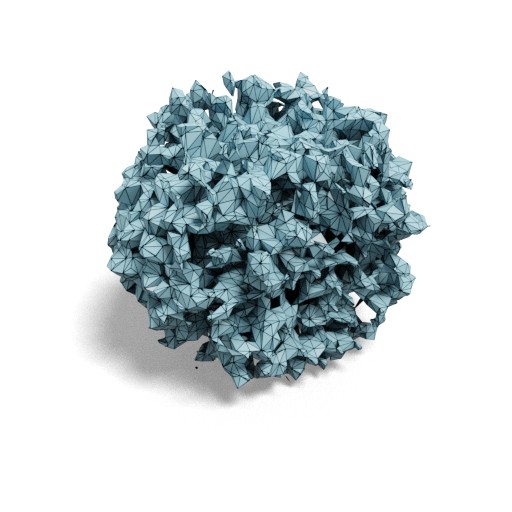}} &
            \raisebox{-0.5\height}{ \includegraphics[width=0.17\columnwidth, trim=0.5cm 2cm 0.5cm 2cm,clip]{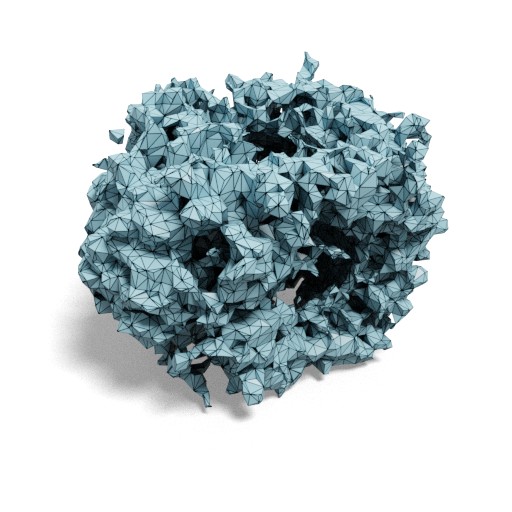}} &
            \raisebox{-0.5\height}{ \includegraphics[width=0.17\columnwidth, trim=0.5cm 2cm 0.5cm 2cm,clip]{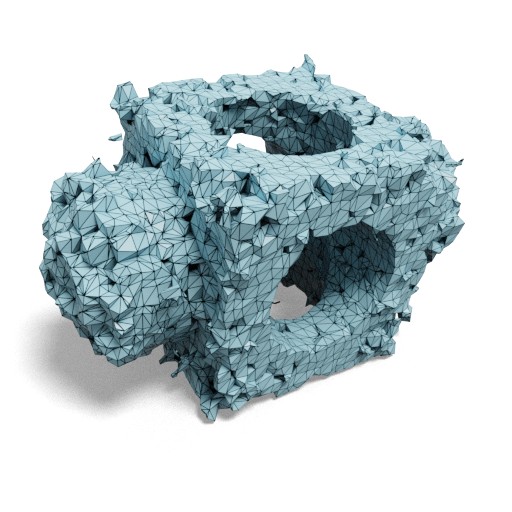}} &
            \raisebox{-0.5\height}{ \includegraphics[width=0.17\columnwidth, trim=0.5cm 2cm 0.5cm 2cm,clip]{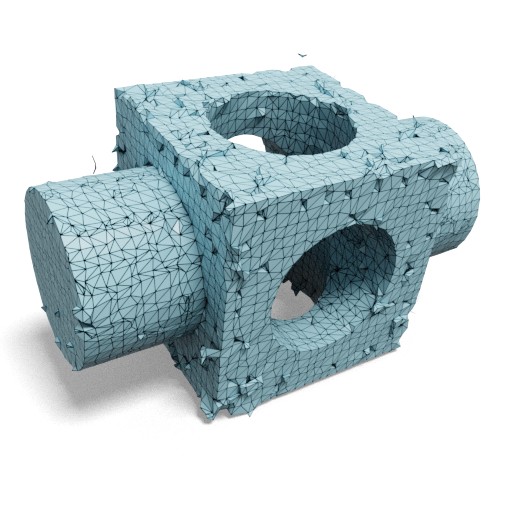}} \\
            \rotatebox[origin=c]{90}{\small{DC$_\mathrm{centroid}$}}  &
            \raisebox{-0.5\height}{ \includegraphics[width=0.17\columnwidth, trim=0.5cm 2cm 0.5cm 2cm,clip]{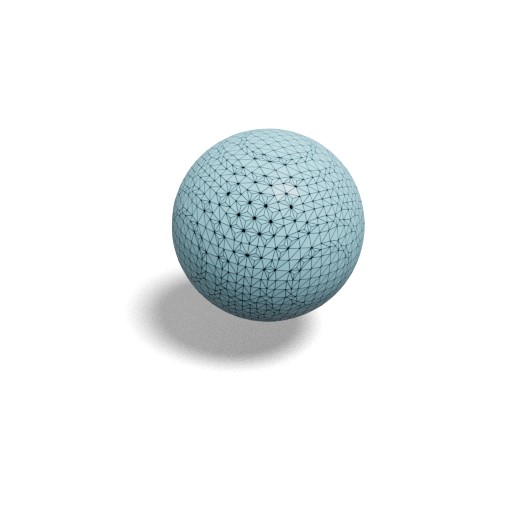}} &
            \raisebox{-0.5\height}{ \includegraphics[width=0.17\columnwidth, trim=0.5cm 2cm 0.5cm 2cm,clip]{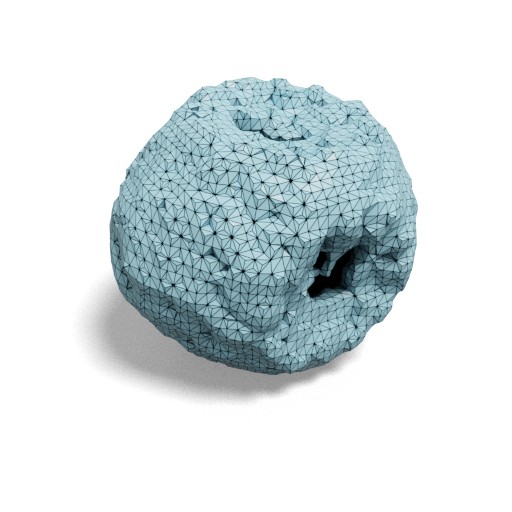}} &
            \raisebox{-0.5\height}{ \includegraphics[width=0.17\columnwidth, trim=0.5cm 2cm 0.5cm 2cm,clip]{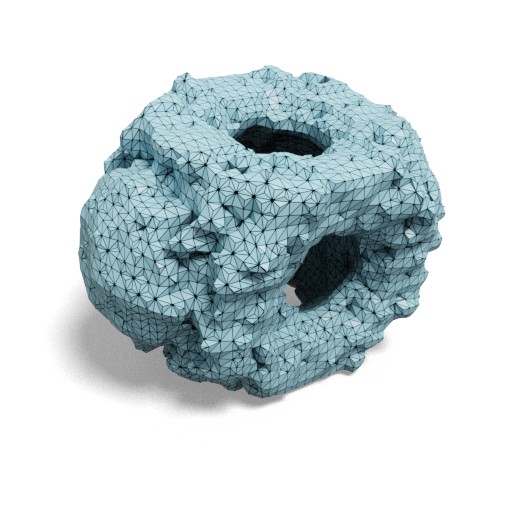}} &
            \raisebox{-0.5\height}{ \includegraphics[width=0.17\columnwidth, trim=0.5cm 2cm 0.5cm 2cm,clip]{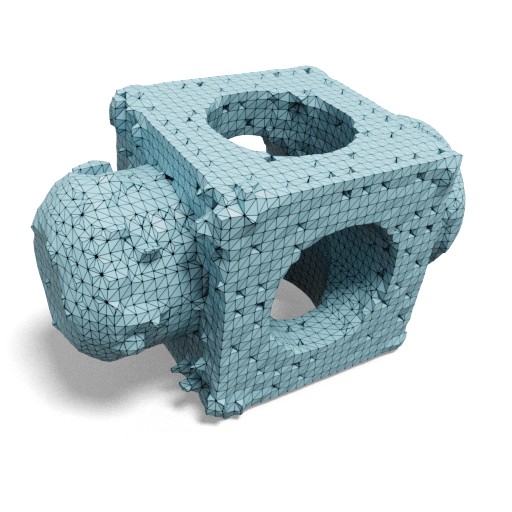}} &
            \raisebox{-0.5\height}{ \includegraphics[width=0.17\columnwidth, trim=0.5cm 2cm 0.5cm 2cm,clip]{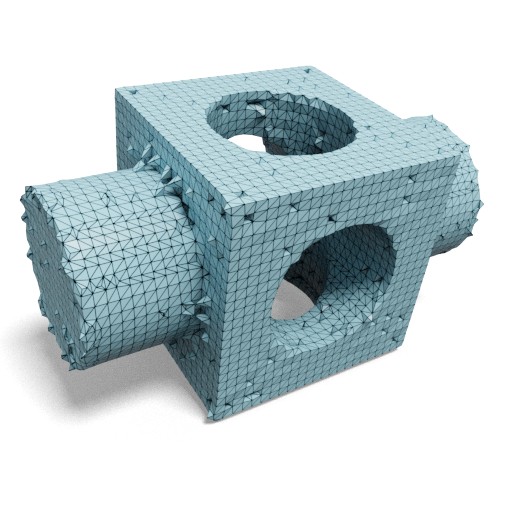}} \\
            \rotatebox[origin=c]{90}{\small{NDC}}  &
            \raisebox{-0.5\height}{ \includegraphics[width=0.17\columnwidth, trim=0.5cm 2cm 0.5cm 2cm,clip]{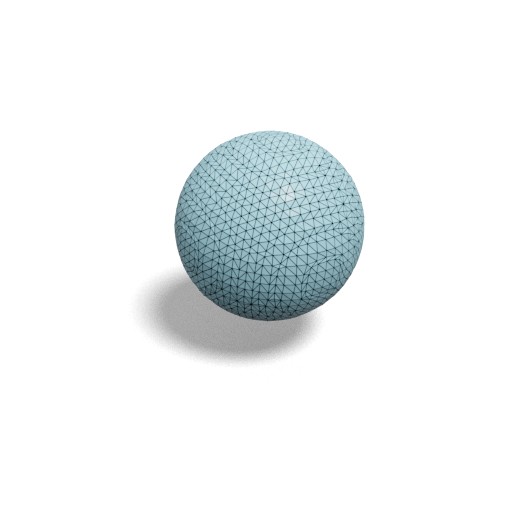}} &
            \raisebox{-0.5\height}{ \includegraphics[width=0.17\columnwidth, trim=0.5cm 2cm 0.5cm 2cm,clip]{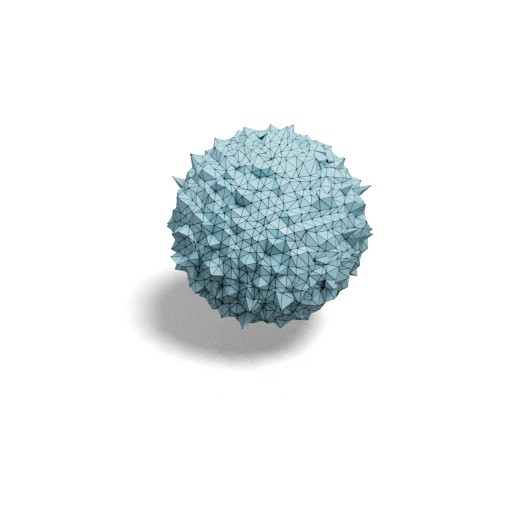}} &
            \raisebox{-0.5\height}{ \includegraphics[width=0.17\columnwidth, trim=0.5cm 2cm 0.5cm 2cm,clip]{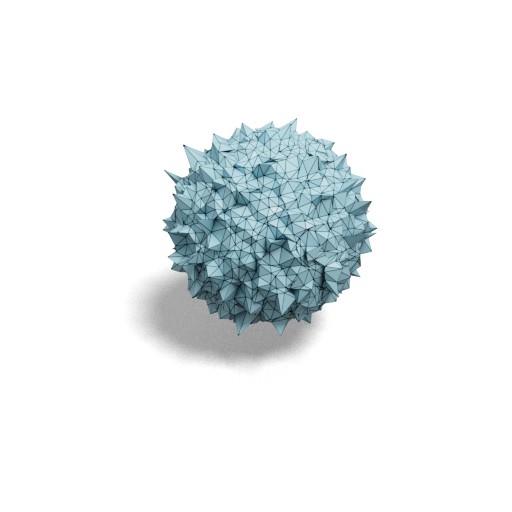}} &
            \raisebox{-0.5\height}{ \includegraphics[width=0.17\columnwidth, trim=0.5cm 2cm 0.5cm 2cm,clip]{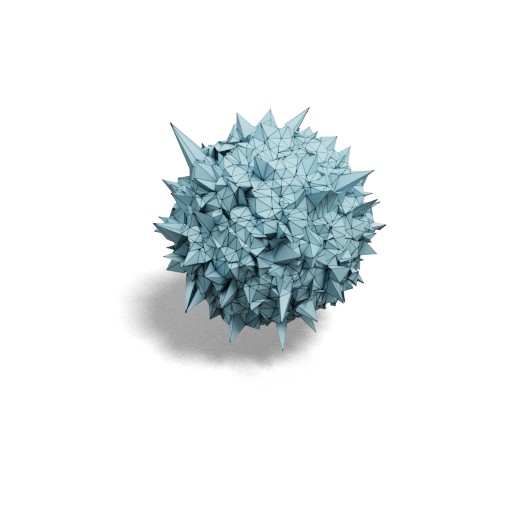}} &
            \raisebox{-0.5\height}{ \includegraphics[width=0.17\columnwidth, trim=0.5cm 2cm 0.5cm 2cm,clip]{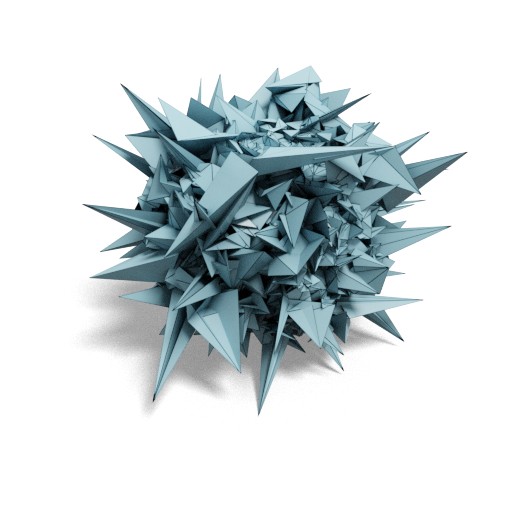}} \\
            \rotatebox[origin=c]{90}{\small{\textsc{DMTet}}}  &
            \raisebox{-0.5\height}{ \includegraphics[width=0.17\columnwidth, trim=0.5cm 2cm 0.5cm 2cm,clip]{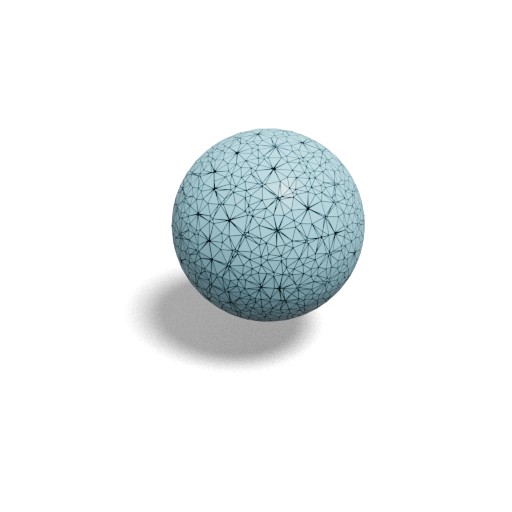}} &
            \raisebox{-0.5\height}{ \includegraphics[width=0.17\columnwidth, trim=0.5cm 2cm 0.5cm 2cm,clip]{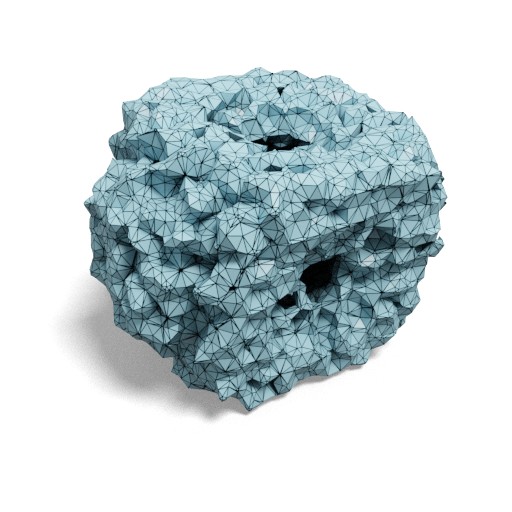}} &
            \raisebox{-0.5\height}{ \includegraphics[width=0.17\columnwidth, trim=0.5cm 2cm 0.5cm 2cm,clip]{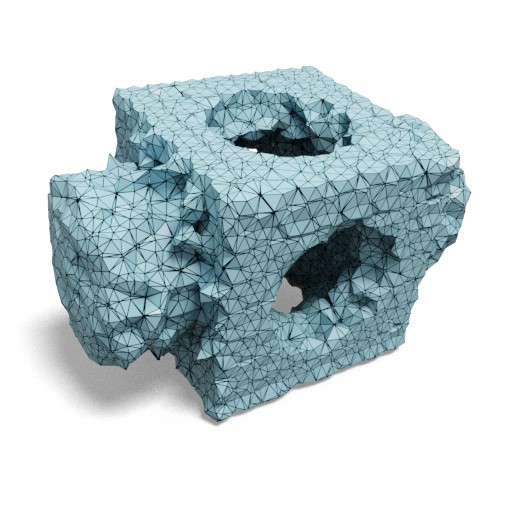}} &
            \raisebox{-0.5\height}{ \includegraphics[width=0.17\columnwidth, trim=0.5cm 2cm 0.5cm 2cm,clip]{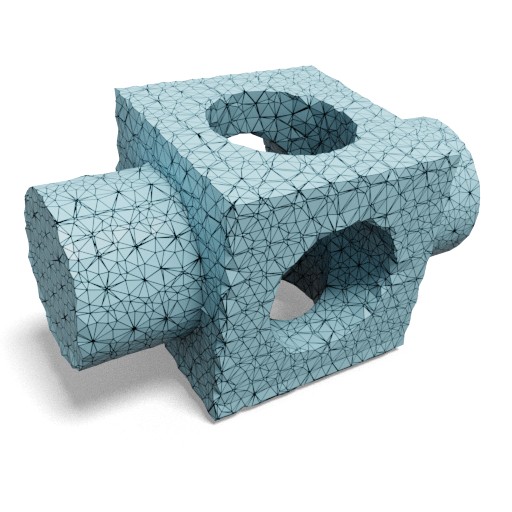}} &
            \raisebox{-0.5\height}{ \includegraphics[width=0.17\columnwidth, trim=0.5cm 2cm 0.5cm 2cm,clip]{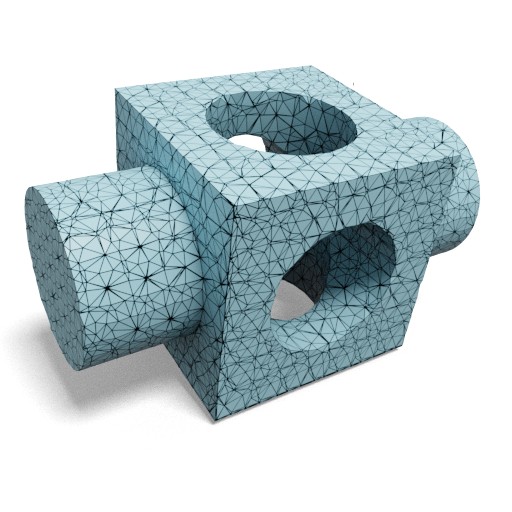}} \\
            \rotatebox[origin=c]{90}{\small{\textsc{FlexiCubes}}}  &
            \raisebox{-0.5\height}{ \includegraphics[width=0.17\columnwidth, trim=0.5cm 2cm 0.5cm 2cm,clip]{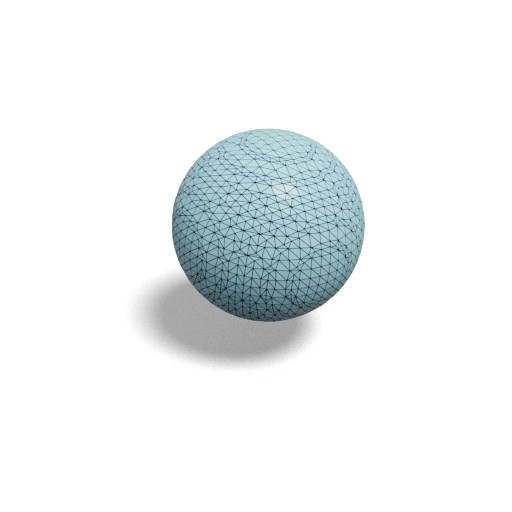}} &
            \raisebox{-0.5\height}{ \includegraphics[width=0.17\columnwidth, trim=0.5cm 2cm 0.5cm 2cm,clip]{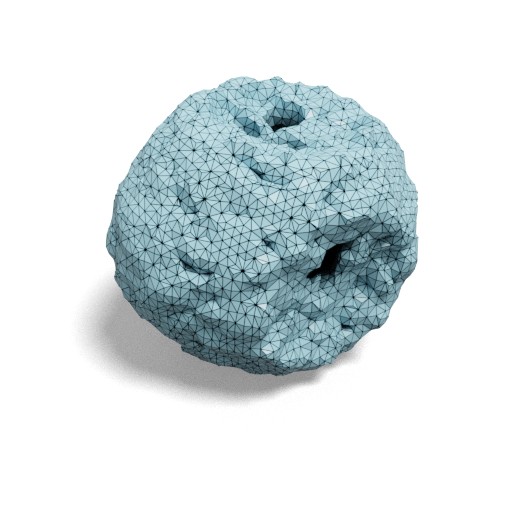}} &
            \raisebox{-0.5\height}{ \includegraphics[width=0.17\columnwidth, trim=0.5cm 2cm 0.5cm 2cm,clip]{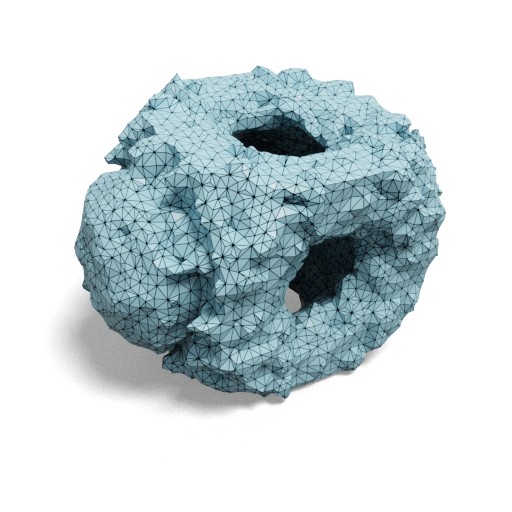}} &
            \raisebox{-0.5\height}{ \includegraphics[width=0.17\columnwidth, trim=0.5cm 2cm 0.5cm 2cm,clip]{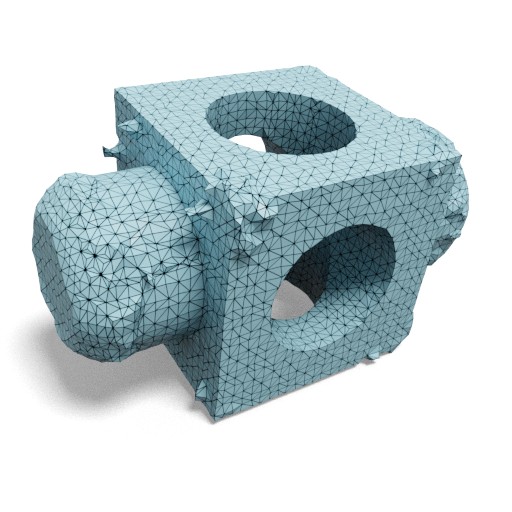}} &
            \raisebox{-0.5\height}{ \includegraphics[width=0.17\columnwidth, trim=0.5cm 2cm 0.5cm 2cm,clip]{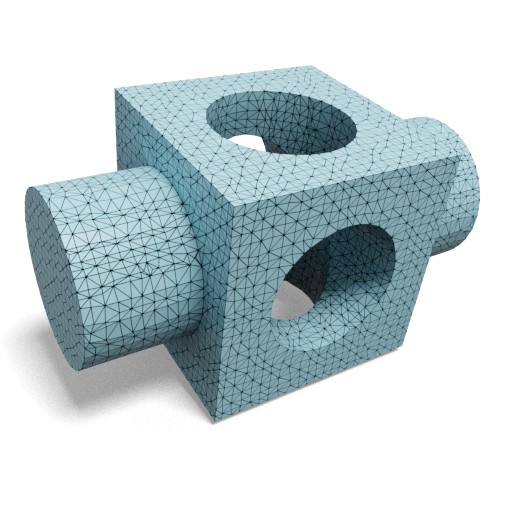}} \\
            & 0 & 50 & 100 & 200 & 1000 \\
            \end{tabular}
            \caption{\textbf{Comparison of different isosurfacing methods for mesh optimization.} Starting from a sphere initialization, we optimize the shapes towards the ground truth mesh using a set of isosurfacing methods. 
            DMTet reconstructs sharp features but produces many sliver triangles. MC~\cite{nielson2003marching} fails to capture sharp features, NDC~\cite{ChenNDC2022} diverges during optimization, DC~\cite{Ju2002} converges with strong regularization, but suffer from artifacts, and \ours{} generates a high quality mesh with details retained. More details about this experiment are provided in the Supplement.}
                \label{fig:differentiability}
            \end{figure}
    }
\else
    \newcommand{\figDifferentiability}{
        \begin{figure}
        \centering
        \includegraphics[width=0.45\textwidth,trim=0 0 0 0,clip]{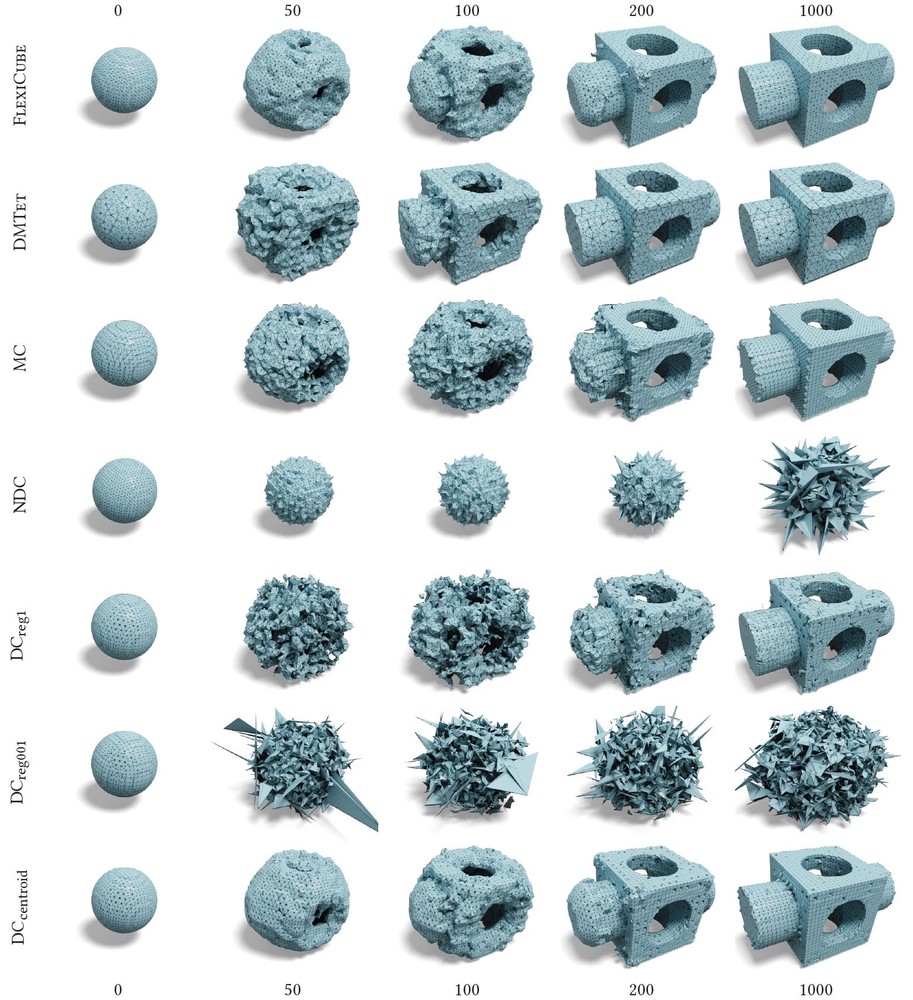}
        \caption{\textbf{Comparison different isosurfacing methods for mesh optimization.} Starting from a sphere initialization, we optimize the shapes towards the ground truth mesh using different iso-surface methods. 
        DMTet reconstructs sharp features but produces many skinning triangles. MC~\cite{nielson2003marching} fails to capture sharp features, NDC~\cite{ChenNDC2022} diverges during optimization, DC~\cite{Ju2002} with regularization suffers a conflict between optimization stability and introducing artifacts, and \ours{} generates best result for both mesh quality and reconstructing geometric details. More details about this experiment are provided in the Supplement.}
        \label{fig:differentiability}
        \end{figure}
    }
\fi

\ifdefined\HQIMAGES
    \newcommand{\figRobustness}{
        \begin{figure}[htb]
        \centering
        \setlength{\tabcolsep}{1pt}
        \begin{tabular}{lccc}
        \rotatebox[origin=c]{90}{MC}  &
        \raisebox{-0.5\height}{\includegraphics[width=0.3\columnwidth, trim=0 0cm 0 5cm, clip]{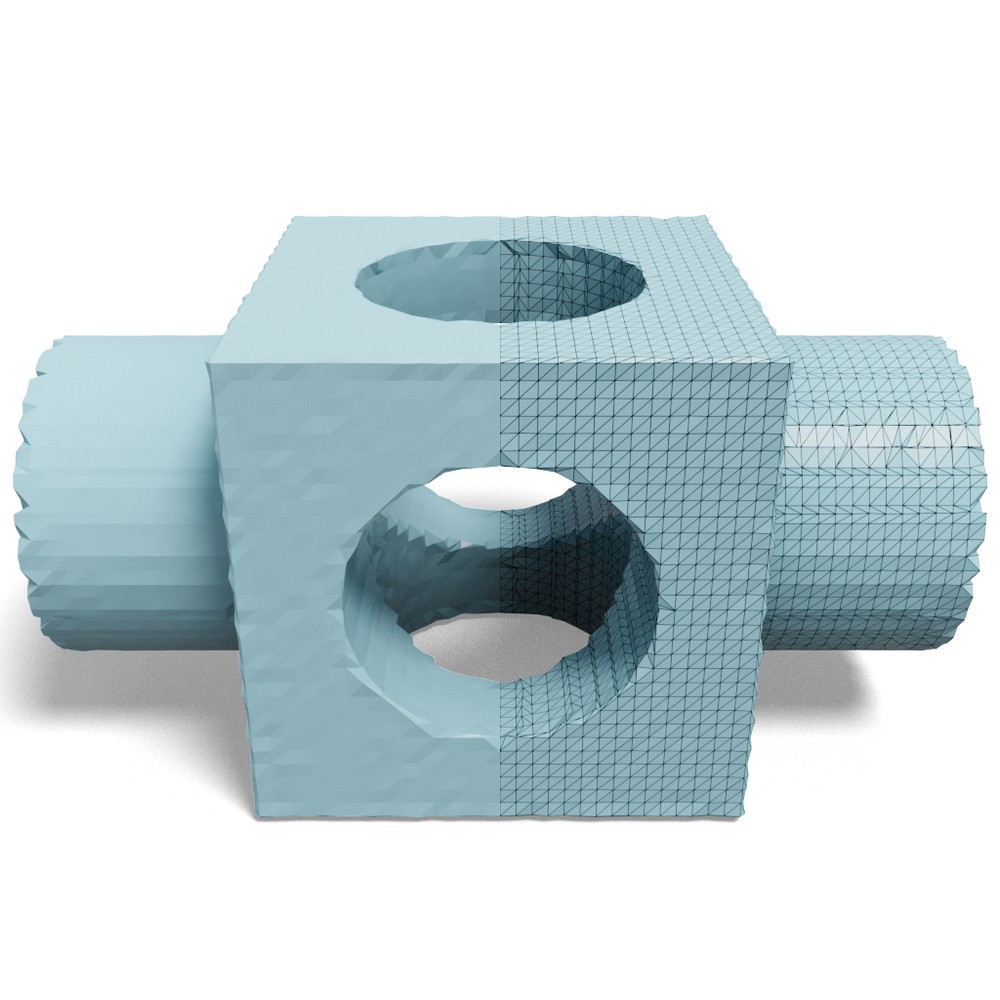}} &
        \raisebox{-0.5\height}{\includegraphics[width=0.3\columnwidth, trim=0 0cm 0 5cm, clip]{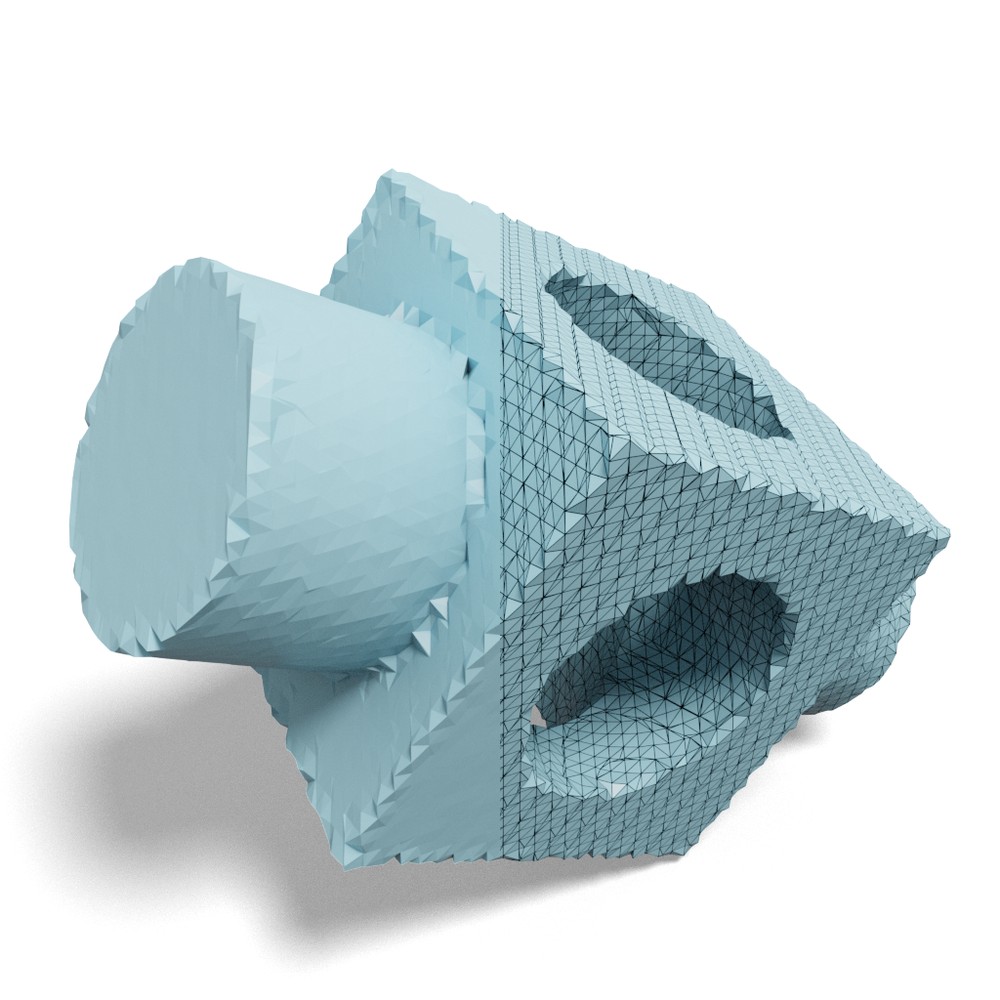}} &
        \raisebox{-0.5\height}{\includegraphics[width=0.3\columnwidth, trim=0 0cm 0 5cm, clip]{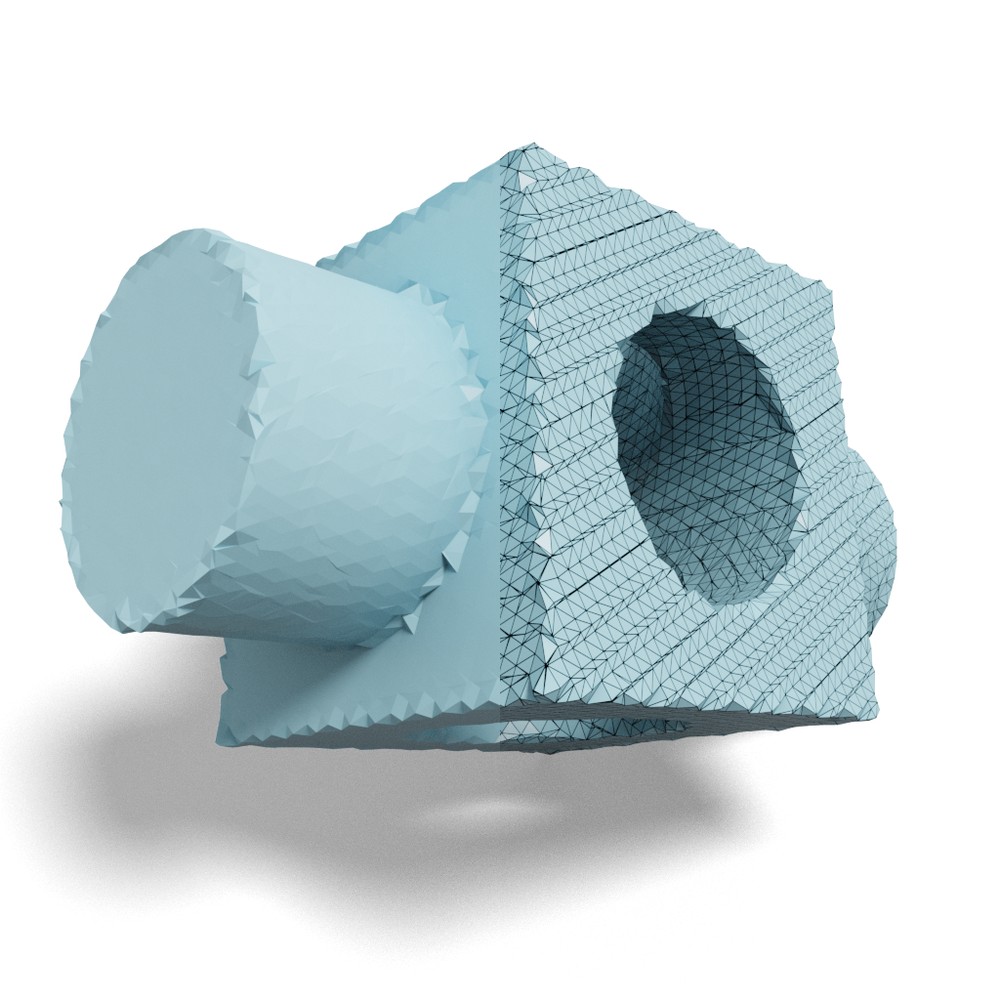}} \\
        \rotatebox[origin=c]{90}{\textsc{DMTet}}  &
        \raisebox{-0.5\height}{\includegraphics[width=0.3\columnwidth, trim=0 0cm 0 5cm, clip]{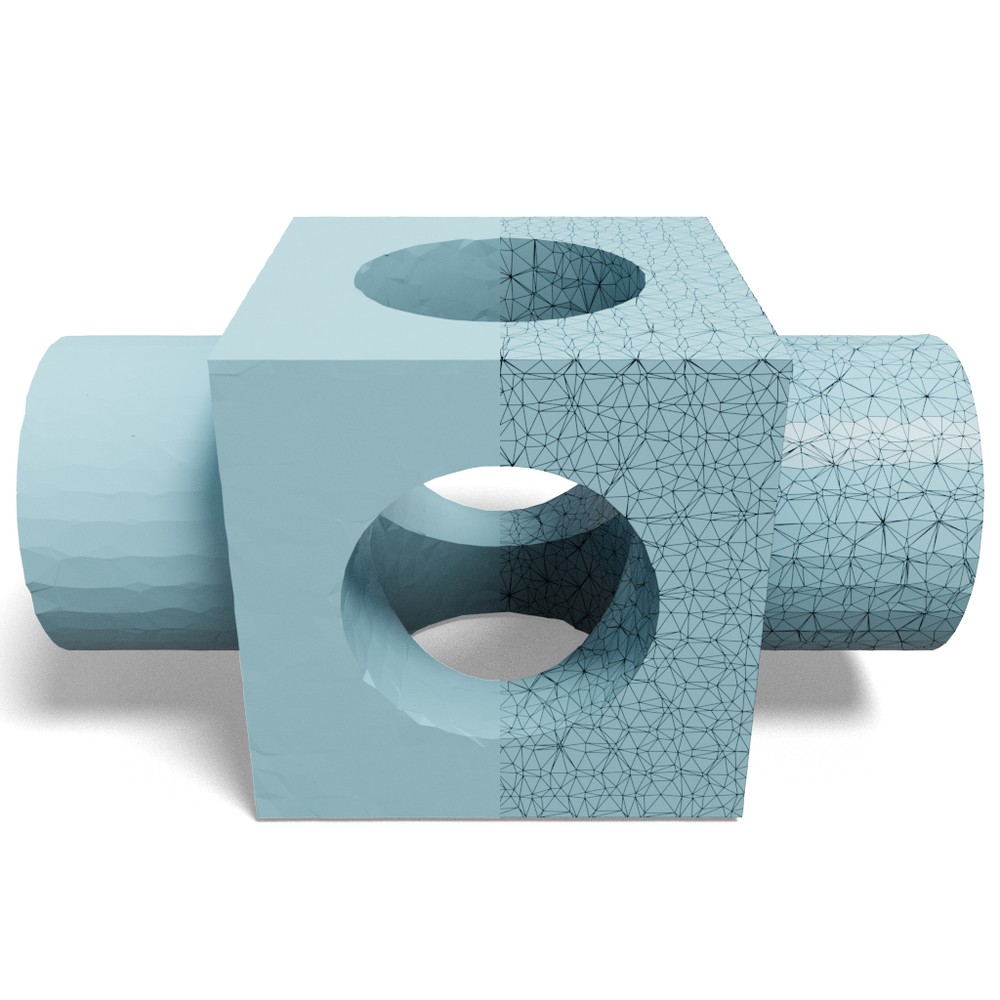}} &
        \raisebox{-0.5\height}{\includegraphics[width=0.3\columnwidth, trim=0 0cm 0 5cm, clip]{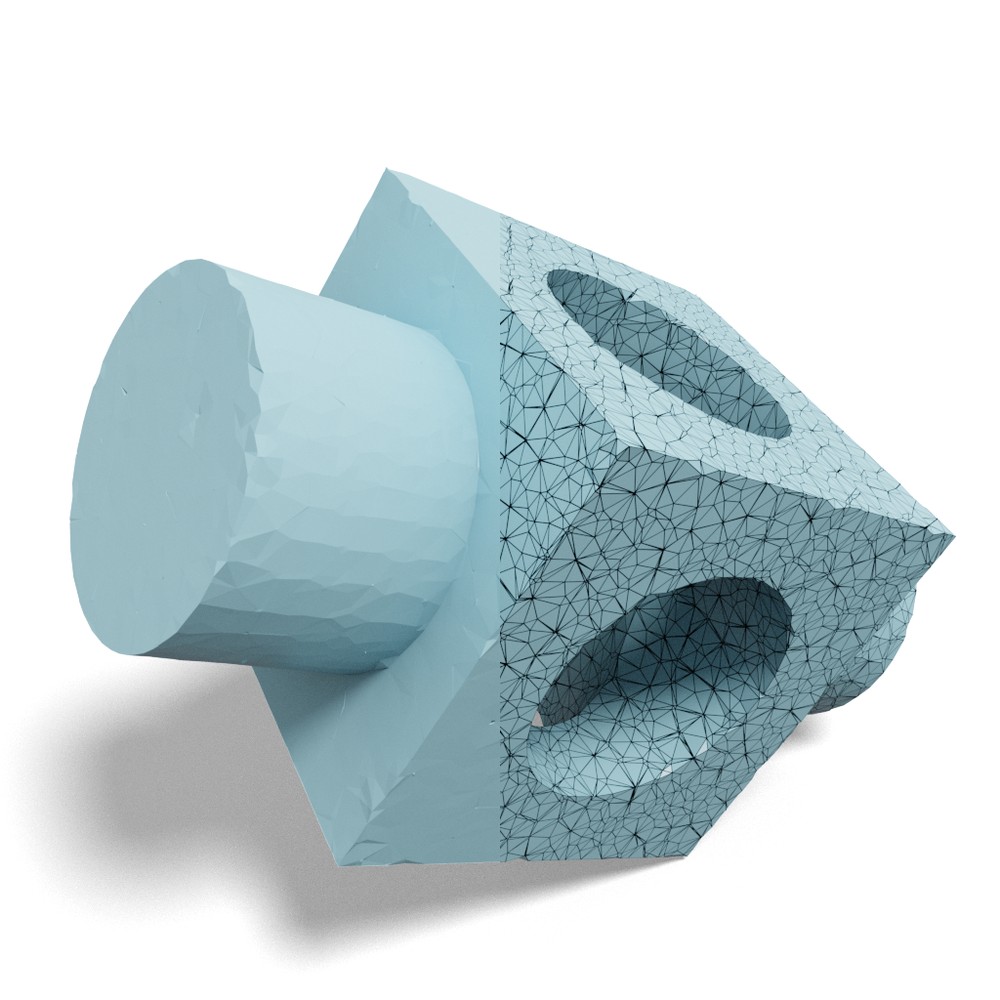}} &
        \raisebox{-0.5\height}{\includegraphics[width=0.3\columnwidth, trim=0 0cm 0 5cm, clip]{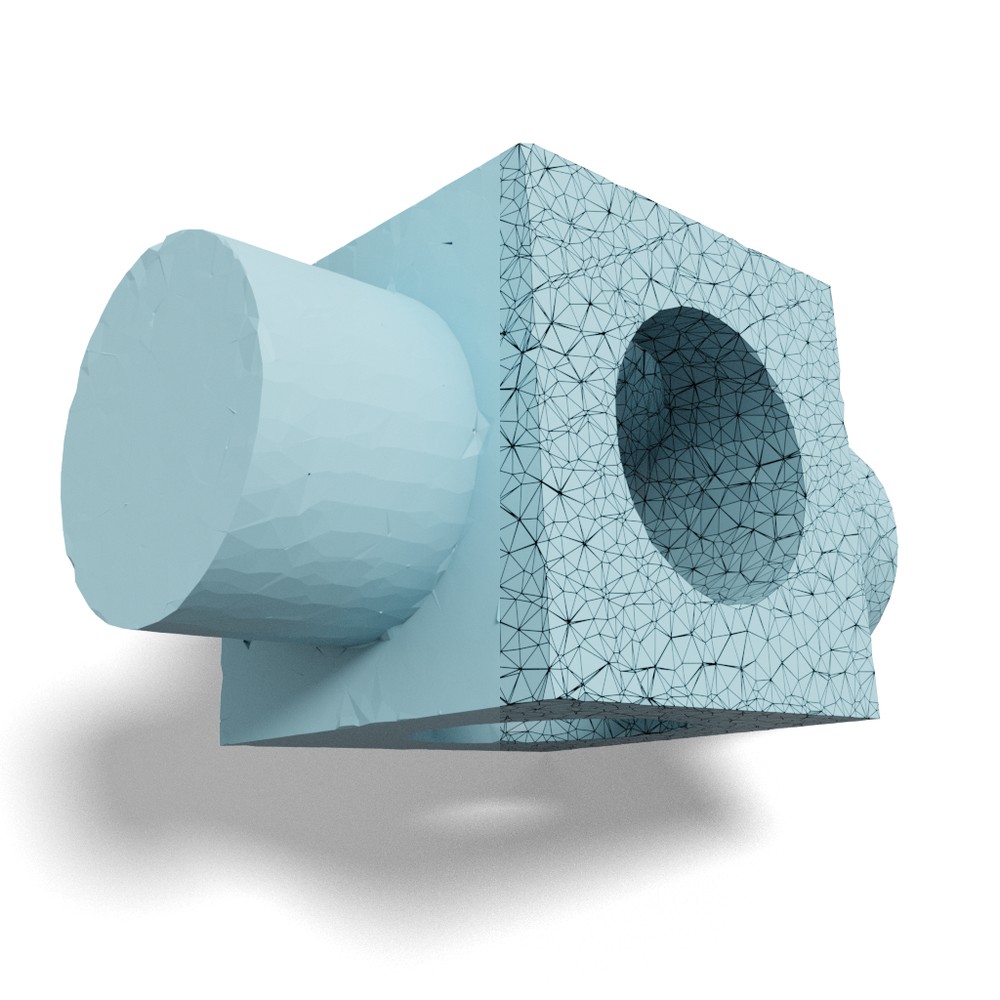}} \\
        \rotatebox[origin=c]{90}{\textsc{FlexiCubes}}  &
        \raisebox{-0.5\height}{\includegraphics[width=0.3\columnwidth, trim=0 0cm 0 5cm, clip]{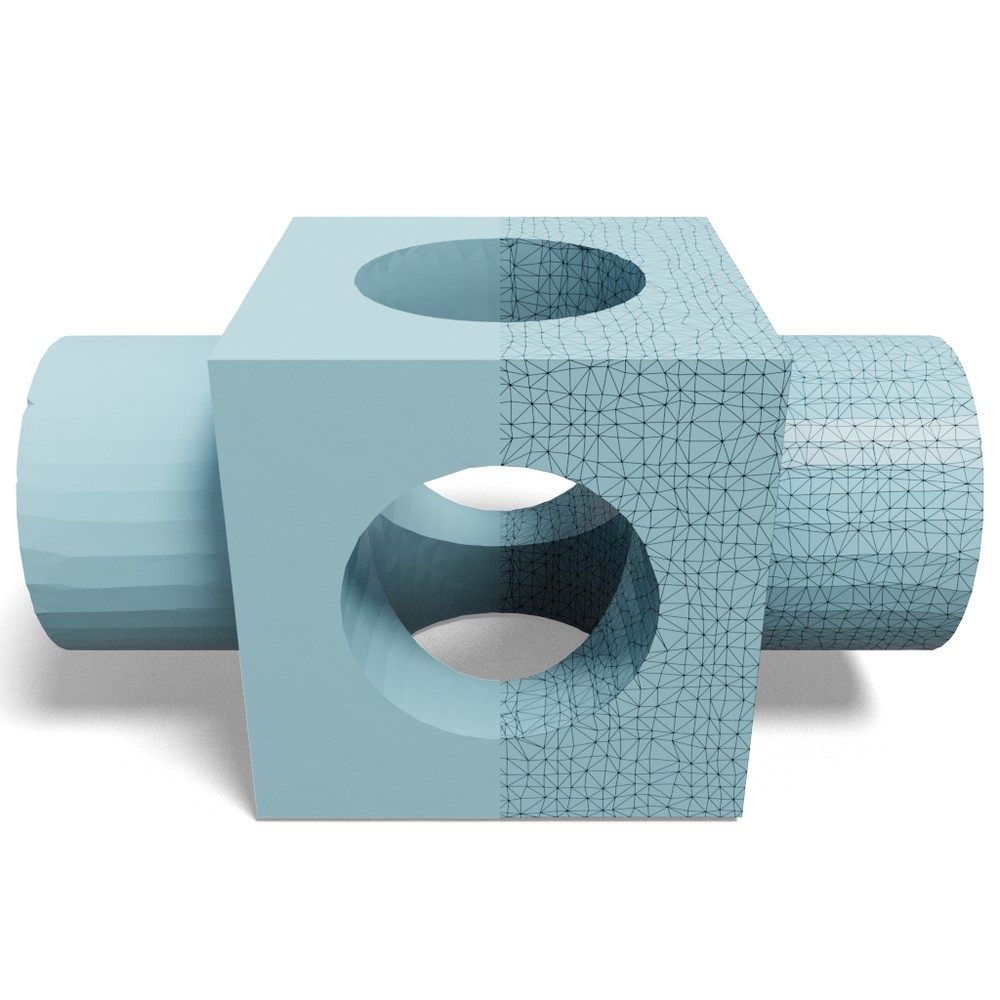}} &
        \raisebox{-0.5\height}{\includegraphics[width=0.3\columnwidth, trim=0 0cm 0 5cm, clip]{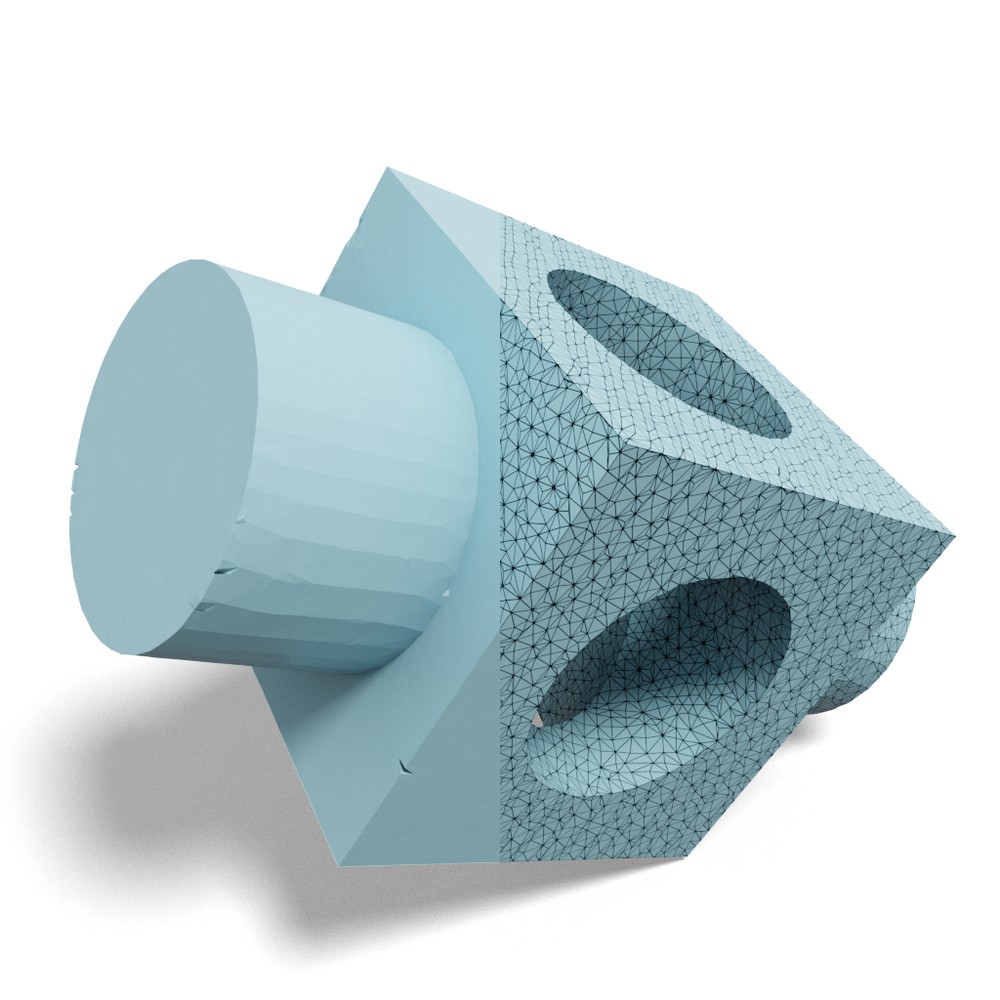}} &
        \raisebox{-0.5\height}{\includegraphics[width=0.3\columnwidth, trim=0 0cm 0 5cm, clip]{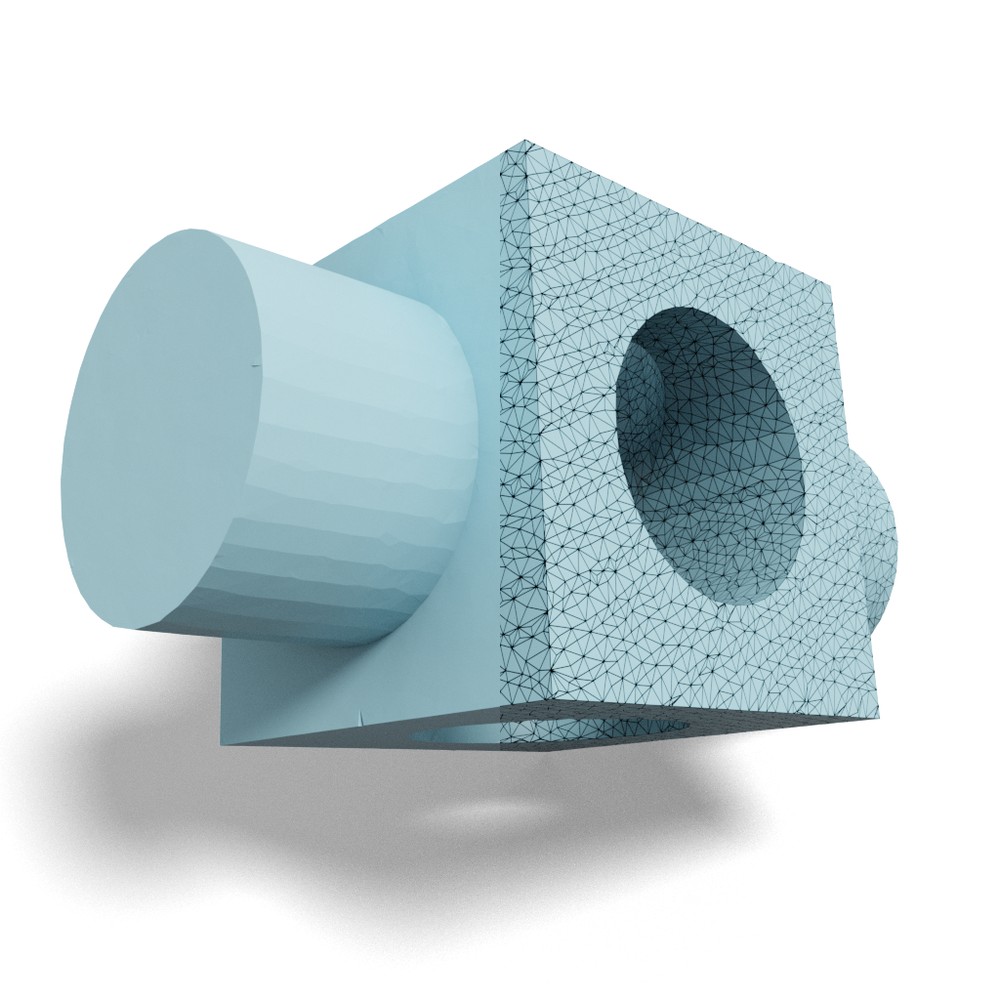}} \\
        & Pose~A & Pose~B & Pose~C \\
        \end{tabular}
        \caption{We reconstruct the same model with three different poses. Marching Cubes performs well when features are axis-aligned, but show stair step artifacts in the other poses. \DMTet{}
        performs more robustly, but produces many sliver triangles. \ours{} has more 
        uniform triangles and no stair step artifacts.}
        \label{fig:robustness}        
        \end{figure}
    }
\else
    \newcommand{\figRobustness}{
        \begin{figure}
        \centering
        \includegraphics[width=0.45\textwidth,trim=0 0 0 0,clip]{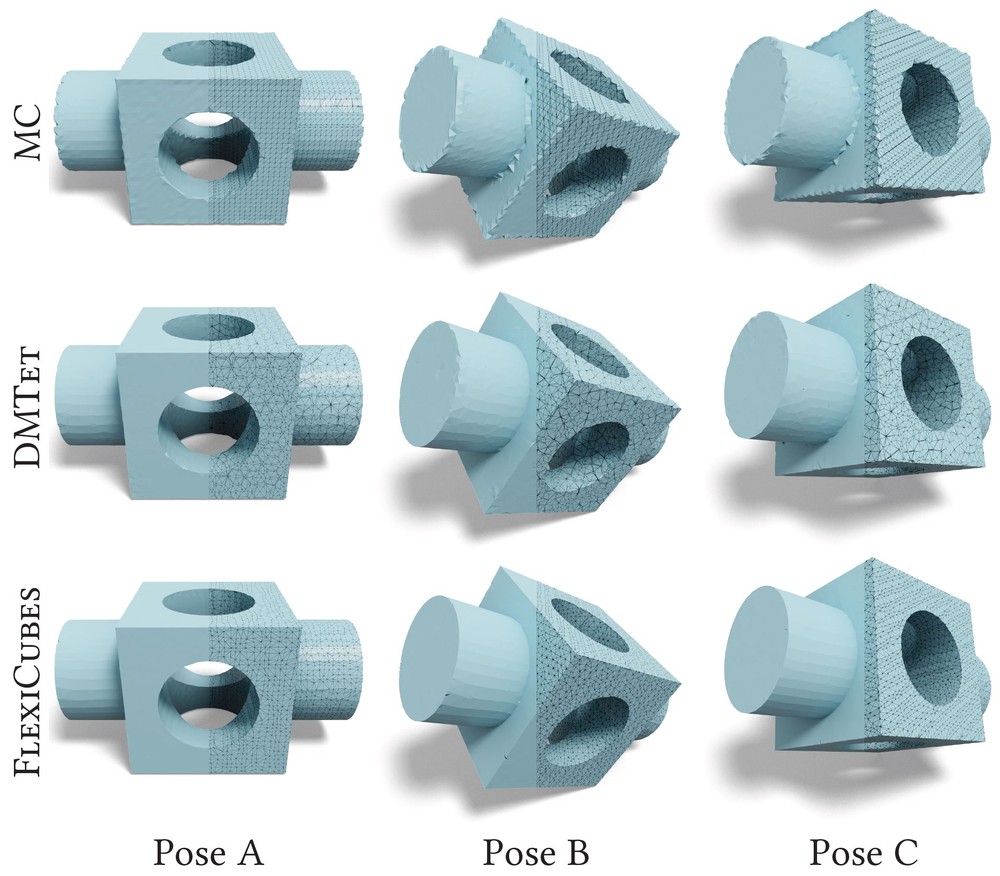}
        \caption{We reconstruct the same model with three different poses. Marching Cubes performs well when features are axis-aligned, but show stair step artifacts in the other poses. \DMTet{}
        performs more robustly, but produces many sliver triangles. \ours{} has more 
        uniform triangles and no stair step artifacts.}
        \label{fig:robustness}
        \end{figure}
    }
\fi

\newcommand{\figQEF}{
    \begin{figure}
    \centering
    \includegraphics[width=0.47\textwidth,trim=0 0 0 0,clip]{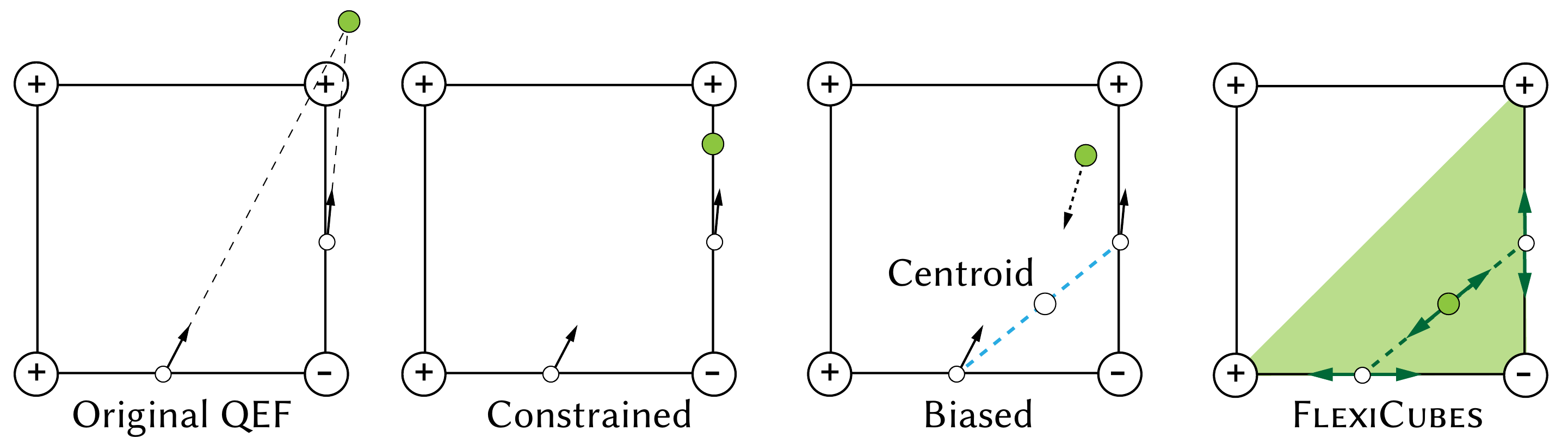}
    \caption{\textbf{Grad. Issue in DC.} \textbf{Left:} When solving a quadratic error function (QEF) the resulting vertex is not guaranteed to be inside the cube. This leads to discrepancies between the geometry and the topological cases. In addition, there exists a singularity in QEF when the normals are coplanar. While there exist techniques to improve the stability of DC - constraining the solution space of QEF or biasing the QEF with regularization loss, they are not readily adaptable in optimization settings. The former (\textbf{Second}) zeros out the gradient in certain directions. The latter (\textbf{Third}) is hard to tune and having a strong regularization will downgrade the advantage of DC in flexibility. Our version (\textbf{Fourth}), \ours{} provides additional degrees of freedom, such that
    the dual vertex can be placed anywhere within the green triangle for this particular configuration.}
    \label{fig:qef}
    \end{figure}
}

\newcommand{\figDMCCompare}{
    \begin{figure}
    \centering
    \includegraphics[width=0.45\textwidth,trim=0 0 0 0,clip]{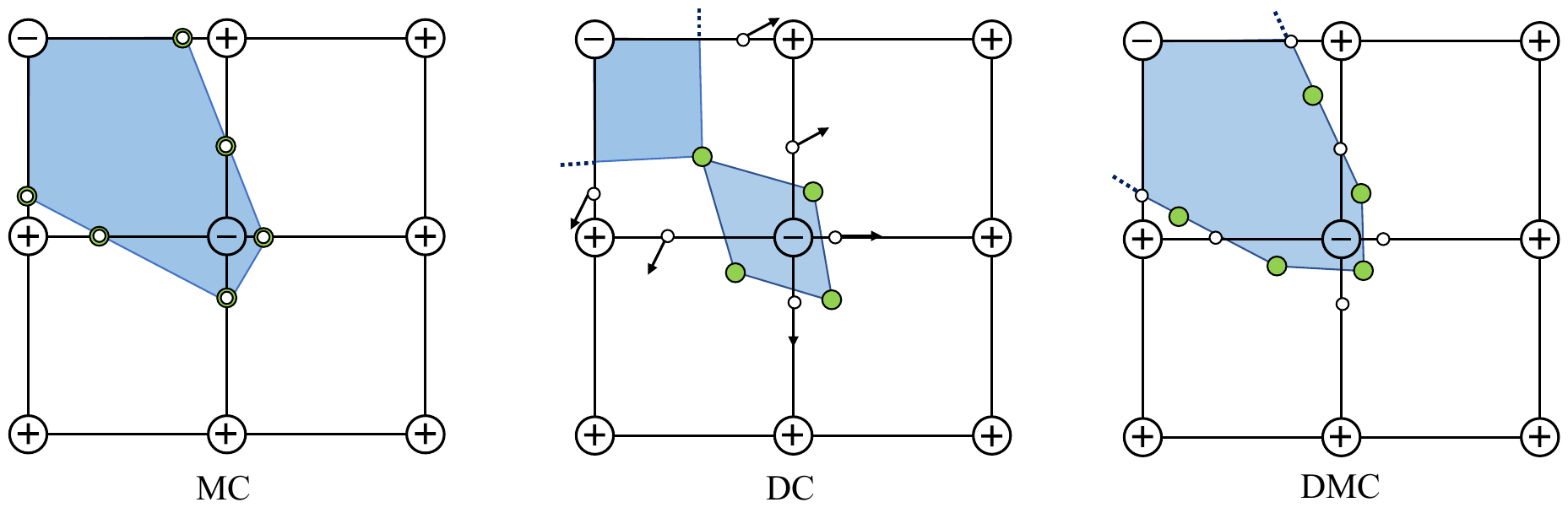}
    \caption{\textbf{Comparison between MC, DC and DMC.} MC fails to capture sharp features. DC captures sharp features but can produce non-manifold vertices in some cases. DMC do not introduce non-manifold vertices, but has less freedom than DC in representing sharp features.}
    \label{fig:dmc_compare}
    \end{figure}
}
\newcommand{\figOctree}{
    \begin{figure}
    \centering
    \includegraphics[width=0.3\textwidth,trim=0 0 0 0,clip]{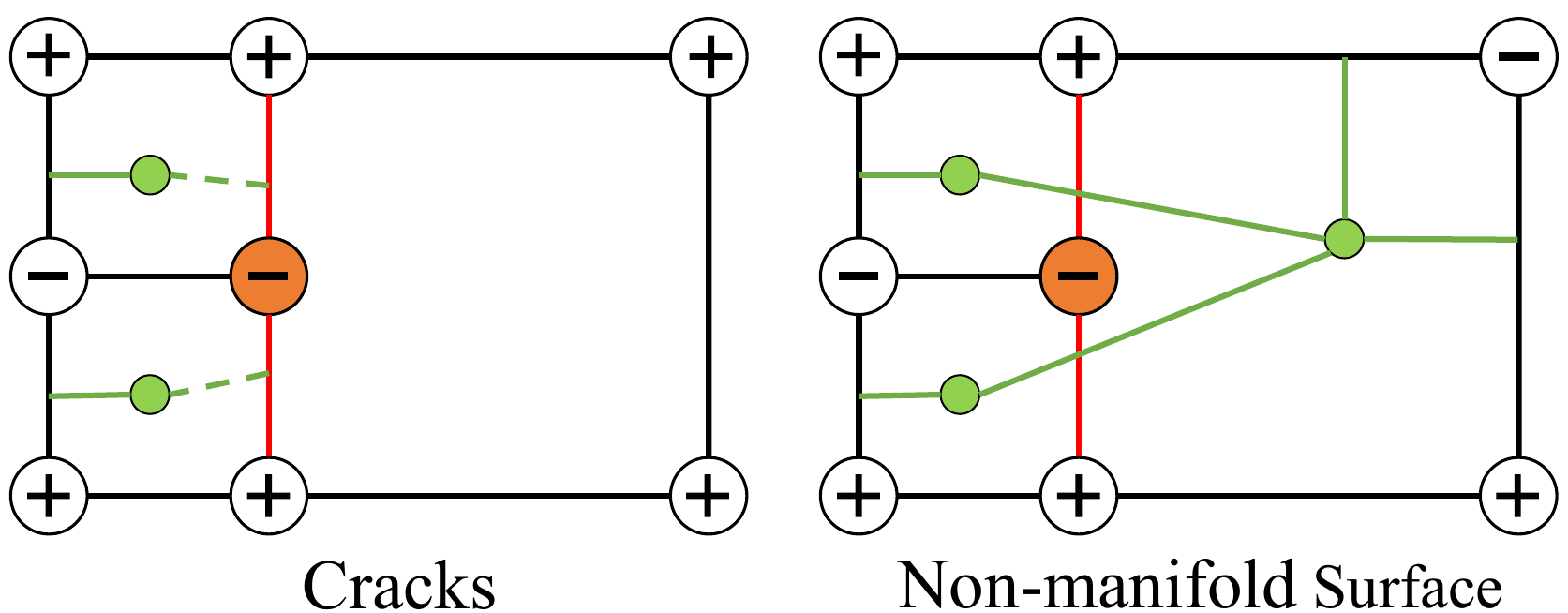}
    \caption{
      \updates{
      If the sign of the SDF is inconsistent on the face shared by cubes at different levels (highlighted in red), it leads to cracks or non-manifold surfaces.
      Two such cases are shown here.
      As such, we restrict the SDF of these vertices (in orange) to be consistent with the face of the larger cube.
      }
    }
    \label{fig:octree}
    \end{figure}
}

\newcommand{\figOctreeFix}{
    \begin{figure}
    \centering
    \setlength{\tabcolsep}{0mm}
    {
    \def\arraystretch{1.3}
    \begin{tabular}{ccc}
        \multirow{2}{*}{\raisebox{-\height}{
            \begin{tikzpicture}
            \node[anchor=south west,inner sep=0] (image) at (0,0) 
            {\includegraphics[width=0.32\columnwidth]{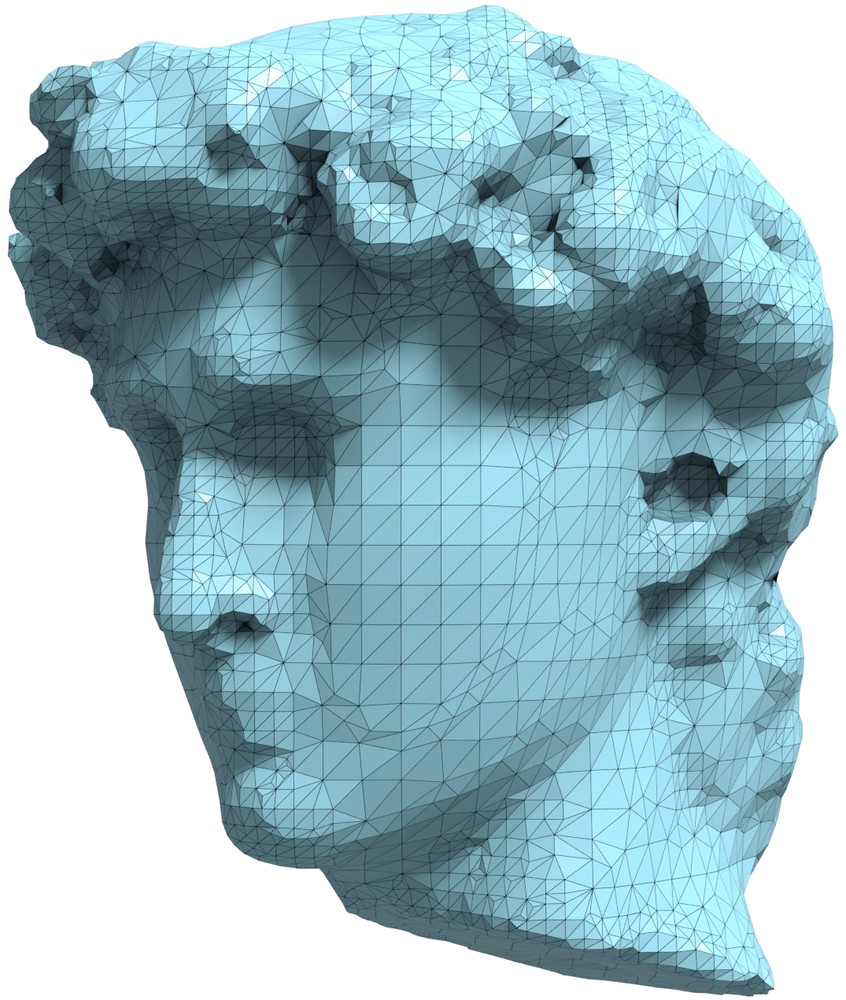}};
            \begin{scope}[x={(image.south east)},y={(image.north west)}]
                \draw[orange,very thick] (0.7, 0.45) rectangle (1.0, .71);
            \end{scope}
            \end{tikzpicture}
        }} &
        \raisebox{-\height}{
            \begin{tikzpicture}
            \node[anchor=south west,inner sep=0] (image) at (0,0) 
            {\includegraphics[width=0.30\columnwidth]{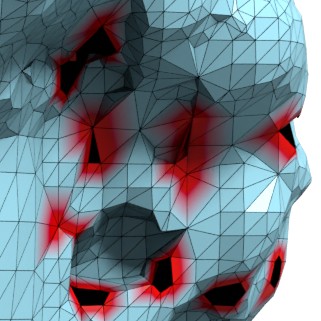}};
            \begin{scope}[x={(image.south east)},y={(image.north west)}]
                \draw[orange,very thick] (0.0, 0.0) rectangle (1.0, 1.0);
            \end{scope}
            \end{tikzpicture}
        } &
        \raisebox{-\height}{
            \begin{tikzpicture}
            \node[anchor=south west,inner sep=0] (image) at (0,0) 
            {\includegraphics[width=0.30\columnwidth]{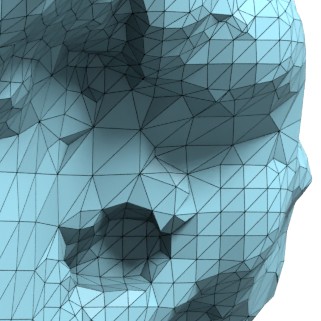}};
            \begin{scope}[x={(image.south east)},y={(image.north west)}]
                \draw[orange,very thick] (0.0, 0.0) rectangle (1.0, 1.0);
            \end{scope}
            \end{tikzpicture}
        } \\
        & Without Constraint & With Constraint
    \end{tabular}   
    }
    \caption{Cracks and non-manifold vertices produced by running \ours{} with an octree representation can be largely resolved by enforcing the constraint on SDF of minimal vertices. For the unconstrained version there are 318 non-manifold vertices in the resulting mesh, and with constraints there is only a single non-manifold vertex remaining. 
    }
    \label{fig:octree_fix}
    \end{figure}
}

\ifdefined\HQIMAGES
    \newcommand{\figMainQual}{
        \begin{figure*}[t]
        \centering
        \setlength{\tabcolsep}{1pt}
        \begin{tabular}{ccc|ccc|c}
        \includegraphics[width=0.14\textwidth, trim=4cm 6cm 4cm 3cm, clip]{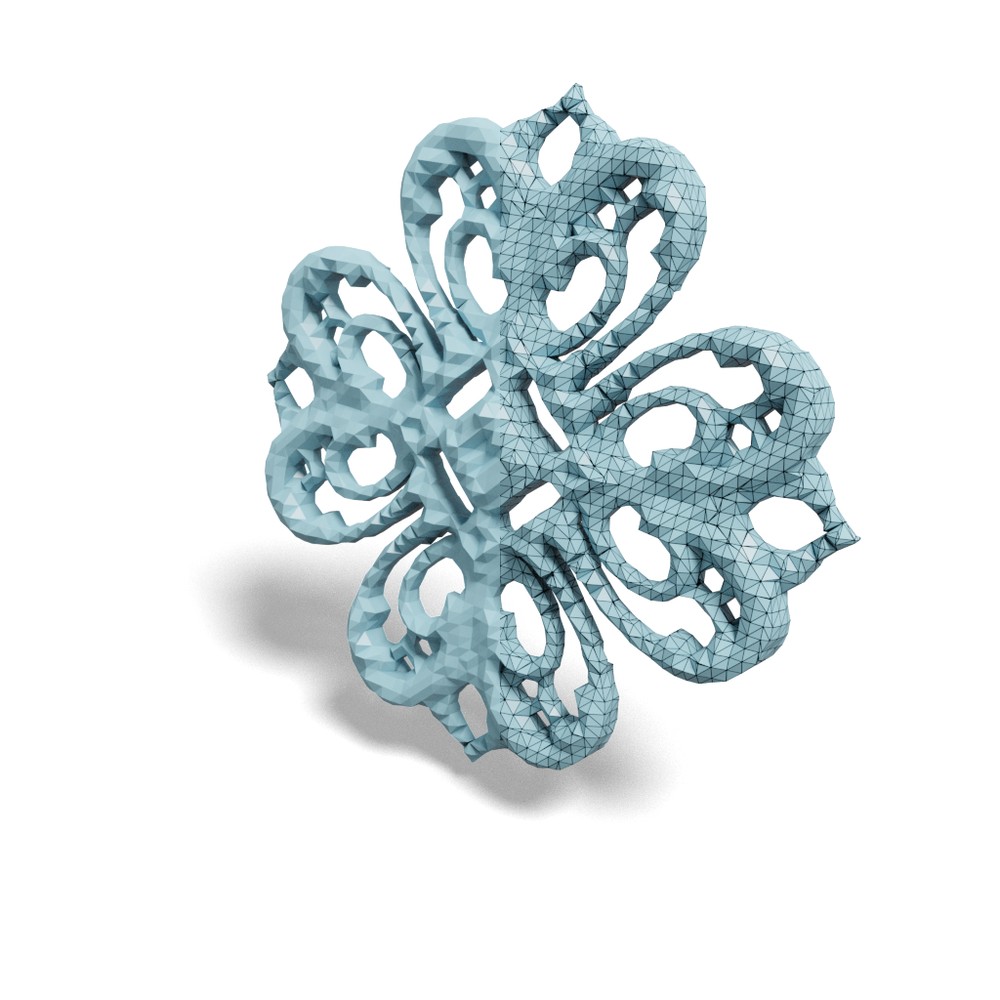} &
        \includegraphics[width=0.14\textwidth, trim=4cm 6cm 4cm 3cm, clip]{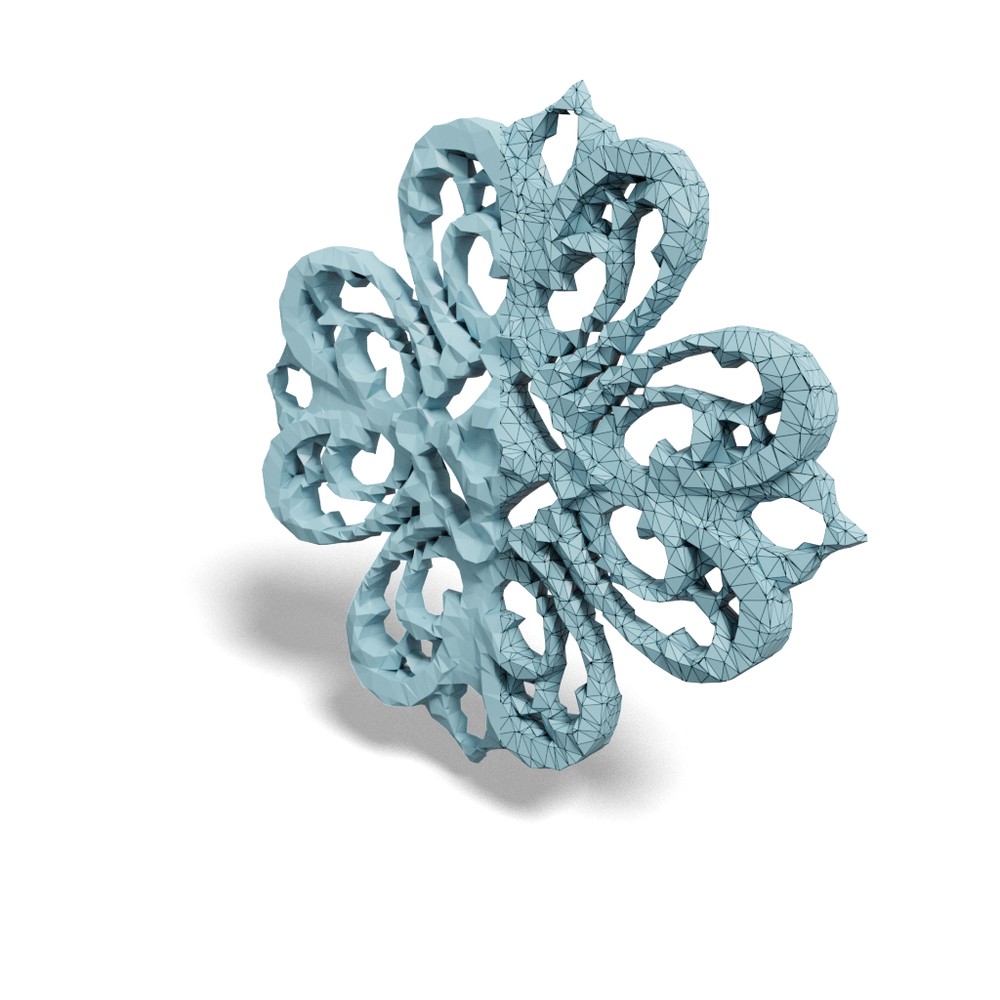} &
        \includegraphics[width=0.14\textwidth, trim=4cm 6cm 4cm 3cm, clip]{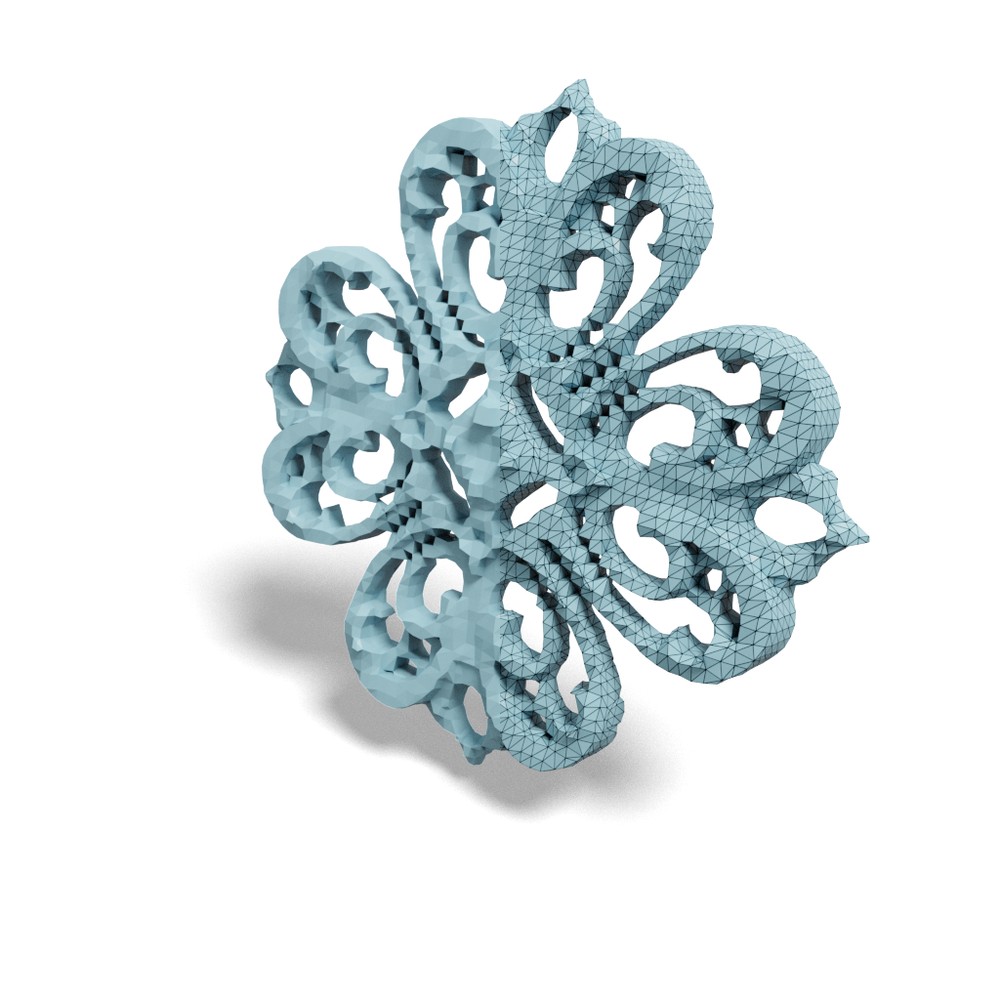} &
        \includegraphics[width=0.14\textwidth, trim=4cm 6cm 4cm 3cm, clip]{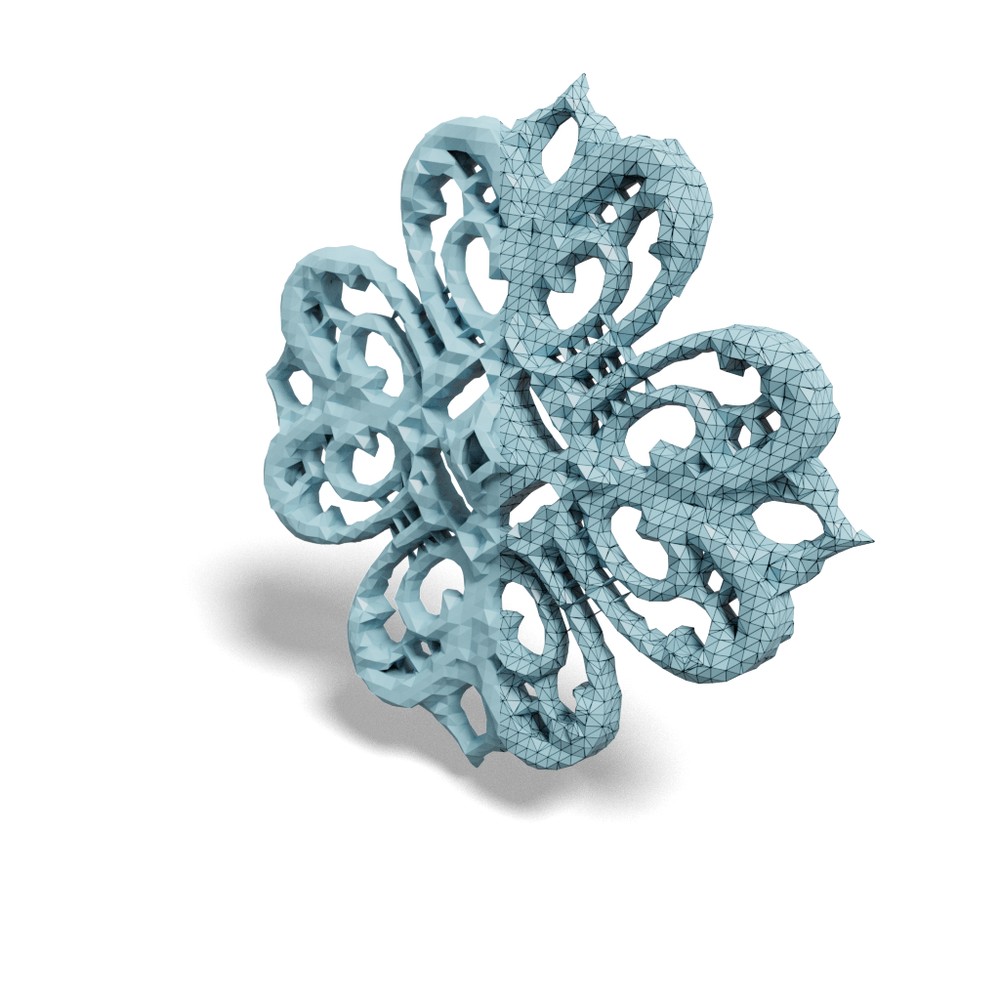} &
        \includegraphics[width=0.14\textwidth, trim=4cm 6cm 4cm 3cm, clip]{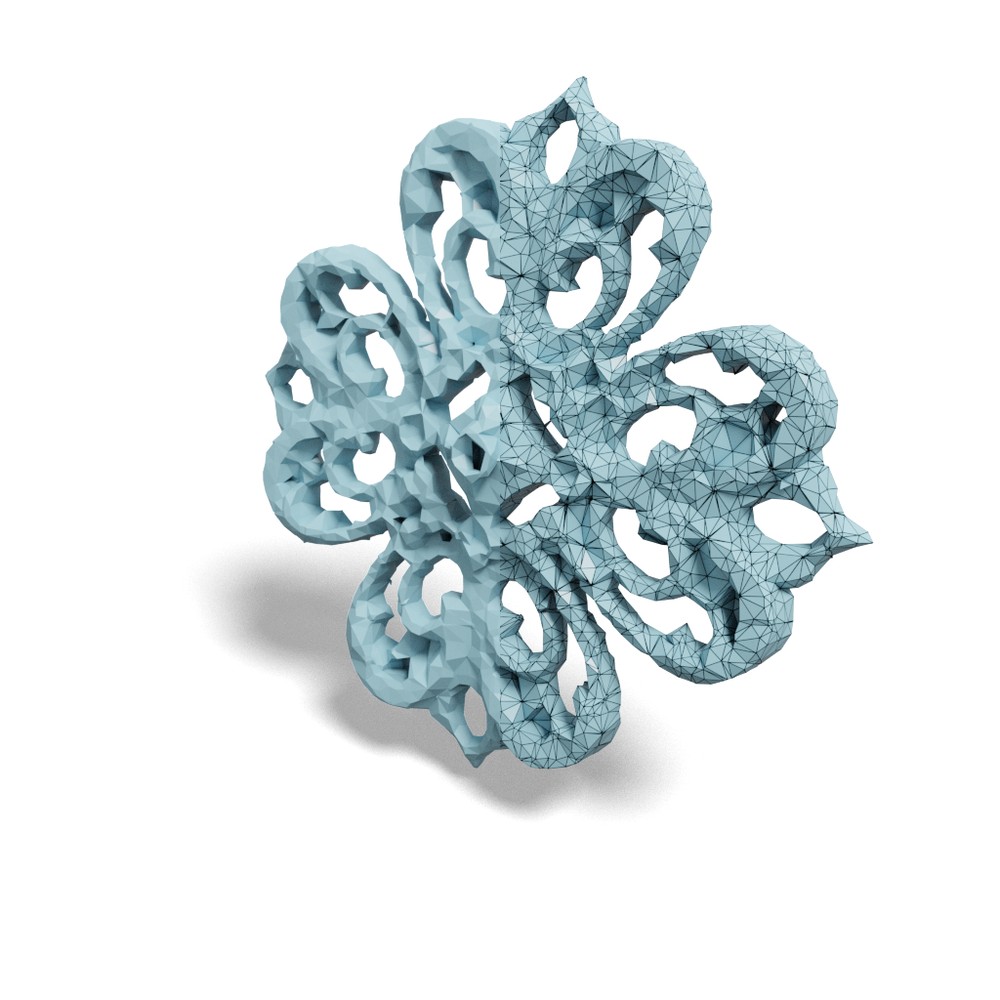} &
        \includegraphics[width=0.14\textwidth, trim=4cm 6cm 4cm 3cm, clip]{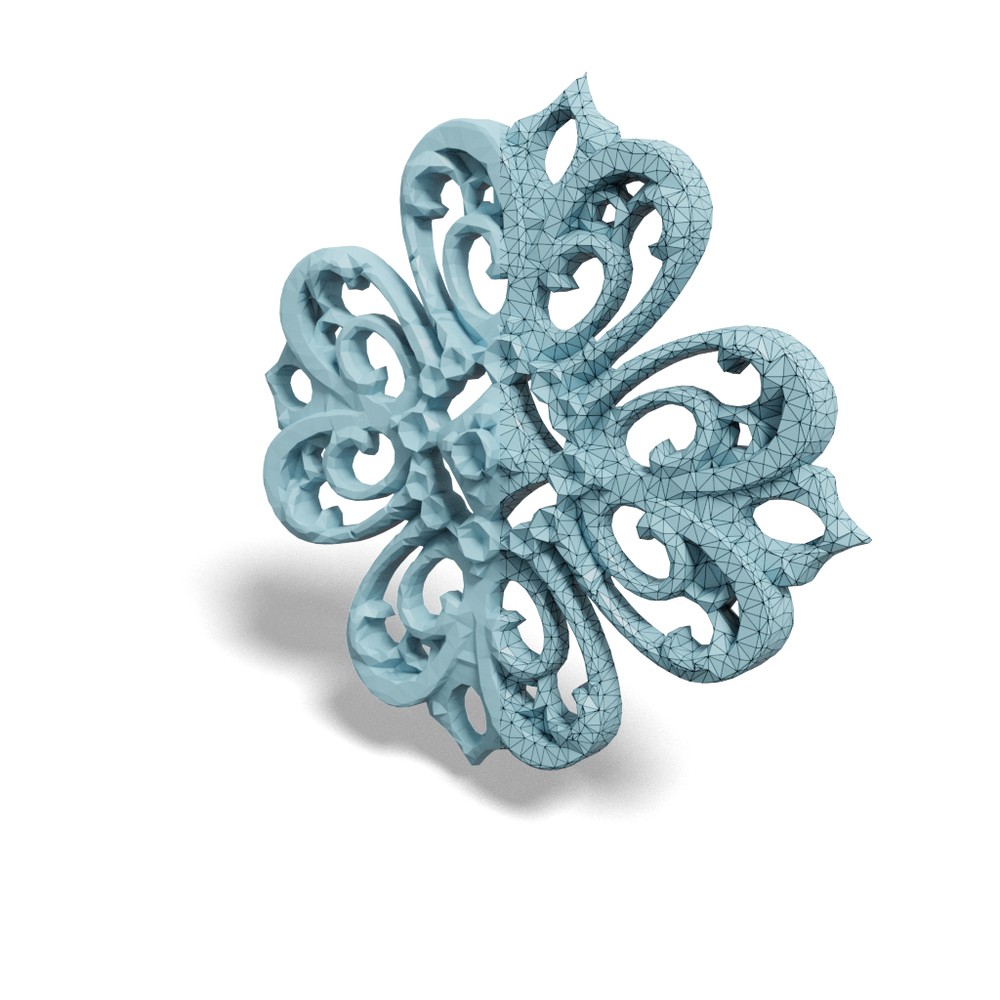} &
        \includegraphics[width=0.14\textwidth, trim=4cm 6cm 4cm 3cm, clip]{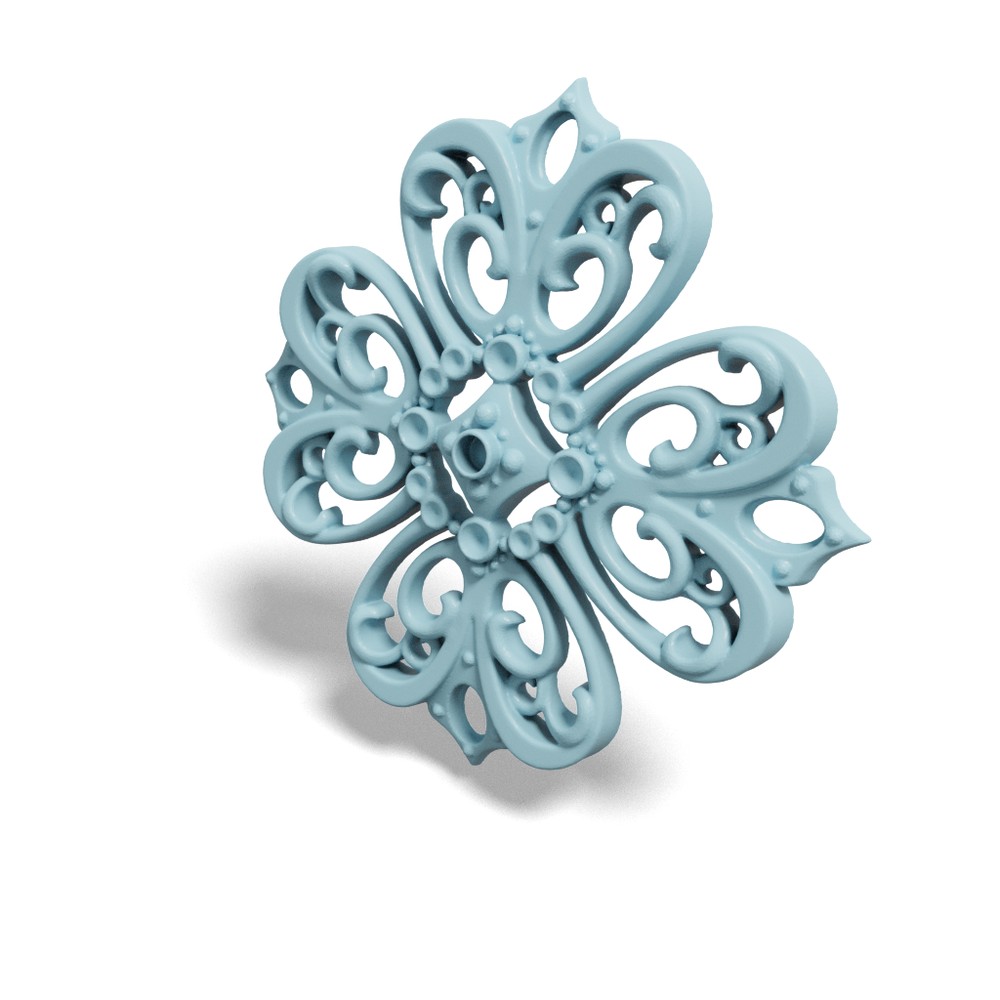} \\
        \includegraphics[width=0.14\textwidth, trim=2cm 2cm 2cm 0cm, clip]{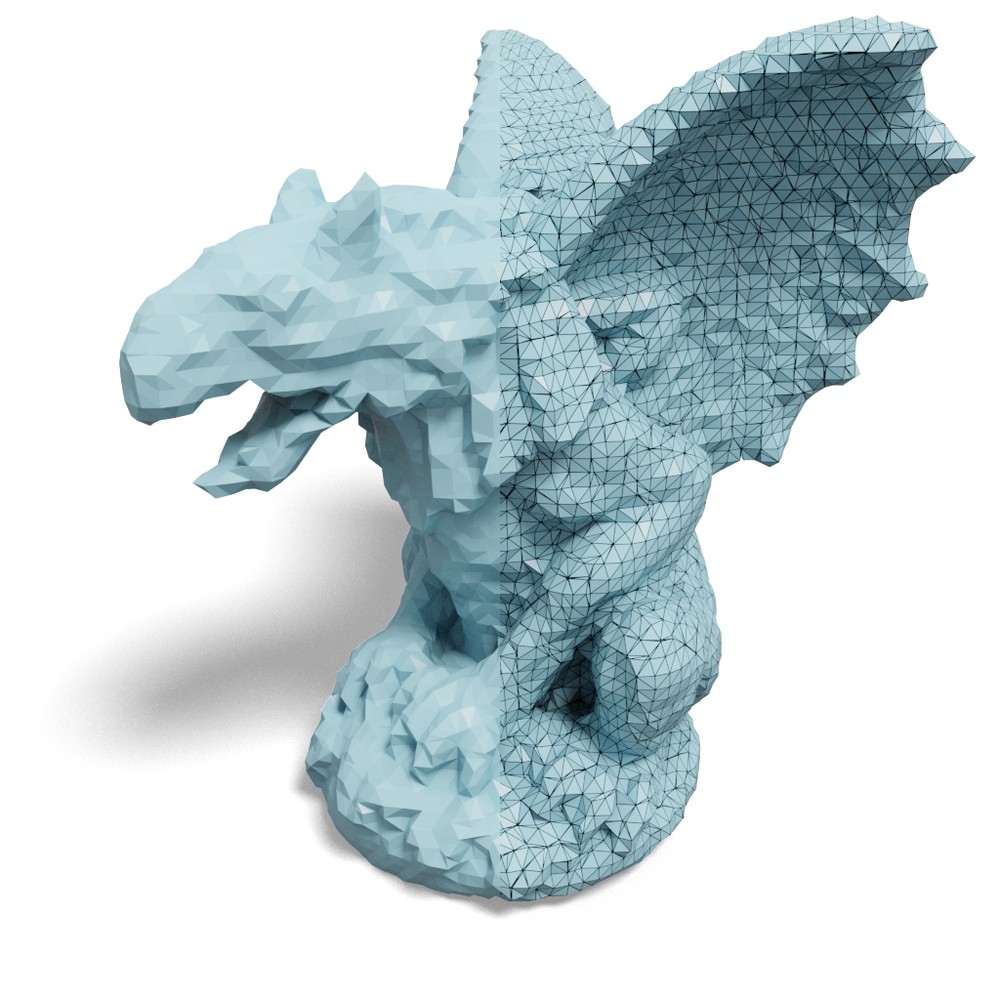} &
        \includegraphics[width=0.14\textwidth, trim=2cm 2cm 2cm 0cm, clip]{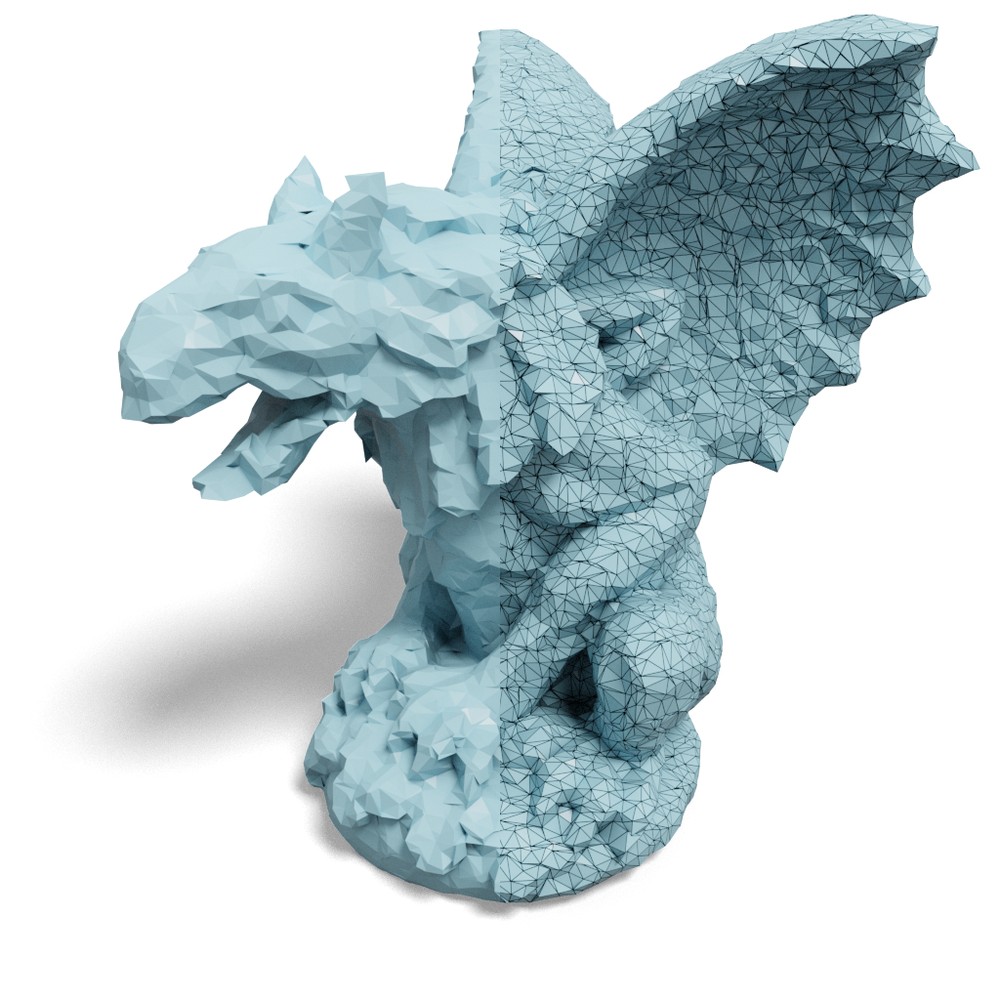} &
        \includegraphics[width=0.14\textwidth, trim=2cm 2cm 2cm 0cm, clip]{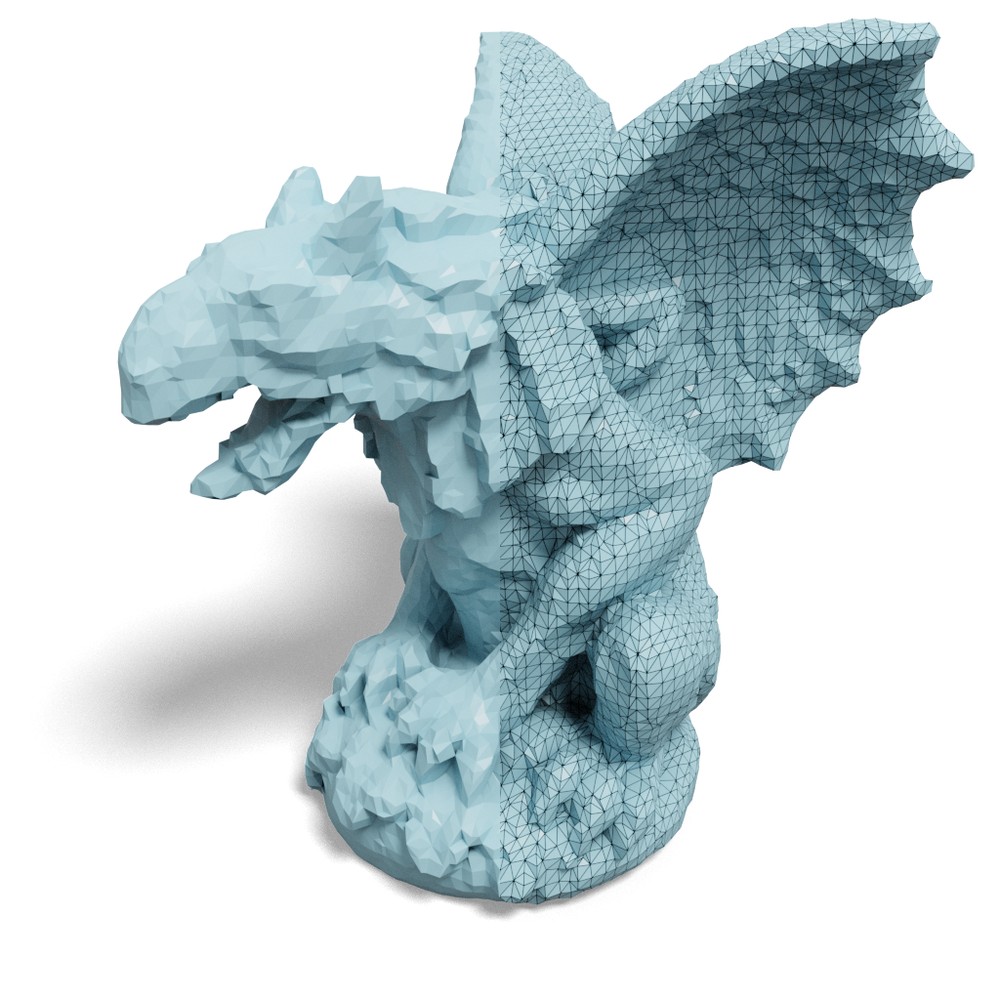} &
        \includegraphics[width=0.14\textwidth, trim=2cm 2cm 2cm 0cm, clip]{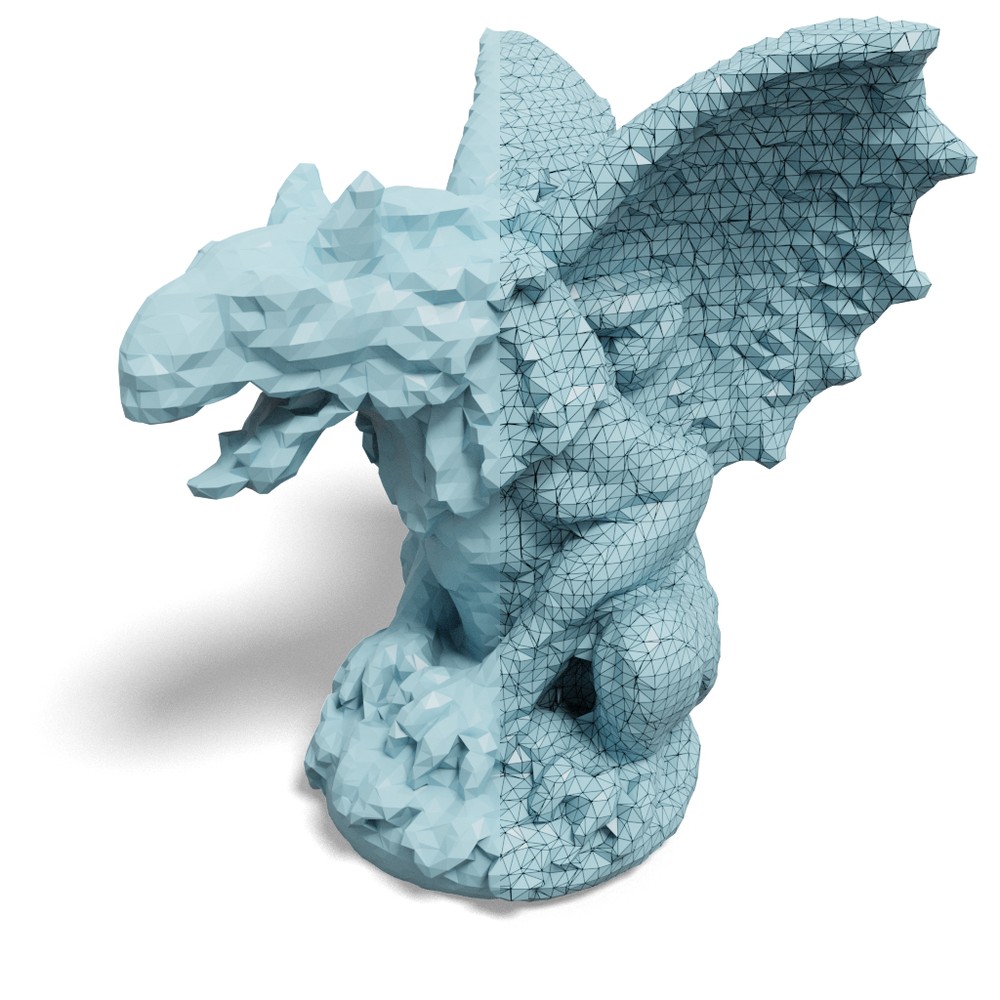} &
        \includegraphics[width=0.14\textwidth, trim=2cm 2cm 2cm 0cm, clip]{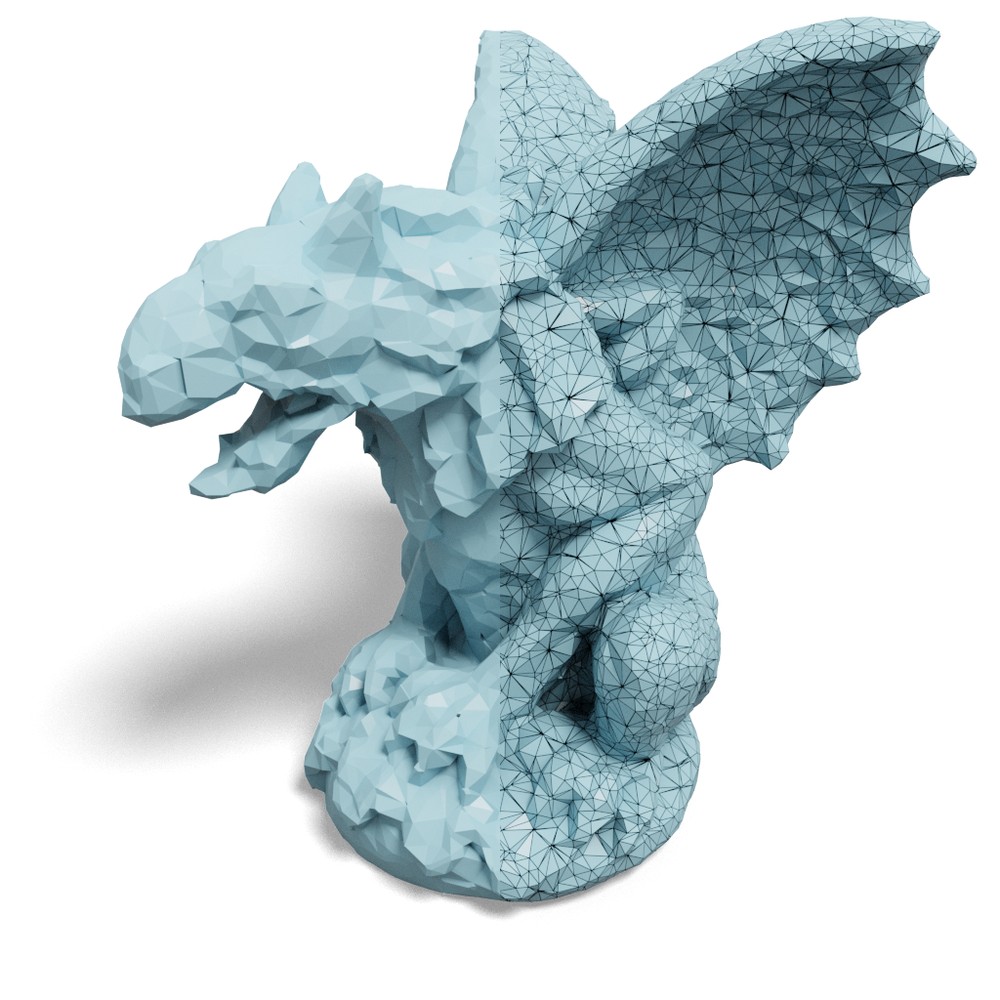} &
        \includegraphics[width=0.14\textwidth, trim=2cm 2cm 2cm 0cm, clip]{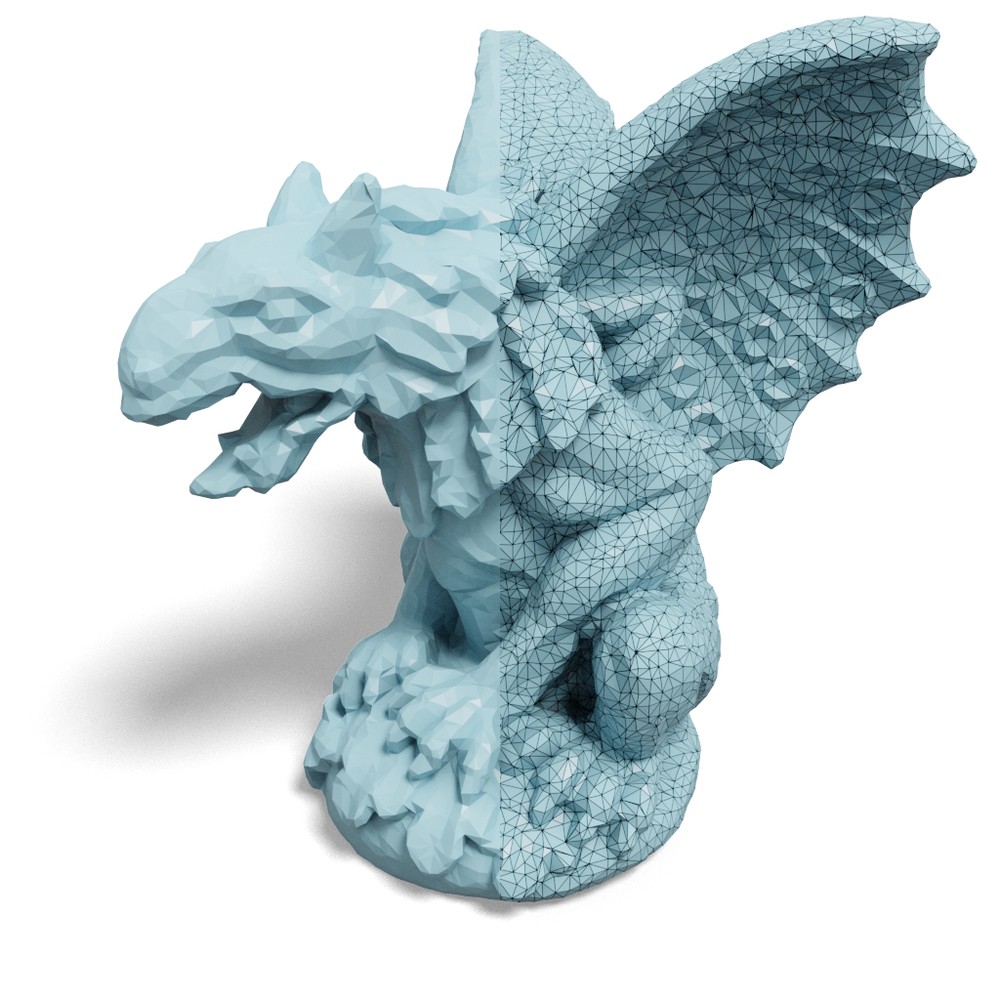} &
        \includegraphics[width=0.14\textwidth, trim=2cm 2cm 2cm 0cm, clip]{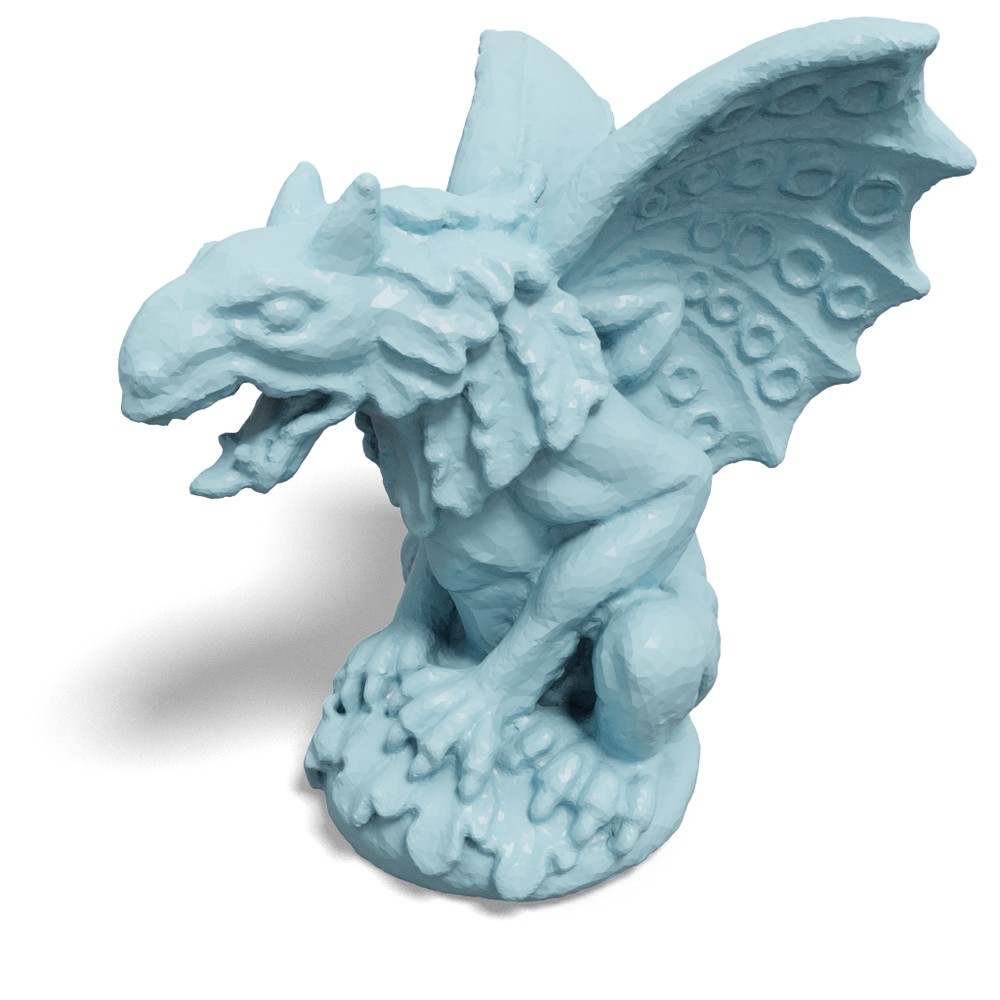} \\
        \includegraphics[width=0.14\textwidth, trim=0cm 2cm 2cm 0cm, clip]{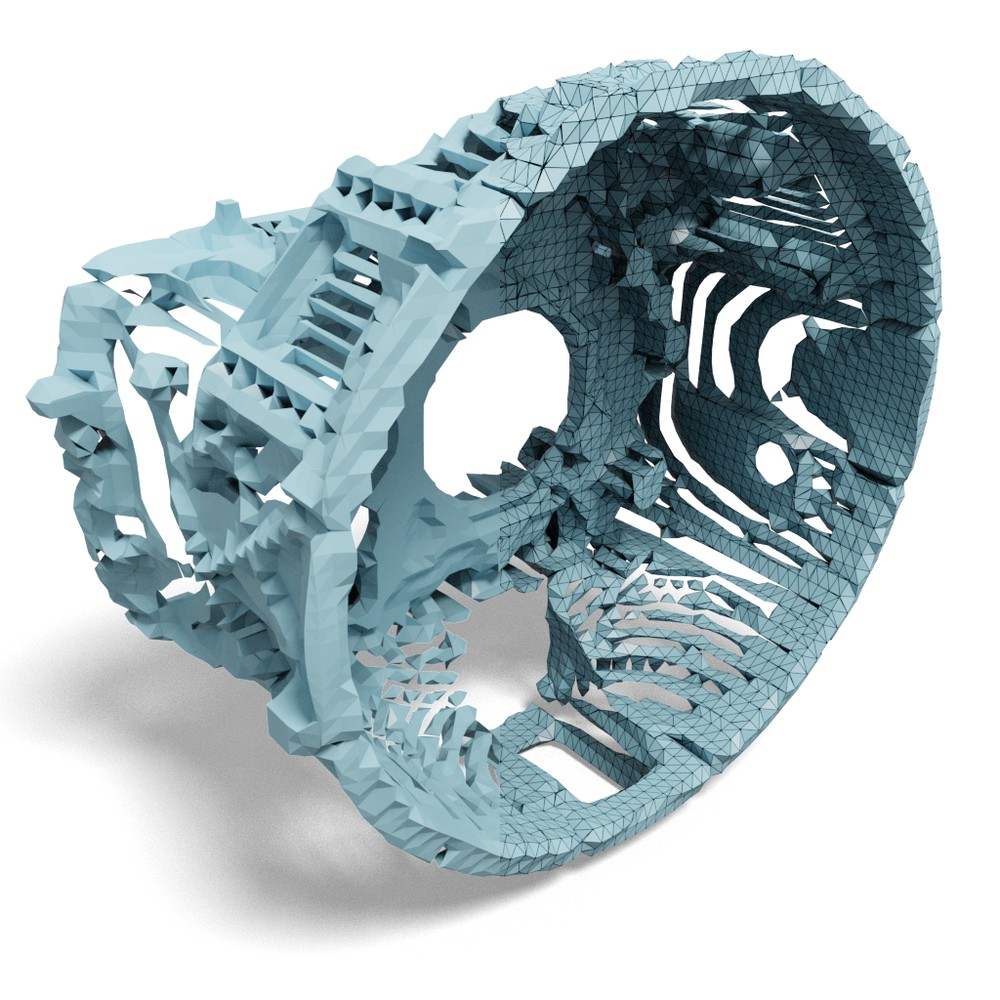} &
        \includegraphics[width=0.14\textwidth, trim=0cm 2cm 2cm 0cm, clip]{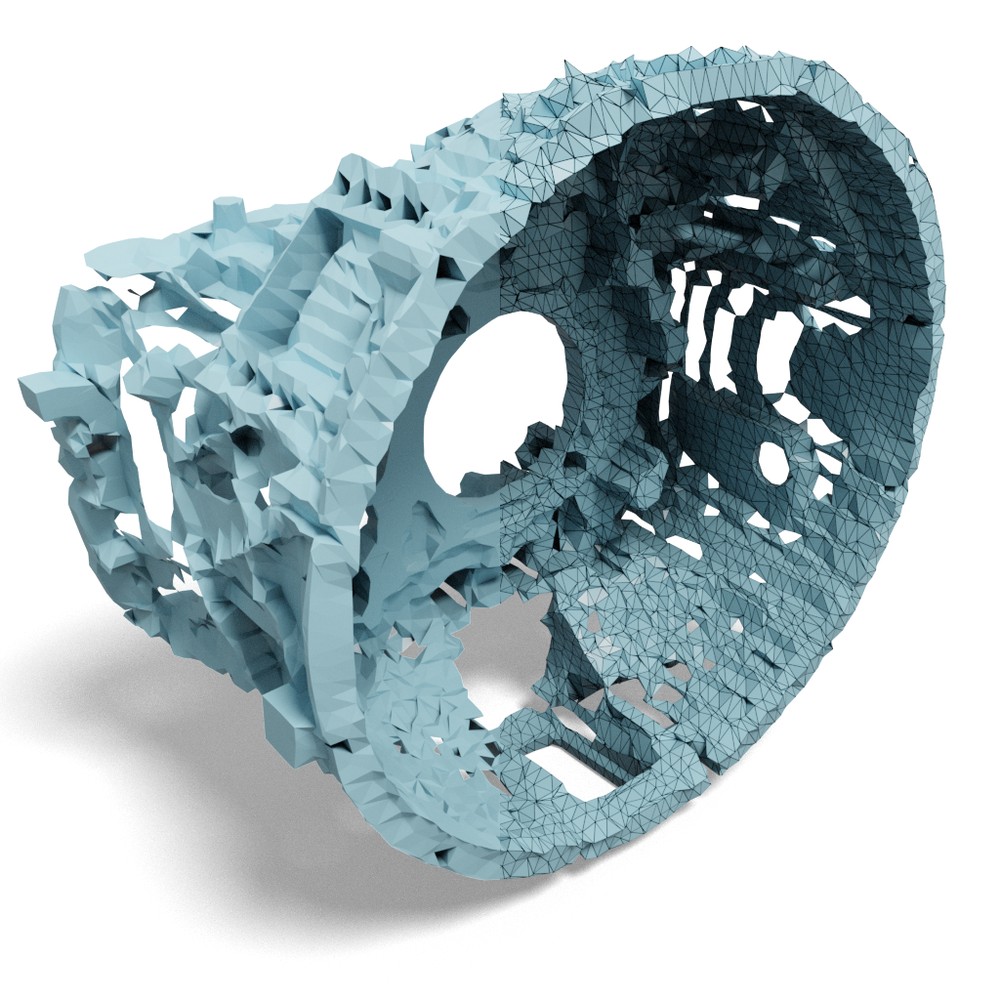} &
        \includegraphics[width=0.14\textwidth, trim=0cm 2cm 2cm 0cm, clip]{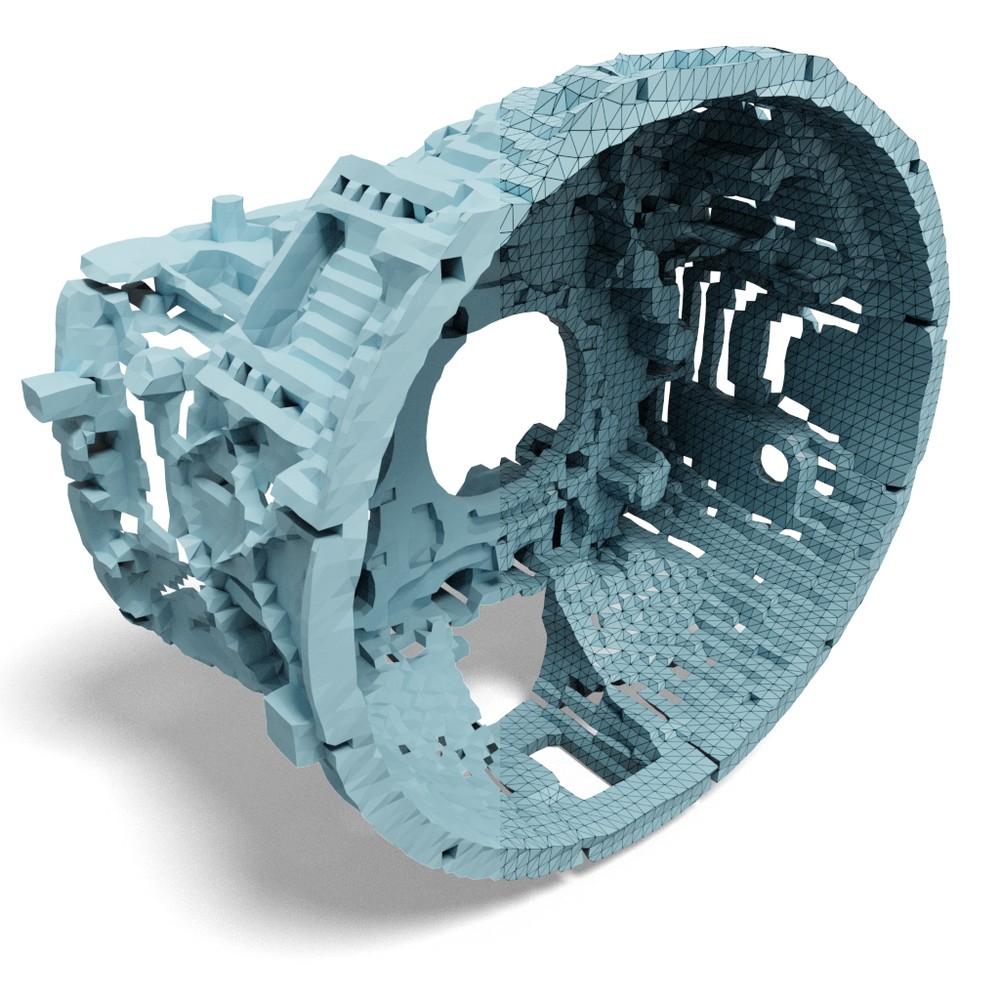} &
        \includegraphics[width=0.14\textwidth, trim=0cm 2cm 2cm 0cm, clip]{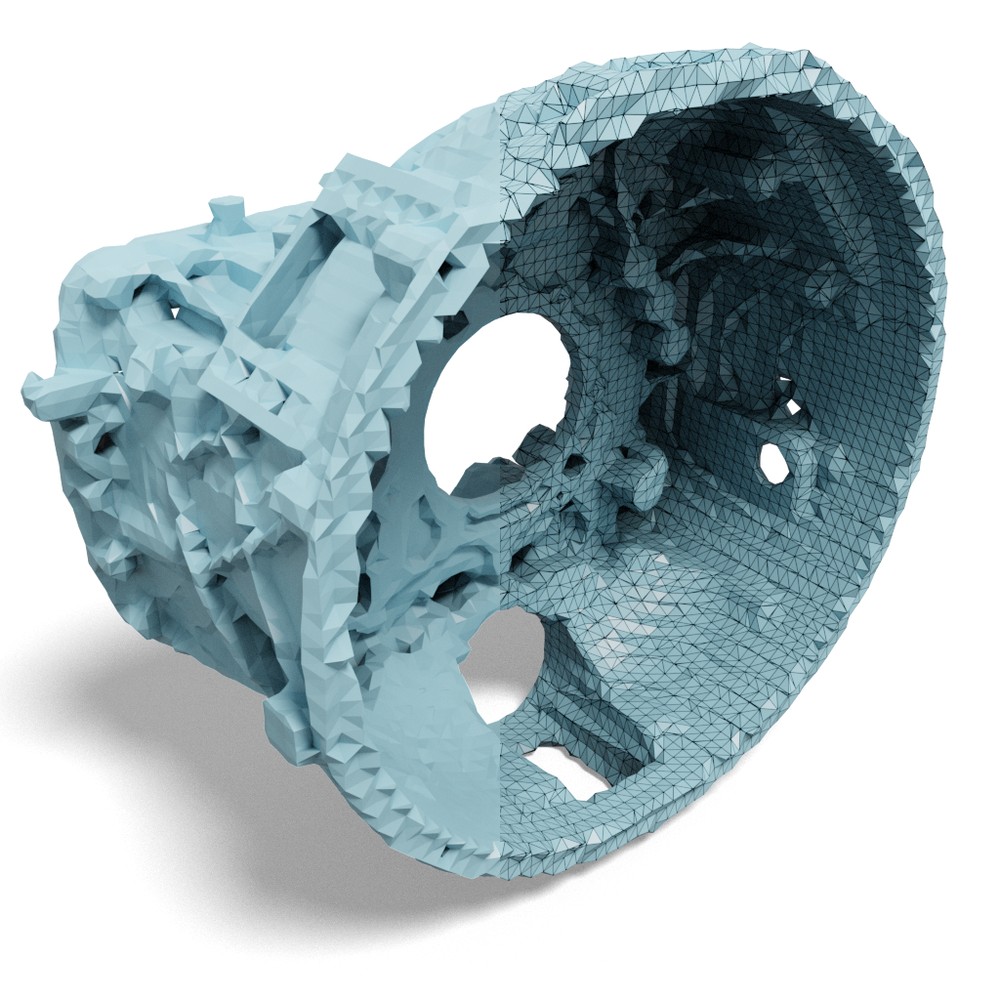} &
        \includegraphics[width=0.14\textwidth, trim=0cm 2cm 2cm 0cm, clip]{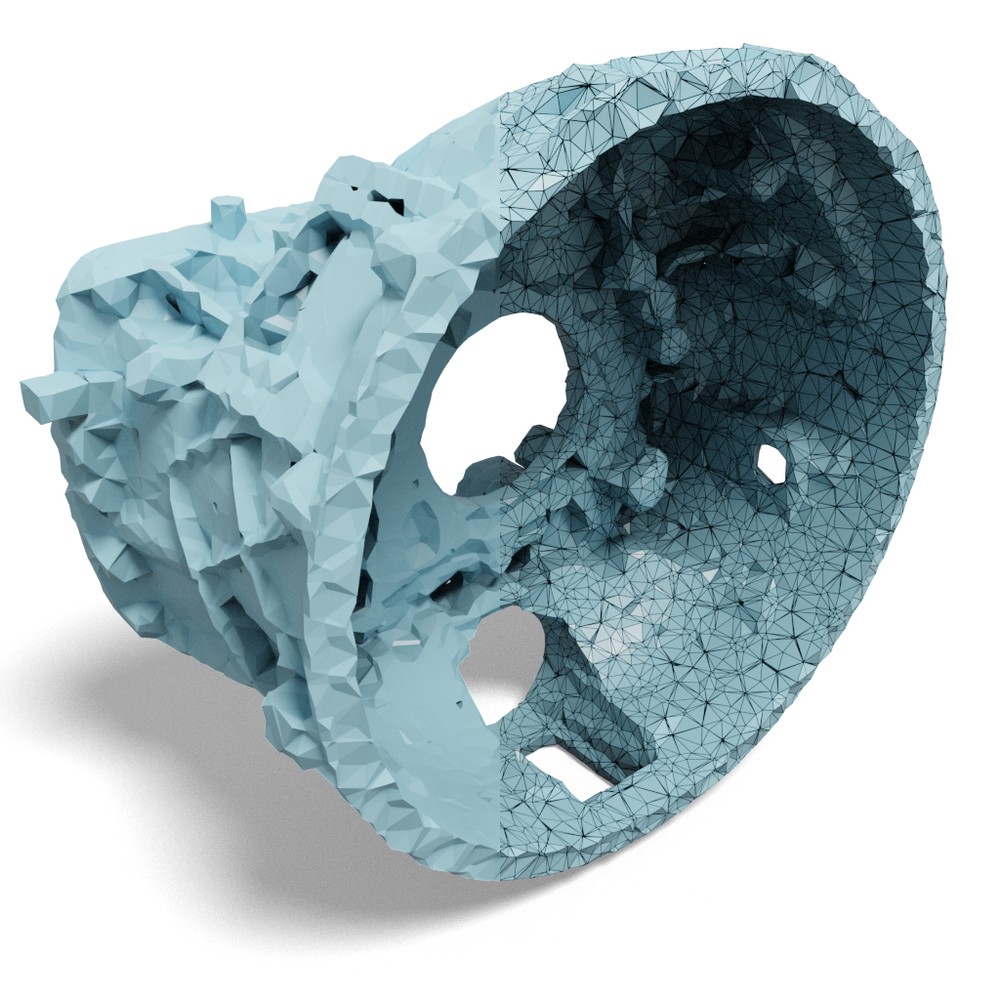} &
        \includegraphics[width=0.14\textwidth, trim=0cm 2cm 2cm 0cm, clip]{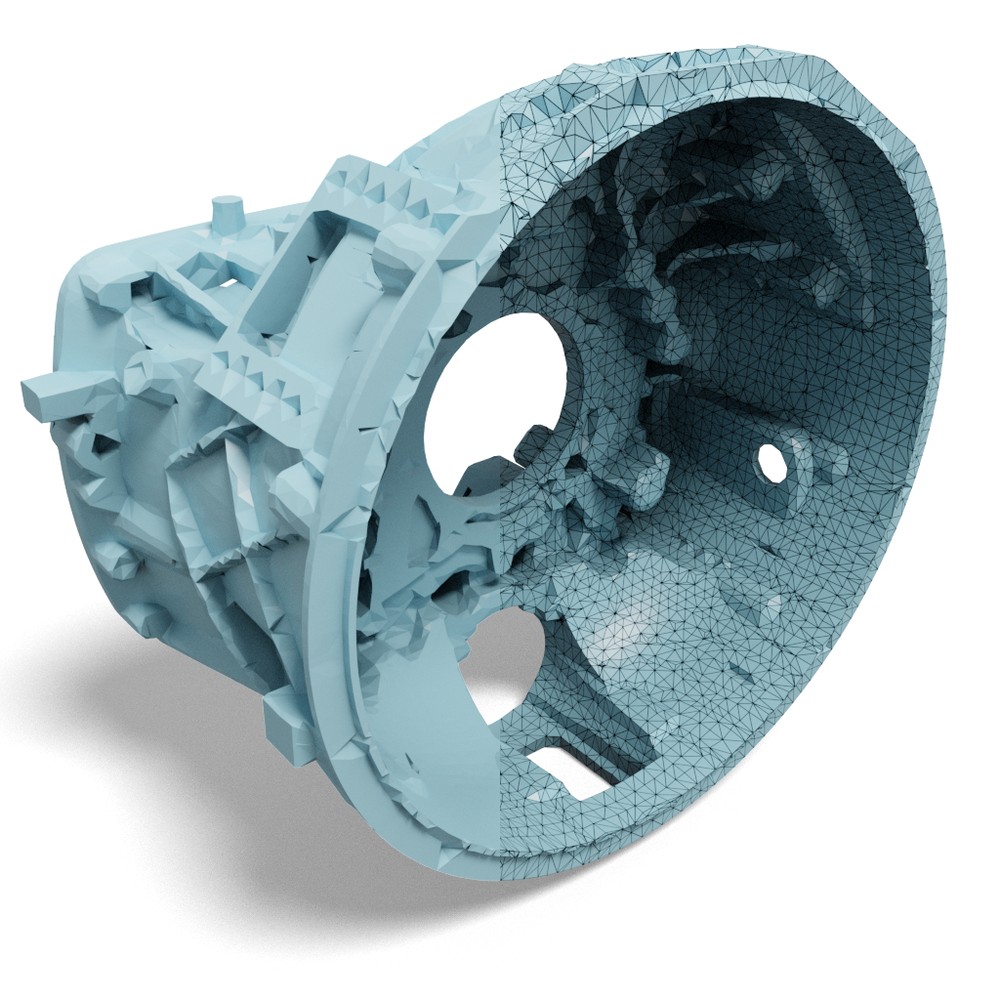} &
        \includegraphics[width=0.14\textwidth, trim=0cm 2cm 2cm 0cm, clip]{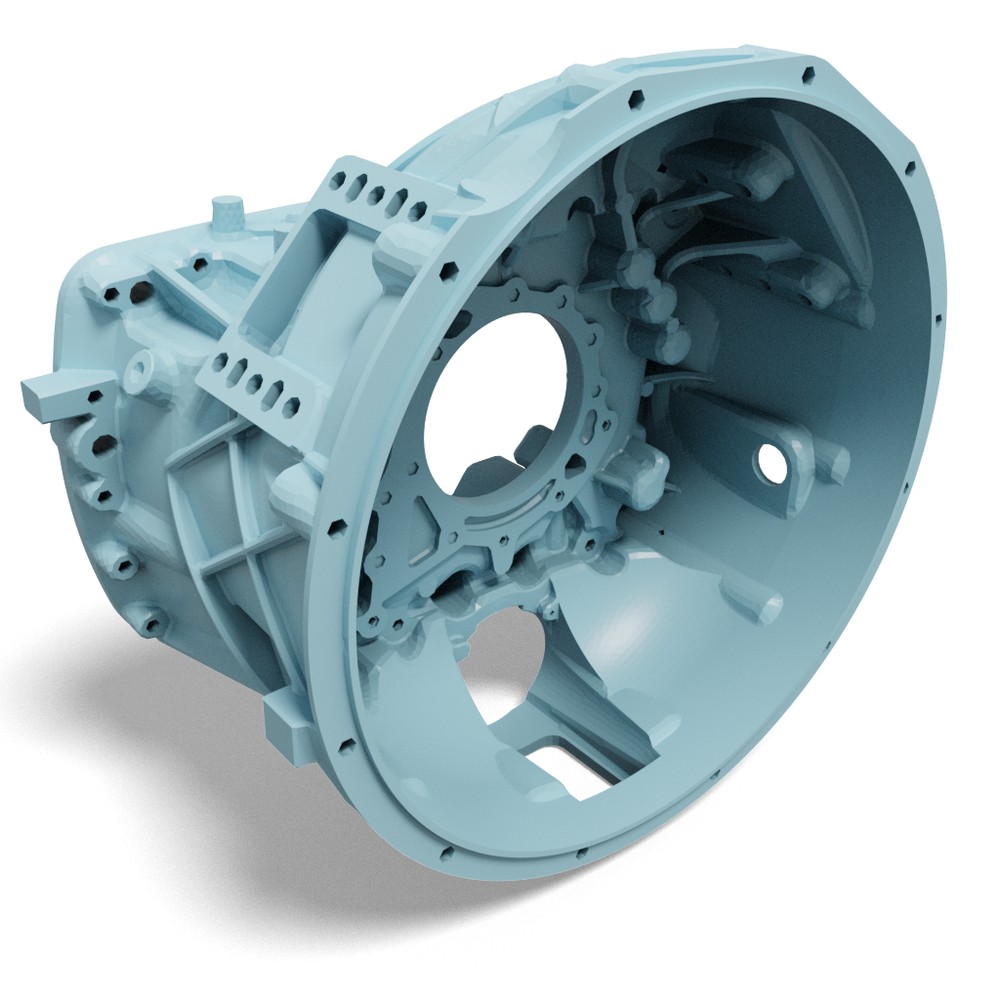} \\
        \includegraphics[width=0.14\textwidth, trim=0cm 2cm 0cm 2cm, clip]{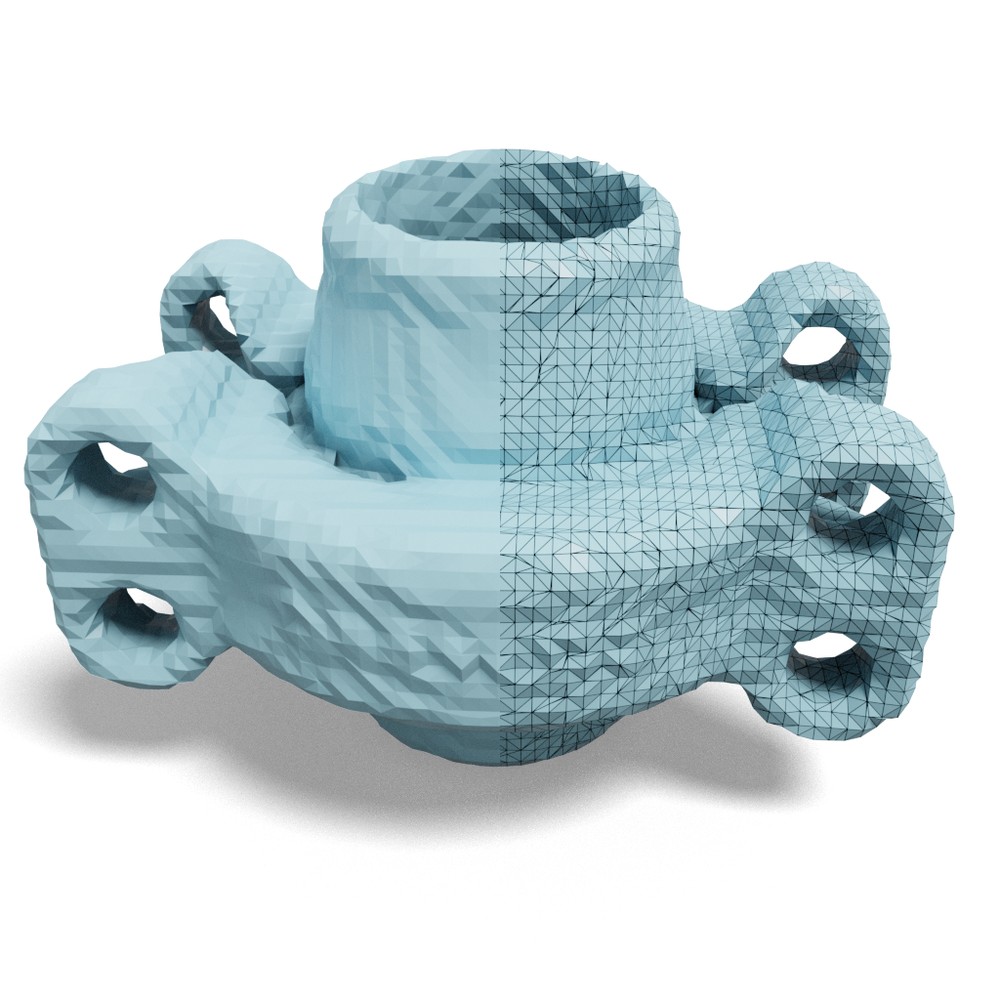} &
        \includegraphics[width=0.14\textwidth, trim=0cm 2cm 0cm 2cm, clip]{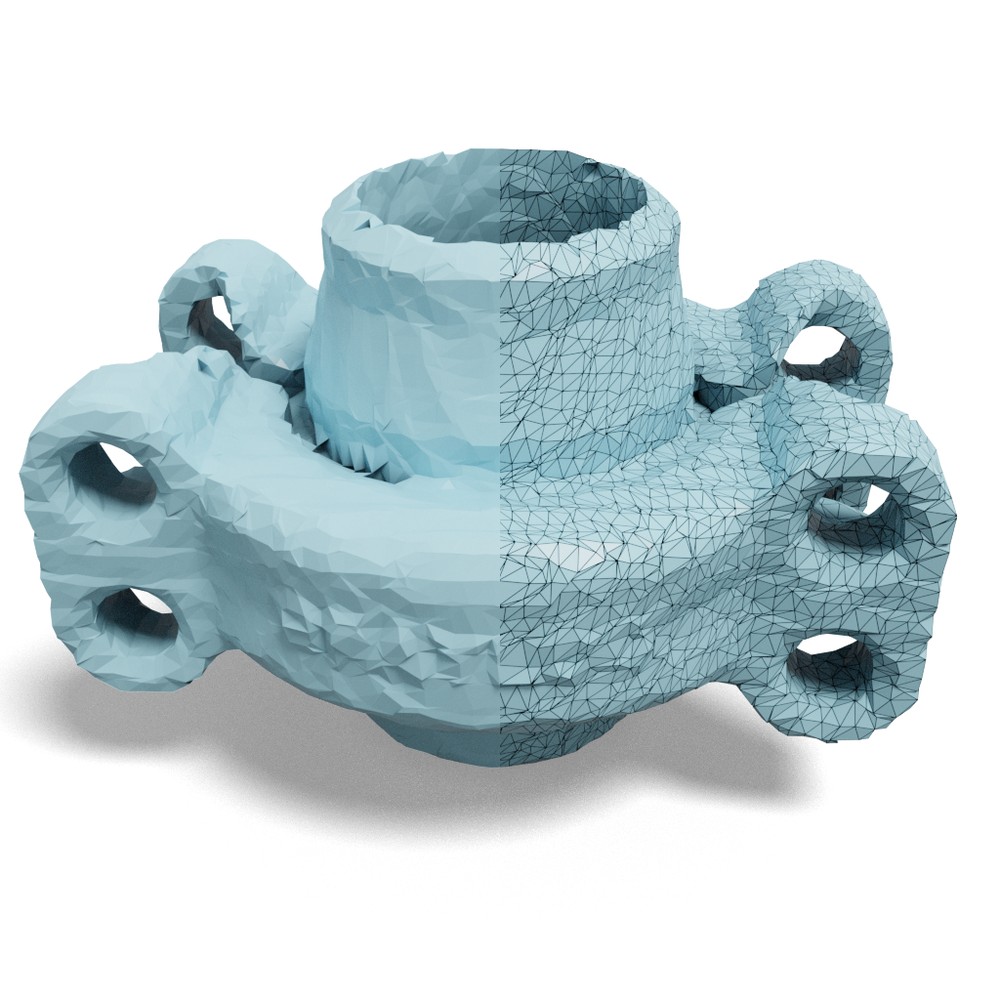} &
        \includegraphics[width=0.14\textwidth, trim=0cm 2cm 0cm 2cm, clip]{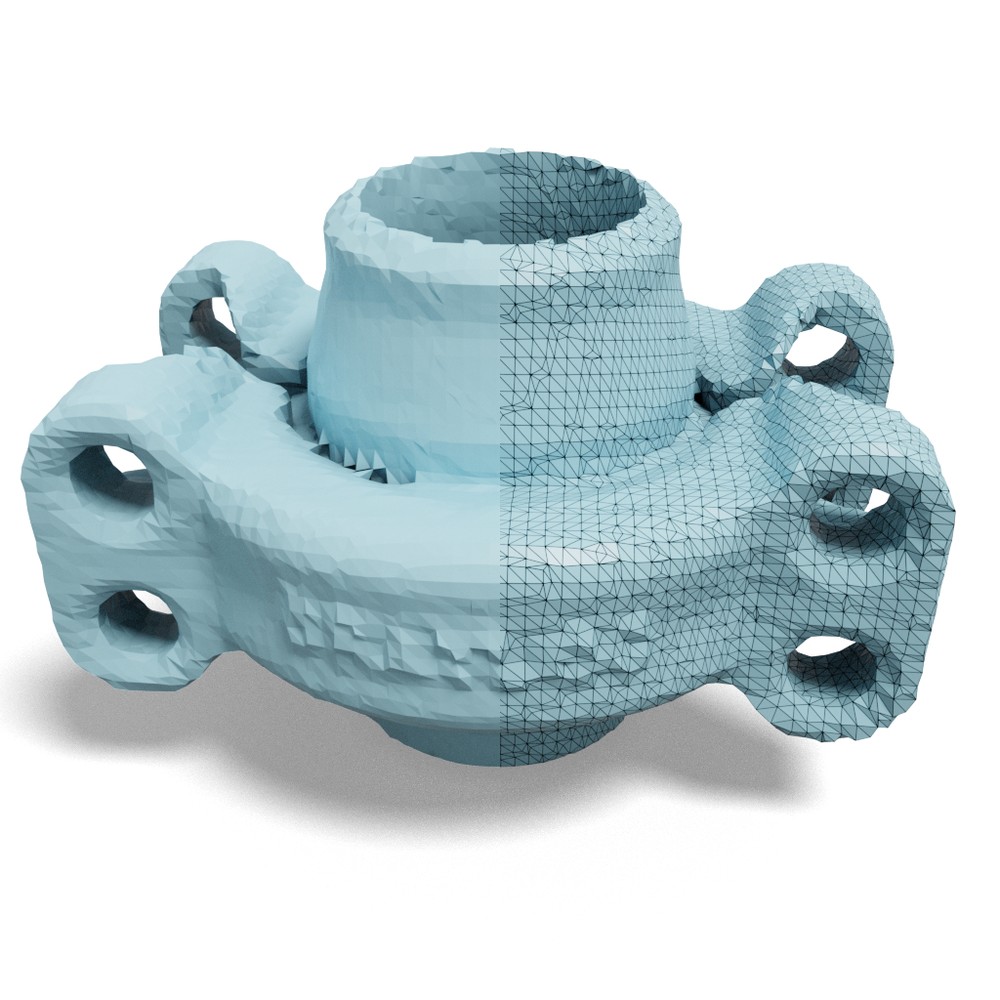} &
        \includegraphics[width=0.14\textwidth, trim=0cm 2cm 0cm 2cm, clip]{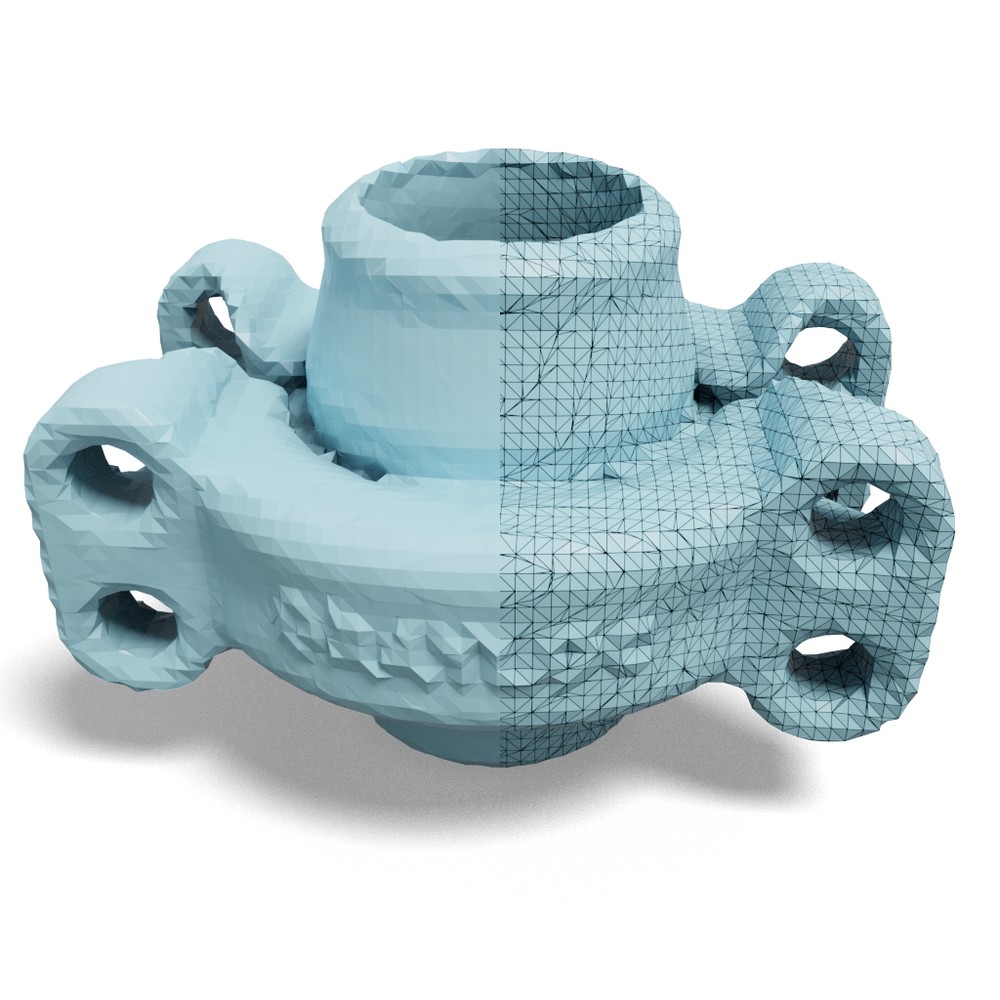} &
        \includegraphics[width=0.14\textwidth, trim=0cm 2cm 0cm 2cm, clip]{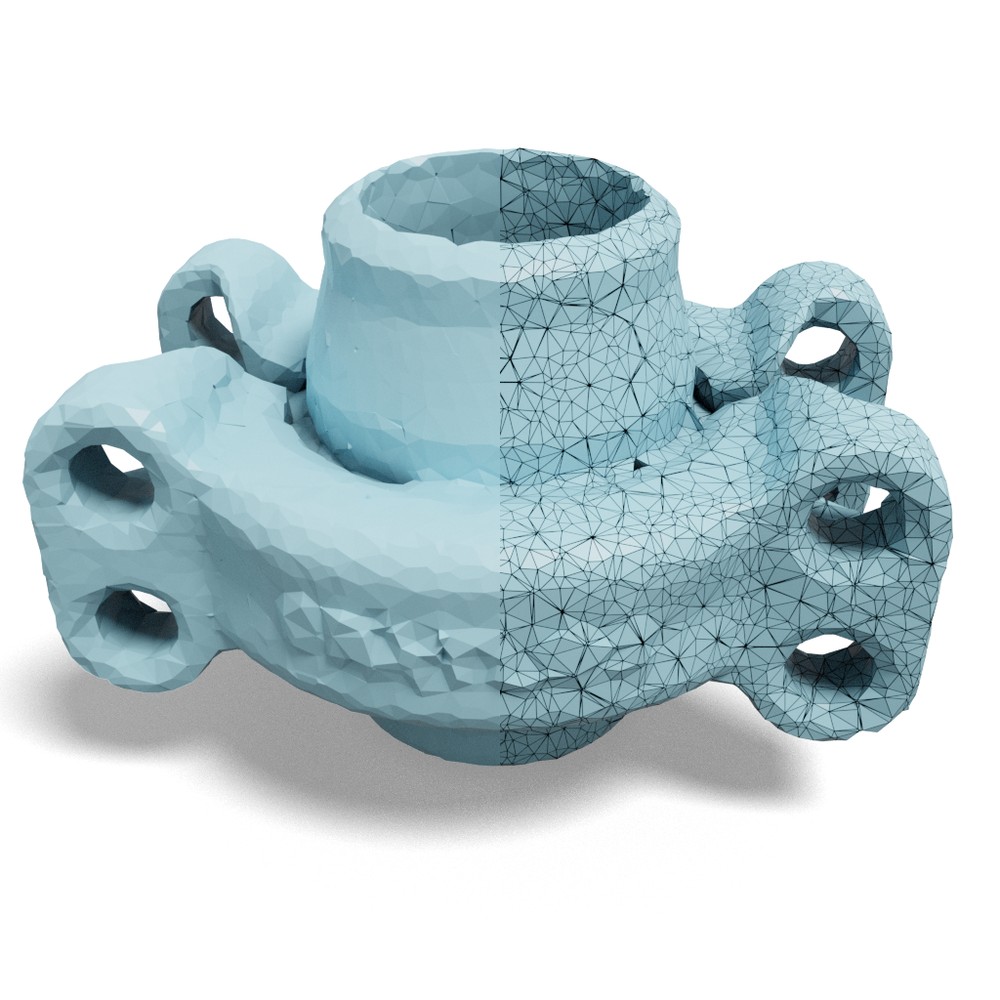} &
        \includegraphics[width=0.14\textwidth, trim=0cm 2cm 0cm 2cm, clip]{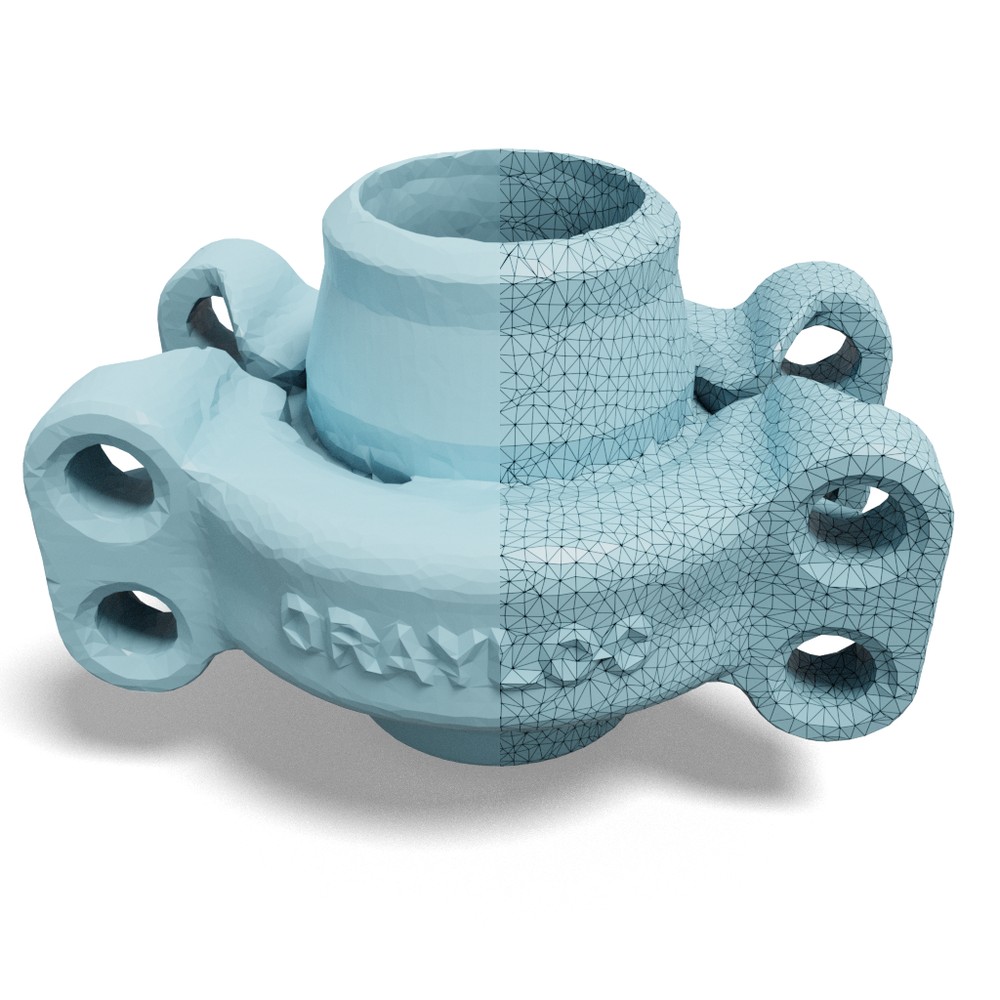} &
        \includegraphics[width=0.14\textwidth, trim=0cm 2cm 0cm 2cm, clip]{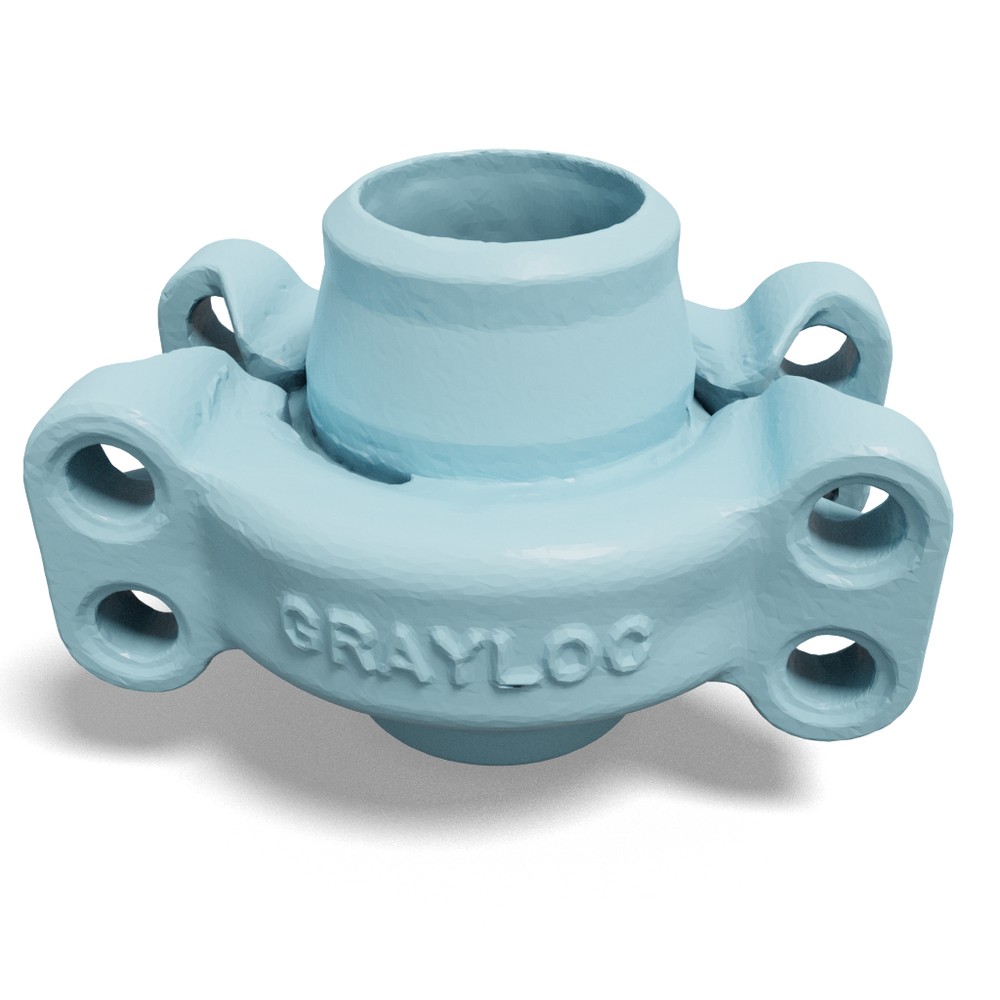} \\
        $MC_{SDF}$ & $DC_{hermite}$ & $NDC_{SDF}$ & MC & \textsc{DMTet} & \textsc{FlexiCubes} & Reference \\
        \multicolumn{3}{c}{Surface extractions from GT SDF} & \multicolumn{3}{c}{Differentiable iso-surfacing} &   %
        \end{tabular}
        \caption{Visual comparison of a set of iso-surfacing techniques. The three leftmost examples: Marching Cubes ($MC_{SDF}$), Dual Contouring, Neural Dual Contouring, use surface extraction from the ground truth SDF. The next three examples: 
        Marching Cubes, Deep Marching Tetrahedra, and \ours{}, use differentiable iso-surfacing. The grid resolution is $64^3$ for all methods except DMTet, which uses $80^3$ tetrahedral grid to match the triangle count in output meshes.}
        \label{fig:main_qual}
        \end{figure*}
    }
\else
   \newcommand{\figMainQual}{
    \begin{figure*}[t]
    \centering
    \includegraphics[width=\textwidth,trim=0 0 0 0,clip]{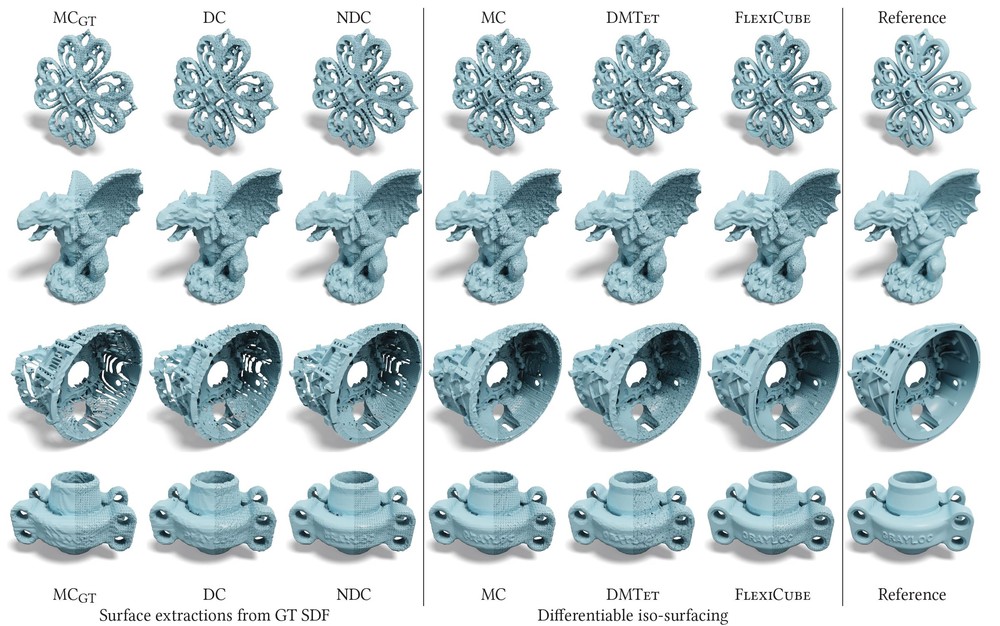}    
    \caption{Visual comparison of different iso-surfacing techniques on mesh reconstruction. The three leftmost examples: Marching Cubes (MC$_\mathrm{SDF}$), Dual Contouring, Neural Dual Contouring, use surface extraction from the GT SDF. The next three examples: 
    Marching Cubes, Deep Marching Tetrahedra, and \ours{}, use differentiable iso-surfacing. \TODO{\textbf{PLACEHOLDER} Labels are fixed in the high quality version of this image.}}
    \label{fig:main_qual}
    \end{figure*}
}
\fi

\ifdefined\HQIMAGES
    \newcommand{\figSupplementalMainQual}{
        \begin{figure*}[t]
        \centering
        \setlength{\tabcolsep}{1pt}
        \begin{tabular}{ccc|ccc|c}
            \includegraphics[width=0.14\textwidth, trim=2cm 2cm 2cm 2cm, clip]{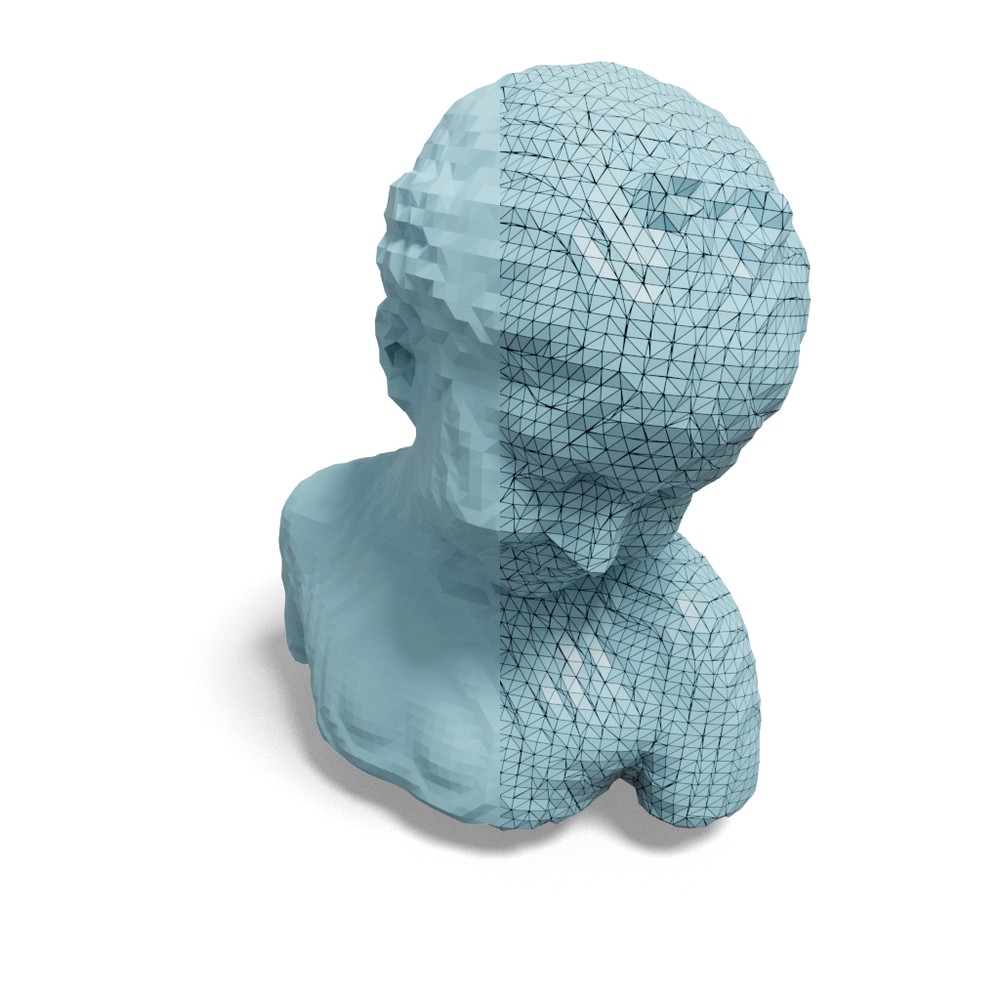} &
            \includegraphics[width=0.14\textwidth, trim=2cm 2cm 2cm 2cm, clip]{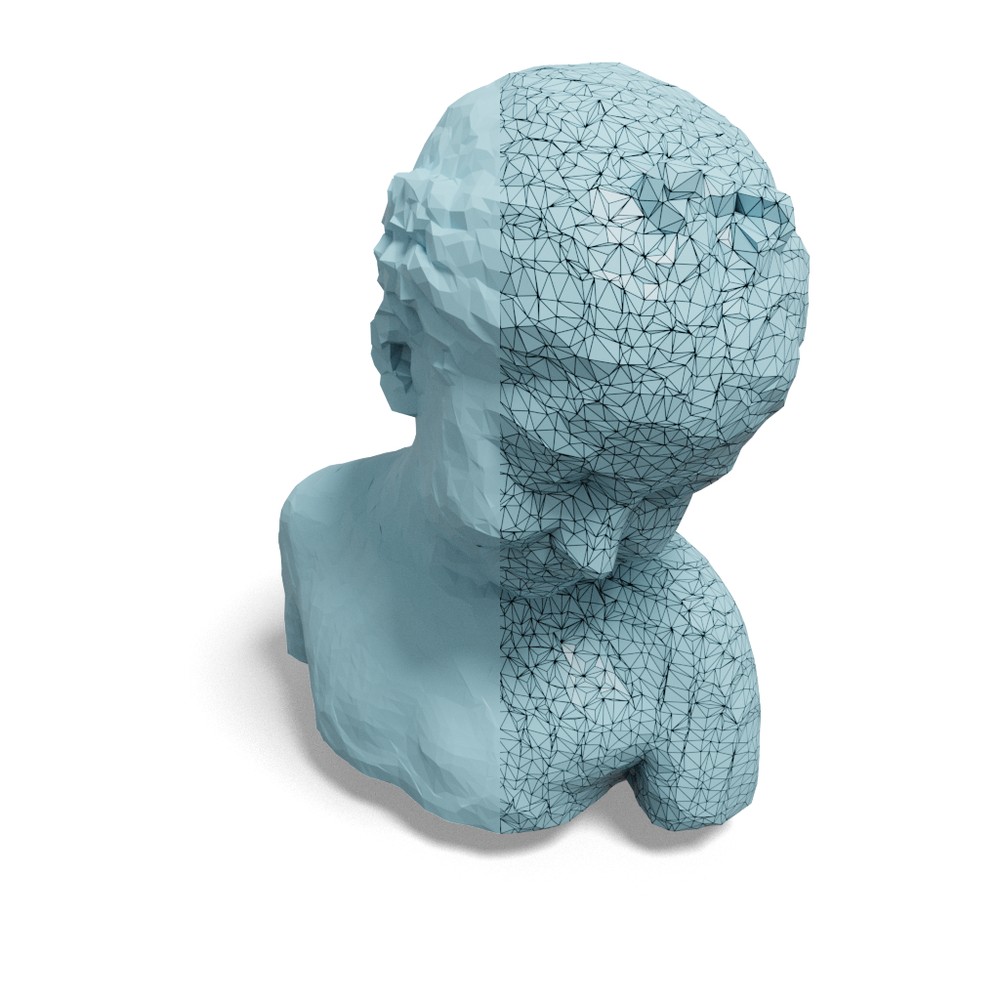} &
            \includegraphics[width=0.14\textwidth, trim=2cm 2cm 2cm 2cm, clip]{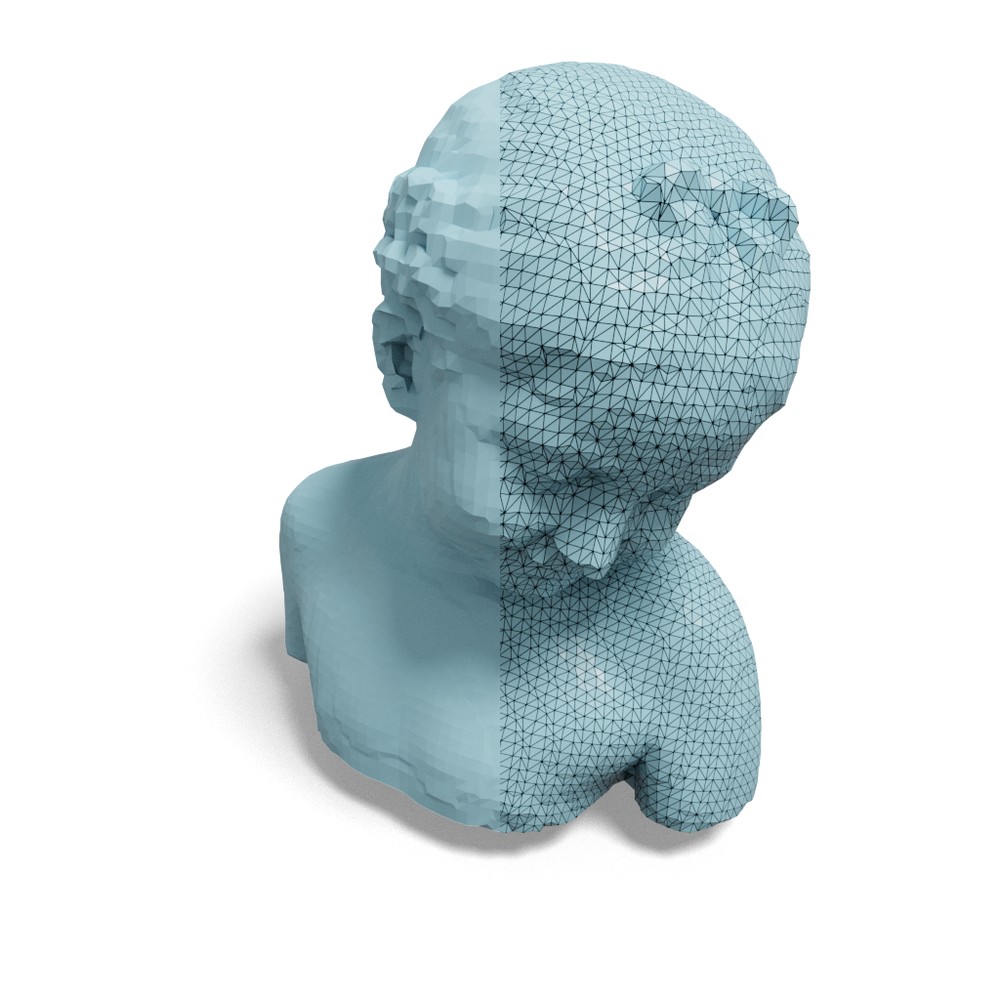} &
            \includegraphics[width=0.14\textwidth, trim=2cm 2cm 2cm 2cm, clip]{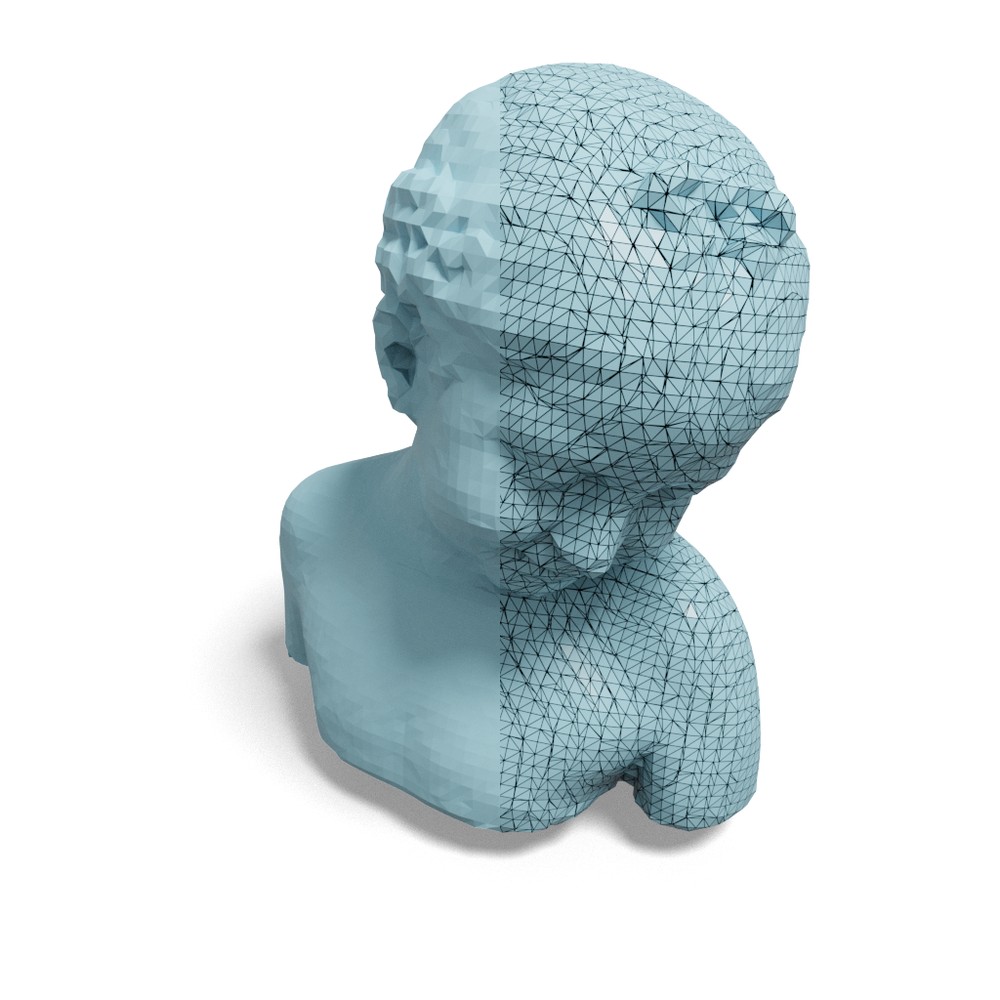} &
            \includegraphics[width=0.14\textwidth, trim=2cm 2cm 2cm 2cm, clip]{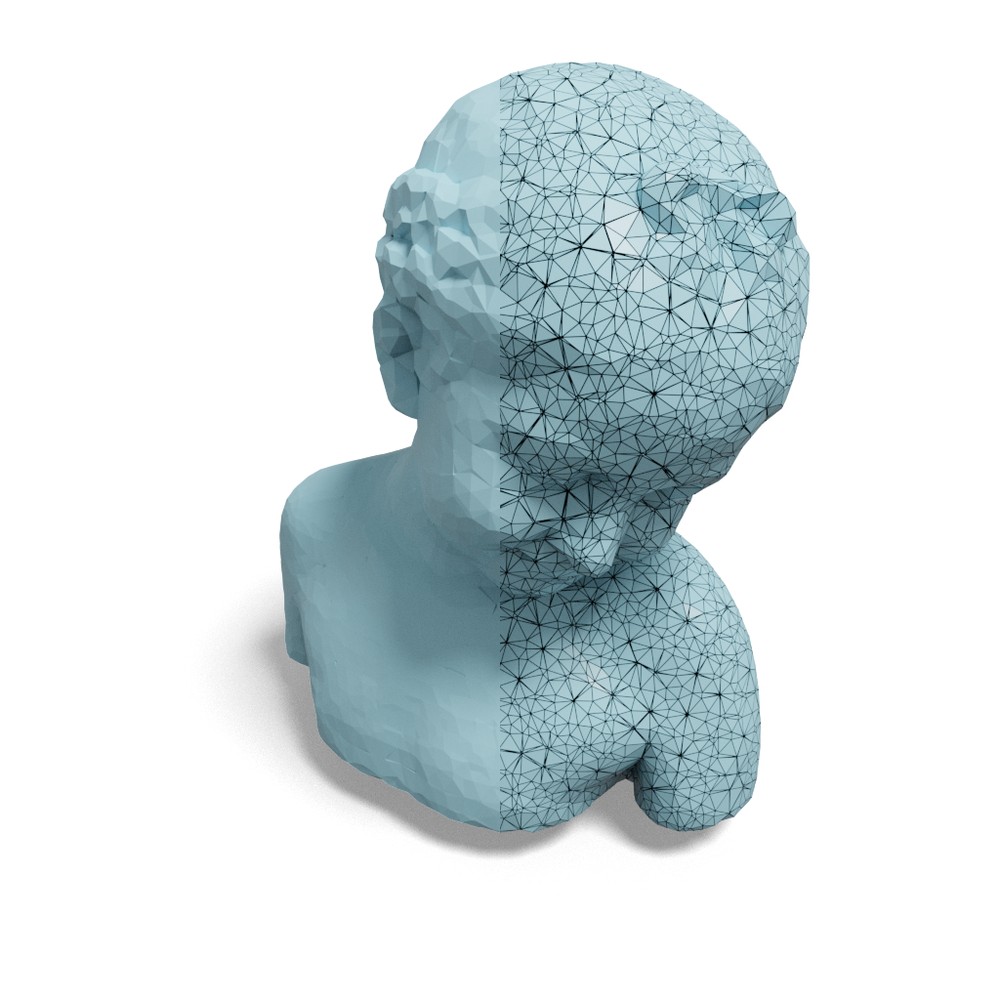} &
            \includegraphics[width=0.14\textwidth, trim=2cm 2cm 2cm 2cm, clip]{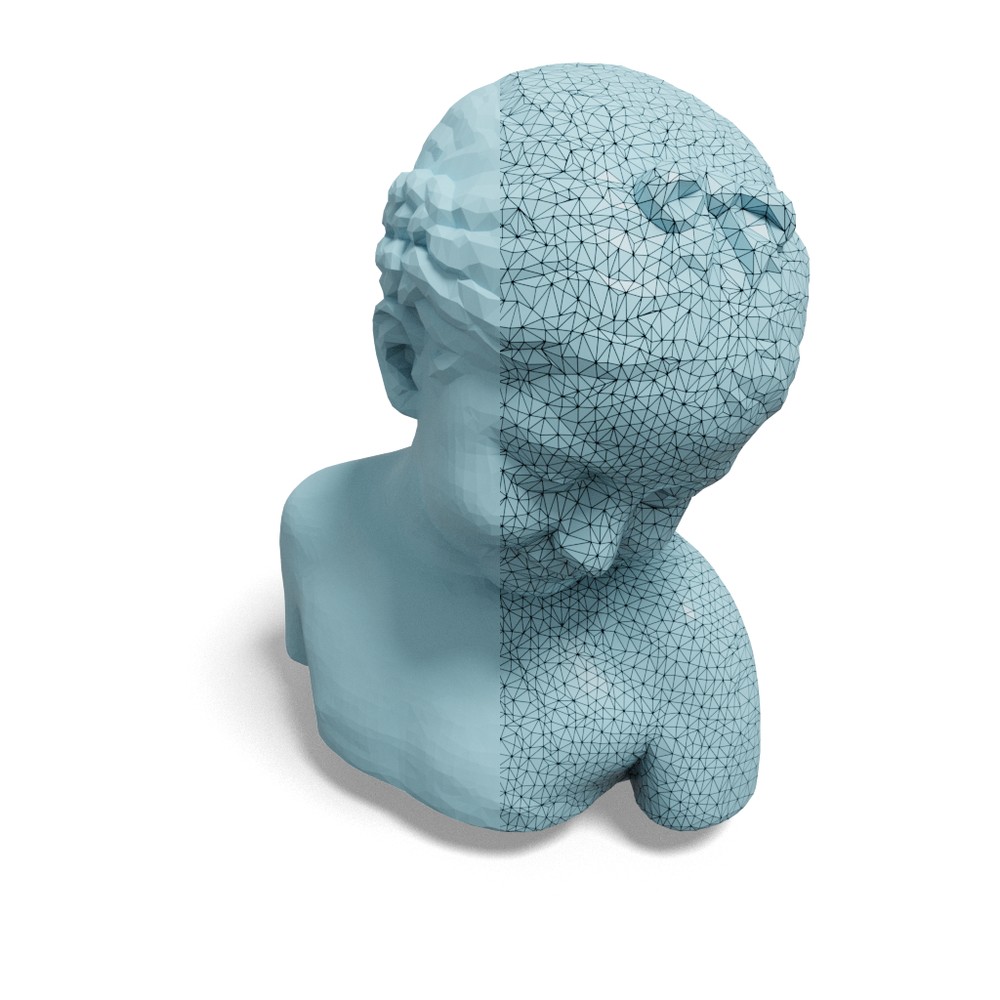} & 
            \includegraphics[width=0.14\textwidth, trim=2cm 2cm 2cm 2cm, clip]{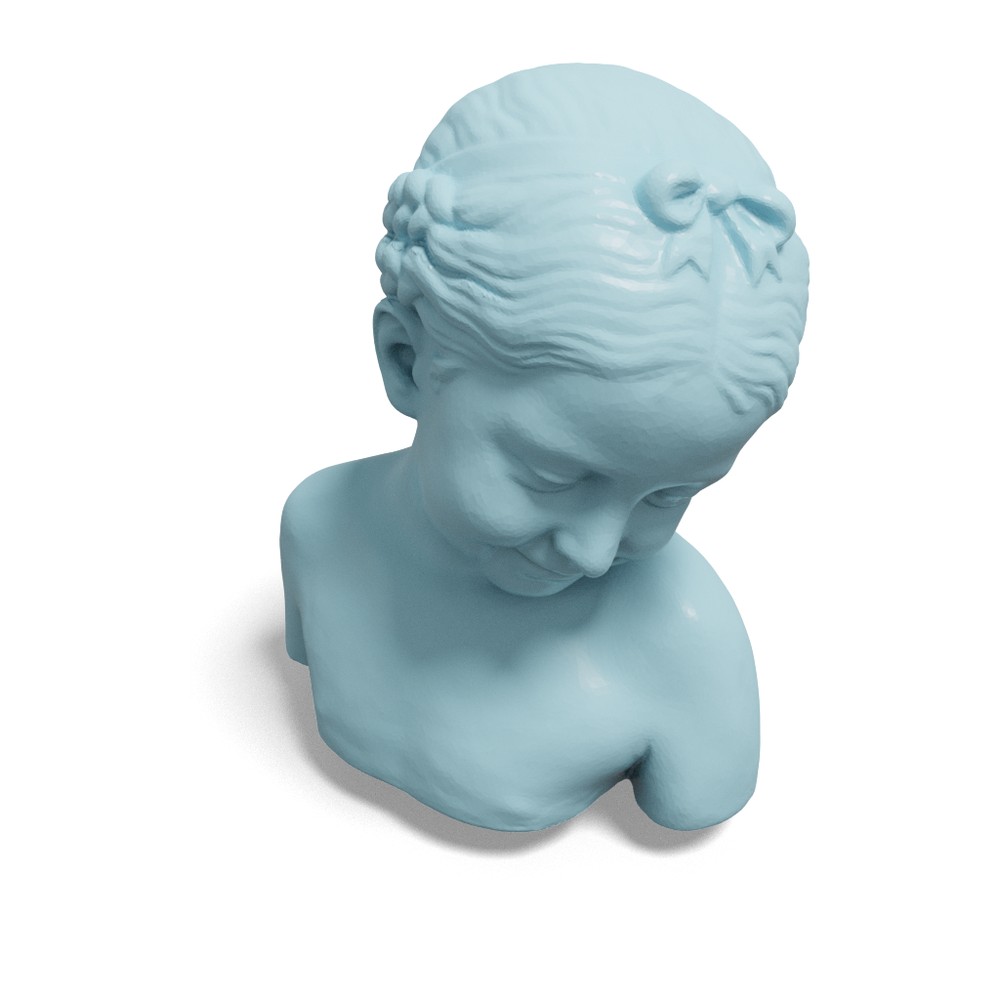} \\ 
            \includegraphics[width=0.14\textwidth, trim=2cm 6cm 2cm 2cm, clip]{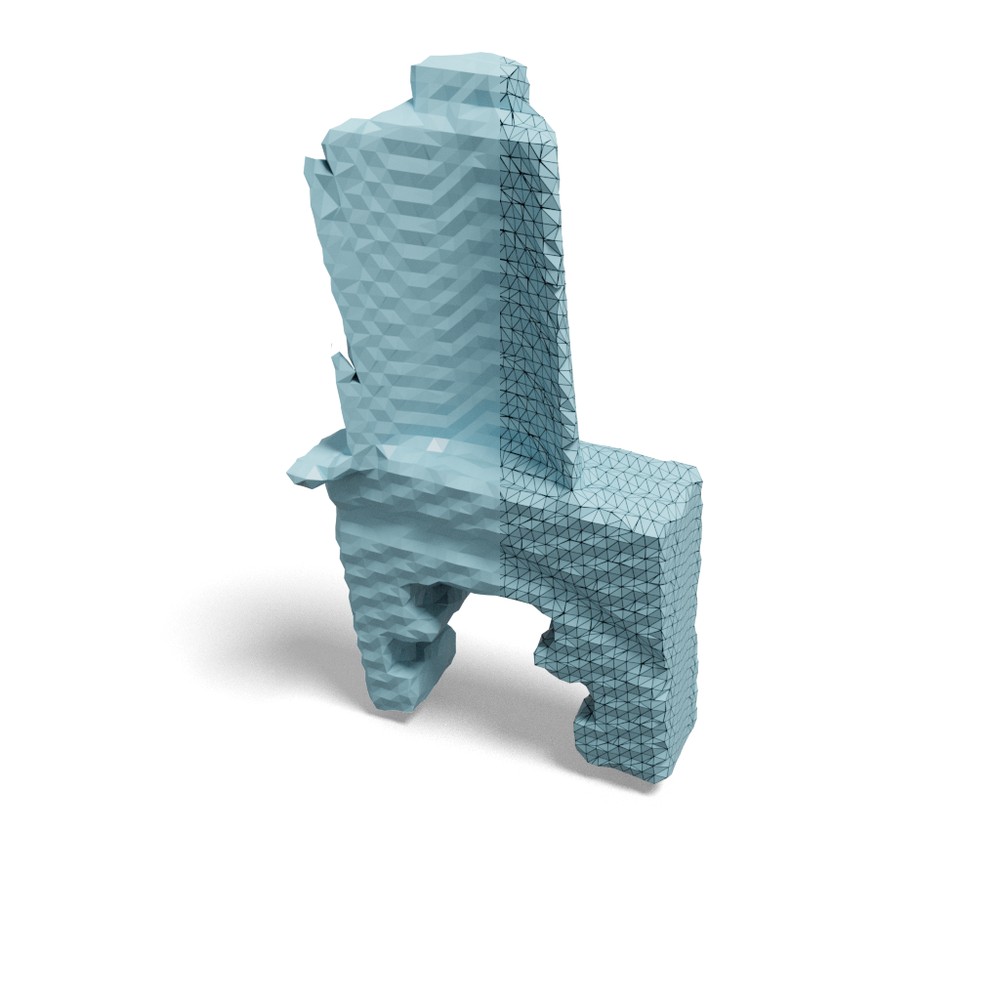} &
            \includegraphics[width=0.14\textwidth, trim=2cm 6cm 2cm 2cm, clip]{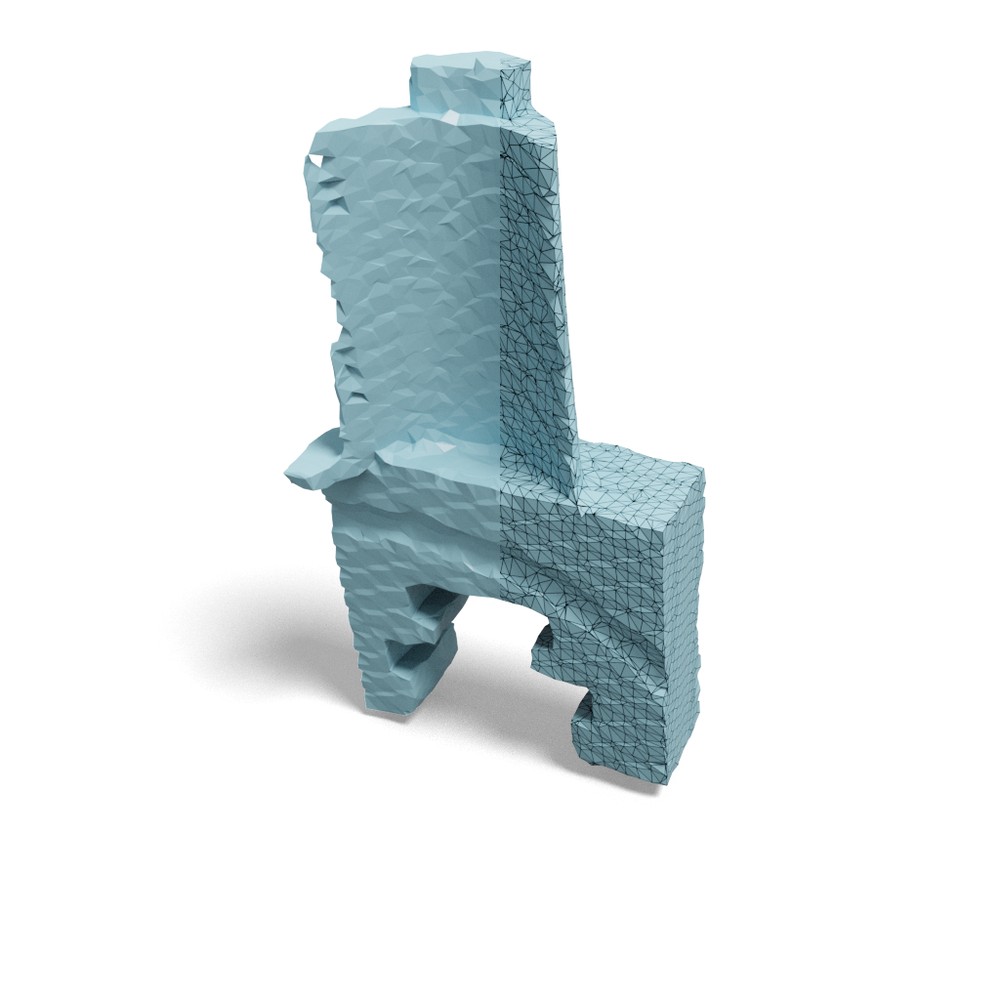} &
            \includegraphics[width=0.14\textwidth, trim=2cm 6cm 2cm 2cm, clip]{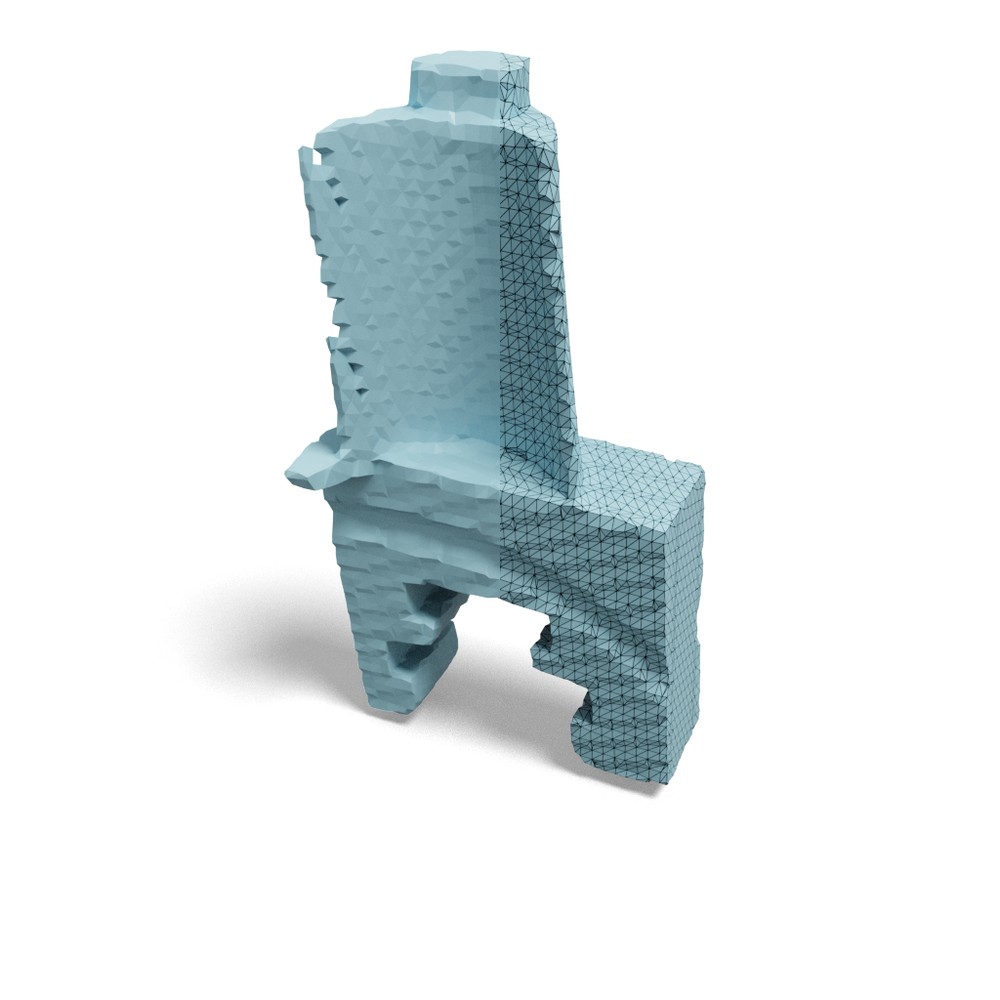} &
            \includegraphics[width=0.14\textwidth, trim=2cm 6cm 2cm 2cm, clip]{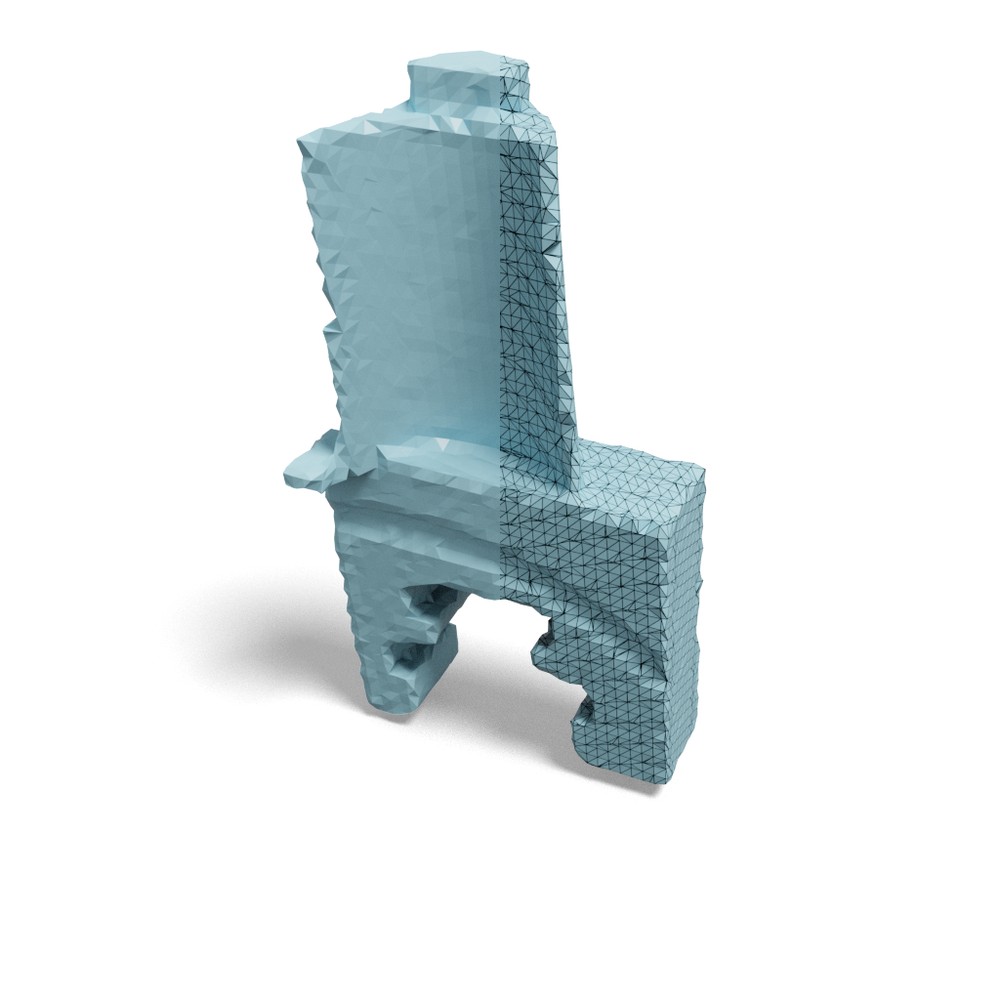} &
            \includegraphics[width=0.14\textwidth, trim=2cm 6cm 2cm 2cm, clip]{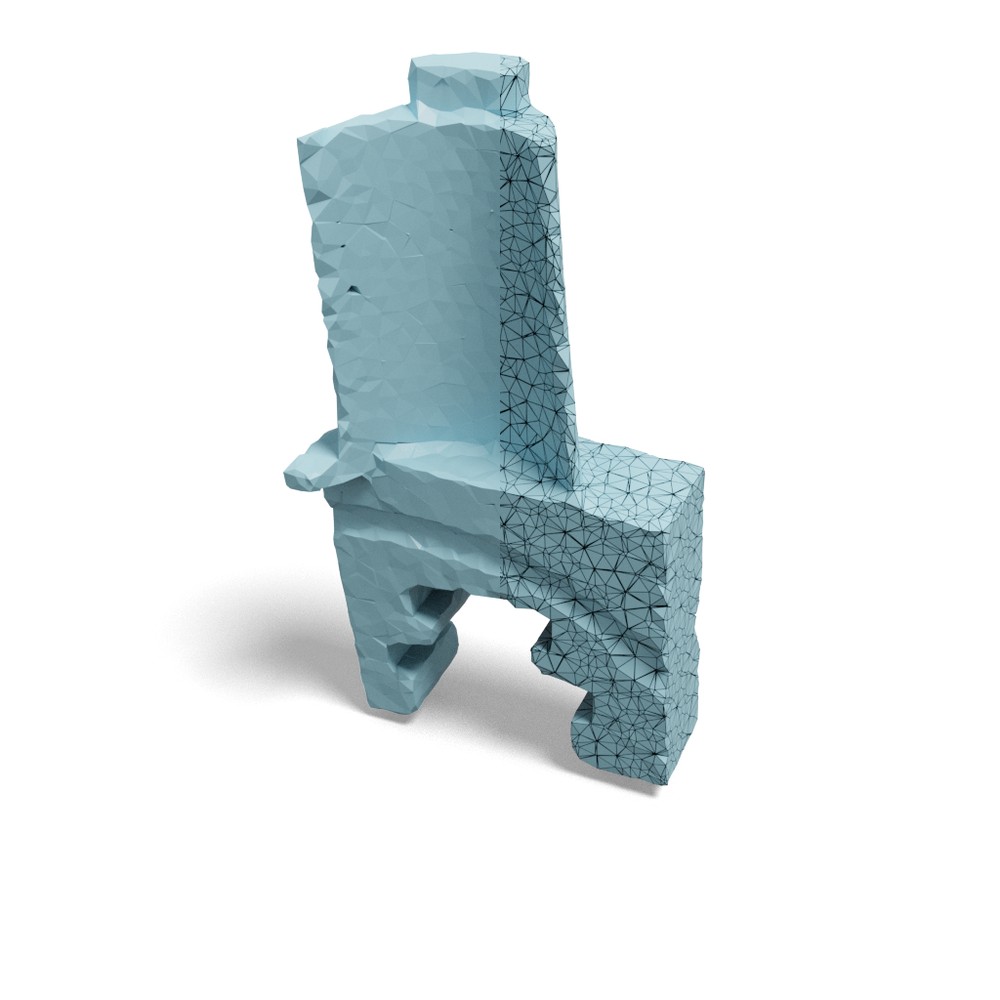} &
            \includegraphics[width=0.14\textwidth, trim=2cm 6cm 2cm 2cm, clip]{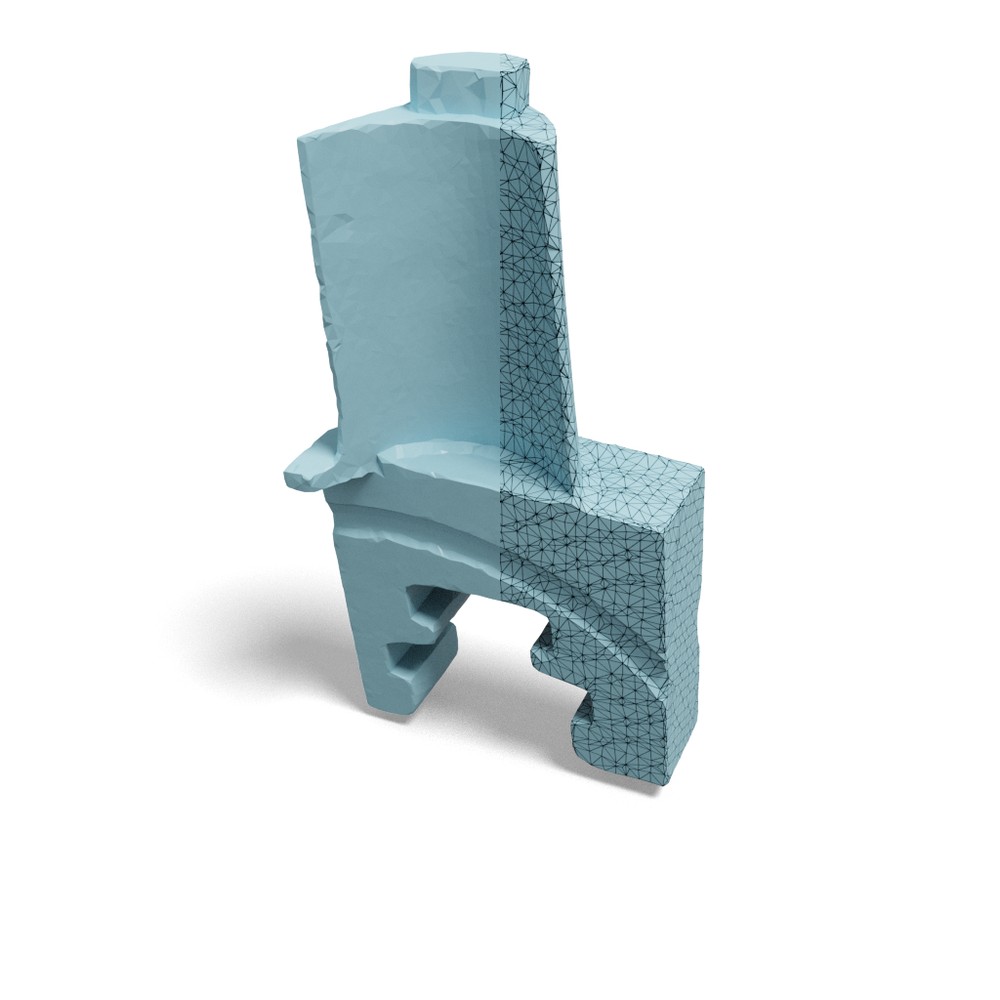} & 
            \includegraphics[width=0.14\textwidth, trim=2cm 6cm 2cm 2cm, clip]{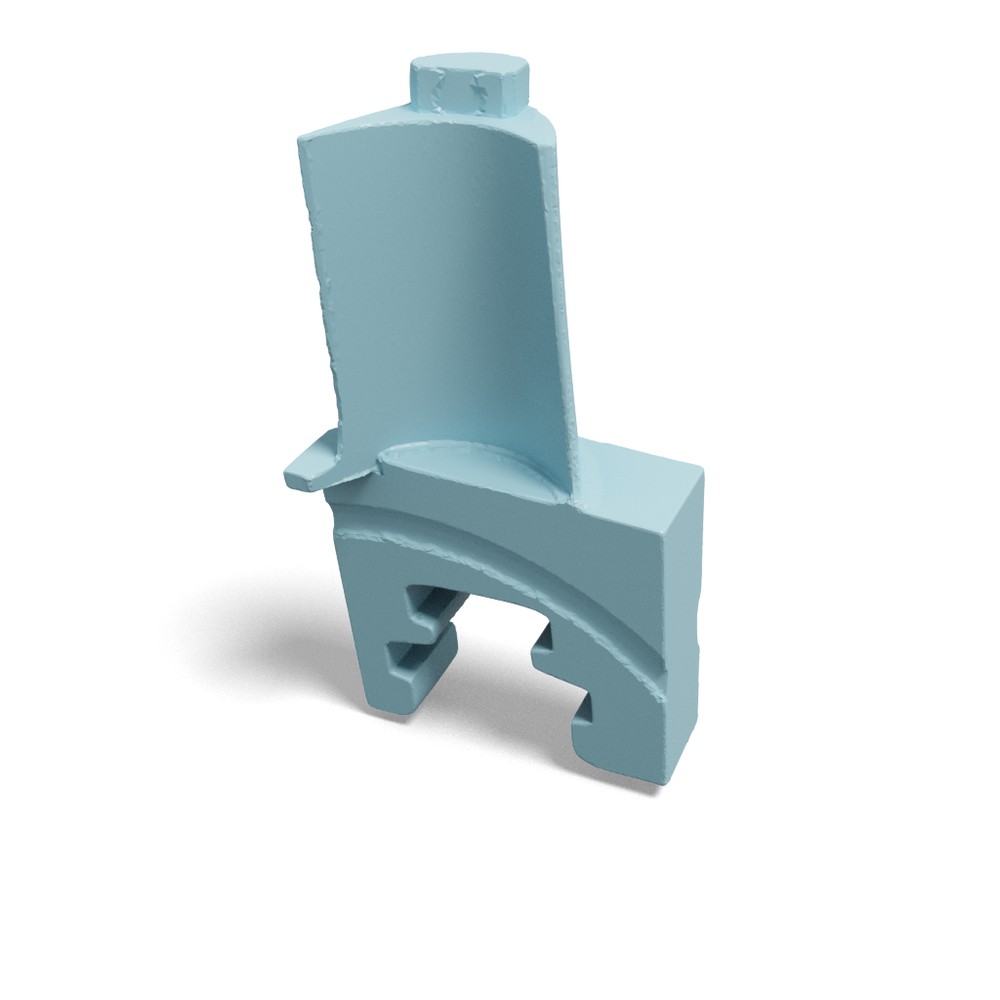} \\ 
            \includegraphics[width=0.14\textwidth, trim=2cm 7cm 2cm 8cm, clip]{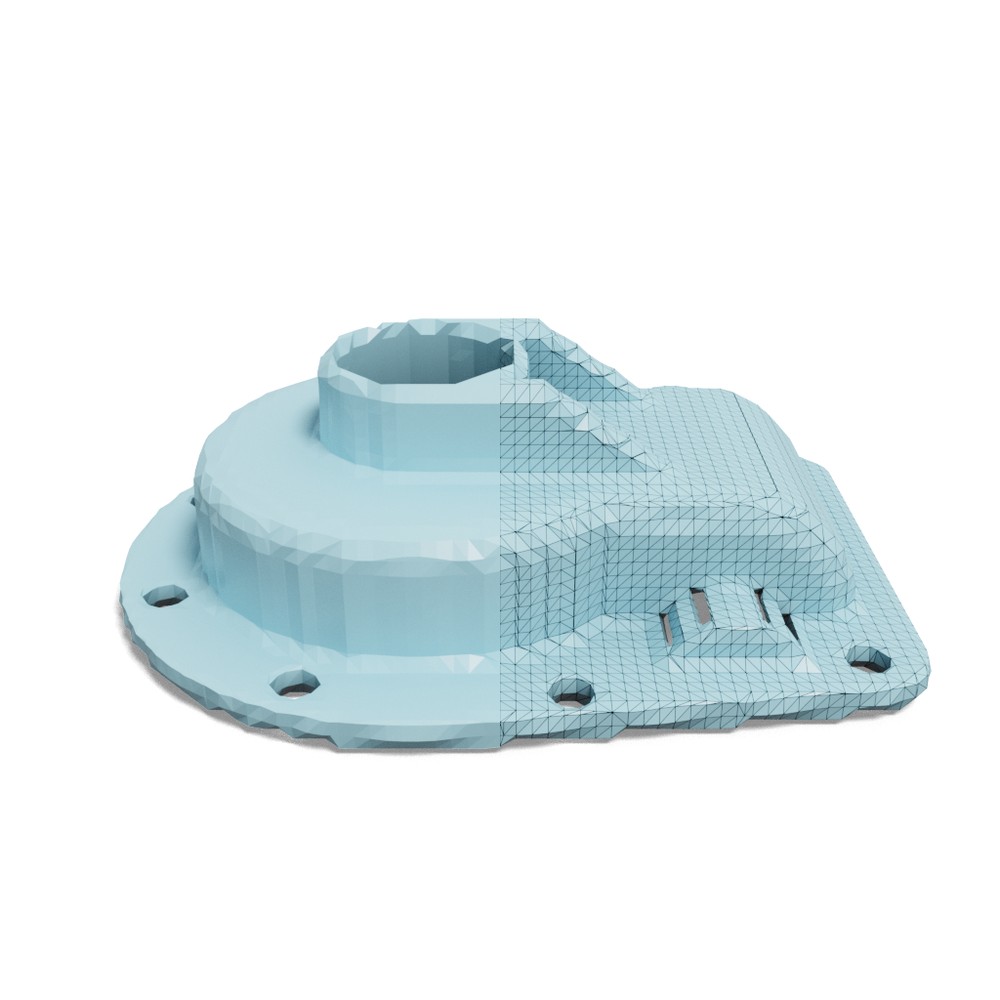} &
            \includegraphics[width=0.14\textwidth, trim=2cm 7cm 2cm 8cm, clip]{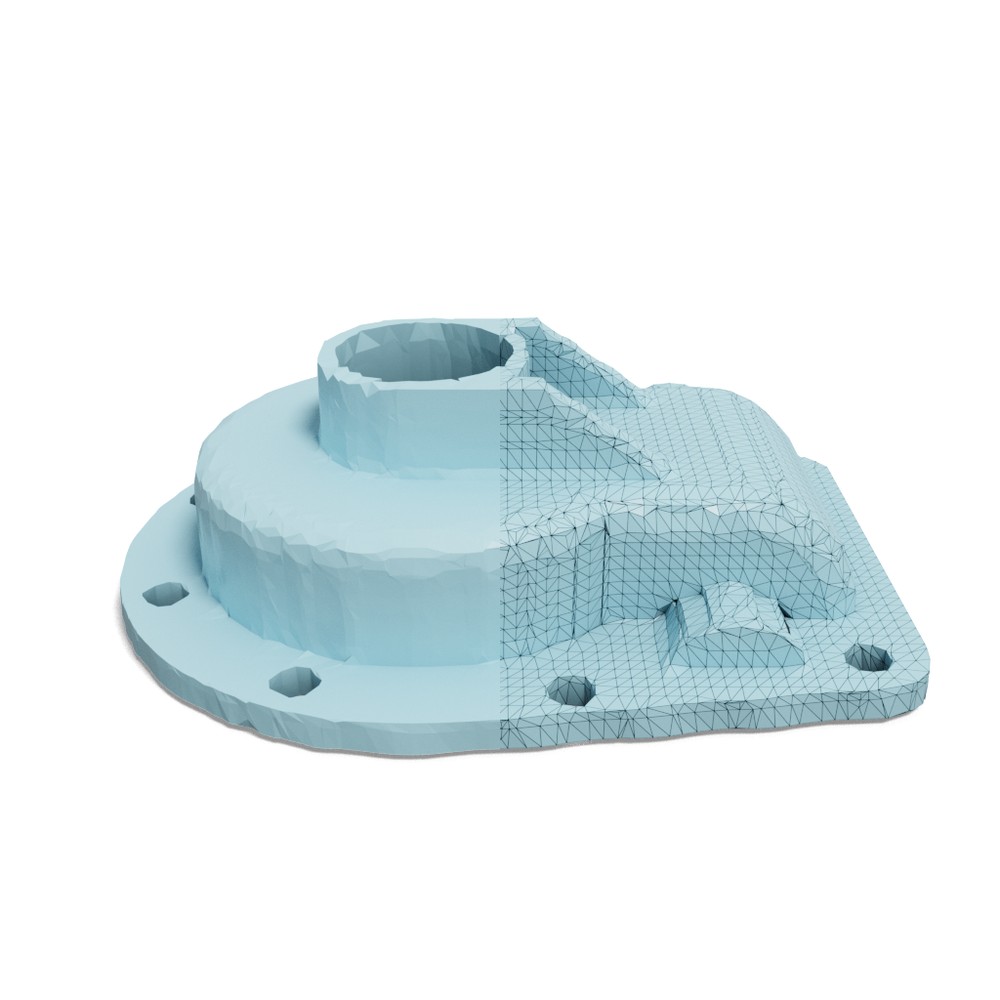} &
            \includegraphics[width=0.14\textwidth, trim=2cm 7cm 2cm 8cm, clip]{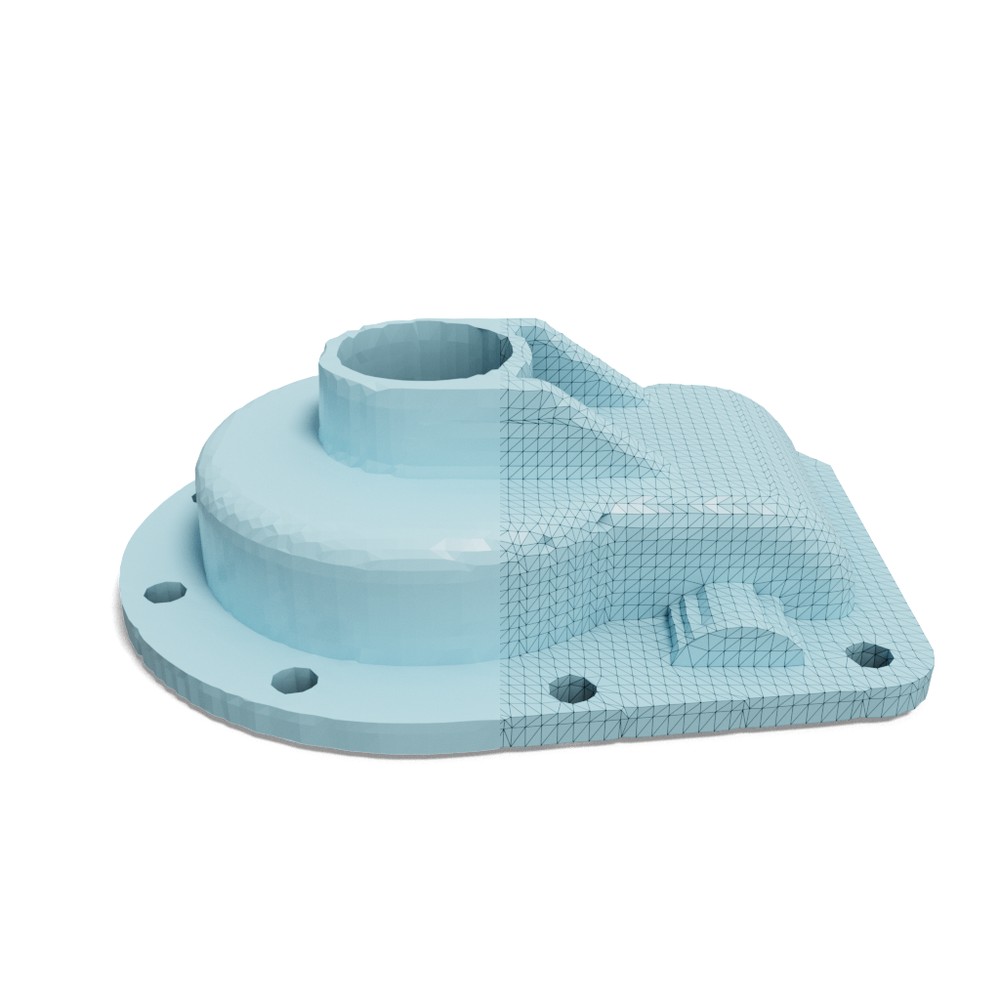} &
            \includegraphics[width=0.14\textwidth, trim=2cm 7cm 2cm 8cm, clip]{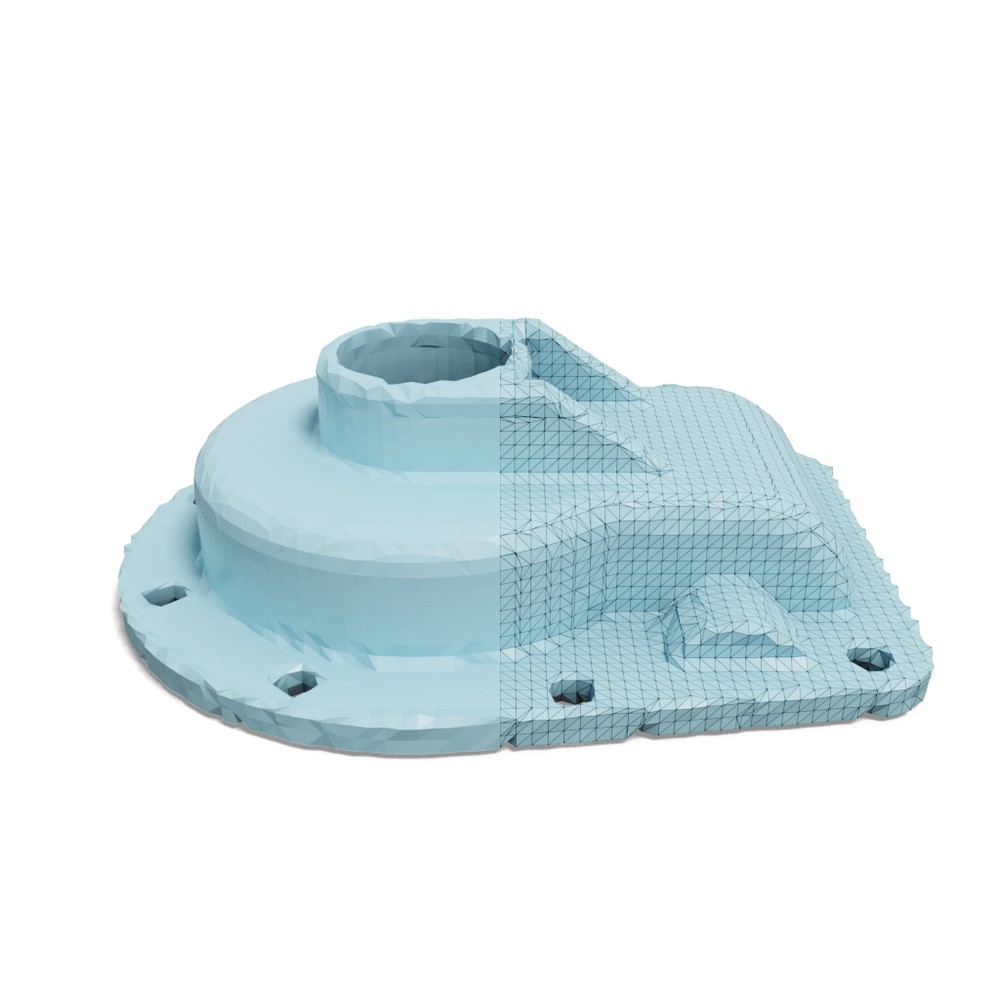} &
            \includegraphics[width=0.14\textwidth, trim=2cm 7cm 2cm 8cm, clip]{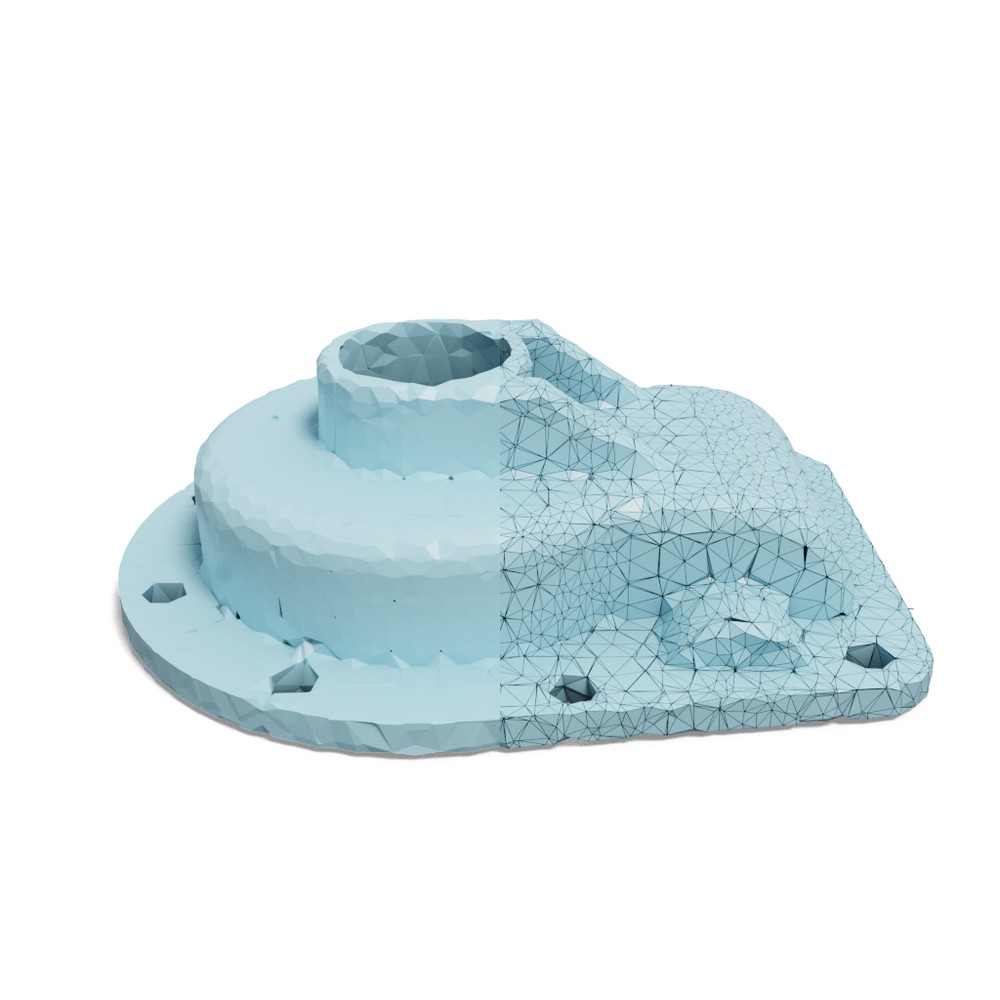} &
            \includegraphics[width=0.14\textwidth, trim=2cm 7cm 2cm 8cm, clip]{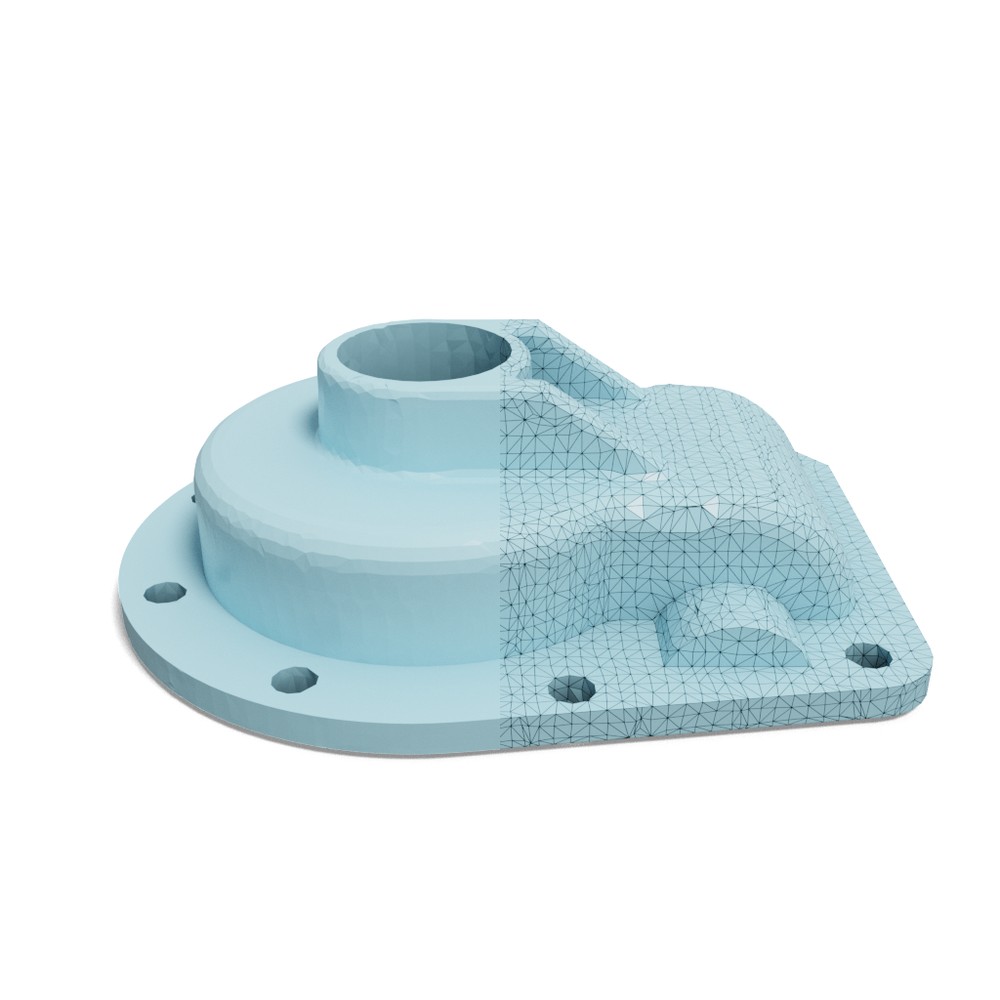} & 
            \includegraphics[width=0.14\textwidth, trim=2cm 7cm 2cm 8cm, clip]{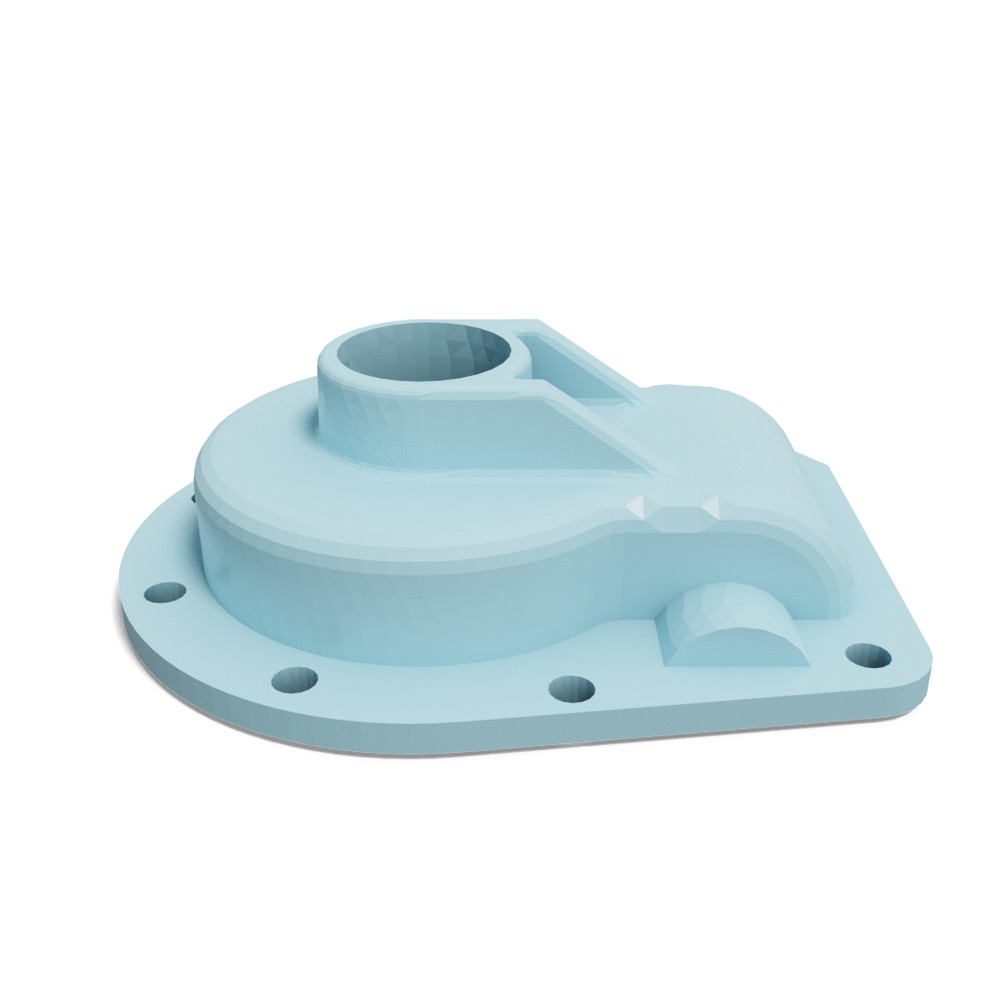} \\ 
            \includegraphics[width=0.14\textwidth, trim=2cm 2cm 2cm 0cm, clip]{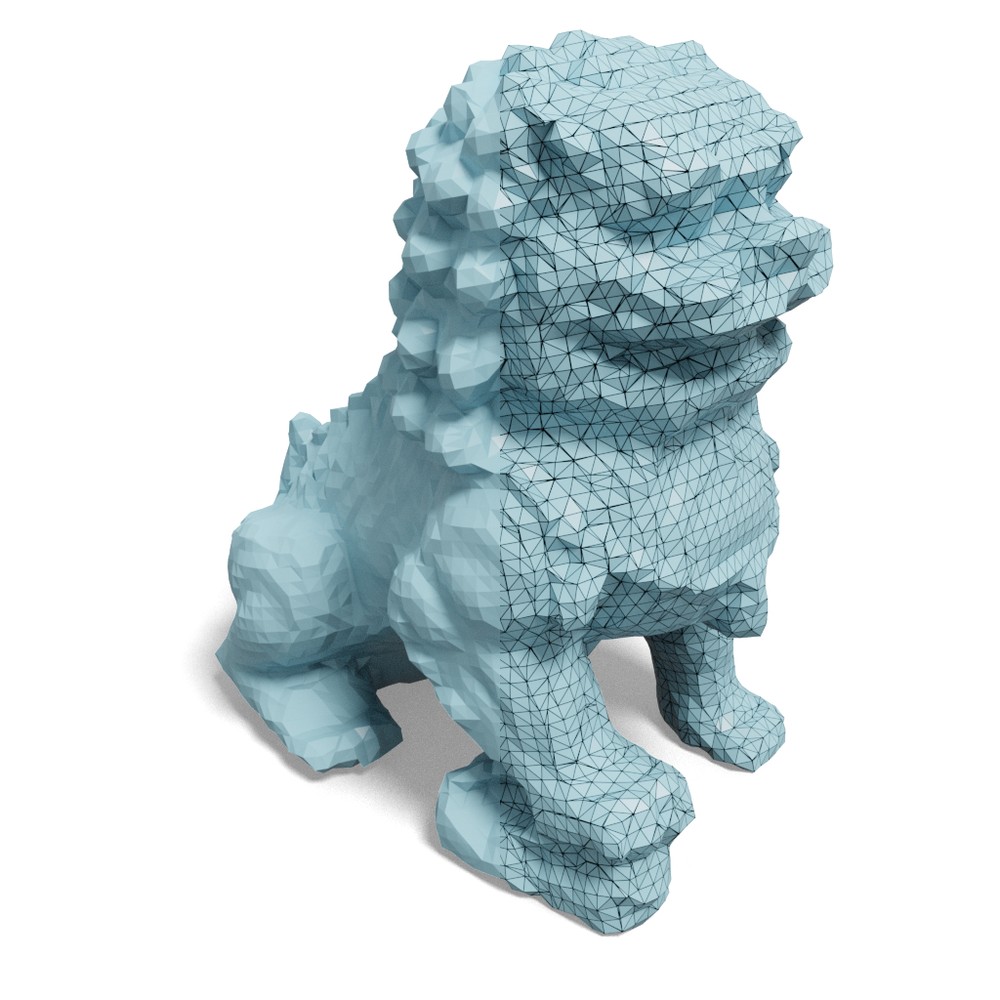} &
            \includegraphics[width=0.14\textwidth, trim=2cm 2cm 2cm 0cm, clip]{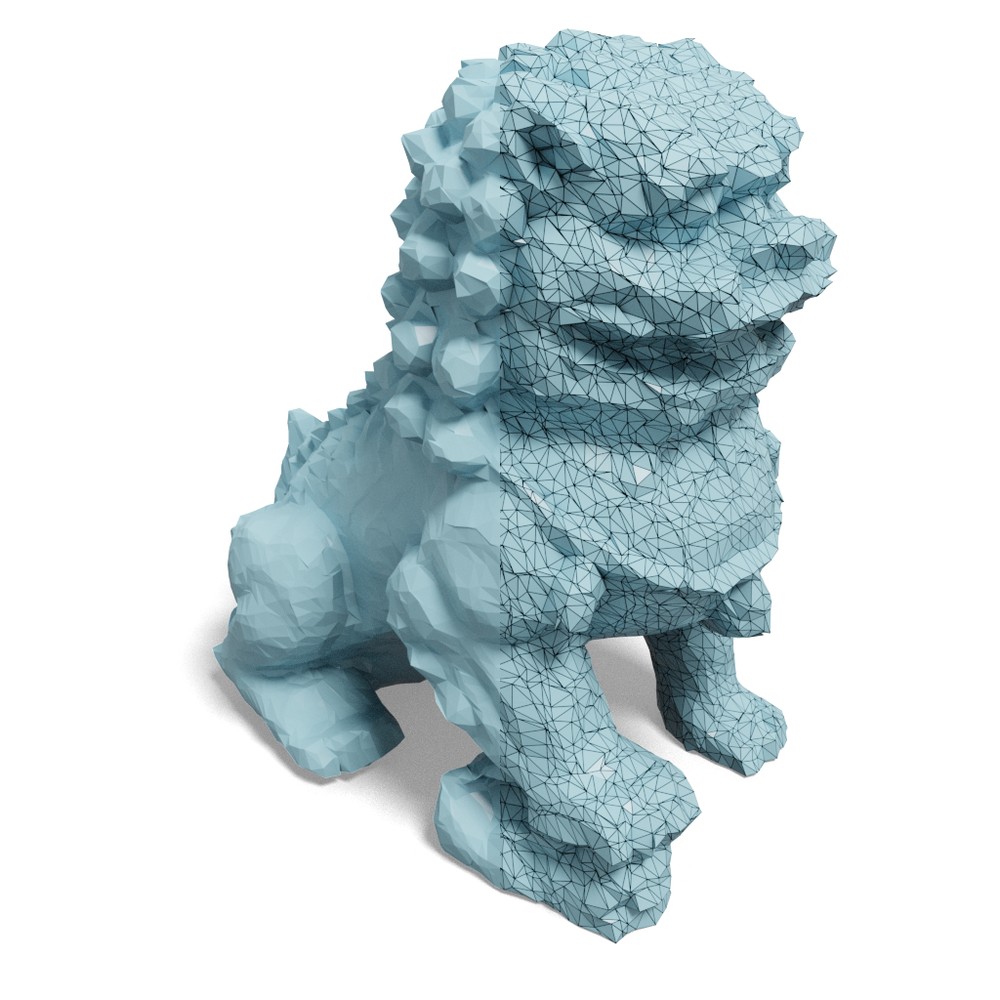} &
            \includegraphics[width=0.14\textwidth, trim=2cm 2cm 2cm 0cm, clip]{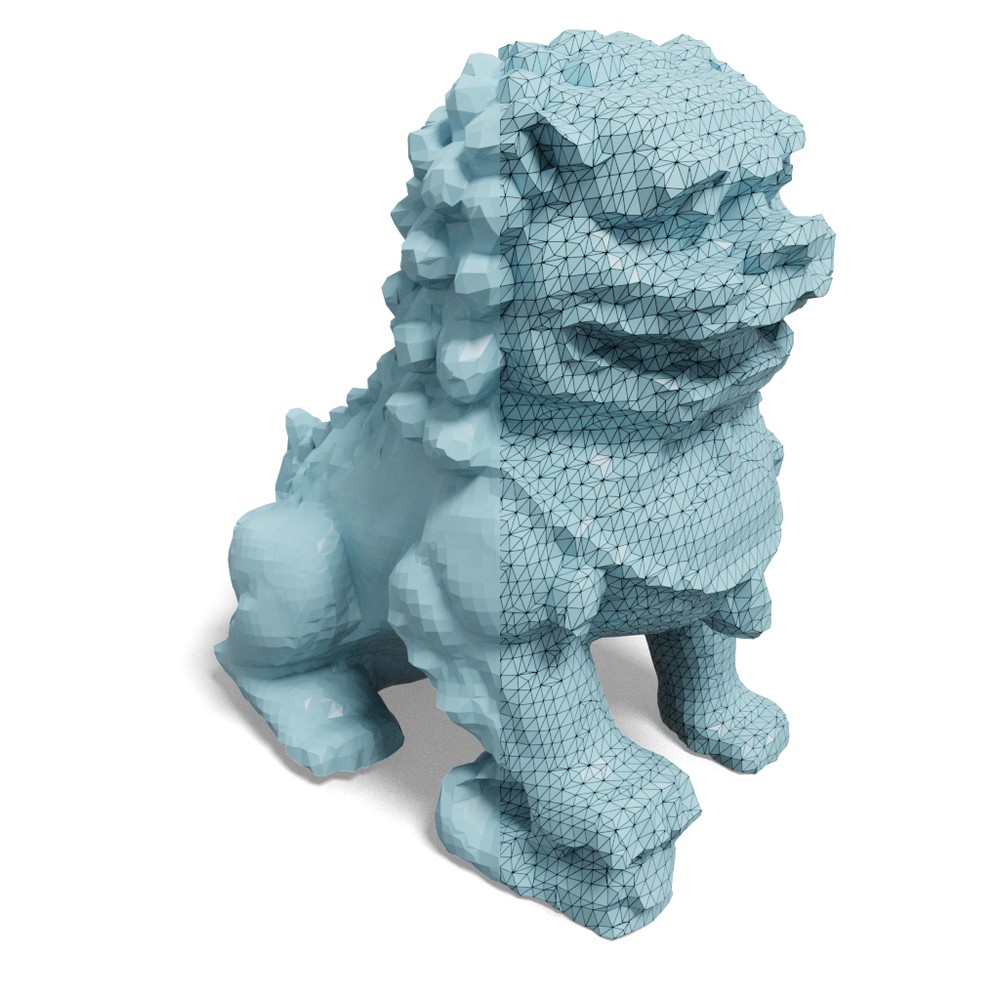} &
            \includegraphics[width=0.14\textwidth, trim=2cm 2cm 2cm 0cm, clip]{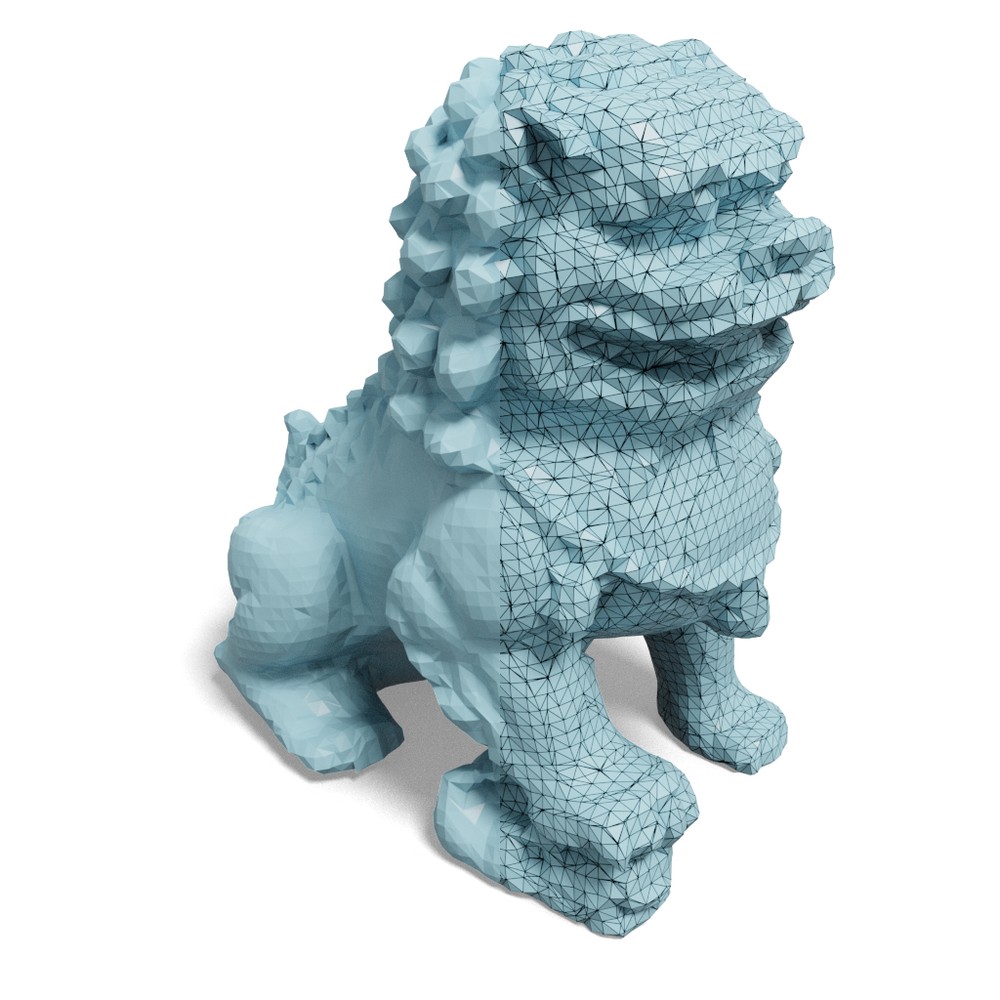} &
            \includegraphics[width=0.14\textwidth, trim=2cm 2cm 2cm 0cm, clip]{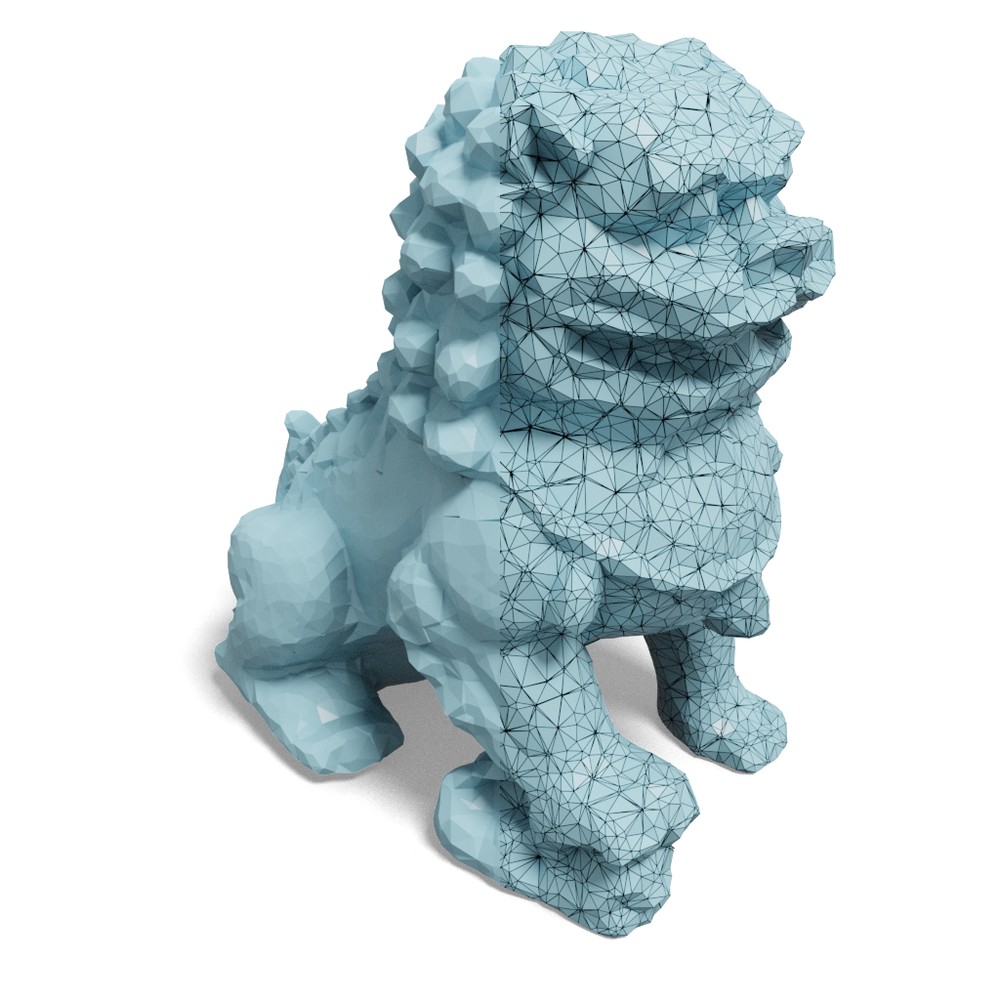} &
            \includegraphics[width=0.14\textwidth, trim=2cm 2cm 2cm 0cm, clip]{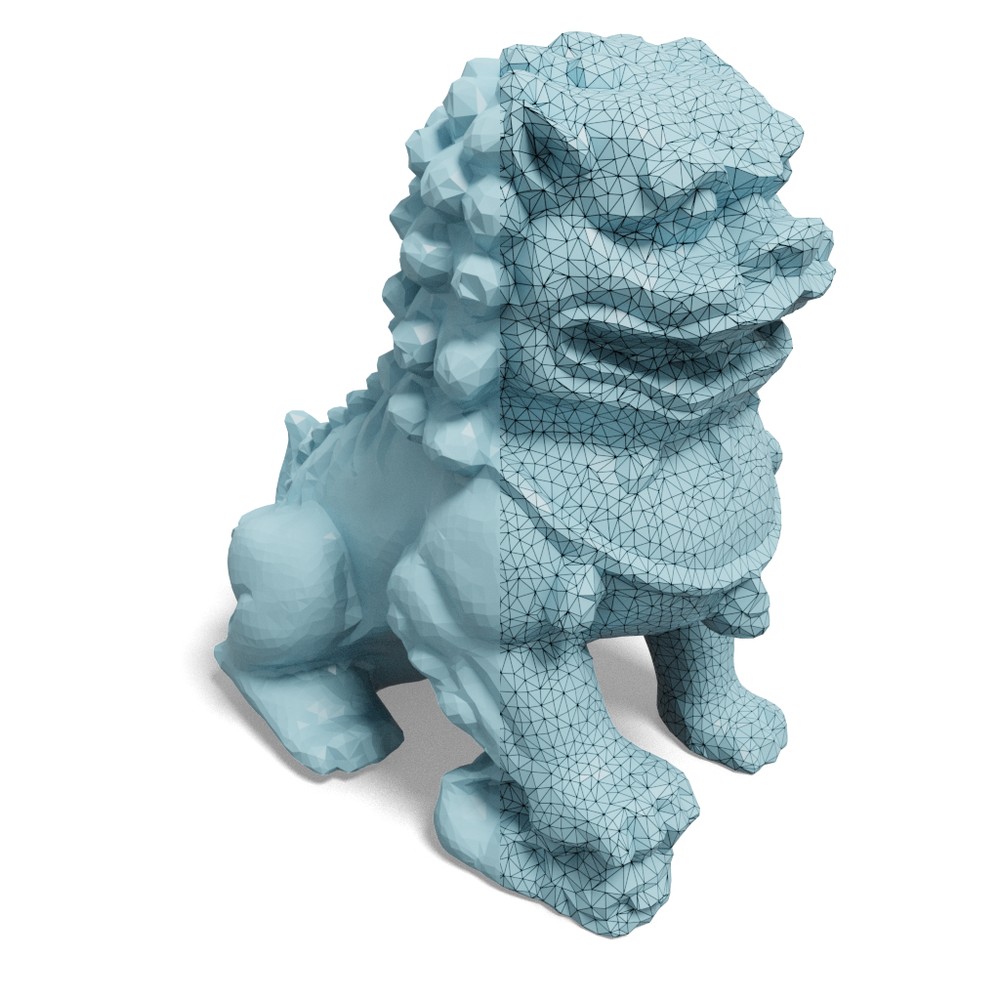} & 
            \includegraphics[width=0.14\textwidth, trim=2cm 2cm 2cm 0cm, clip]{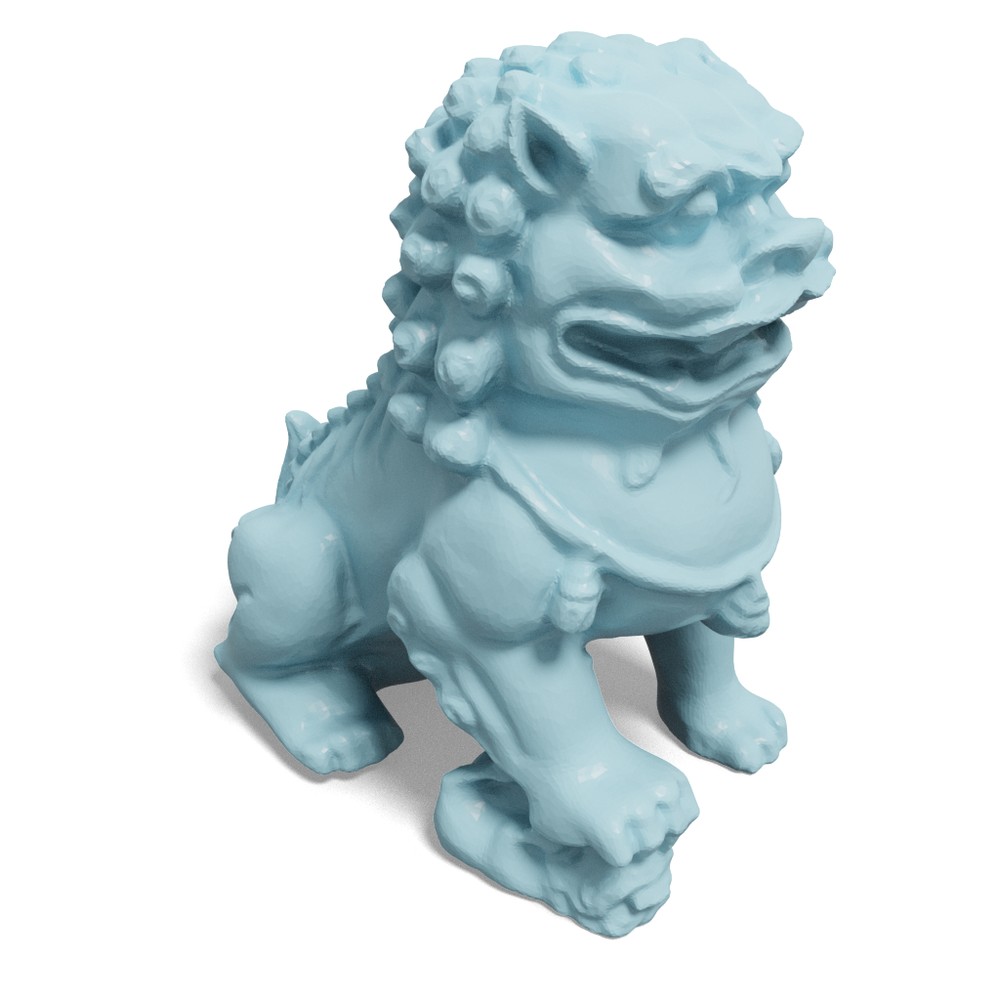} \\ 
            \includegraphics[width=0.14\textwidth, trim=2cm 2cm 2cm 0cm, clip]{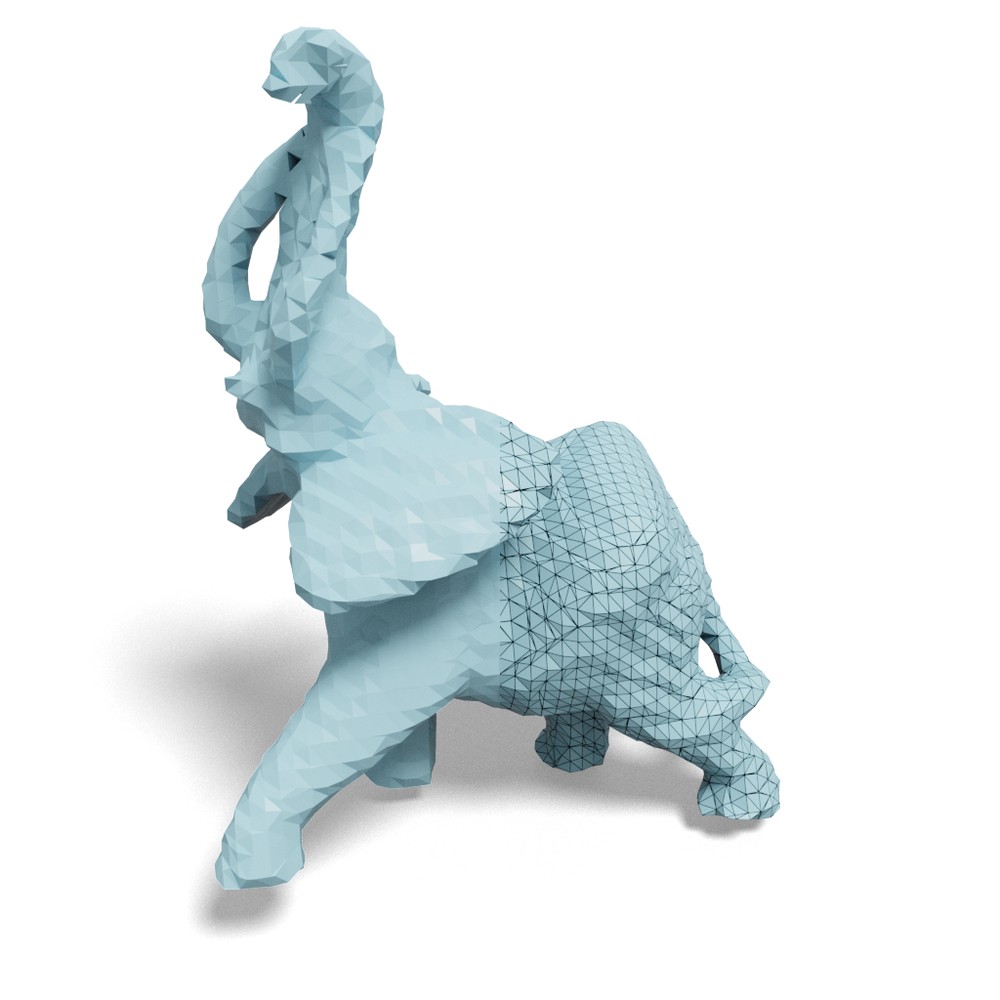} &
            \includegraphics[width=0.14\textwidth, trim=2cm 2cm 2cm 0cm, clip]{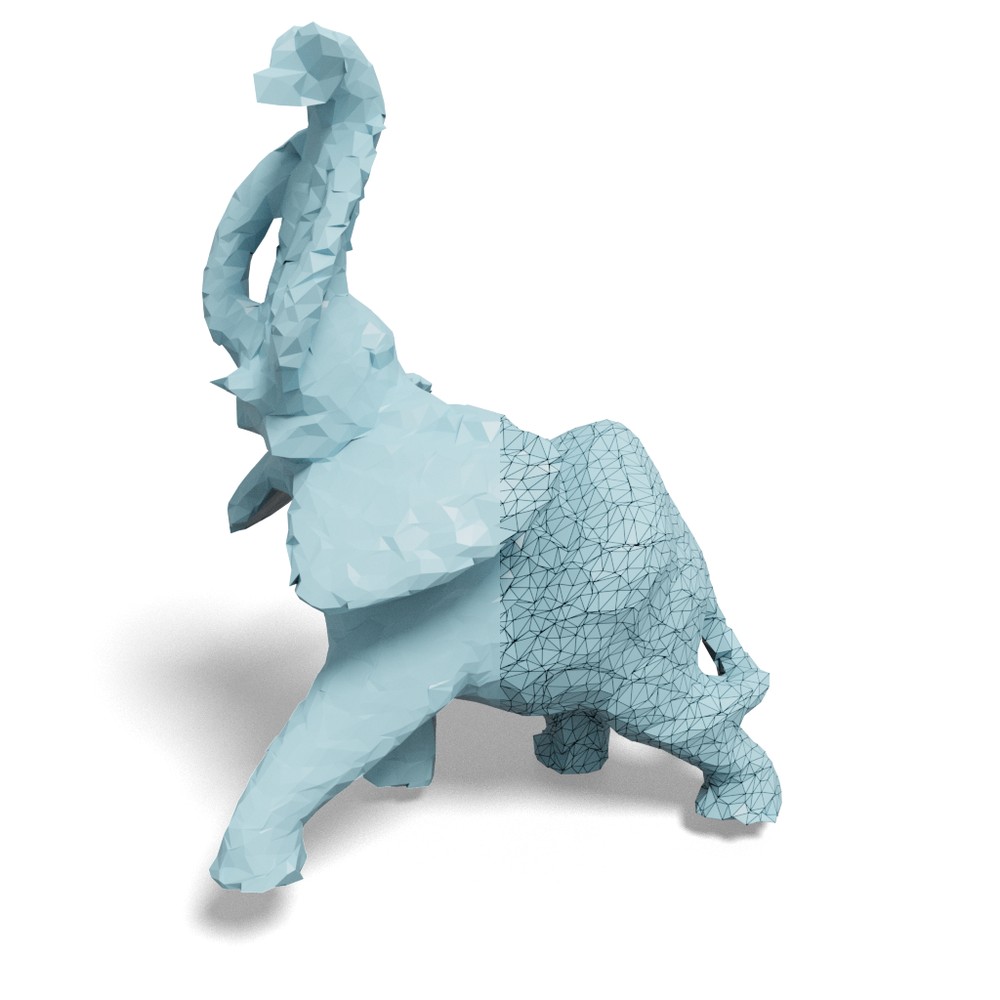} &
            \includegraphics[width=0.14\textwidth, trim=2cm 2cm 2cm 0cm, clip]{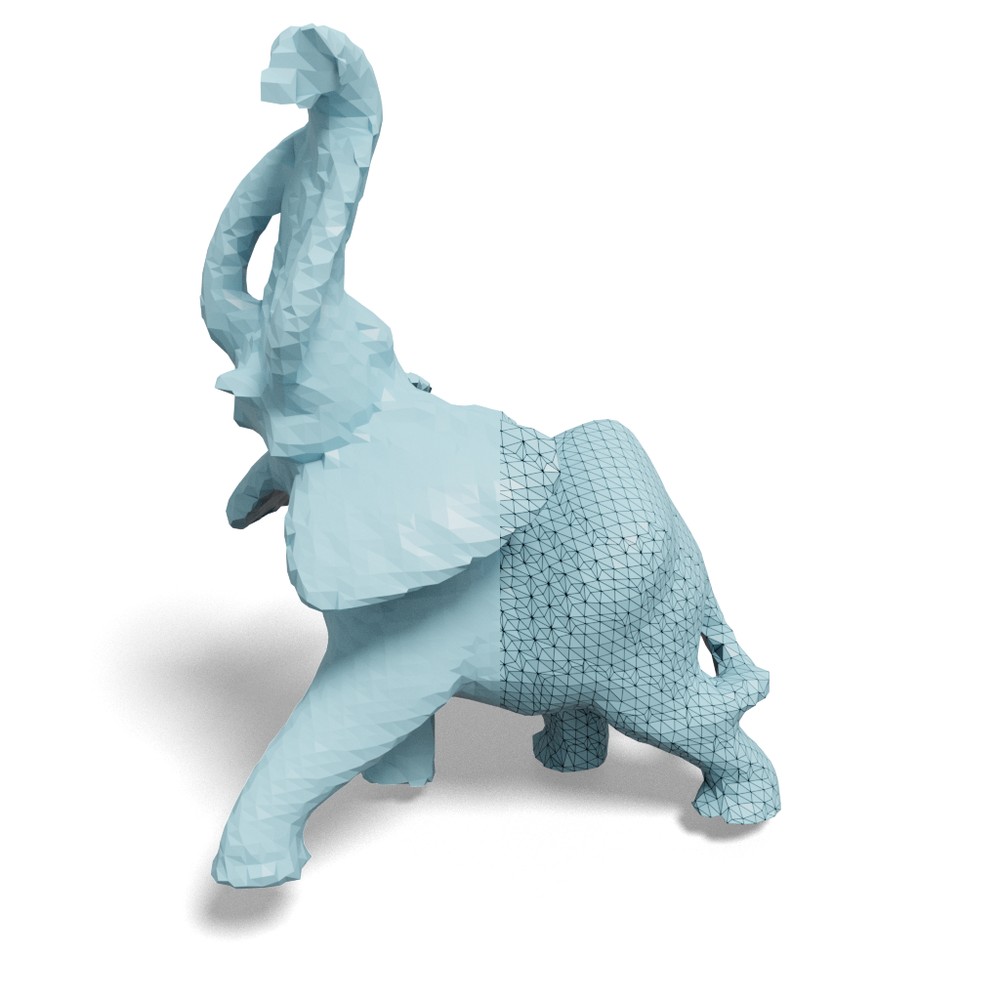} &
            \includegraphics[width=0.14\textwidth, trim=2cm 2cm 2cm 0cm, clip]{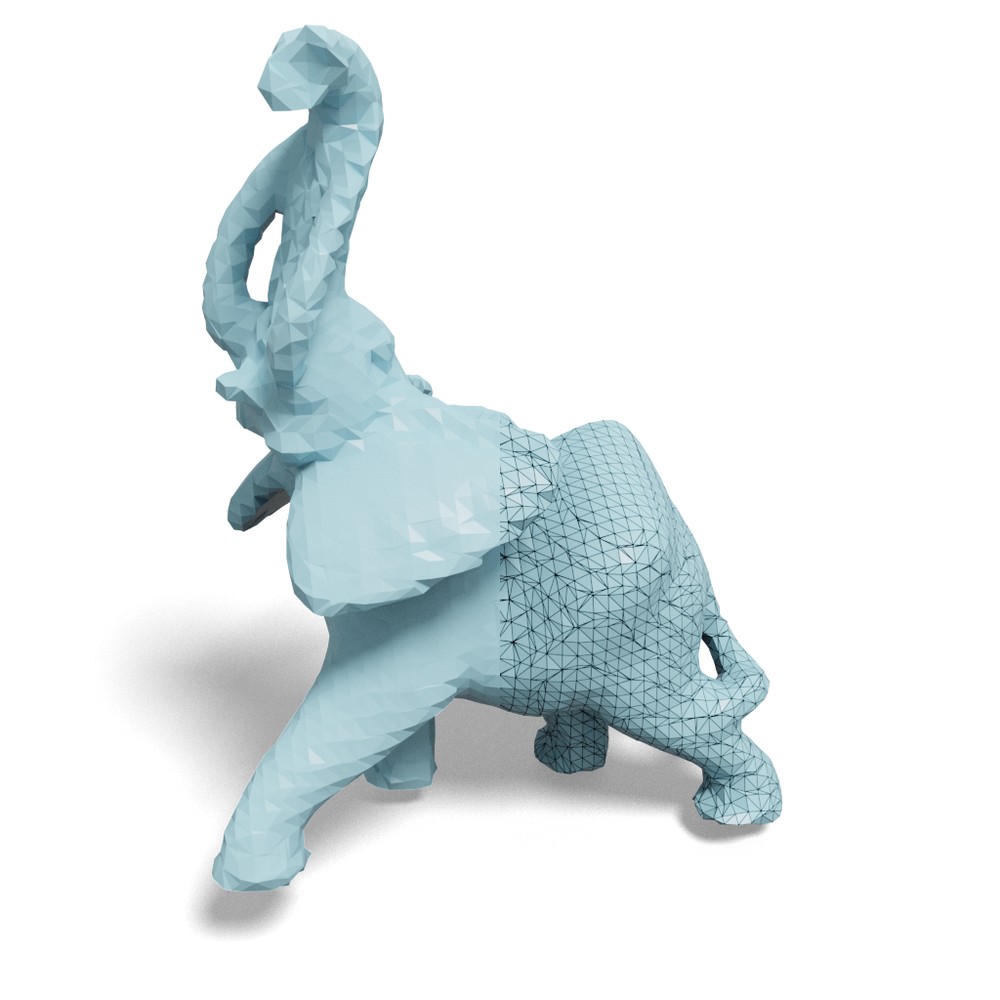} &
            \includegraphics[width=0.14\textwidth, trim=2cm 2cm 2cm 0cm, clip]{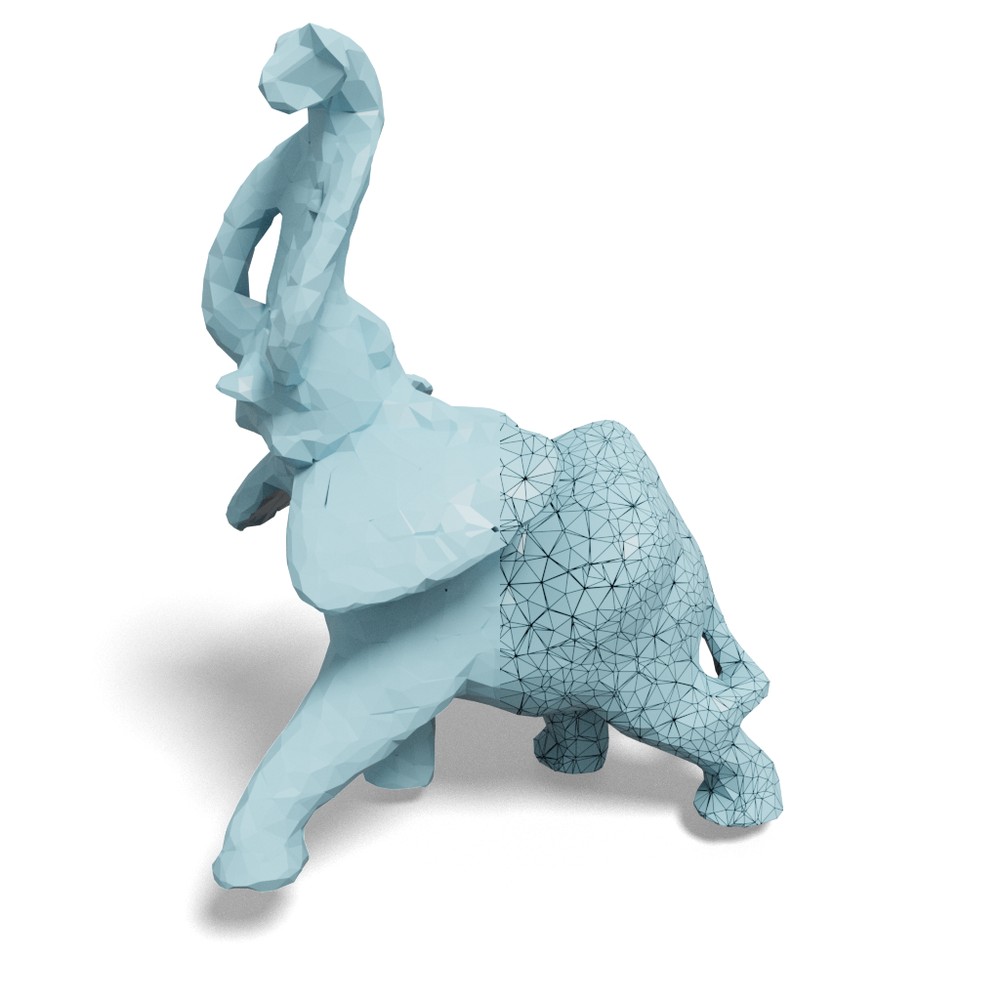} &
            \includegraphics[width=0.14\textwidth, trim=2cm 2cm 2cm 0cm, clip]{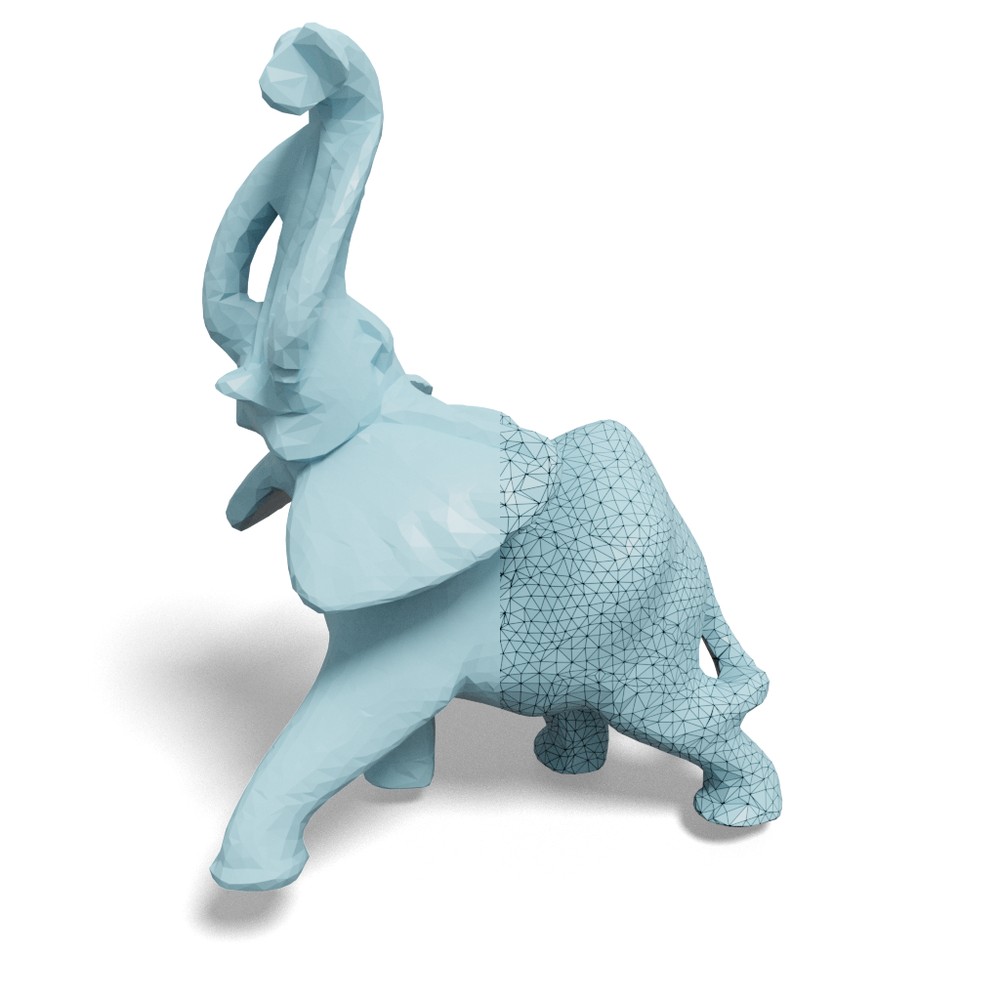} & 
            \includegraphics[width=0.14\textwidth, trim=2cm 2cm 2cm 0cm, clip]{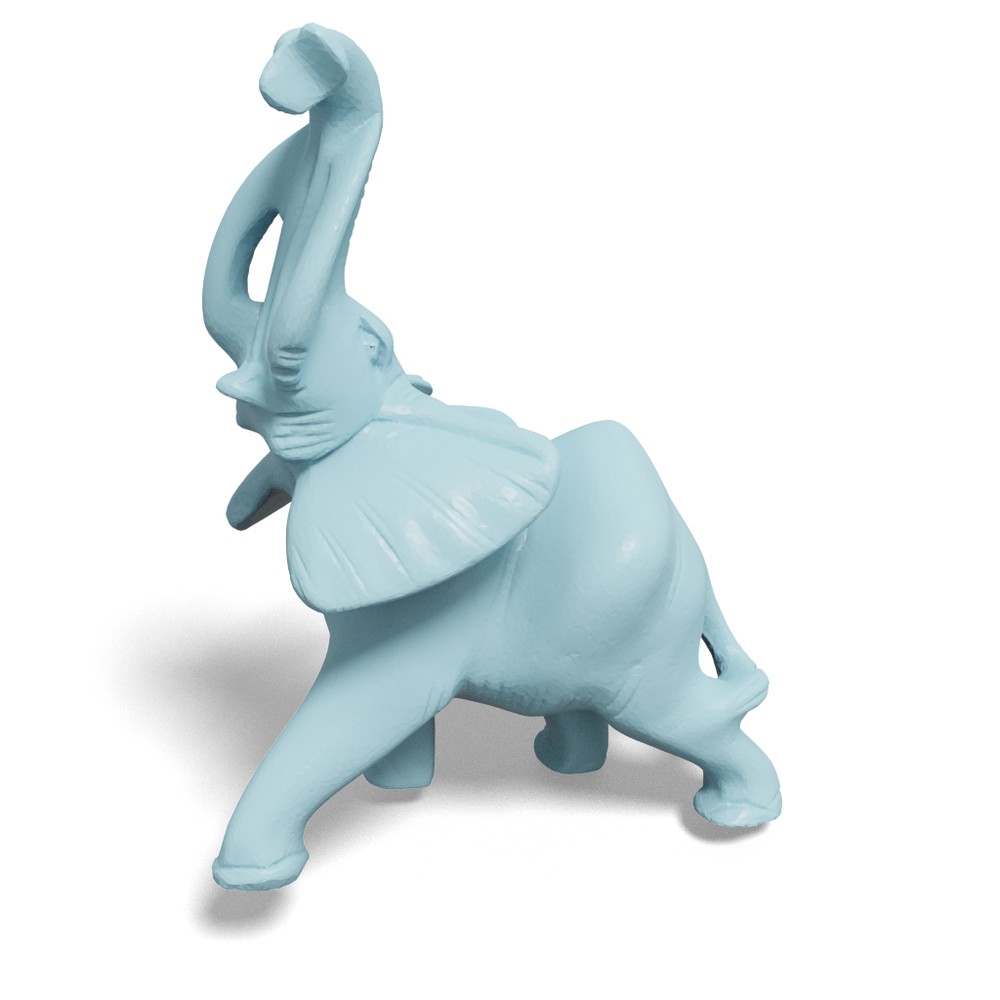} \\ 
            \includegraphics[width=0.14\textwidth, trim=2cm 2cm 2cm 2cm, clip]{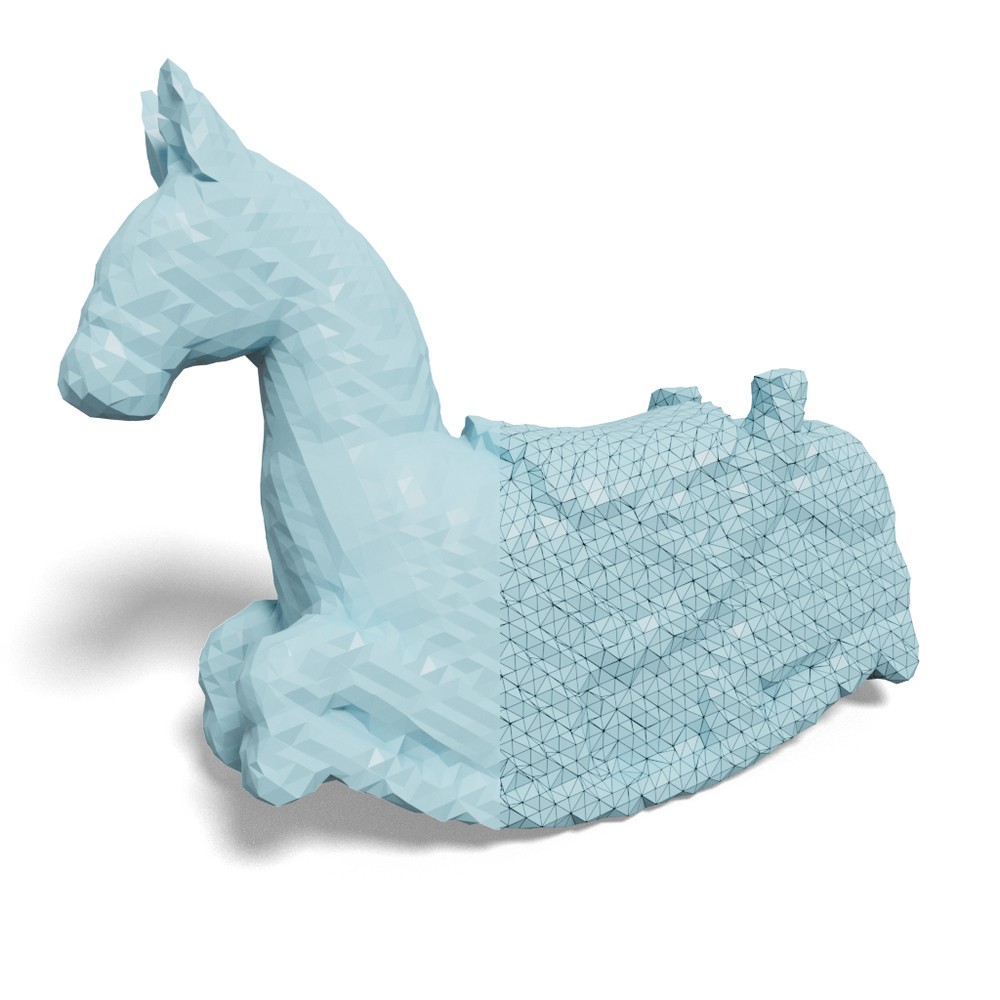} &
            \includegraphics[width=0.14\textwidth, trim=2cm 2cm 2cm 2cm, clip]{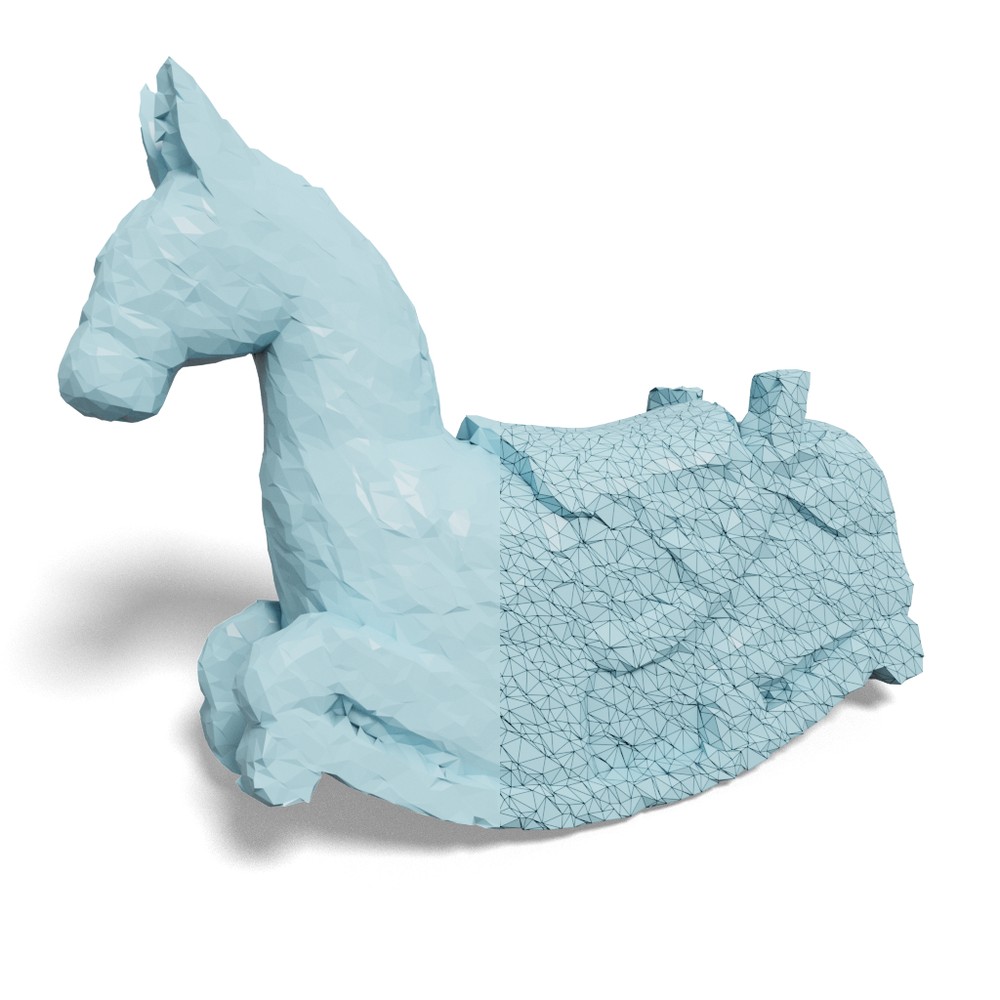} &
            \includegraphics[width=0.14\textwidth, trim=2cm 2cm 2cm 2cm, clip]{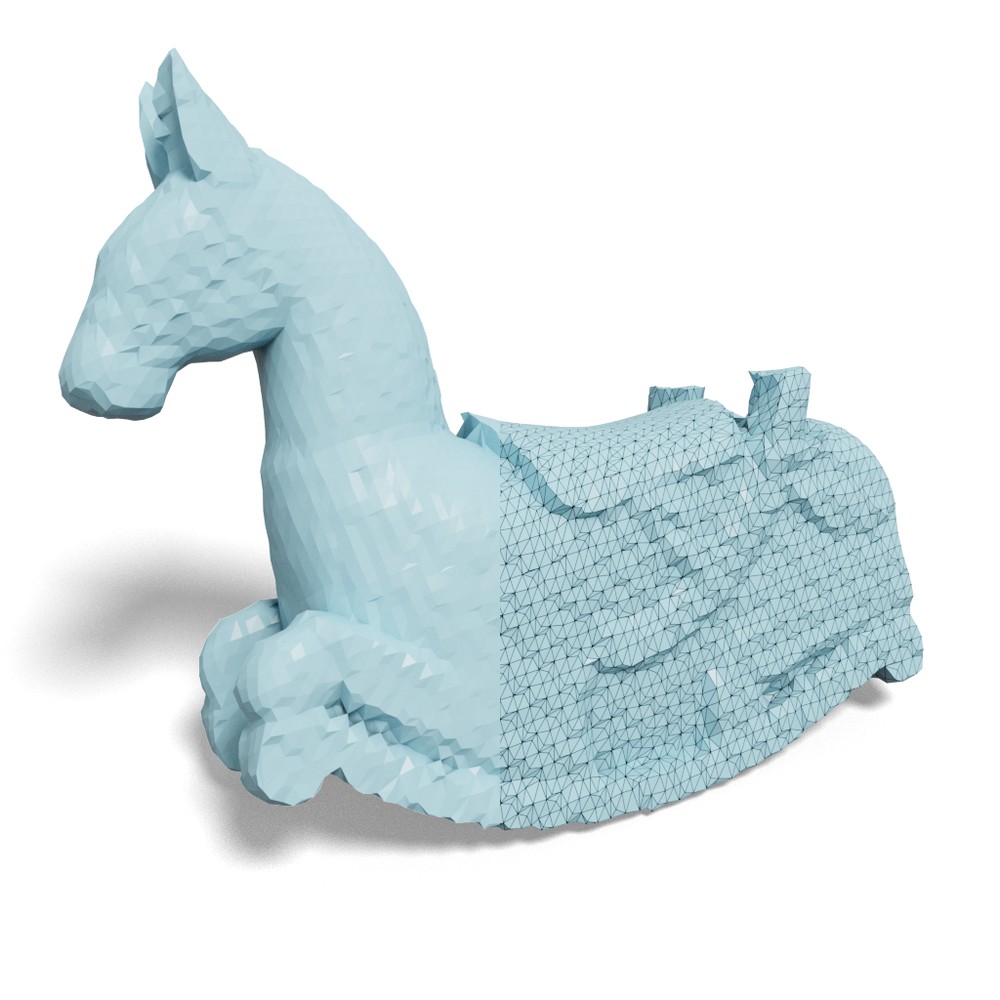} &
            \includegraphics[width=0.14\textwidth, trim=2cm 2cm 2cm 2cm, clip]{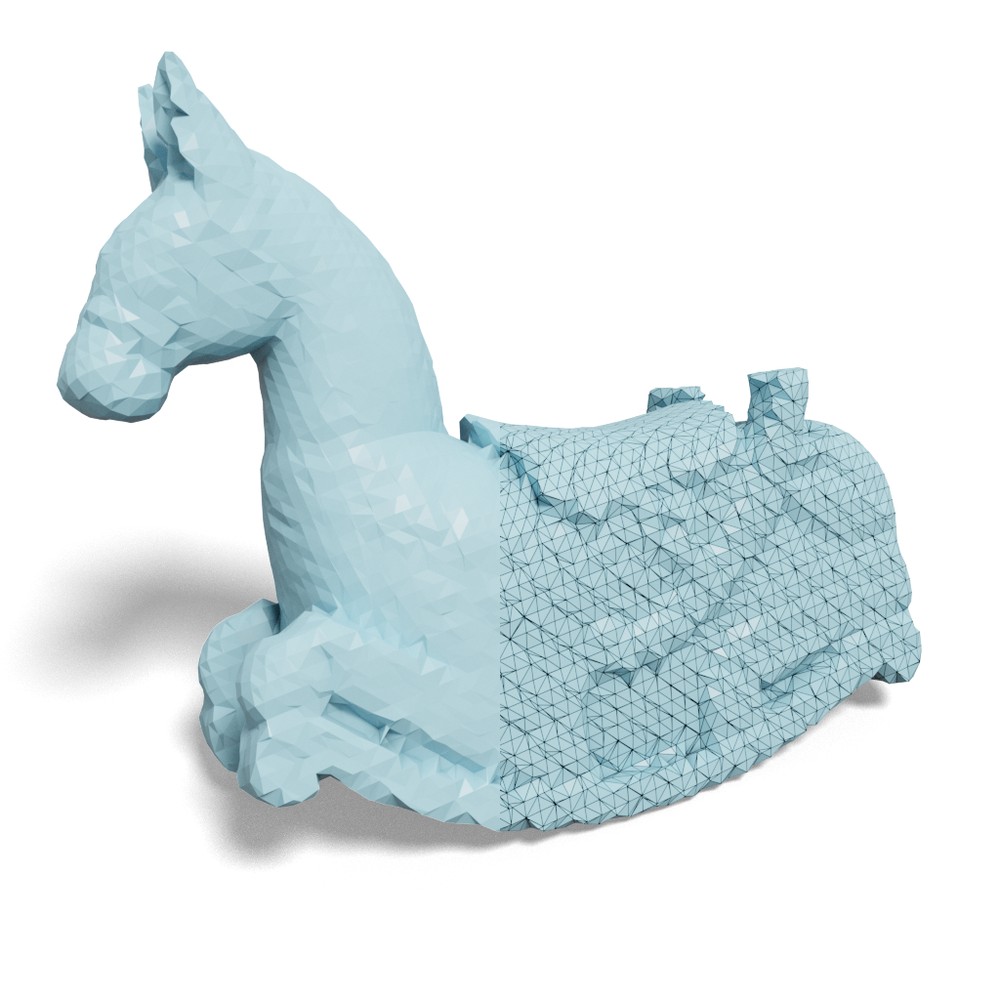} &
            \includegraphics[width=0.14\textwidth, trim=2cm 2cm 2cm 2cm, clip]{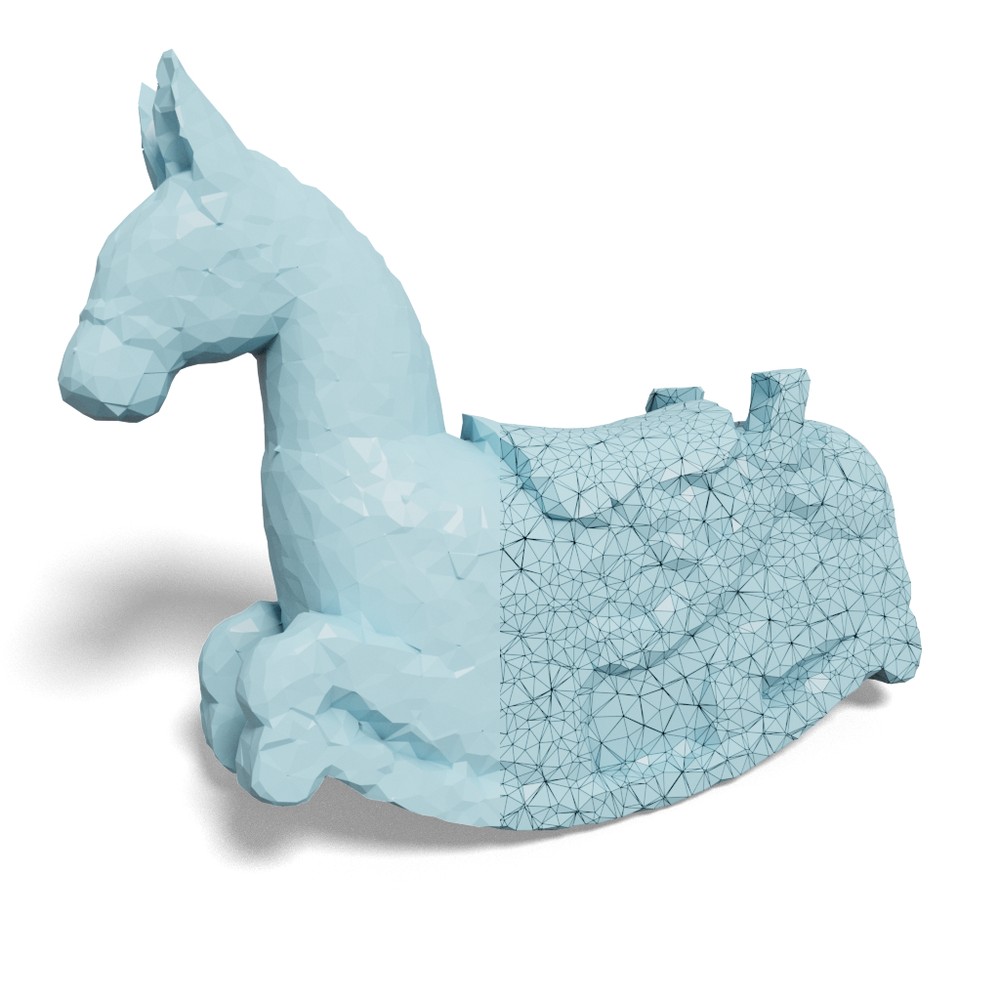} &
            \includegraphics[width=0.14\textwidth, trim=2cm 2cm 2cm 2cm, clip]{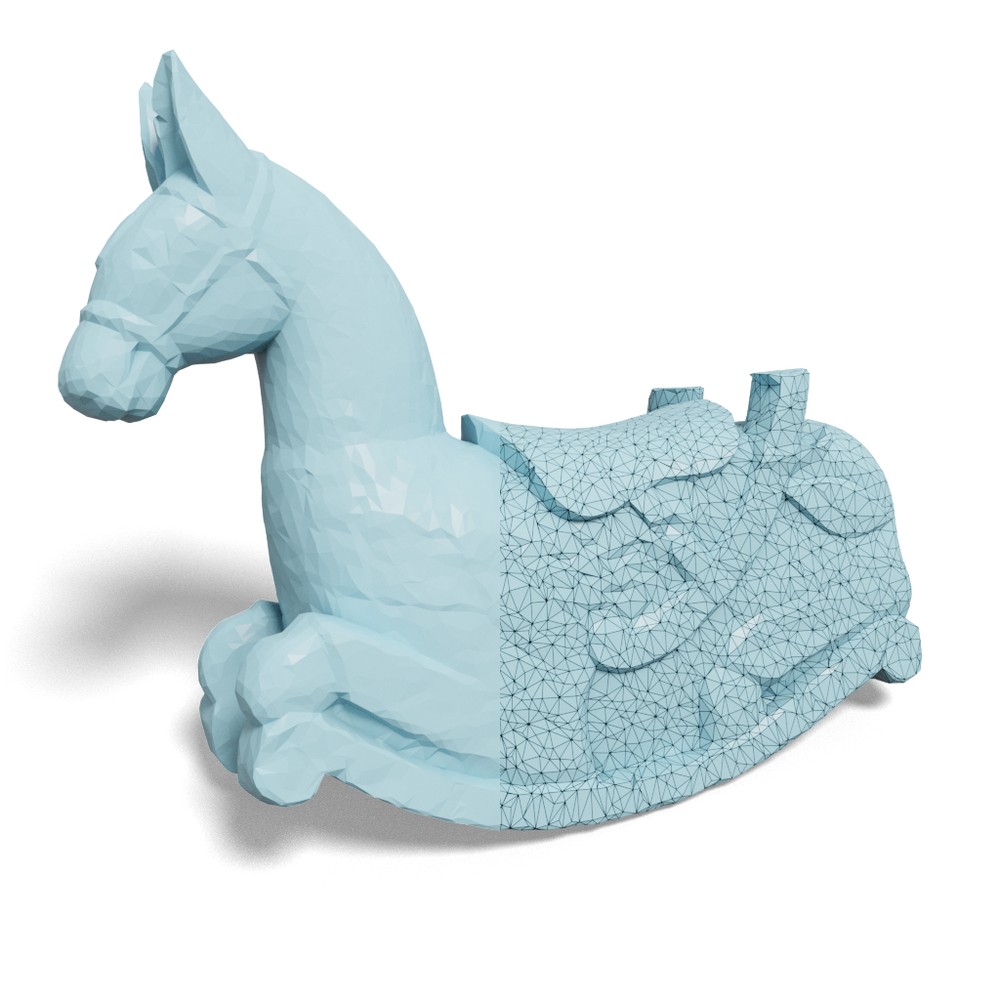} & 
            \includegraphics[width=0.14\textwidth, trim=2cm 2cm 2cm 2cm, clip]{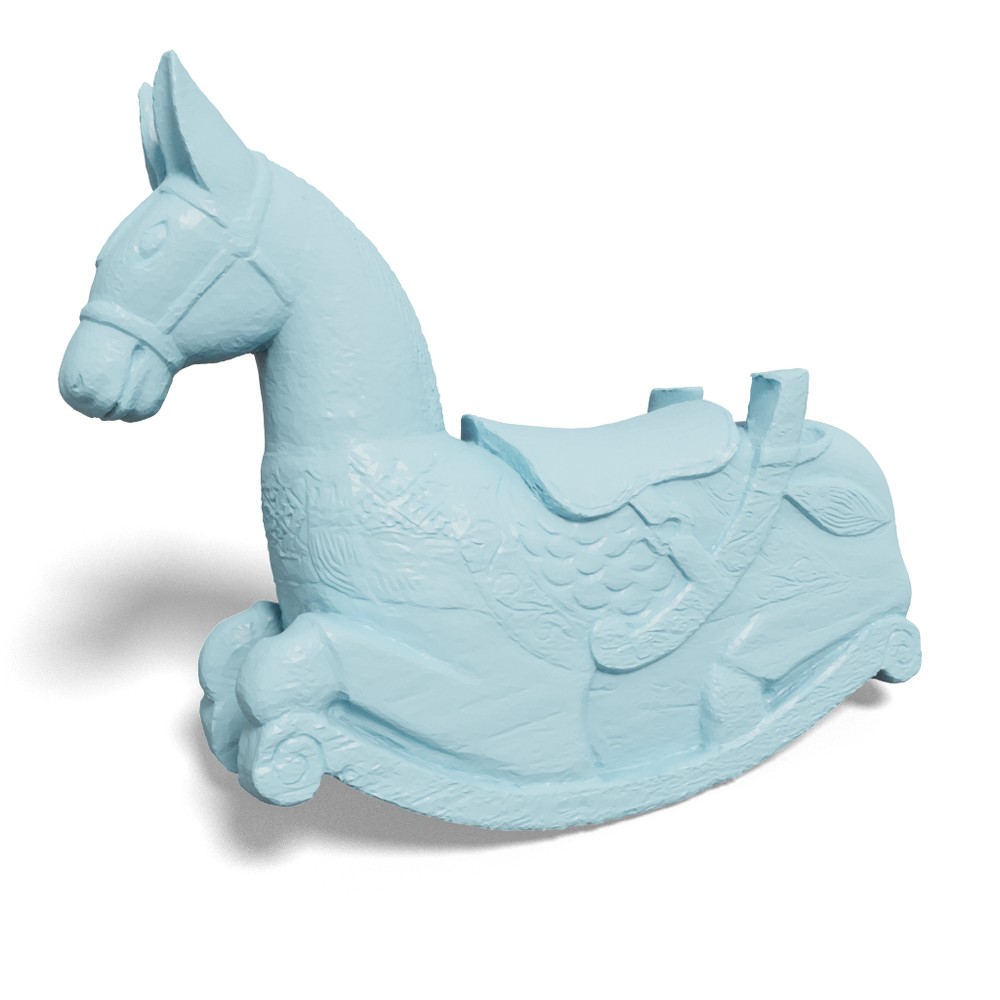} \\ 
            \includegraphics[width=0.14\textwidth, trim=2cm 2cm 2cm 2cm, clip]{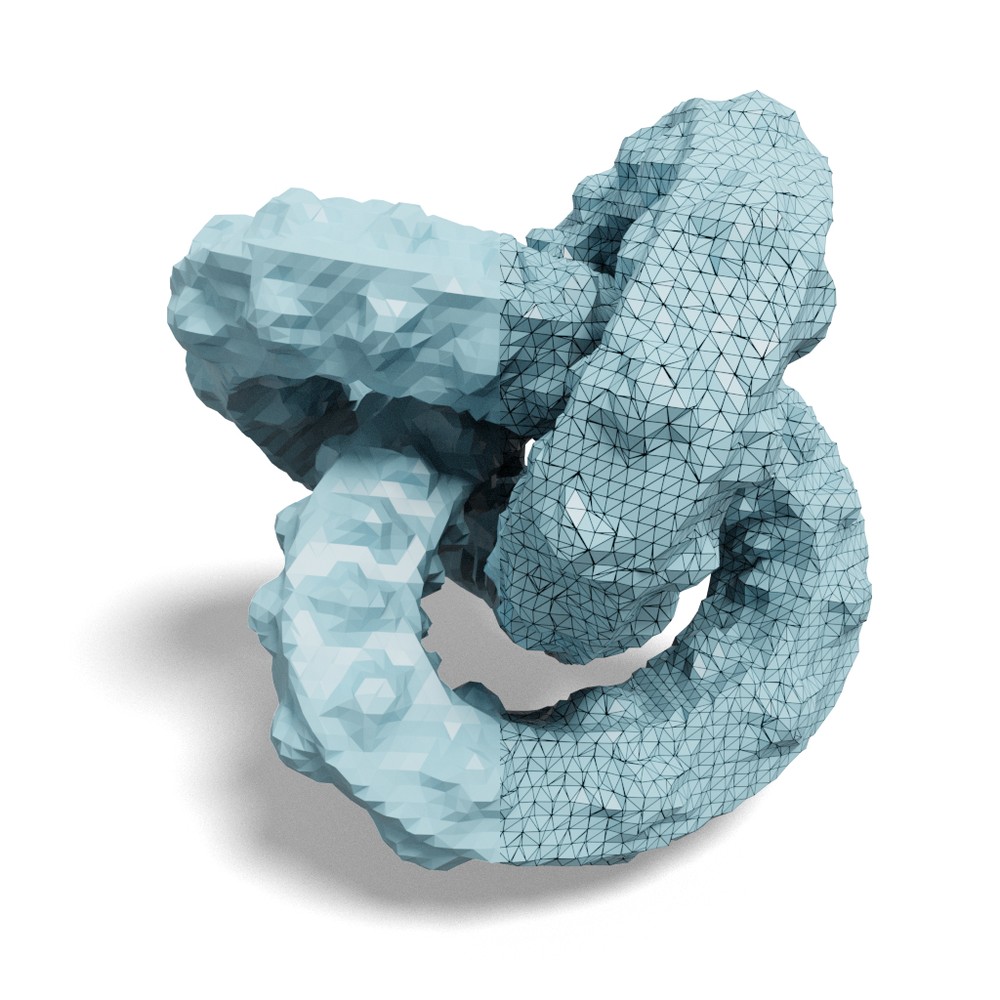} &
            \includegraphics[width=0.14\textwidth, trim=2cm 2cm 2cm 2cm, clip]{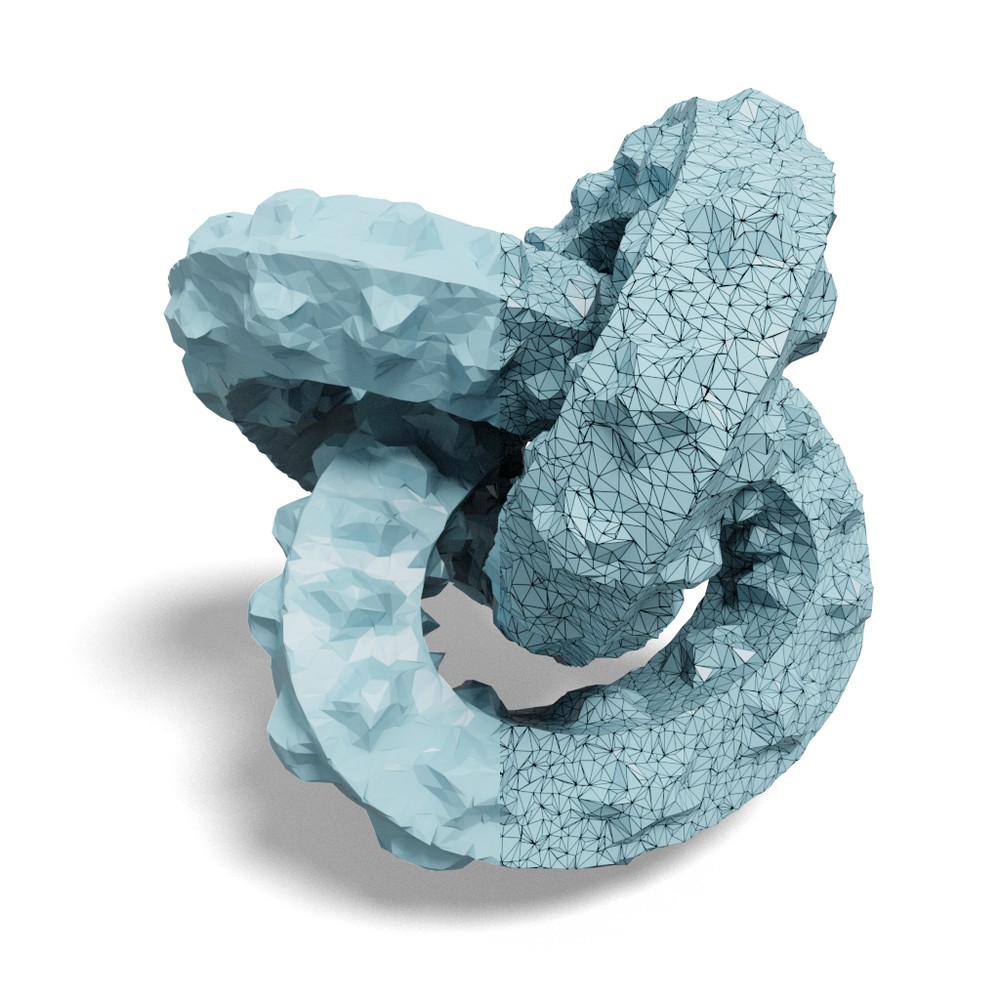} &
            \includegraphics[width=0.14\textwidth, trim=2cm 2cm 2cm 2cm, clip]{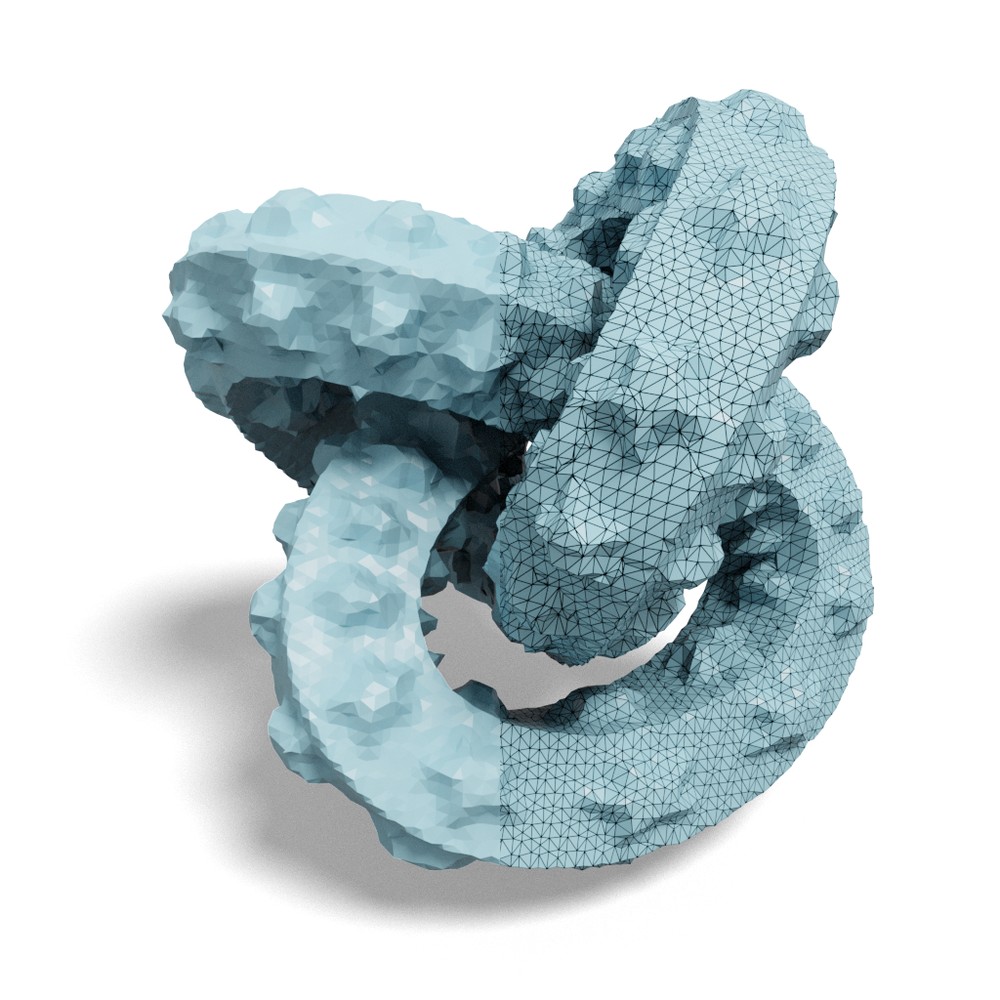} &
            \includegraphics[width=0.14\textwidth, trim=2cm 2cm 2cm 2cm, clip]{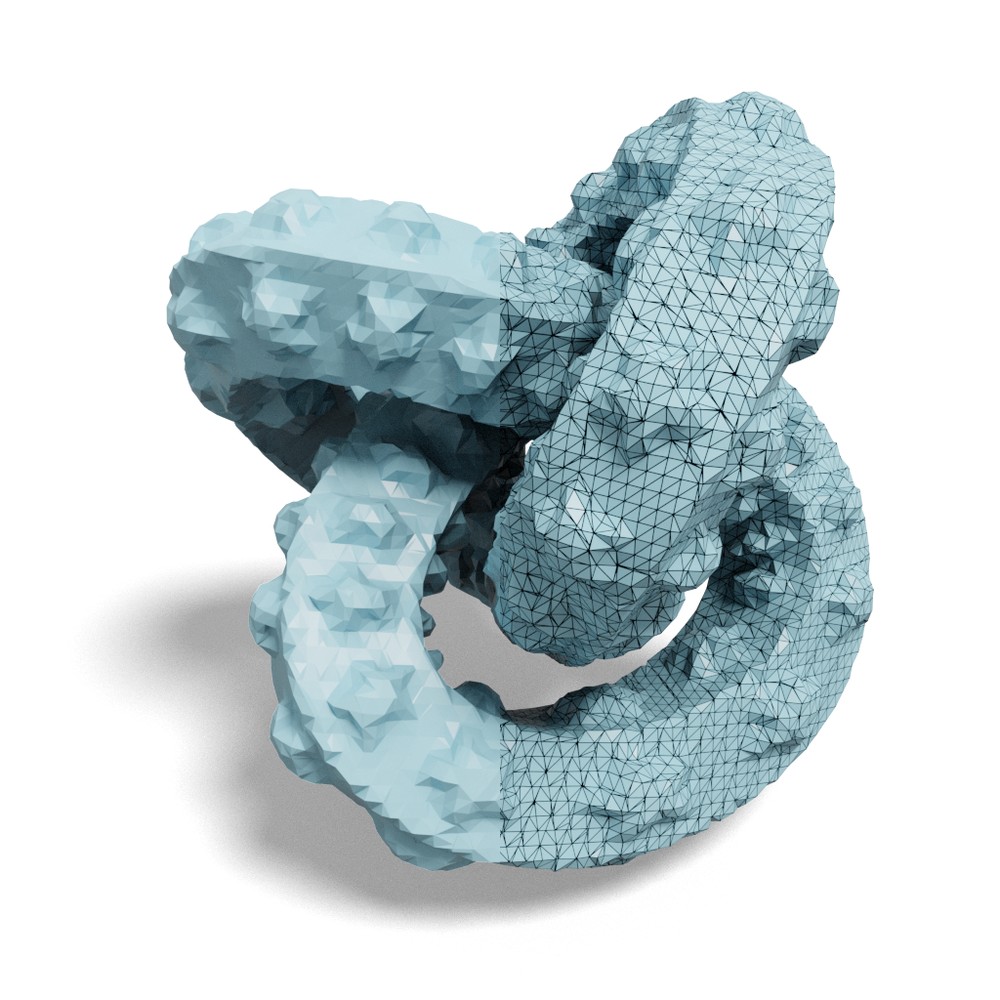} &
            \includegraphics[width=0.14\textwidth, trim=2cm 2cm 2cm 2cm, clip]{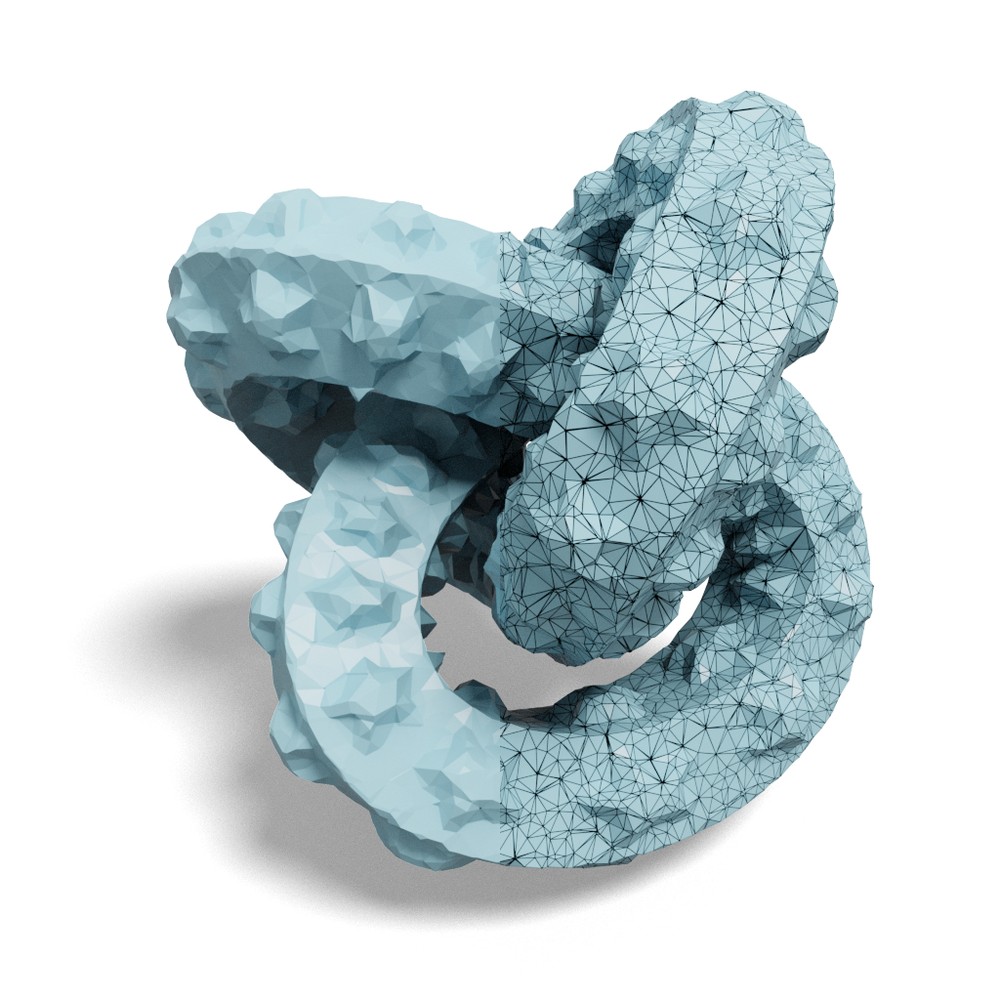} &
            \includegraphics[width=0.14\textwidth, trim=2cm 2cm 2cm 2cm, clip]{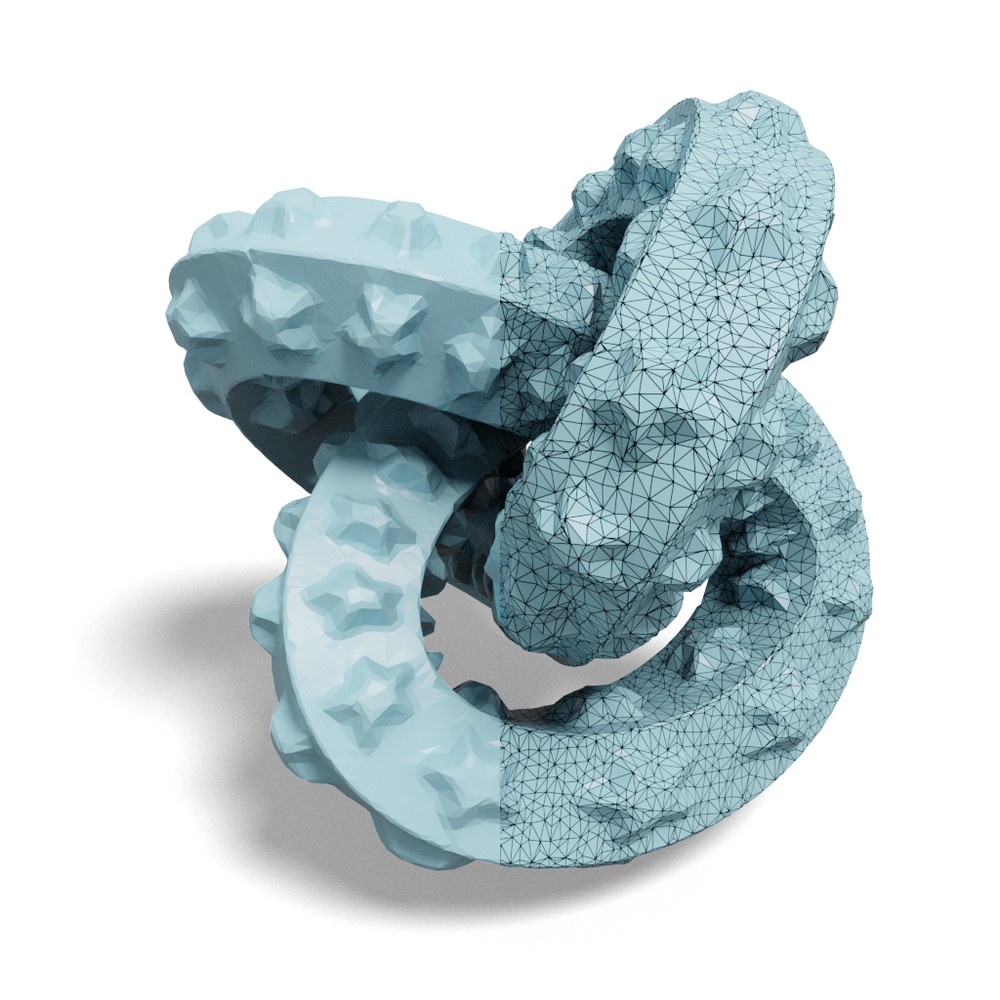} & 
            \includegraphics[width=0.14\textwidth, trim=2cm 2cm 2cm 2cm, clip]{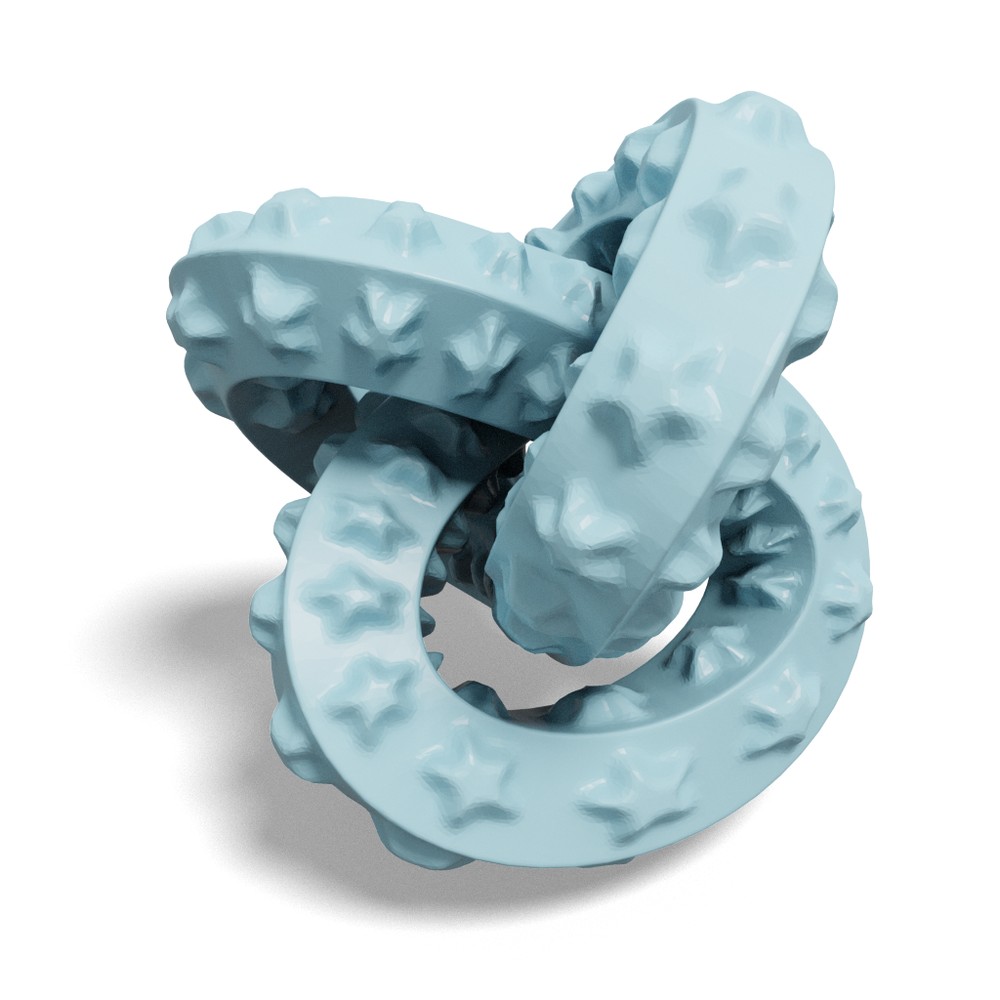} \\ 
            \includegraphics[width=0.14\textwidth, trim=2cm 2cm 2cm 2cm, clip]{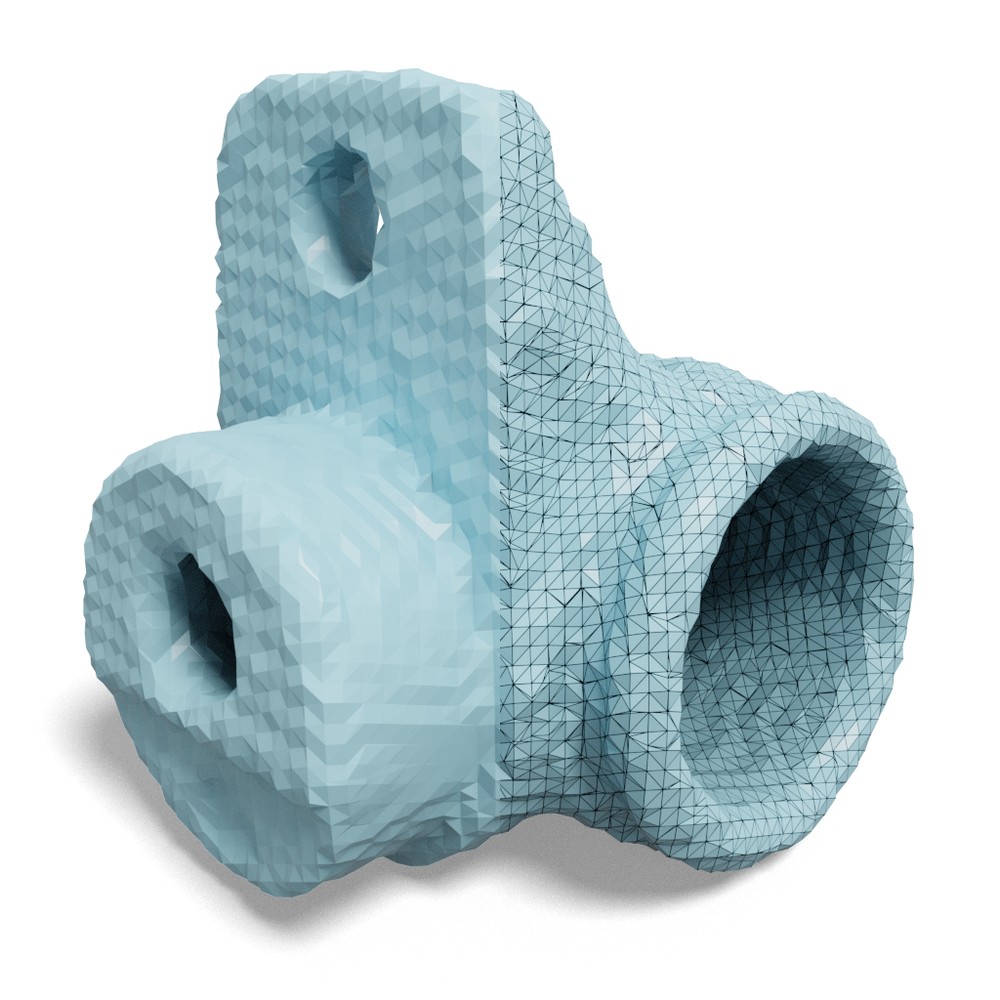} &
            \includegraphics[width=0.14\textwidth, trim=2cm 2cm 2cm 2cm, clip]{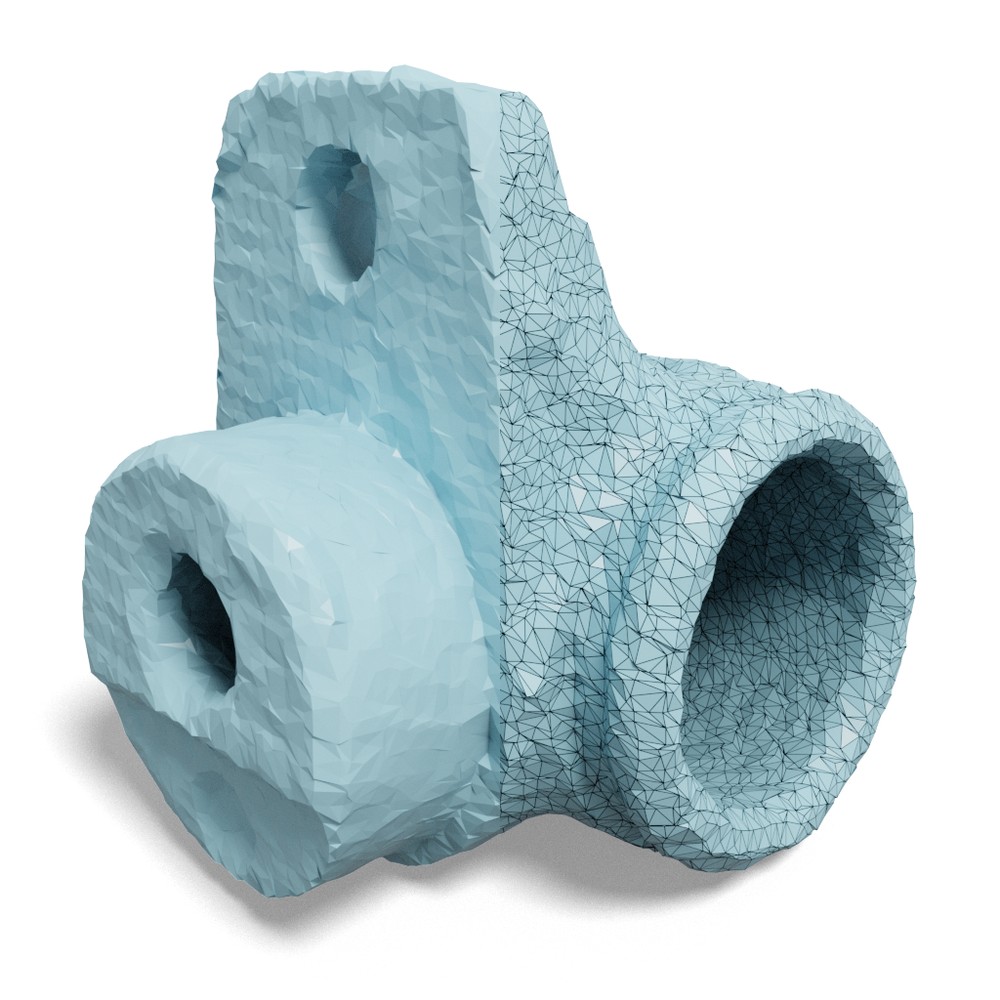} &
            \includegraphics[width=0.14\textwidth, trim=2cm 2cm 2cm 2cm, clip]{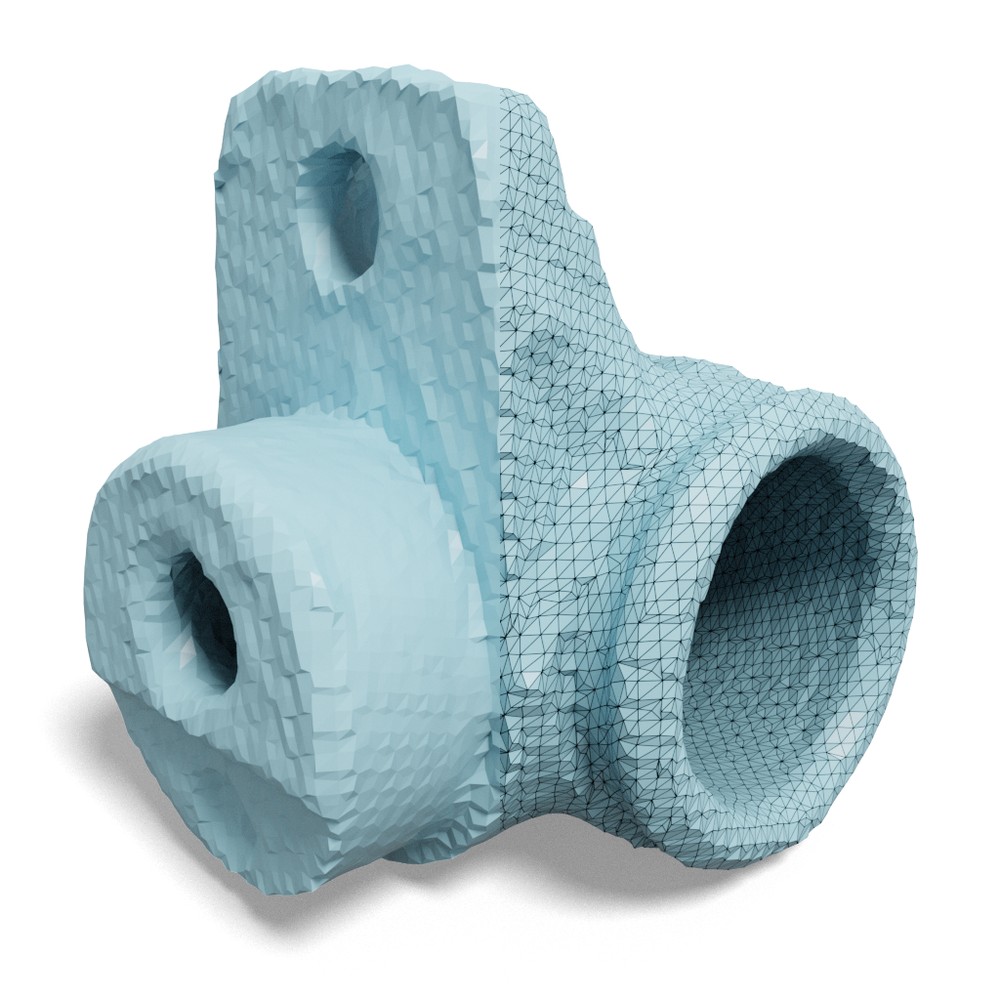} &
            \includegraphics[width=0.14\textwidth, trim=2cm 2cm 2cm 2cm, clip]{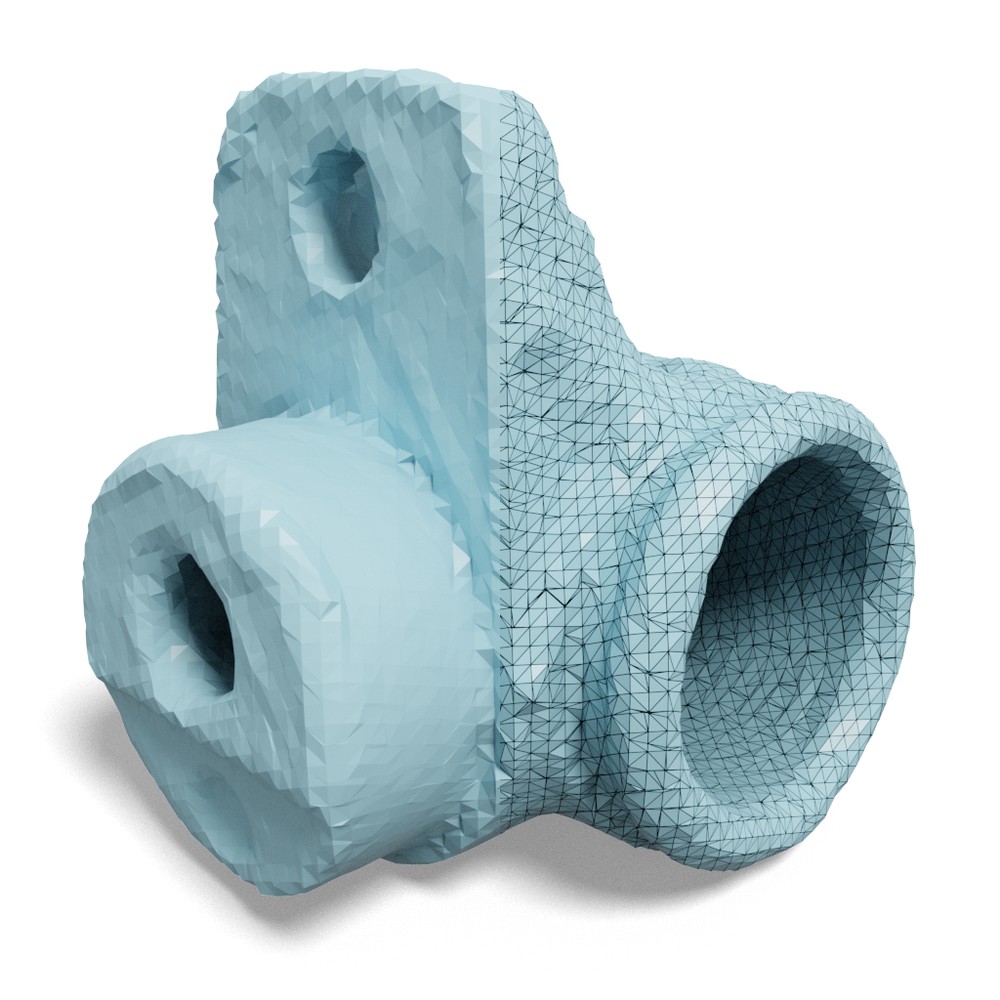} &
            \includegraphics[width=0.14\textwidth, trim=2cm 2cm 2cm 2cm, clip]{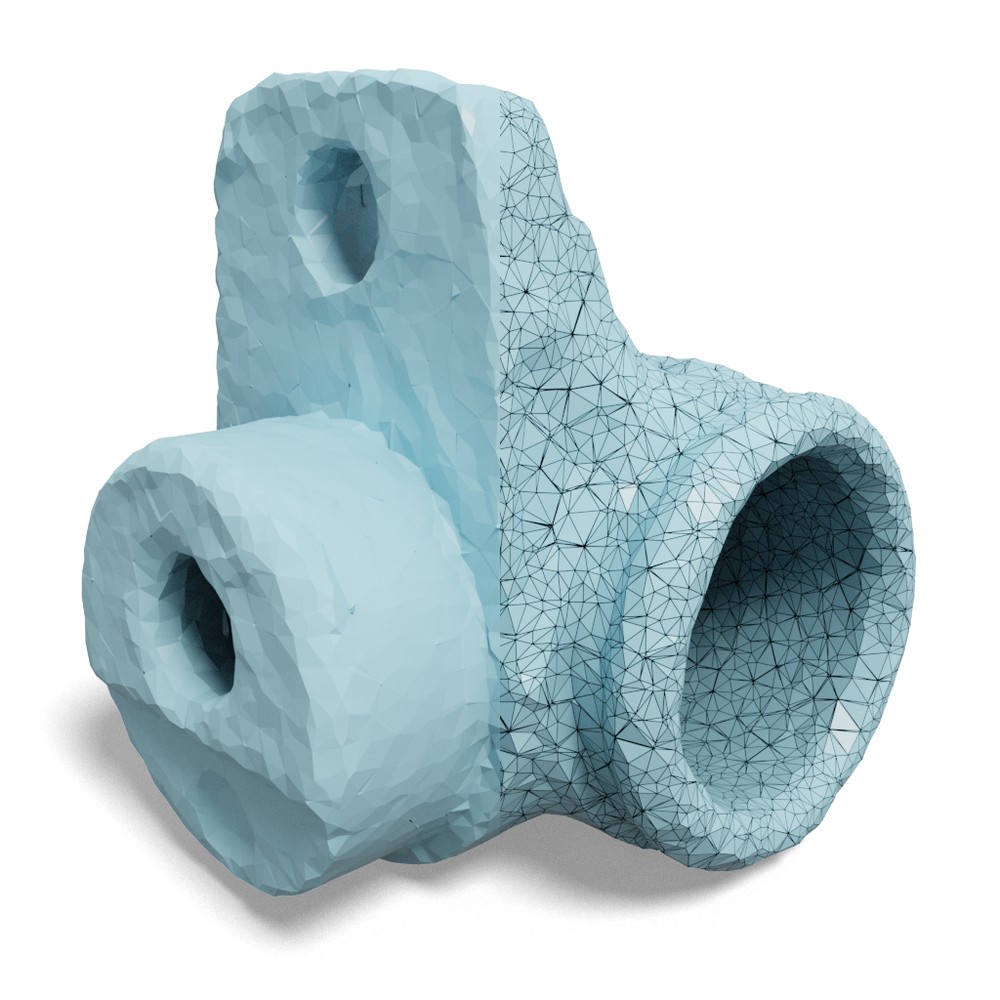} &
            \includegraphics[width=0.14\textwidth, trim=2cm 2cm 2cm 2cm, clip]{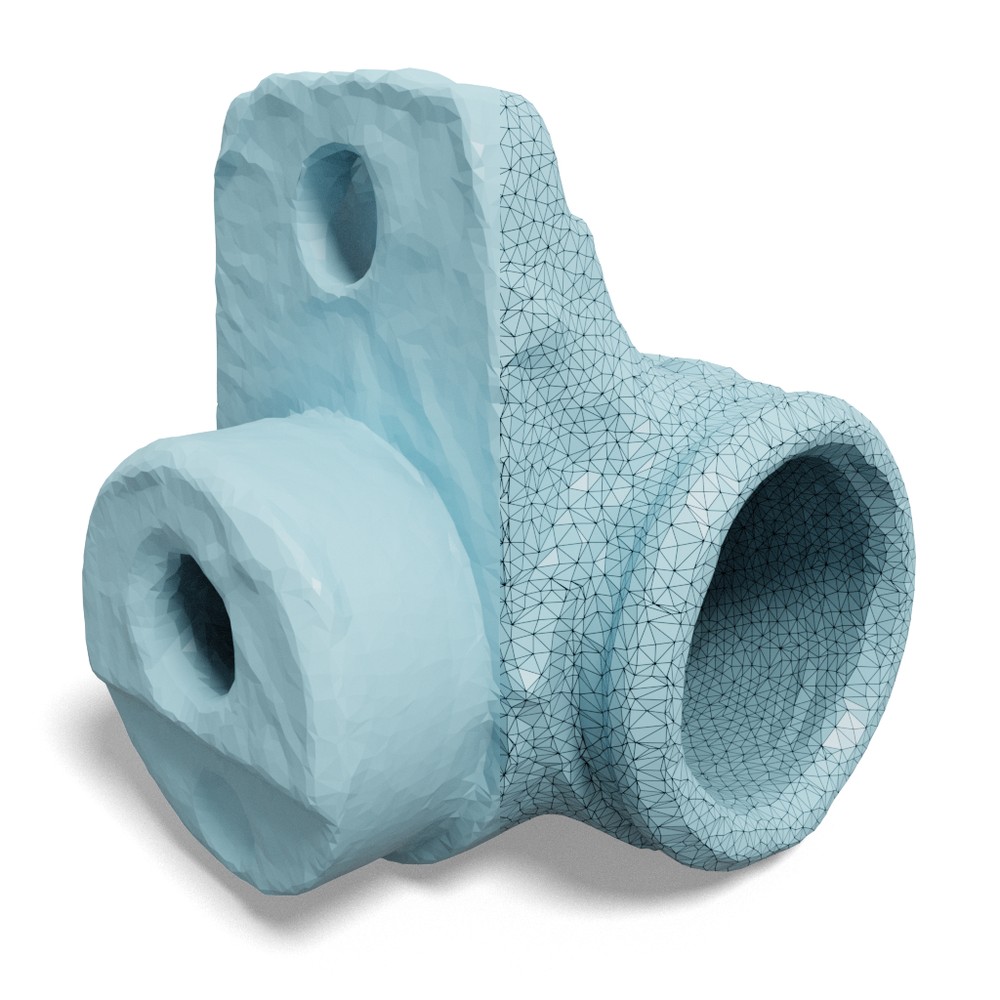} & 
            \includegraphics[width=0.14\textwidth, trim=2cm 2cm 2cm 2cm, clip]{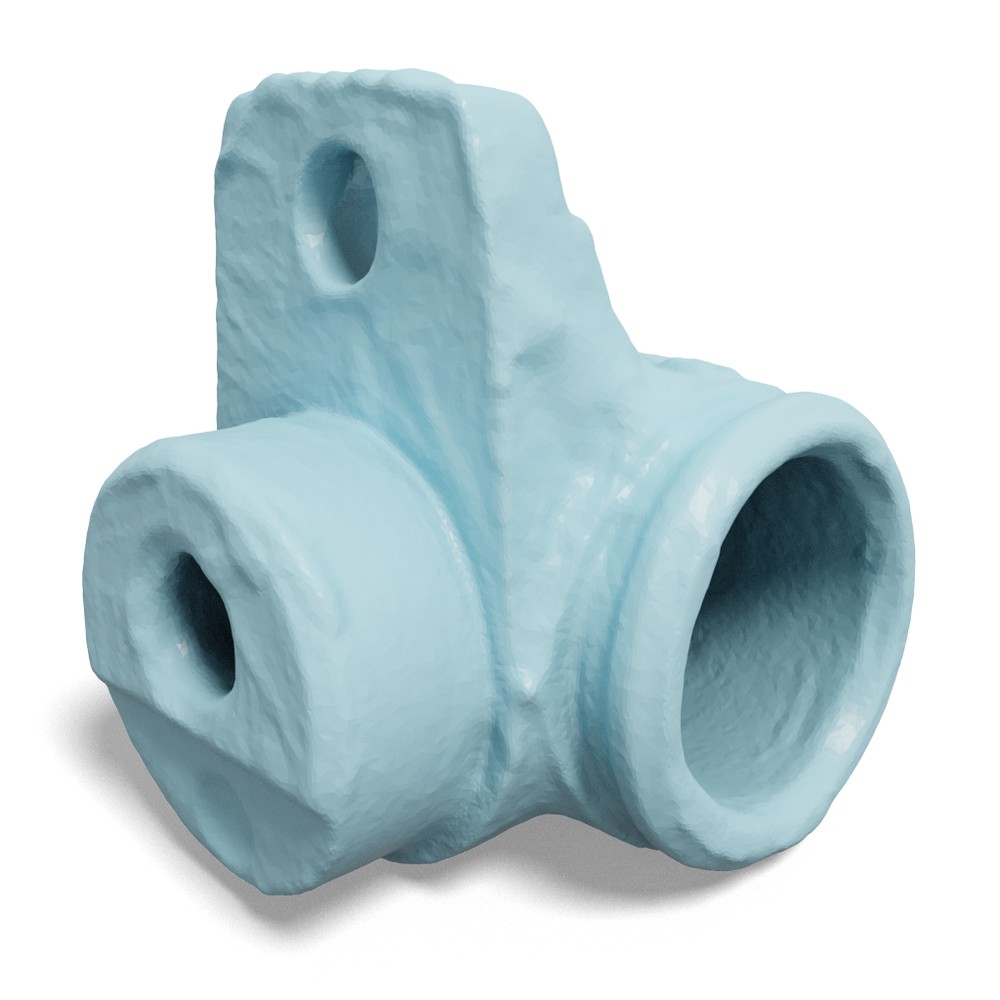} \\ 
            MC$_\mathrm{SDF}$ & DC$_\mathrm{hermite}$ & NDC$_\mathrm{SDF}$ & MC & \textsc{DMTet} & \textsc{FlexiCubes} & Reference \\
            \multicolumn{3}{c}{Surface extractions from GT SDF} & \multicolumn{3}{c}{Differentiable iso-surfacing} &
        \end{tabular}
        \vspace{-2mm}
        \caption{Visual comparison of a set of iso-surfacing techniques. The three leftmost examples: Marching Cubes (MC$_\mathrm{SDF}$), Dual Contouring, Neural Dual Contouring, use surface extraction from the GT SDF. The next three examples: 
        Marching Cubes, Deep Marching Tetrahedra, and \ours{}, use differentiable iso-surfacing. The grid resolution is $64^3$ for all methods except DMTet, which uses $80^3$ tetrahedral grid to match the triangle count in output meshes.}
        \label{fig:sup_main_qual}
        \end{figure*}
    }
\else
   \newcommand{\figSupplementalMainQual}{
    \begin{figure*}[t]
    \centering
    \includegraphics[width=\textwidth,trim=0 0 0 0,clip]{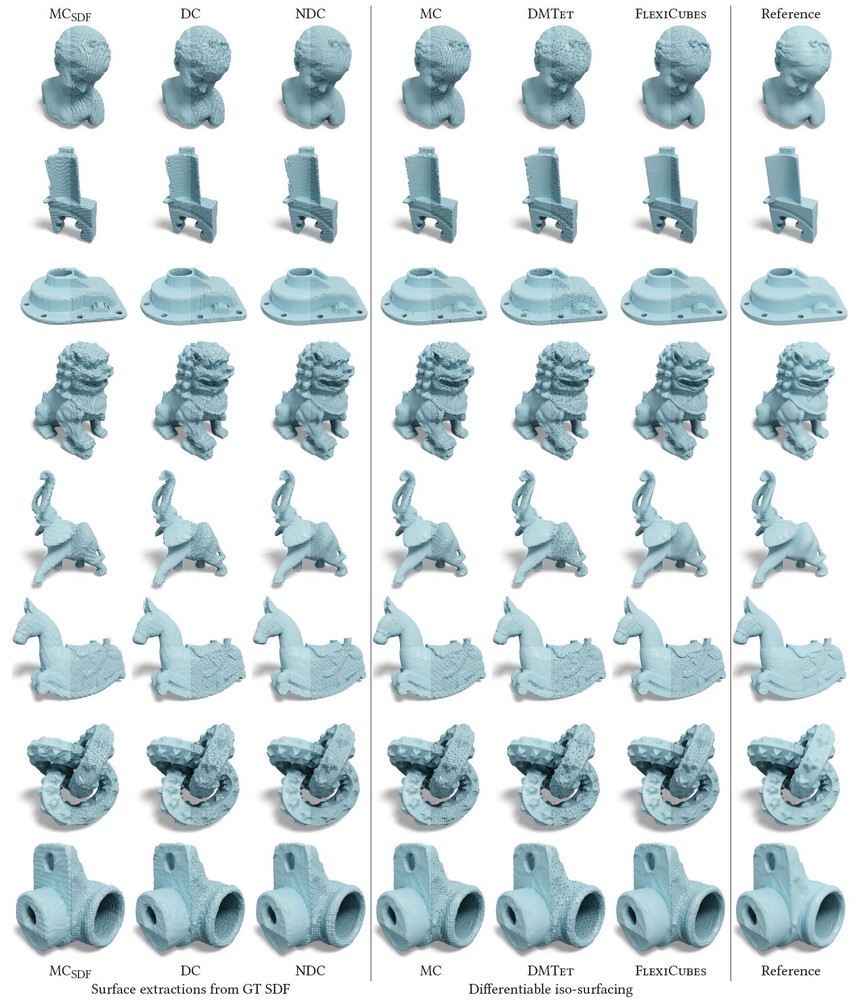}    
    \caption{ Visual comparison of a set of iso-surfacing techniques. The three leftmost examples: Marching Cubes (MC$_\mathrm{SDF}$), Dual Contouring, Neural Dual Contouring, use surface extraction from the GT SDF. The next three examples: 
    Marching Cubes, Deep Marching Tetrahedra, and \ours{}, use differentiable iso-surfacing.}
    \label{fig:sup_main_qual}
    \end{figure*}
}
\fi

\newcommand{\figAdaptiveQual}{
    \begin{figure}
    \centering
    \includegraphics[width=0.45\textwidth,trim=0 0 0 0,clip]{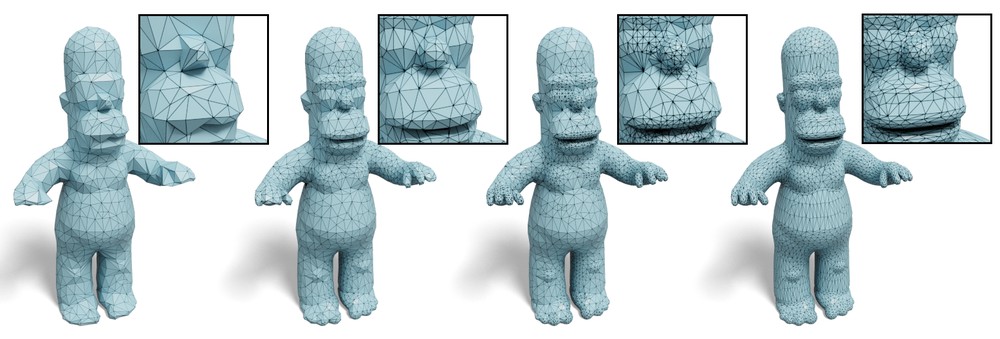}
    \setlength{\tabcolsep}{1mm}
    \begin{tabularx}{\columnwidth}{XXXX}
    \small{Uniform $32^3$} & \small{Adaptive $64^3$} & \small{Adaptive $128^3$} & \small{GT} \\
    \small{2.1k~tris, CD:4.3} & \small{4.5k~tris, CD:2.5} & \small{7.6k~tris, CD: 2.4} & \small{10k~tris}
    \end{tabularx}
    \includegraphics[width=0.45\textwidth,trim=0 0 0 0,clip]{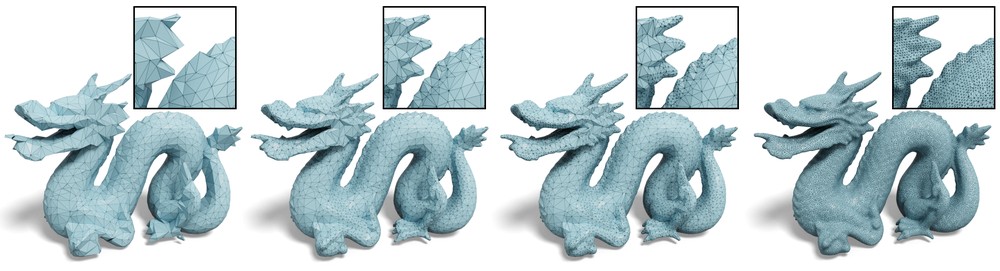}
    \begin{tabularx}{\columnwidth}{XXXX}
    \small{Uniform $32^3$} & \small{Adaptive $64^3$} & \small{Adaptive $128^3$} & \small{GT} \\
    \small{2.1k~tris, CD:5.7} & \small{5.7k~tris, CD:4.7} & \small{22k~tris, CD: 4.4} & \small{104k~tris}
    \end{tabularx}
    \caption{Adaptive meshing with \ours{} optimizing mesh topology and the octree structure jointly. CD denotes Chamfer distance ($10^{-5}$). %
    }. 
    \label{fig:adaptive_qual}
    \end{figure}
}

\newcommand{\figTriQualityHist}{
    \begin{figure*}[t]
    \centering
    \includegraphics[width=\textwidth,trim=0 0 0 0,clip]{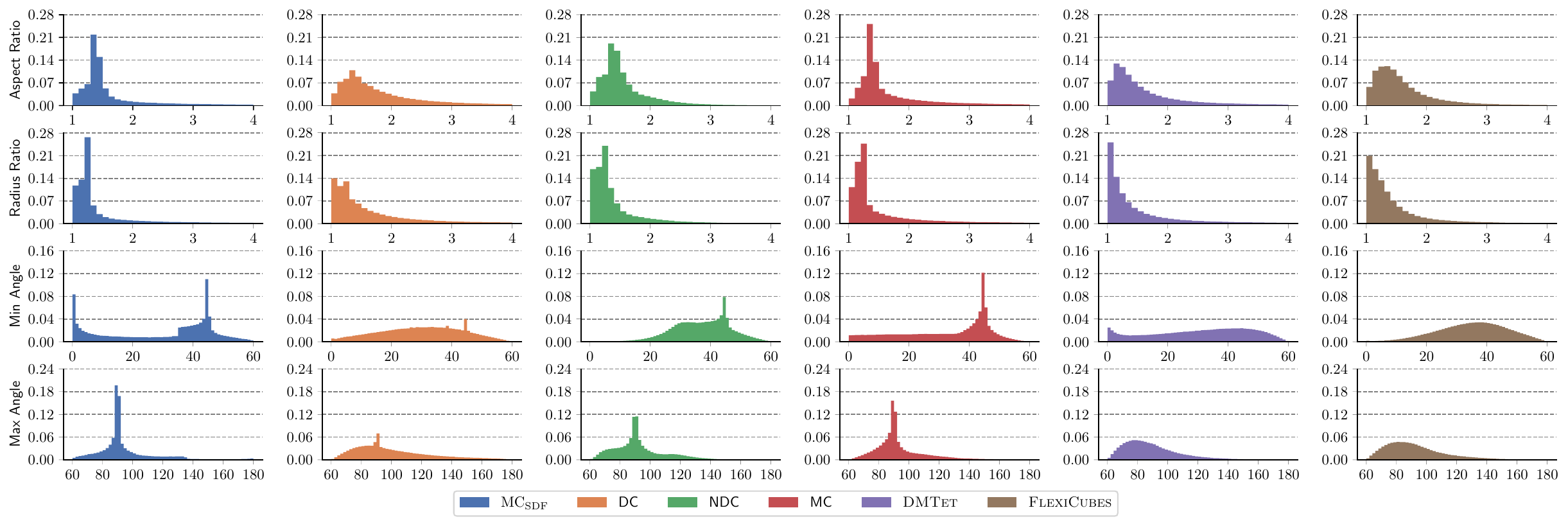}
    \caption{Quantitative comparison of the intrinsic quality of extracted meshes. \ours{} extracts high-quality meshes with a significantly smaller amount of sliver triangles (row 3) and more equilateral sides (row 1). The smooth distributions of Min and Max Angle (rows 3 and 4) show that \ours{} can adjust the triangles to better fit the geometry in slightly non-planar regions.   
    }
    \label{fig:tri_quality_hist}
    \end{figure*}
}

\newcommand{\figRoller}{
    \begin{figure}
    \centering
    \includegraphics[width=0.99\columnwidth,trim=0 0 0 0,clip]{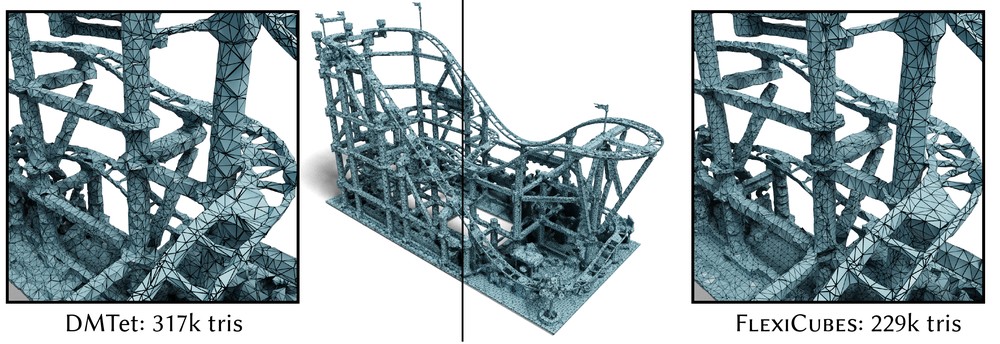}
    \caption{Extracting a 3D model of the Roller scene from LDraw \protect resources~\cite{Lasser2022} using \textsc{nvdiffrecmc}~\shortcite{hasselgren2022nvdiffrecmc} unmodified (DMTet) and \textsc{nvdiffrecmc} with \ours{} for 
    the topology extraction step. We note more uniform triangulation and higher detail.}
    \label{fig:roller}
    \end{figure}
}

\newcommand{\figUV}{
    \begin{figure}
    \centering
    \setlength{\tabcolsep}{1mm}
    \begin{tabular}{cccc}
        \includegraphics[width=0.45\columnwidth]{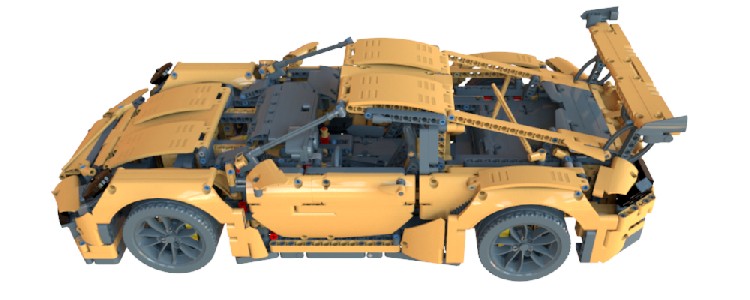} &
        \includegraphics[width=0.23\columnwidth]{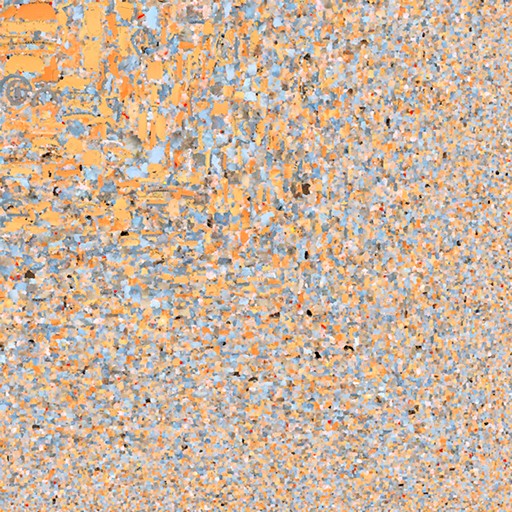} &
        \includegraphics[width=0.23\columnwidth]{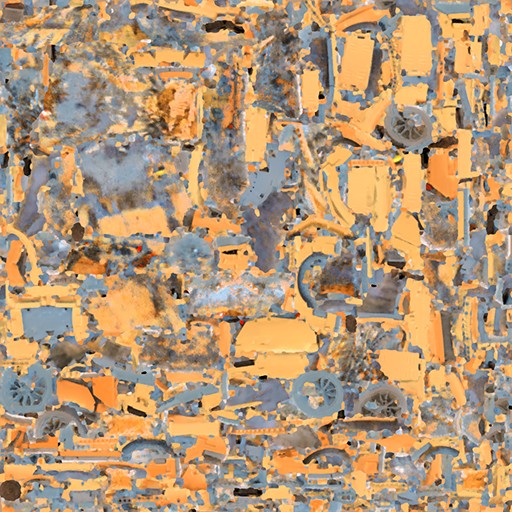} \\
        3D Model & DMTet & \ours{} 
    \end{tabular}

    \caption{Extracting a 3D model of the Porsche scene from LDraw \protect resources~\cite{Lasser2022} using \textsc{nvdiffrecmc}~\shortcite{hasselgren2022nvdiffrecmc}. We visualize the diffuse 
    texture, and note that \ours{} simplifies the UV unwrapping step (performed using xatlas~\shortcite{xatlas}). The resulting 
    textures have larger regions, which improves texture filtering.}
    \label{fig:uv}
    \end{figure}
}

\newcommand{\figNerfVisu}{
    \begin{figure}
    \centering
    \includegraphics[width=0.45\textwidth,trim=0 0 0 0,clip]{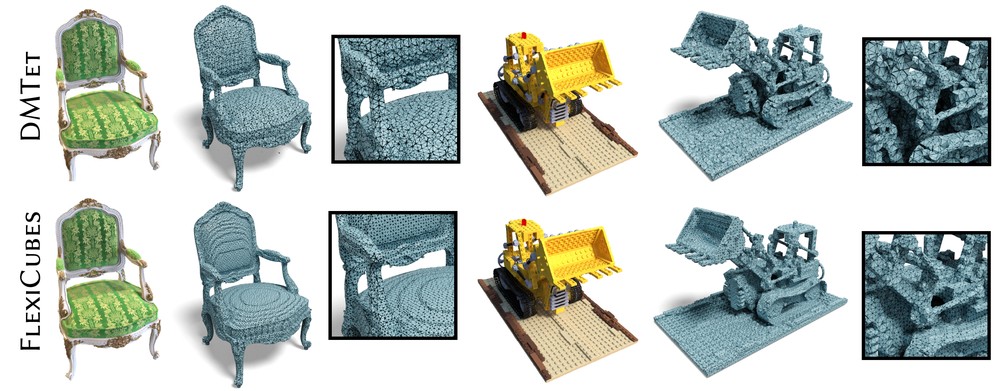}
    \includegraphics[width=0.45\textwidth,trim=0 0 0 0,clip]{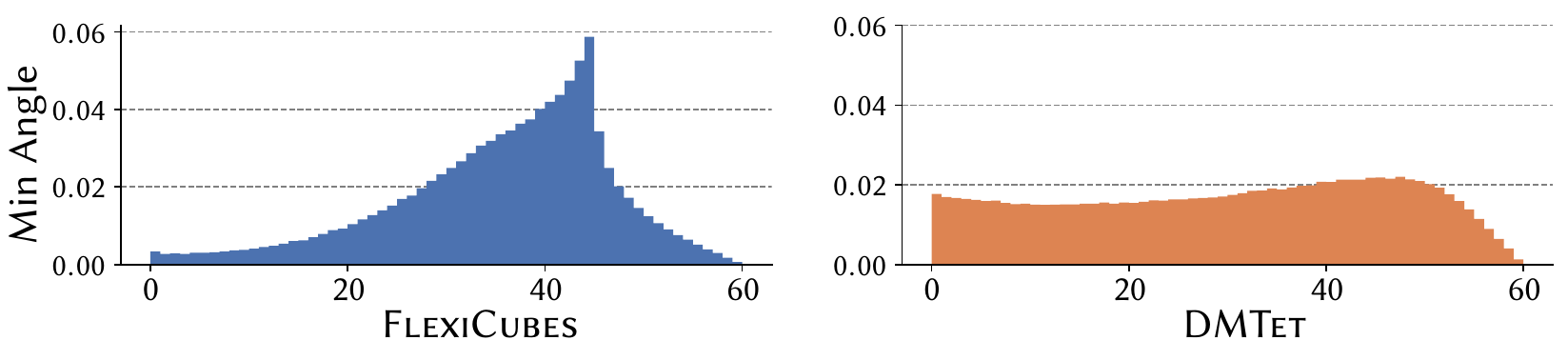}    
    \caption{Visualization of \nvdiffrec{} reconstructions for two scenes in the NeRF synthetic dataset. We compare \textsc{DMTet} and \ours{} for the topology extraction step. We note fewer sliver triangles for \ours{}.  We illustate this by including min angle histogram for \nvdiffrec{} reconstructions for all eight scenes in the NeRF synthetic dataset. Fewer triangles with small angles means less sliver triangles for \ours{}. 
    }
    \label{fig:nerf_visu}
    \end{figure}
}

\newcommand{\figSkining}{
    \begin{figure}
    \centering
    \setlength{\tabcolsep}{0mm}
    \begin{tabular}{ccc}
        \multicolumn{3}{c}{\begin{tabular}{cc}
            \raisebox{-0.5\height}{\includegraphics[width=0.49\columnwidth]{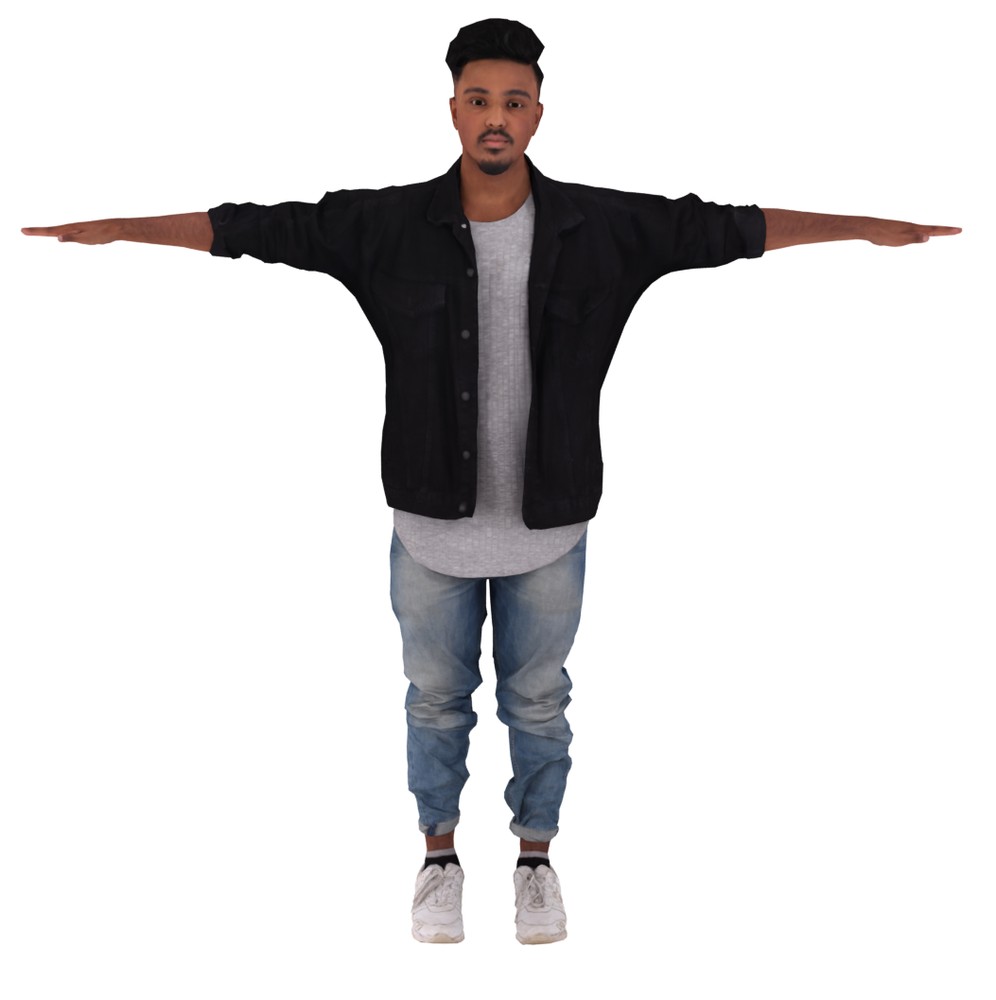}} &
            \raisebox{-0.5\height}{
                \begin{tikzpicture}
    			\node[anchor=south west,inner sep=0, outer sep=0] (image) at (0,0) 
                {\includegraphics[width=0.45\columnwidth]{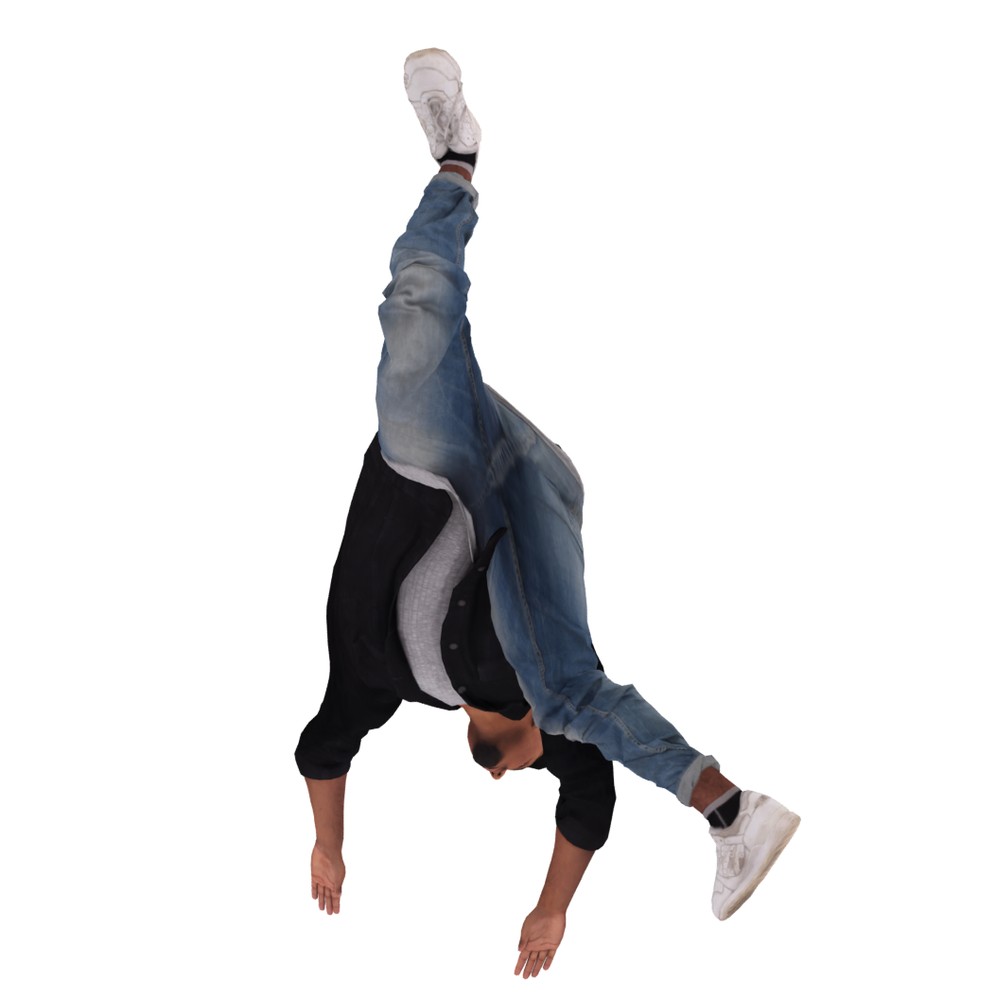}};
        		\begin{scope}[x={(image.south east)},y={(image.north west)}]
        			\draw[orange,very thick] (0.30, 0.33) rectangle (.63, .66);
        		\end{scope}
                \end{tikzpicture}
            } \\
            T-pose & Frame 47
        \end{tabular}} \\
        \raisebox{-0.5\height}{
            \includegraphics[width=0.32\columnwidth]{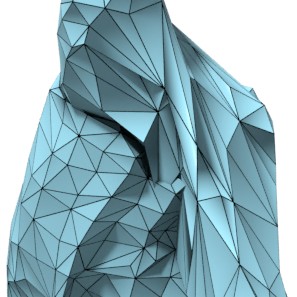}
        } &
        \raisebox{-0.5\height}{
            \includegraphics[width=0.32\columnwidth]{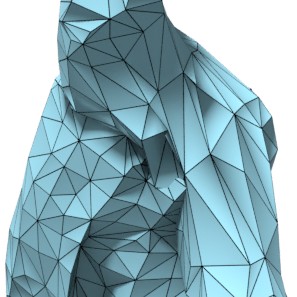}
        } &
        \raisebox{-0.5\height}{
            \includegraphics[width=0.32\columnwidth]{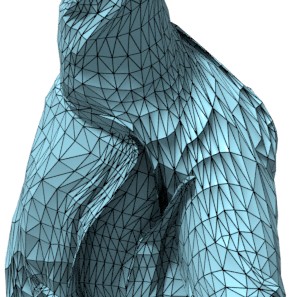}
        } \\
        \ours{} (T-pose)  & \ours{} (e2e) & Reference\\
        1.7k~tris & 1.7k~tris & 23k~tris
    \end{tabular}
    \caption{Mesh extraction of a skinned animation, showing the benefits of end-to-end optimization using an explicit, differentiable, mesh representation. We learn the topology and appearance of a low-poly mesh through image supervision using nvdiffrec. The top row 
    shows the reference mesh T-pose and an animated frame. The bottom row shows a baseline of \ours{} optimized for the T-pose 
    and re-skinned in a post pass, \ours{} with end-to-end optimization, where we re-skin the mesh in each optimization step and use a randomized viewpoint and animation frame, and the reference. 
    }
    \label{fig:skinning}
    \end{figure}
}

\newcommand{\figApplicationMeshGeneration}{
    \begin{figure}
    \centering
    \includegraphics[width=0.5\textwidth,trim=0 0 0 0,clip]{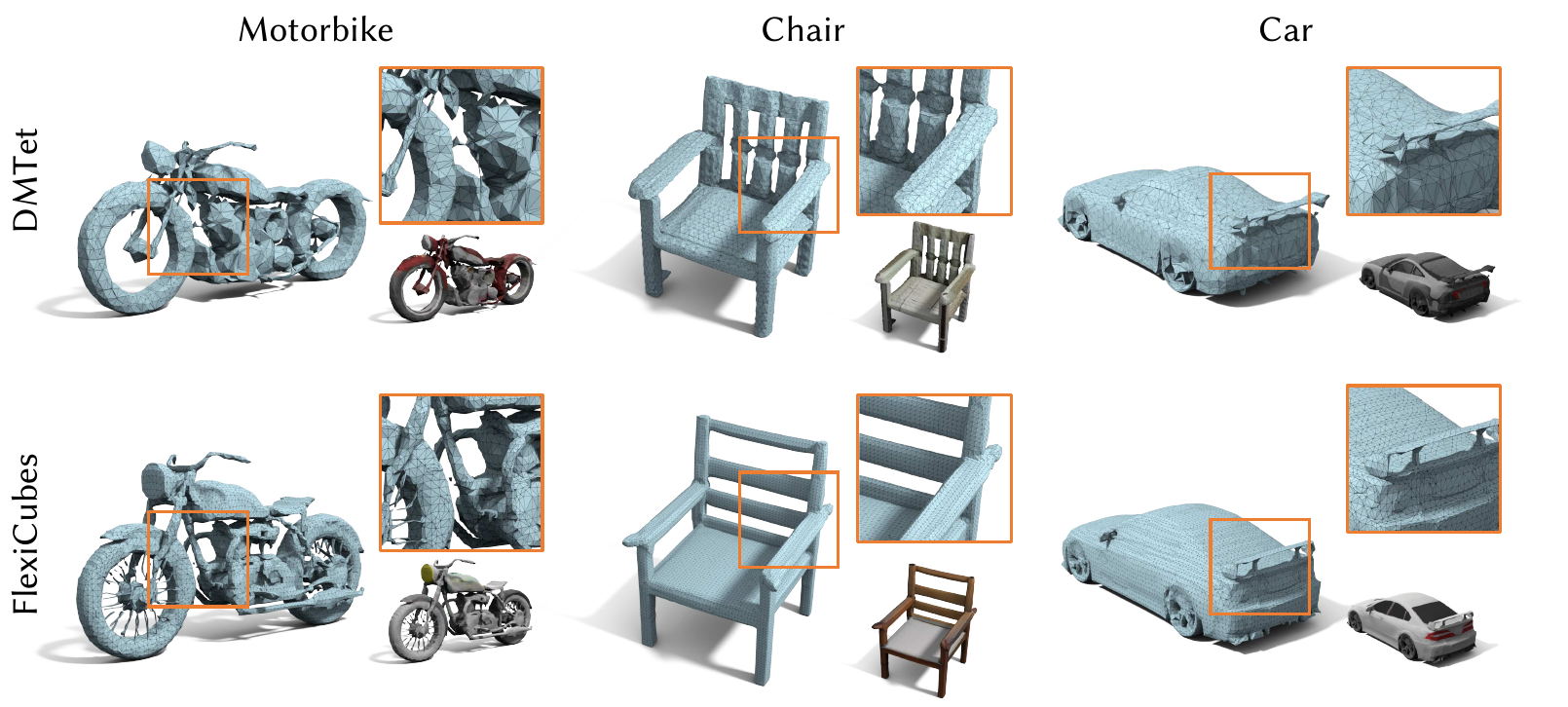}    
    \caption{
    Qualitative results for 3D generative modeling with meshes in GET3D~\cite{gao2022get3d}. \ours{} produces significantly improved mesh quality with detailed thin structures and more uniform surfaces. 
    }
    \label{fig:ApplicationMeshGeneration}
    \end{figure}
}

\newcommand{\figApplicationMeshGenerationSupp}{
    \begin{figure*}
    \centering
    \includegraphics[width=\textwidth,trim=0 0 0 0,clip]{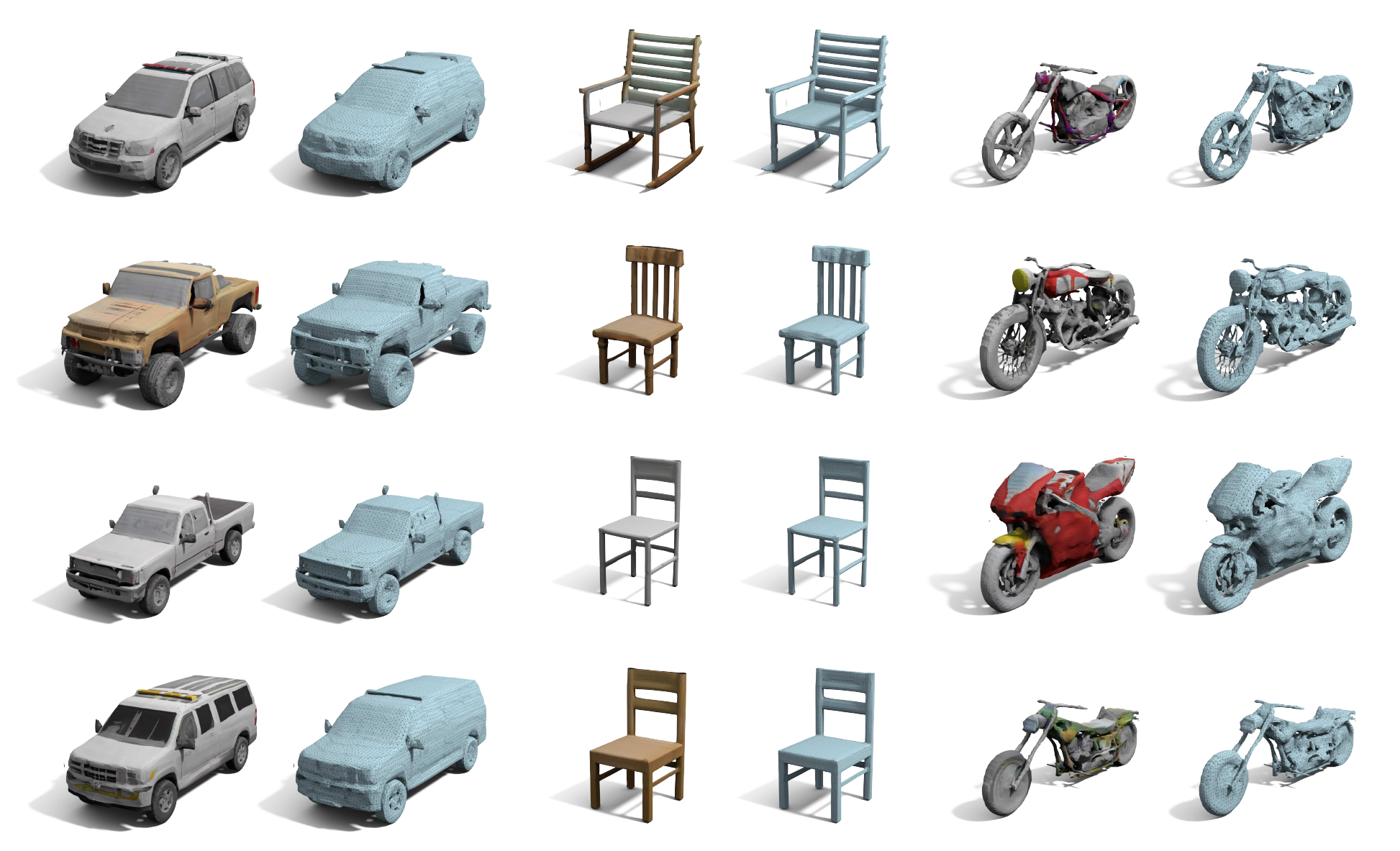}    
    \caption{
    Qualitative textured mesh generation combining \ours{} with GET3D~\cite{gao2022get3d}. 
    }
    \label{fig:ApplicationMeshGenerationSupp}
    \end{figure*}
}

\newcommand{\figEdgeReg}{
    \begin{figure}
    \centering
    \includegraphics[width=0.46\textwidth]{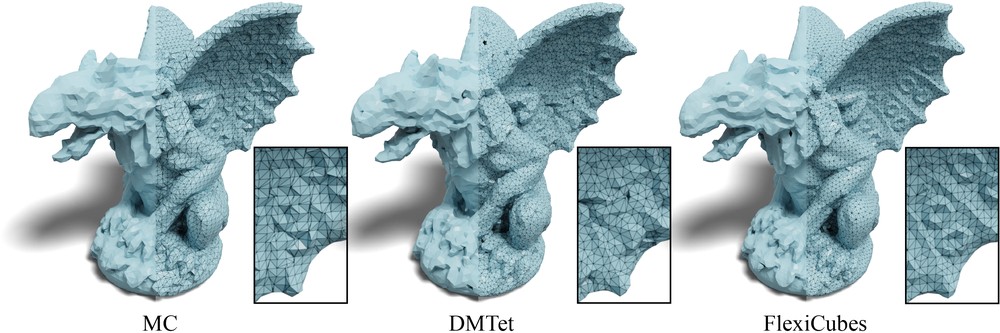}
    \caption{ Adding equilateral edge regularization to enhance triangle quality. Comparing with results in Figure~\ref{fig:main_qual}, which does not have equilateral edge regularization loss, \ours{} only has slight degradation of the reconstruction quality, while MC and DMTet lose details in the local geometry.  Please zoom in to see the mesh details.
    }
    \label{fig:EdgeReg}
    \end{figure}
}

\newcommand{\figdevelopable}{
    \begin{figure}
    \centering
    \includegraphics[width=0.46\textwidth]{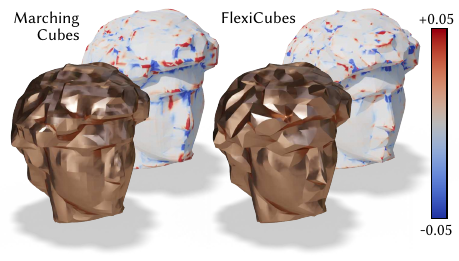}    
    \vspace*{-0.8em}
    \caption{
      To demonstrate mesh-based regularizers, we combine synthetic photogrammetry with a \emph{developability} term~\cite{Stein:2018:DSF}, which encourages fabricability from panels.
      \ours{} has the freedom to fit the shape and satisfy the regularizer, whereas Marching Cubes does not. 
      \emph{Top:} Gaussian curvature in radians, \emph{Bottom:} stylized render to show panelization.
    }
    \label{fig:developable}
    \end{figure}
}

\newcommand{\figphysics}{
    \begin{figure}
    \centering
    \setlength{\tabcolsep}{0mm}
    \begin{tabular}{cccccc}
   & && $t=0s$ & $t=1.0s$ & $t=1.8s$ \\
   \rotatebox{90}{ \quad  Initialization} & &
\rotatebox{90}{ \quad  $D= 1.500$}
         &
  \includegraphics[width=0.30\columnwidth,trim={0cm 0cm 0cm 0cm},clip]{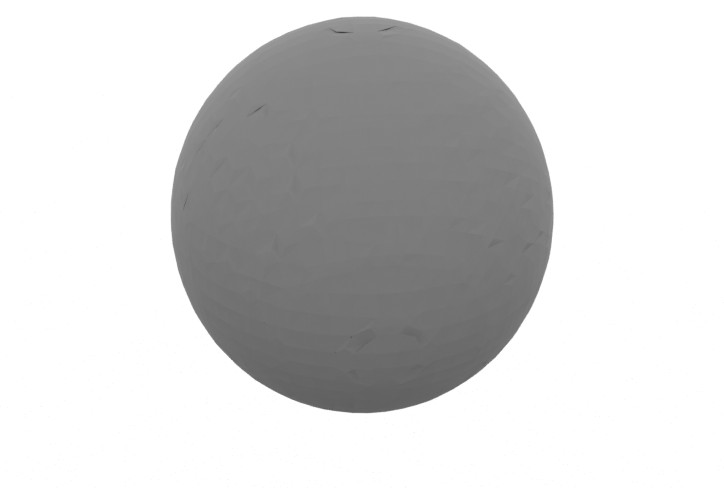} &
       \includegraphics[width=0.30\columnwidth,trim={0cm 0cm 0cm 0cm},clip]{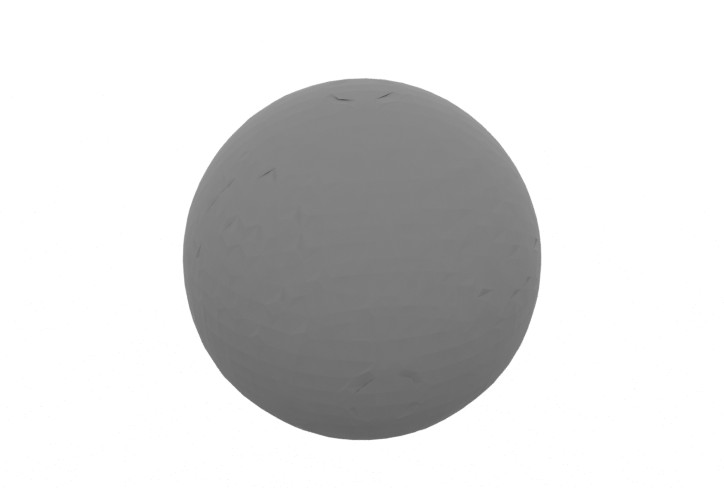} &
   \includegraphics[width=0.30\columnwidth,trim={0cm 0cm 0cm 0cm},clip]{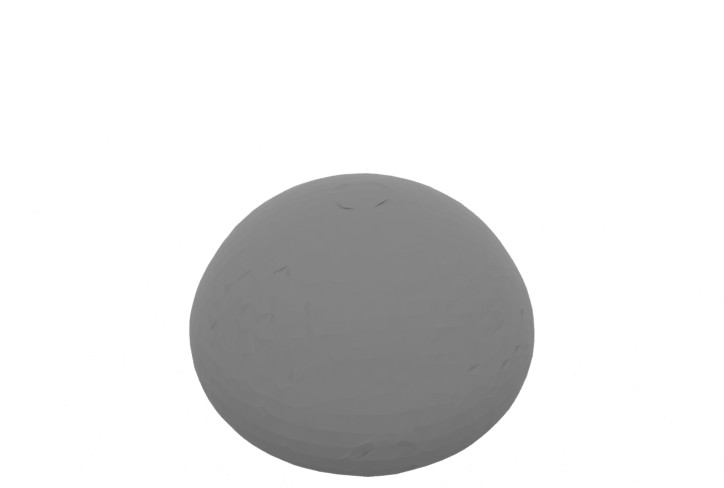} \\
   \rotatebox{90}{ \quad  Optimized} & &
   \rotatebox{90}{ \quad  $D= 0.311$}
        &
    \includegraphics[width=0.30\columnwidth,trim={0cm 0cm 0cm 1cm},clip]{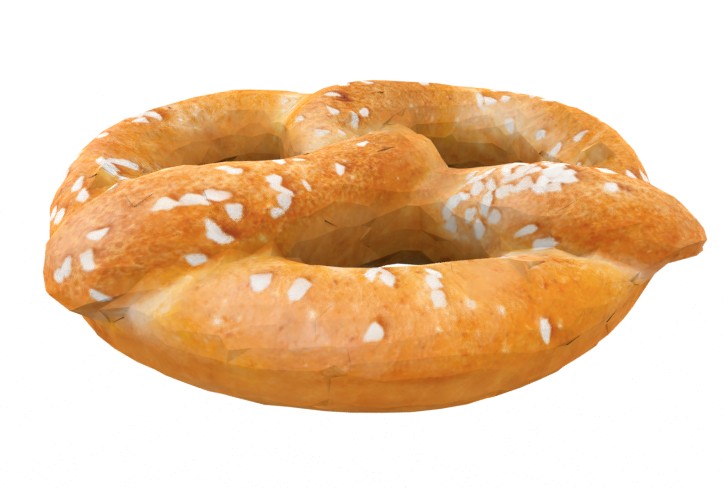} &
        \includegraphics[width=0.30\columnwidth,trim={0cm 0cm 0cm 1cm},clip]{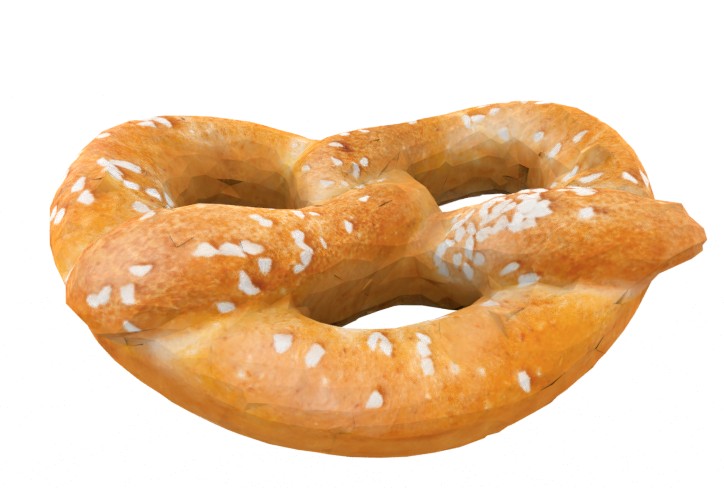} &
        \includegraphics[width=0.30\columnwidth,trim={0cm 0cm 0cm 1cm},clip]{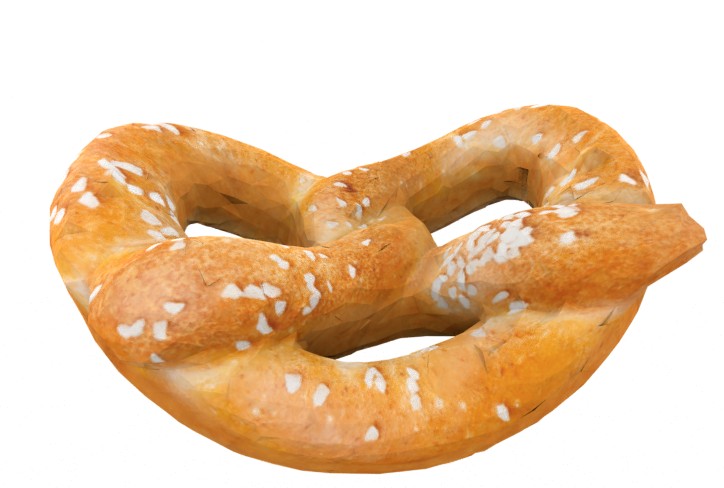} \\
    \rotatebox{90}{\quad   Reference} & &
    	\rotatebox{90}{\quad   $D= 0.300$}
    &
    \includegraphics[width=0.30\columnwidth,trim={0cm 0cm 0cm 1cm},clip]{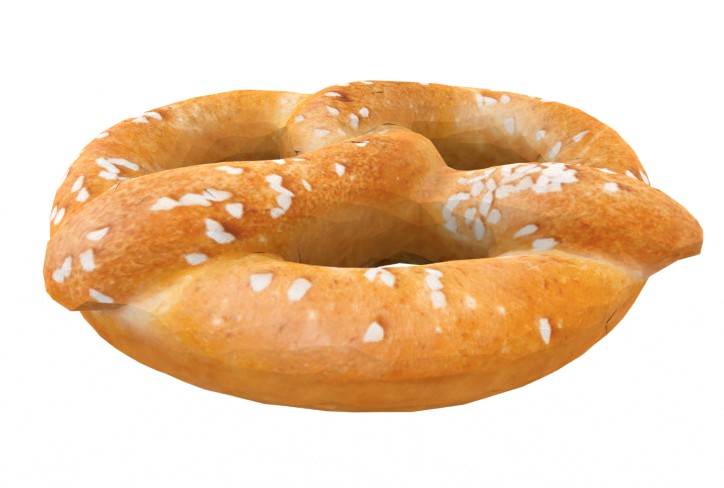} &
     \includegraphics[width=0.30\columnwidth,trim={0cm 0cm 0cm 1cm},clip]{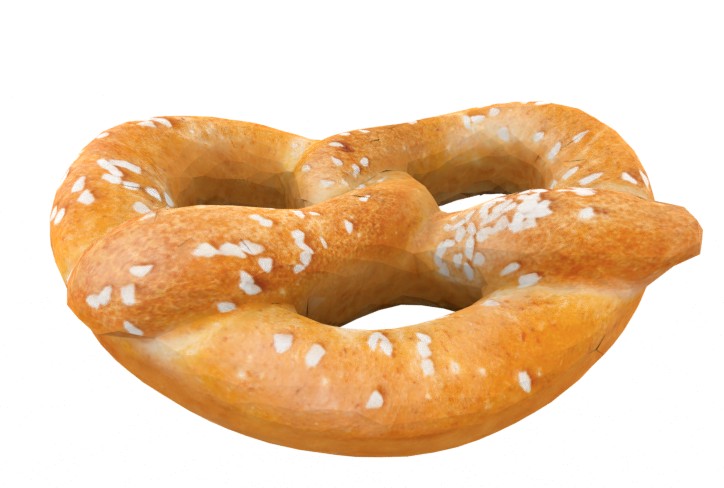} &
\includegraphics[width=0.30\columnwidth,trim={0cm 0cm 0cm 1cm},clip]{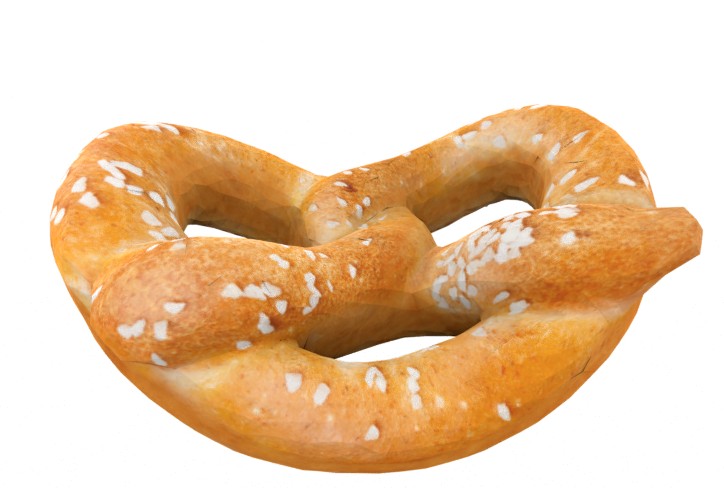} \\
    \end{tabular}
    \caption{We optimize shape, texture, and physical properties of an object from multi-view videos by feeding the extracted tetrahedral mesh from \ours{} into differentiable physics simulation~\cite{gradsim} and differentiable rendering~\cite{Laine2020diffrast} pipeline.
    In the bottom row (Reference shape), we set the mass density $D=0.300$ and fix the two ending points of an object, letting it drop and deform under gravity. We show the output at three timesteps (three columns).
    We initialize an SDF with a sphere geometry and randomly-guessed mass density ($D=1.500$, top row), and optimize in a two-stage manner. In the first stage, we utilize the beginning frame to optimize the shape and texture. In the second stage, we  extract a tetrahedral mesh (Section~\ref{sec:tetmesh}) using \ours{} and apply differentiable physics simulation pipeline to optimize the physical parameters. 
     The optimized physical parameters ($D=0.311$, middle row) are close to the ground truth ($D=0.300$, bottom row). 
    }
    \label{fig:physicus}
    \end{figure}
}

\newcommand{\figRegAbl}{
    \begin{figure}
    \centering
    \setlength{\tabcolsep}{0mm}
    \begin{tabular}{cc}
        \includegraphics[width=0.45\columnwidth]{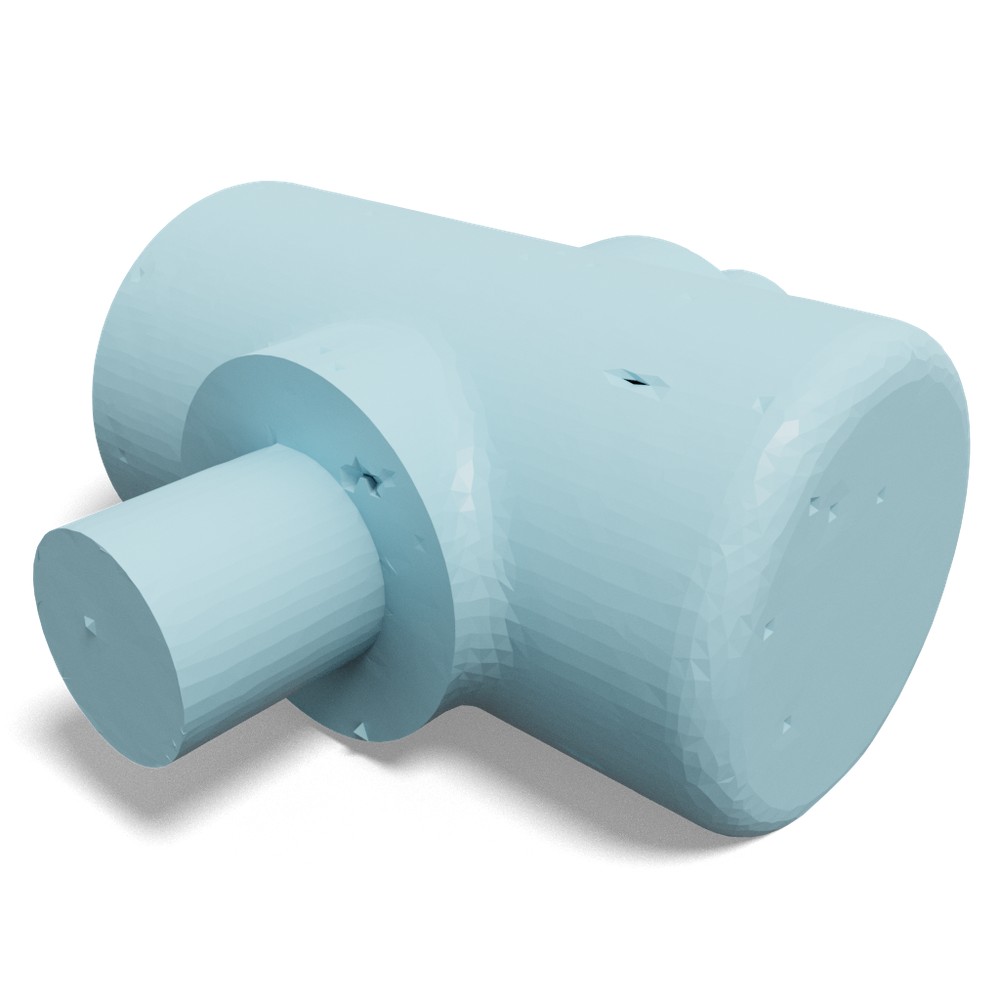} &
        \includegraphics[width=0.45\columnwidth]{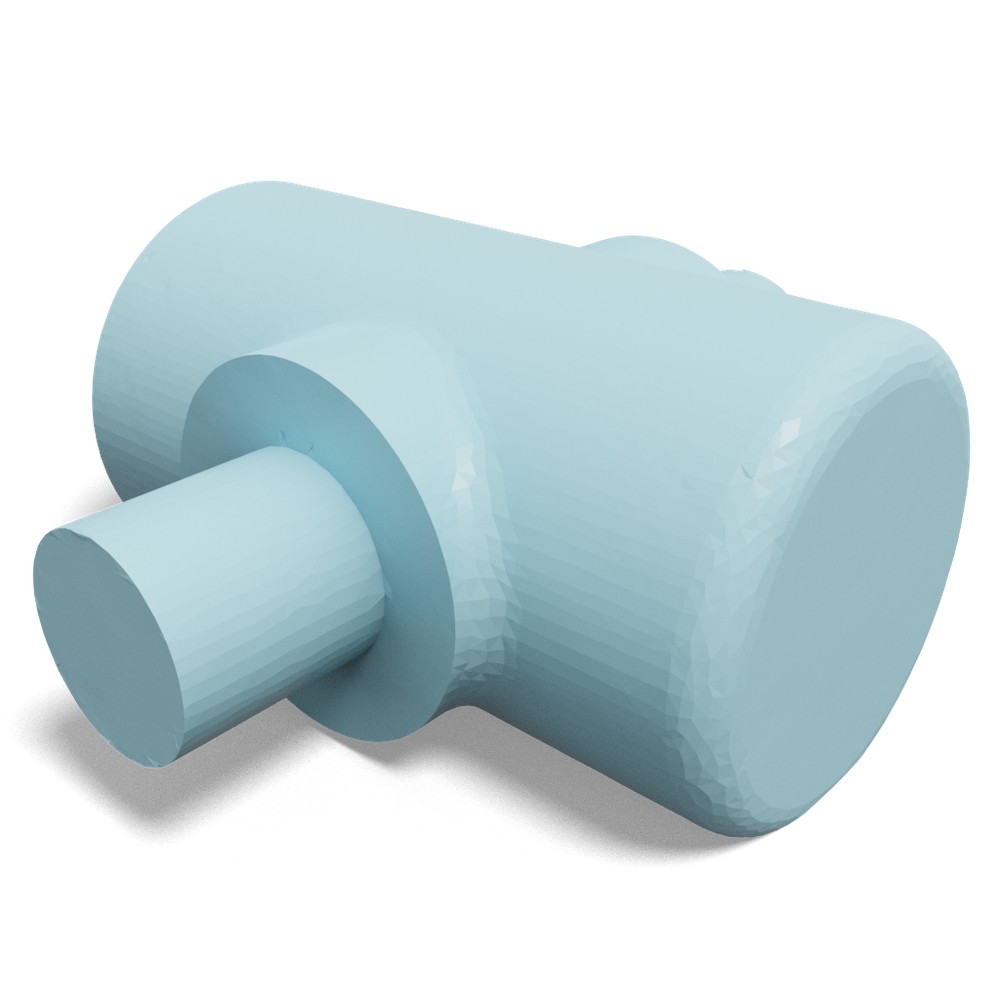} \\
        $\mathcal{L}_{\textrm{dev}}$ off & $\mathcal{L}_{\textrm{dev}}$ on
    \end{tabular}
    \caption{Influence of the regularizer $\mathcal{L}_{\textrm{dev}}$. The reconstructed shape with $\mathcal{L}_{\textrm{dev}}$ has less visible seams compared to the result without $\mathcal{L}_{\textrm{dev}}$ applied.}
    \label{fig:reg_abl}
    \end{figure}
}

\newcommand{\figNVDIFFRECNerfSupplemental}{
	\begin{figure*}
		\centering
		\setlength{\tabcolsep}{0mm}
		\begin{tabular}{cccccccccc}
			\rotatebox[origin=c]{90}{Chair} &
			\raisebox{-0.5\height}{\includegraphics[width=0.24\columnwidth]{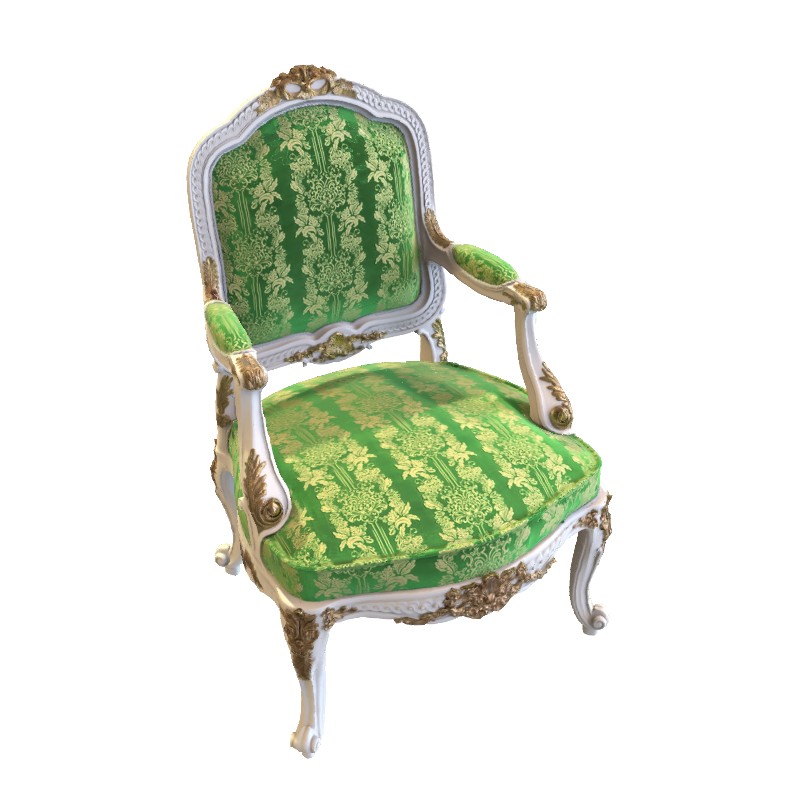}} &
			\raisebox{-0.5\height}{\includegraphics[width=0.24\columnwidth]{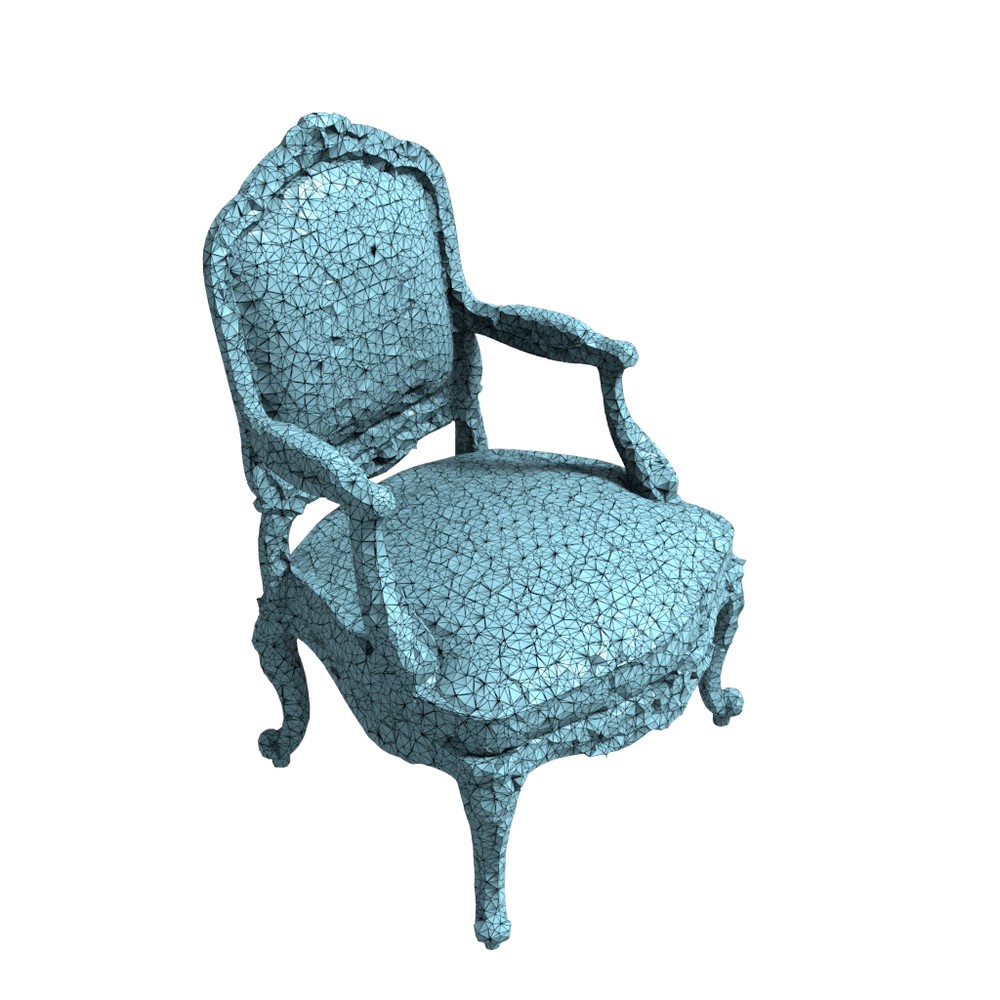}} &
			\raisebox{-0.5\height}{\includegraphics[width=0.24\columnwidth]{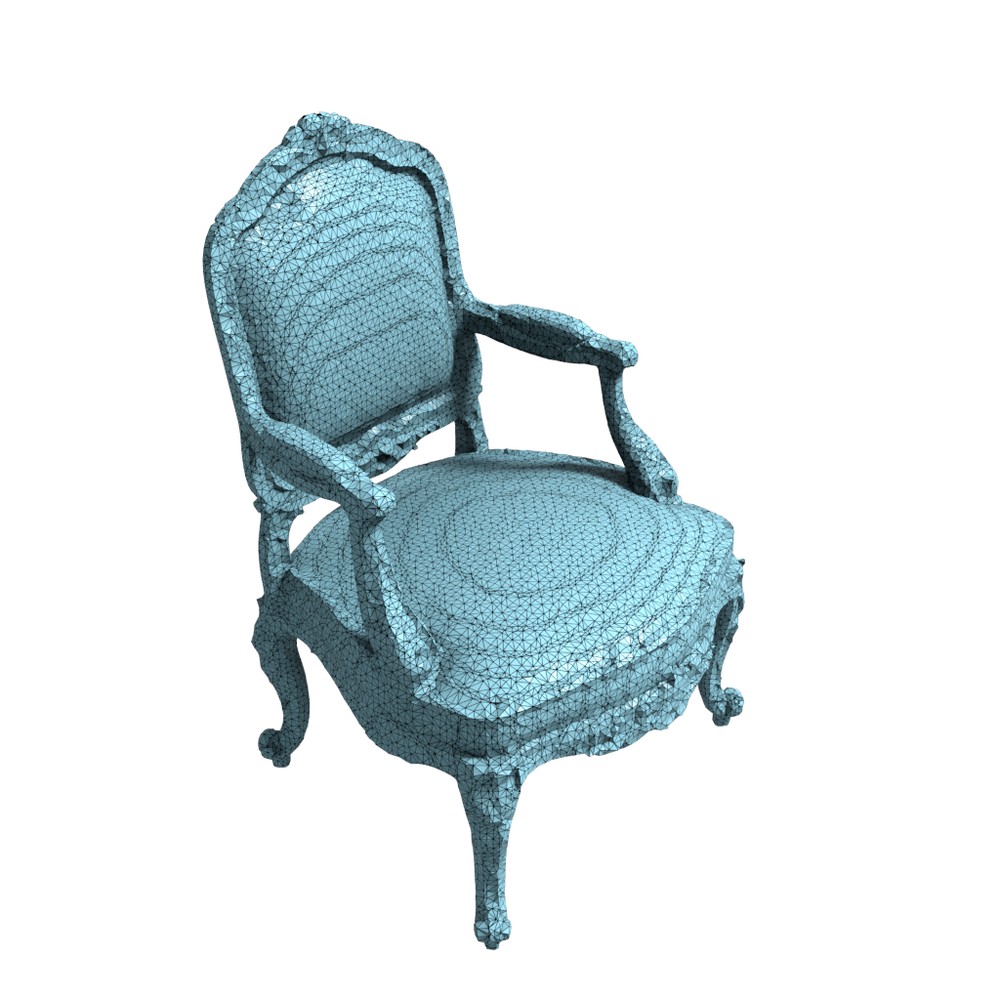}} &
			\raisebox{-0.5\height}{\includegraphics[width=0.24\columnwidth]{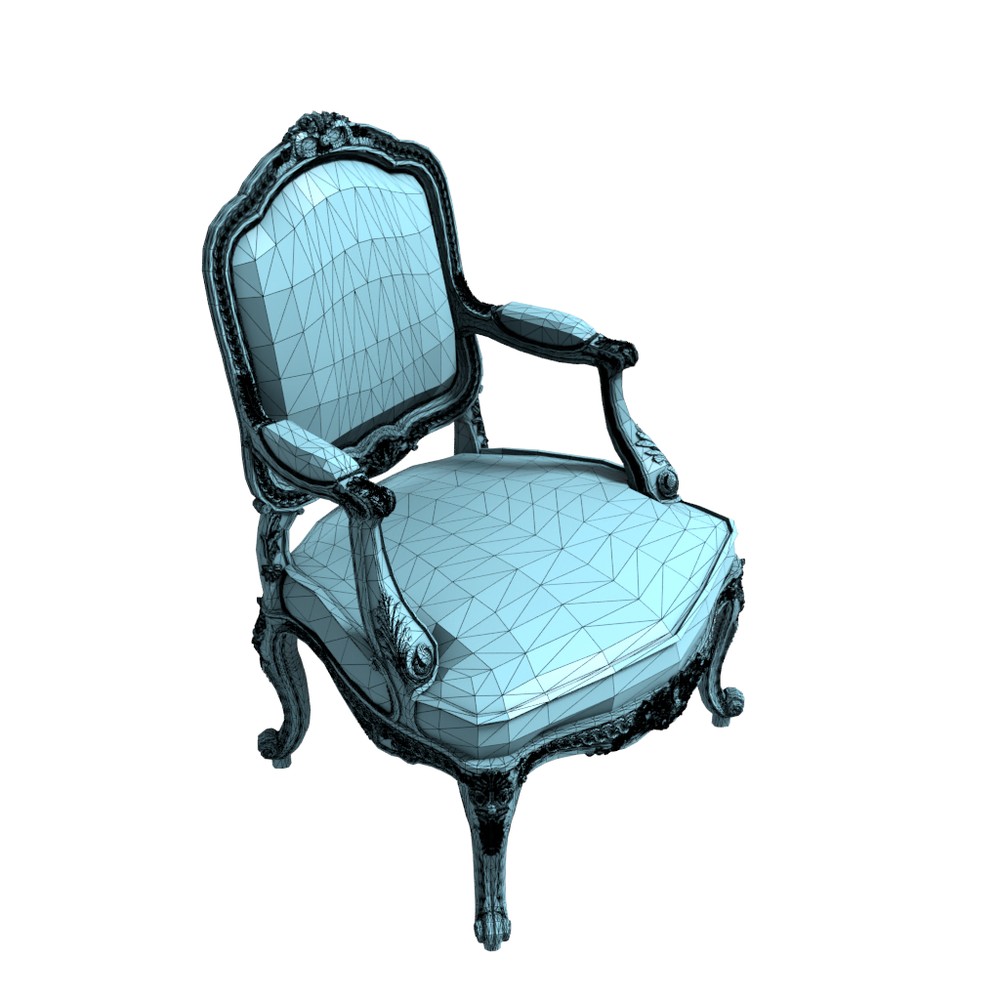}} &
						
			\rotatebox[origin=c]{90}{Drums} &
			\raisebox{-0.5\height}{\includegraphics[width=0.24\columnwidth]{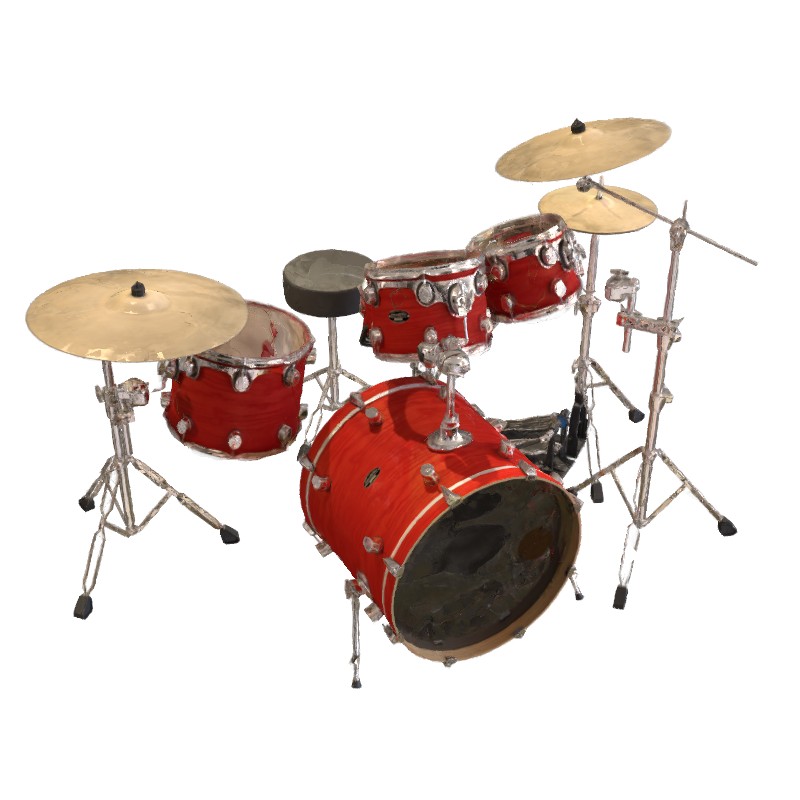}} &
			\raisebox{-0.5\height}{\includegraphics[width=0.24\columnwidth]{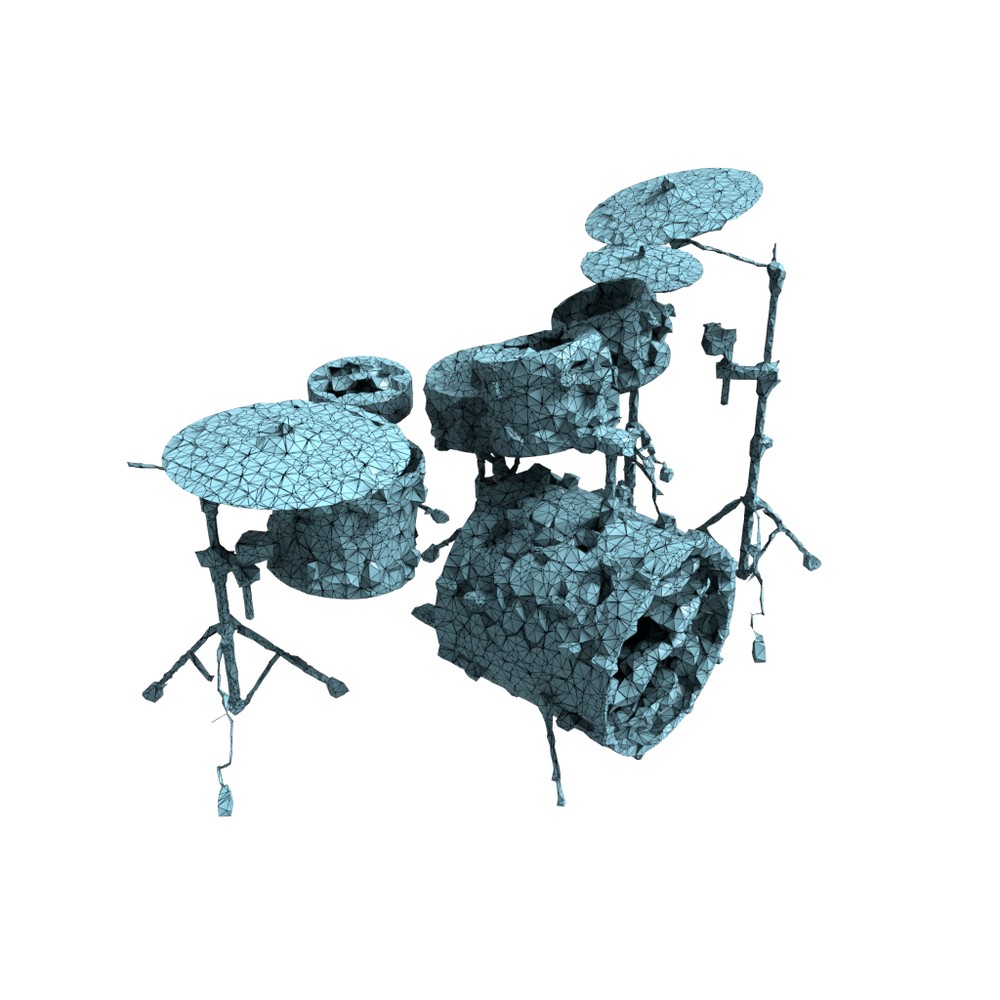}} &
			\raisebox{-0.5\height}{\includegraphics[width=0.24\columnwidth]{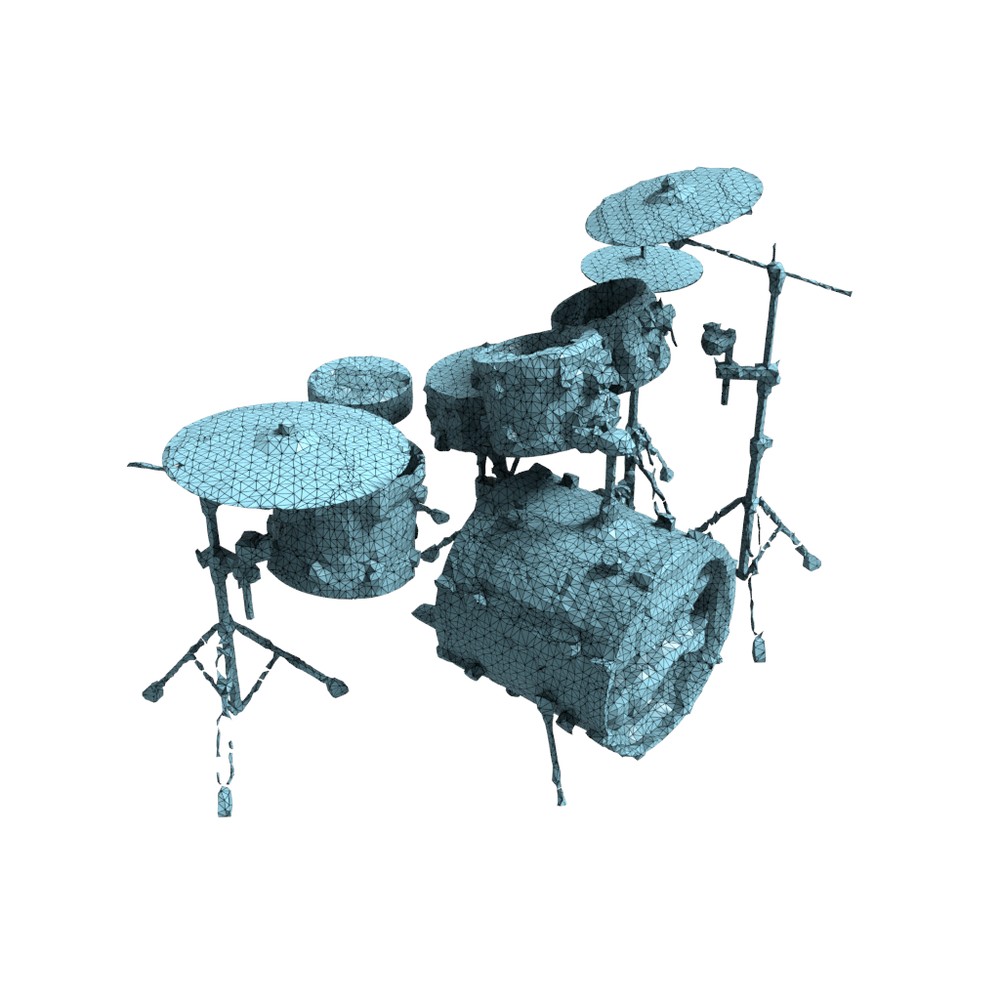}} &
			\raisebox{-0.5\height}{\includegraphics[width=0.24\columnwidth]{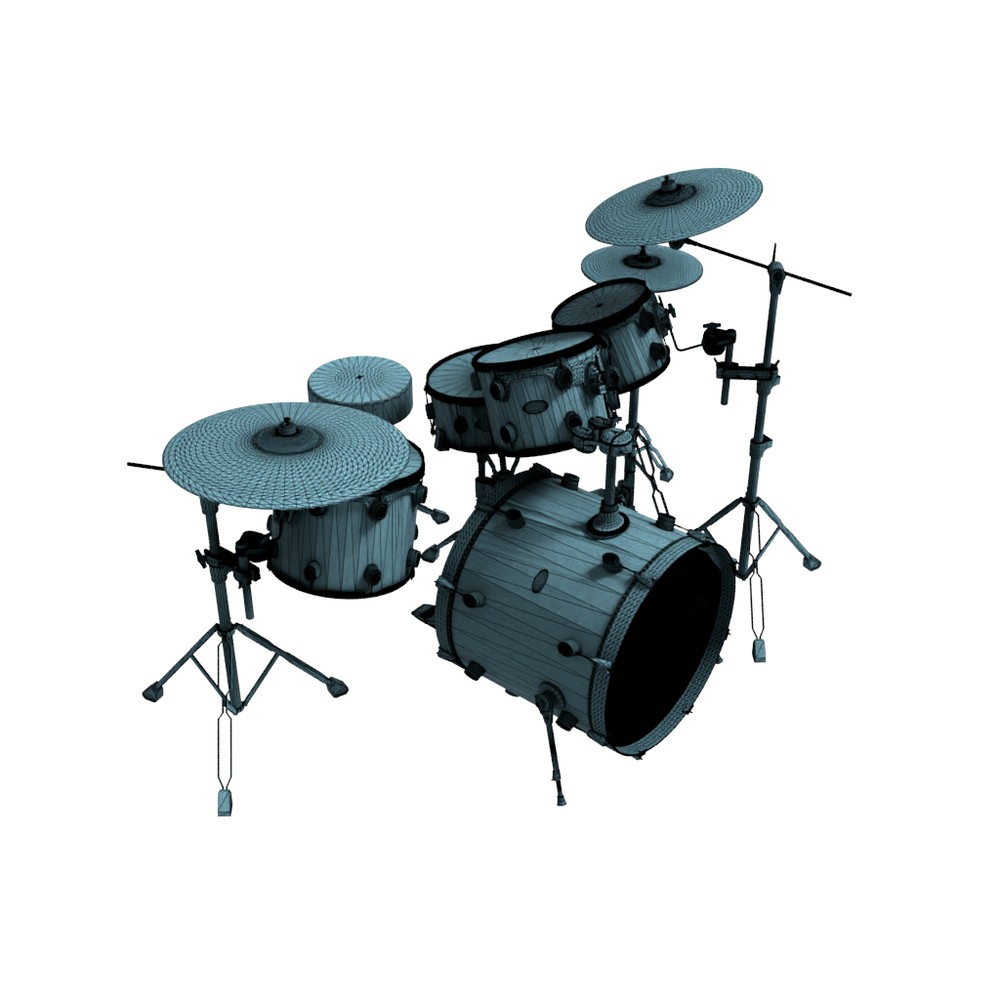}} \\

			\rotatebox[origin=c]{90}{Ficus} &
			\raisebox{-0.5\height}{\includegraphics[width=0.24\columnwidth]{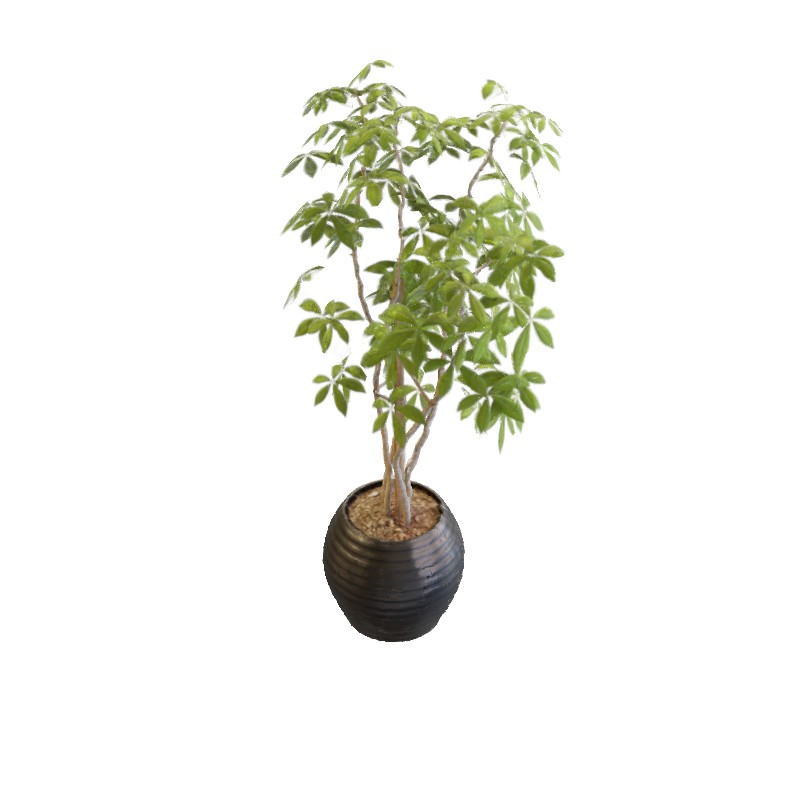}} &
			\raisebox{-0.5\height}{\includegraphics[width=0.24\columnwidth]{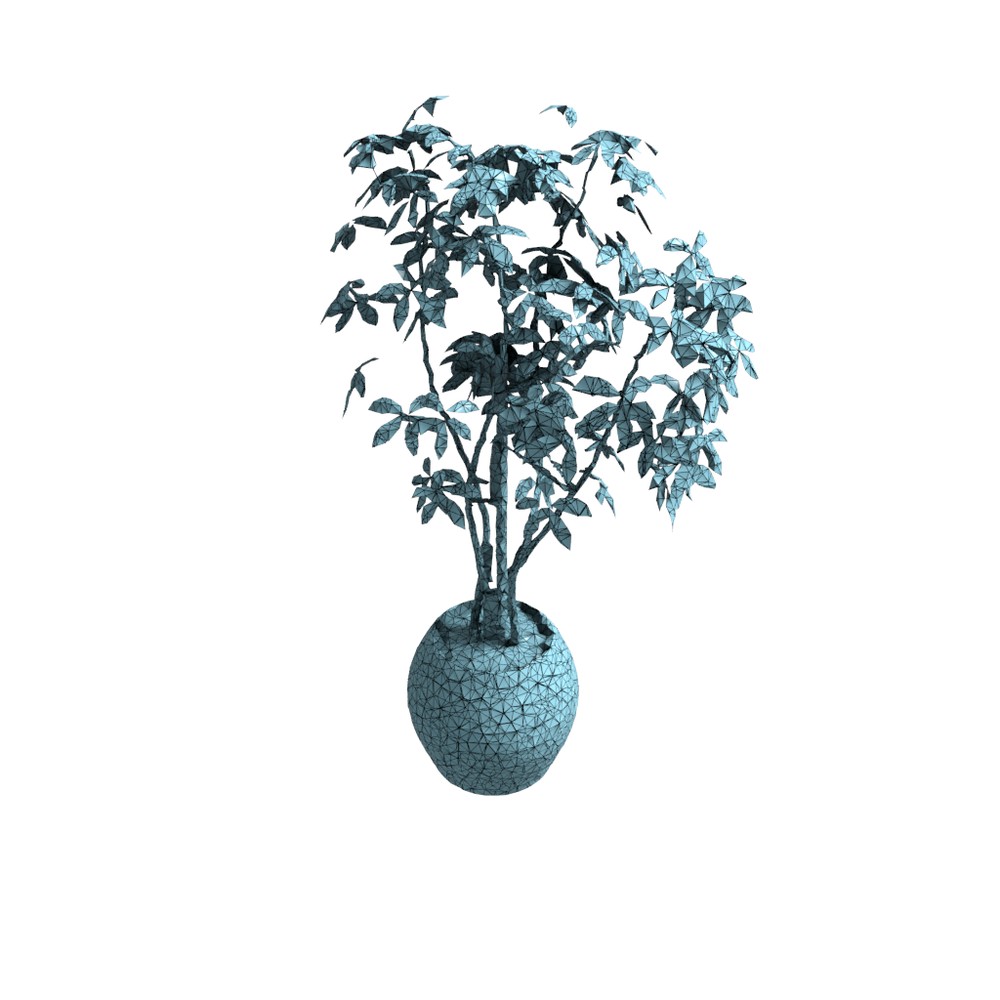}} &
			\raisebox{-0.5\height}{\includegraphics[width=0.24\columnwidth]{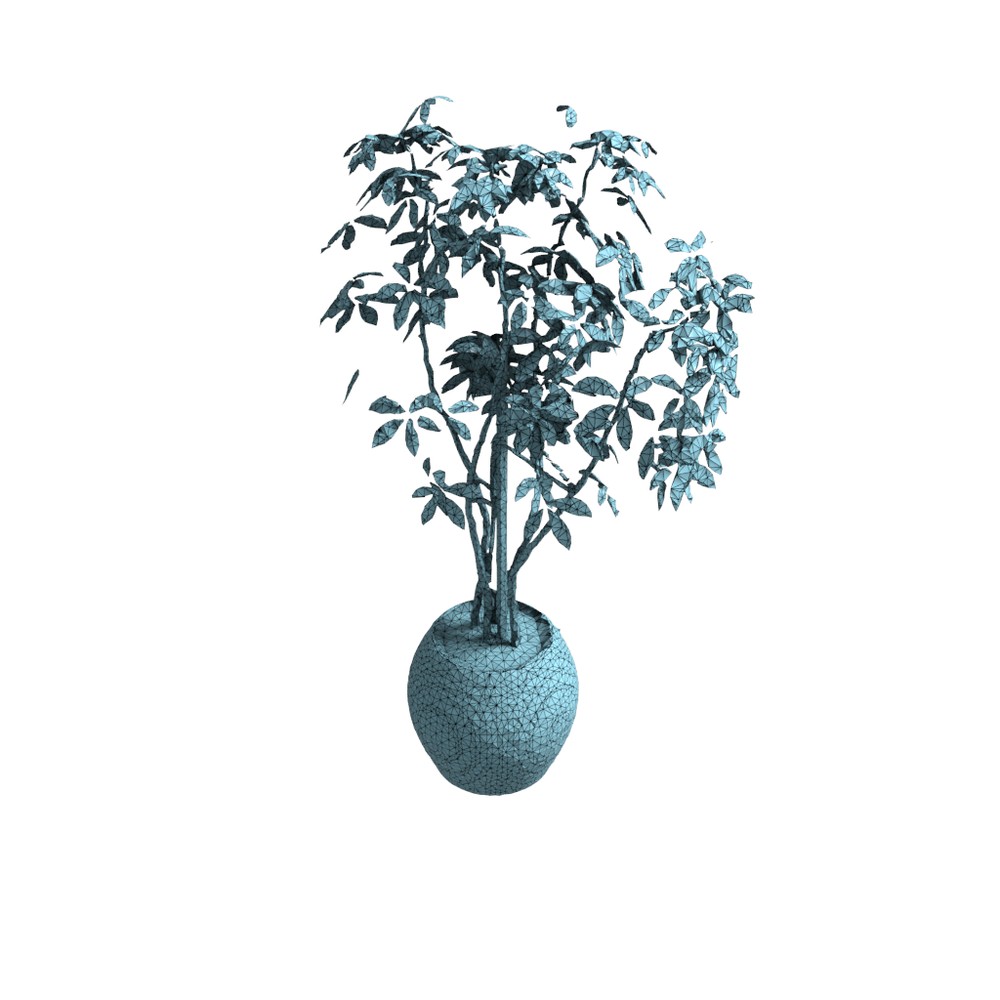}} &
			\raisebox{-0.5\height}{\includegraphics[width=0.24\columnwidth]{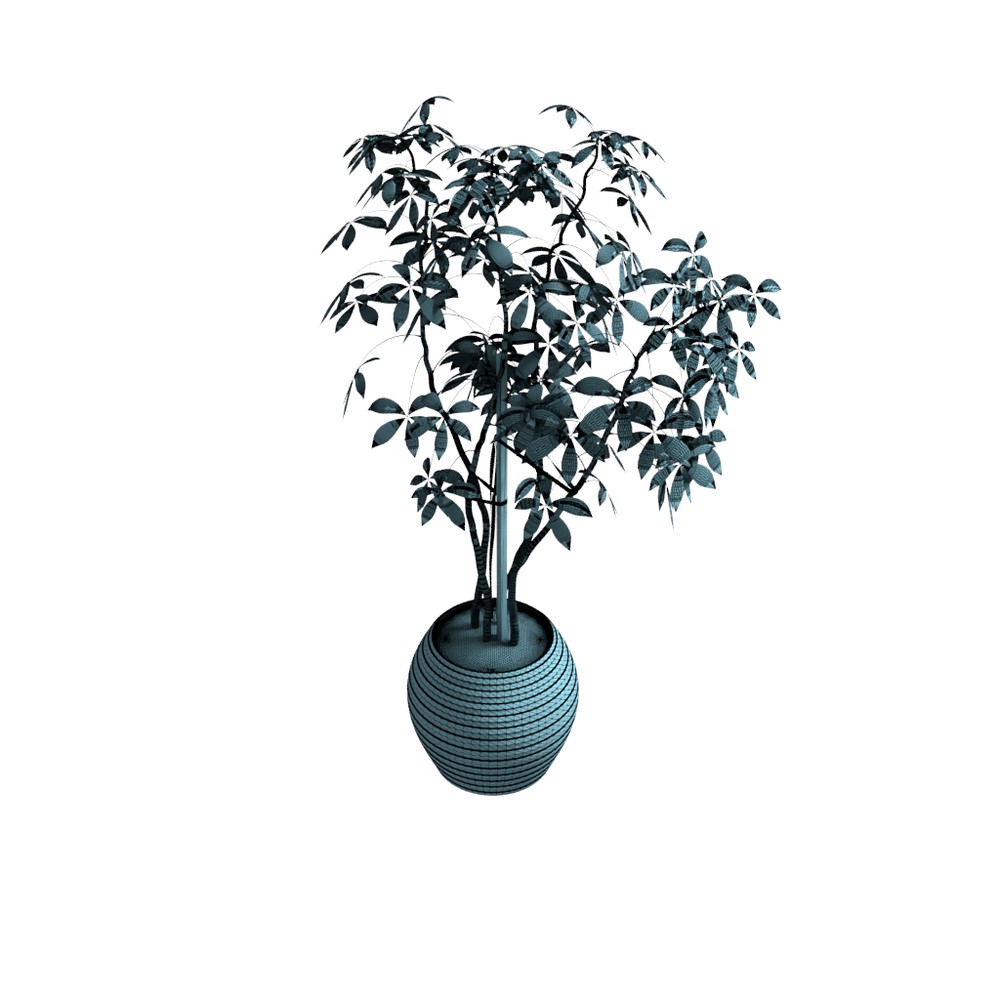}} &

			\rotatebox[origin=c]{90}{Hotdog} &
			\raisebox{-0.5\height}{\includegraphics[width=0.24\columnwidth]{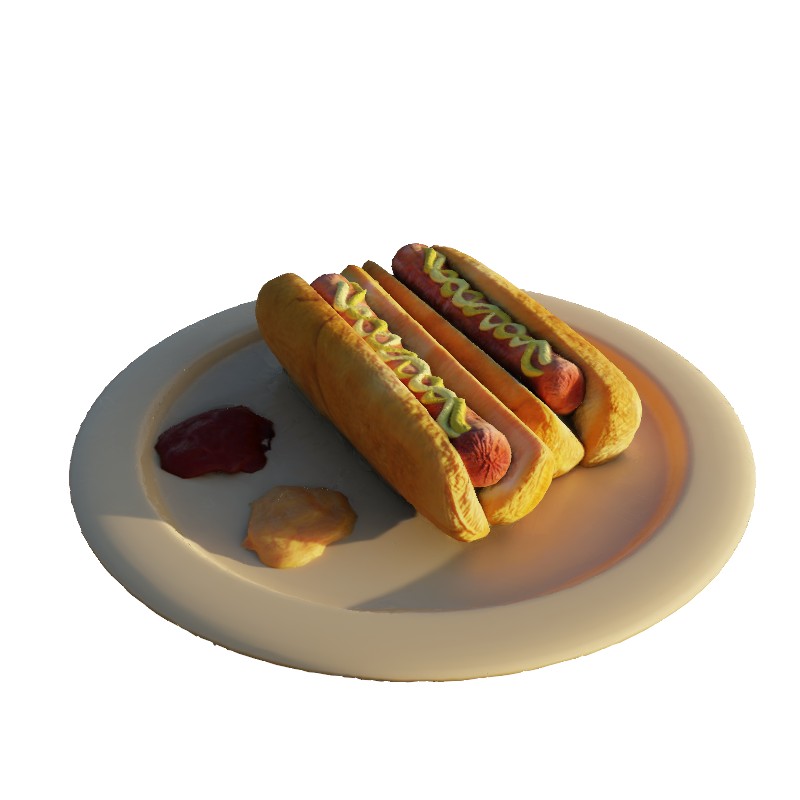}} &
			\raisebox{-0.5\height}{\includegraphics[width=0.24\columnwidth]{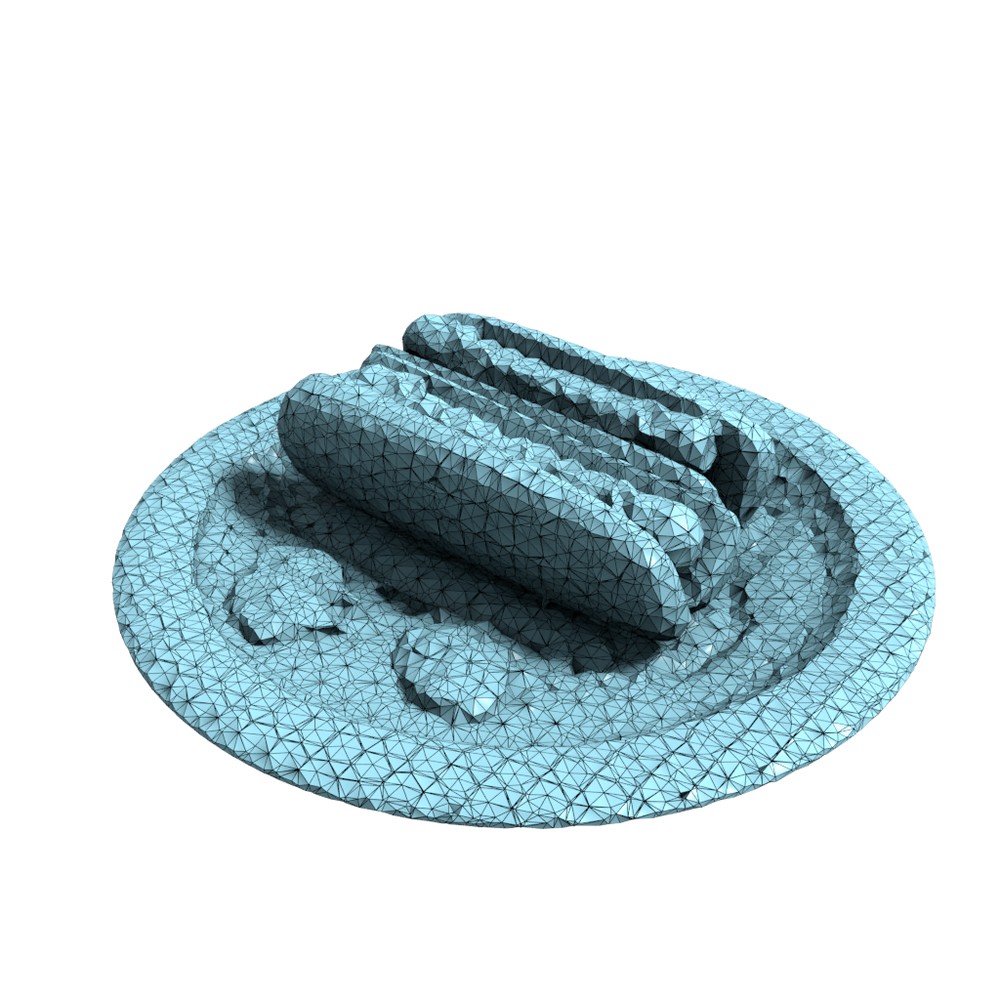}} &
			\raisebox{-0.5\height}{\includegraphics[width=0.24\columnwidth]{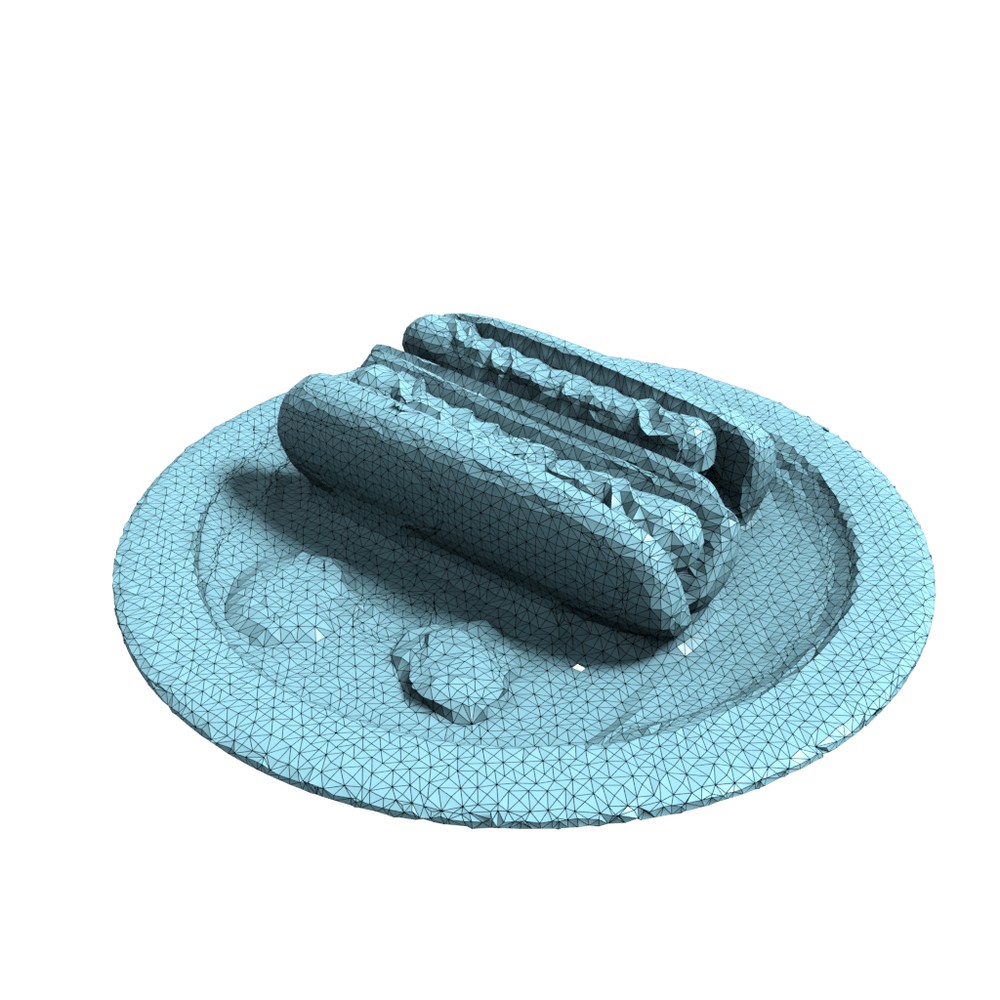}} &
			\raisebox{-0.5\height}{\includegraphics[width=0.24\columnwidth]{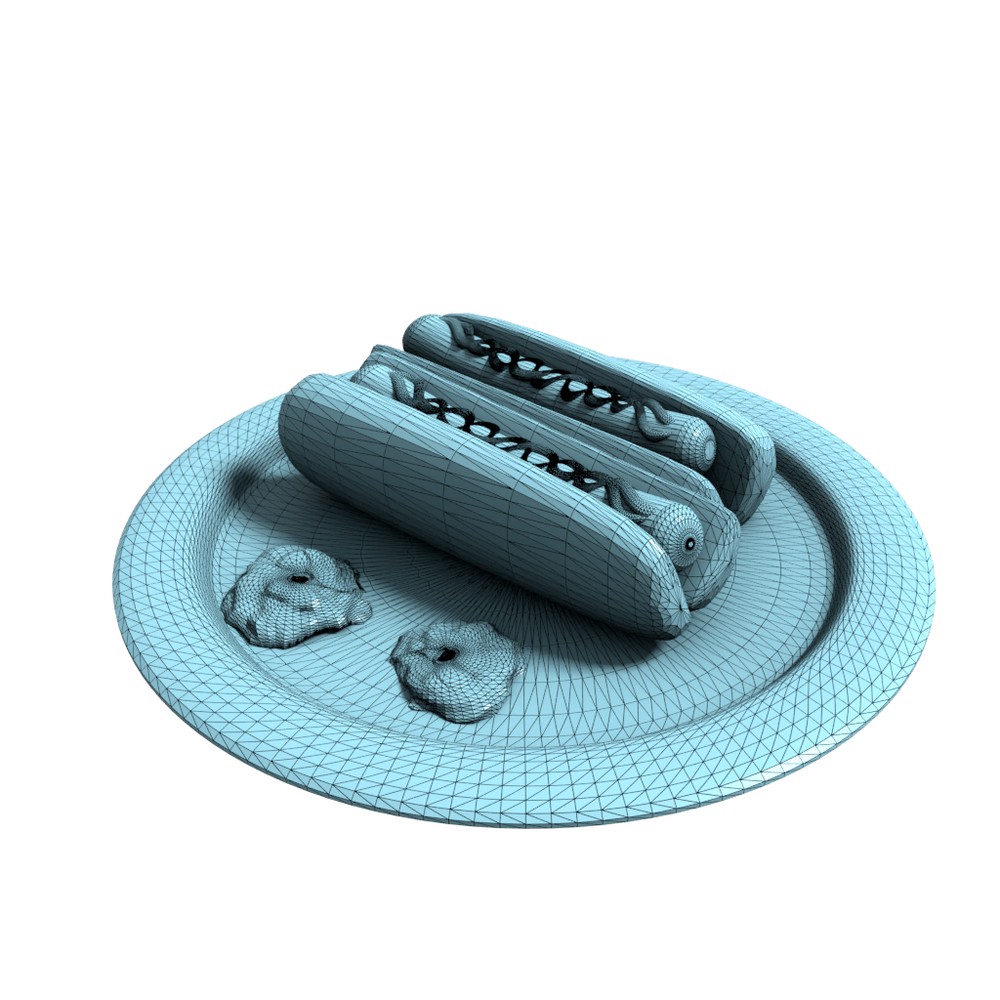}} \\

			\rotatebox[origin=c]{90}{Lego} &
			\raisebox{-0.5\height}{\includegraphics[width=0.24\columnwidth]{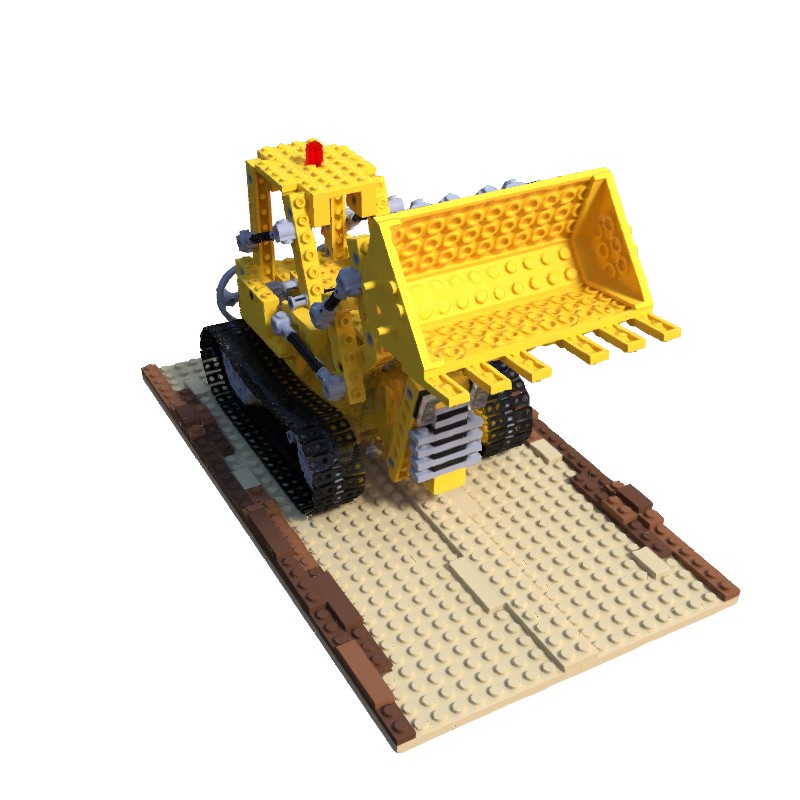}} &
			\raisebox{-0.5\height}{\includegraphics[width=0.24\columnwidth]{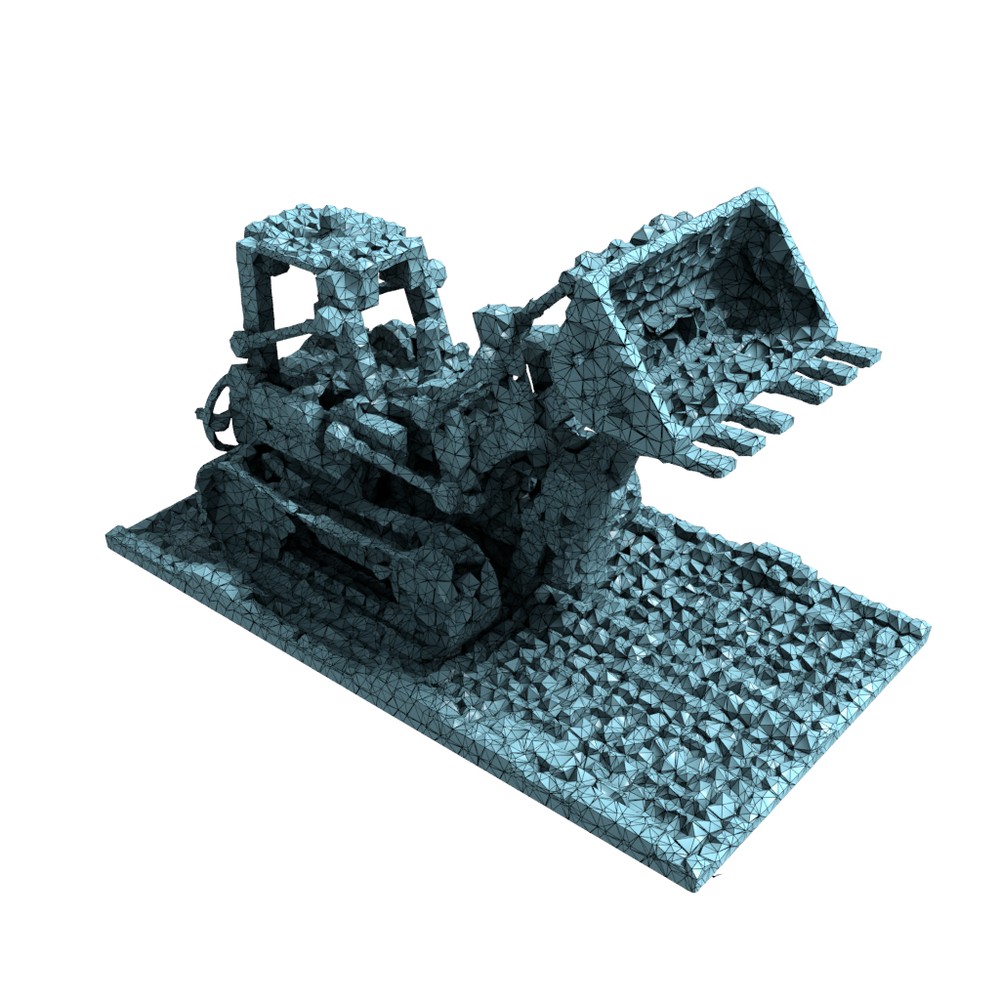}} &
			\raisebox{-0.5\height}{\includegraphics[width=0.24\columnwidth]{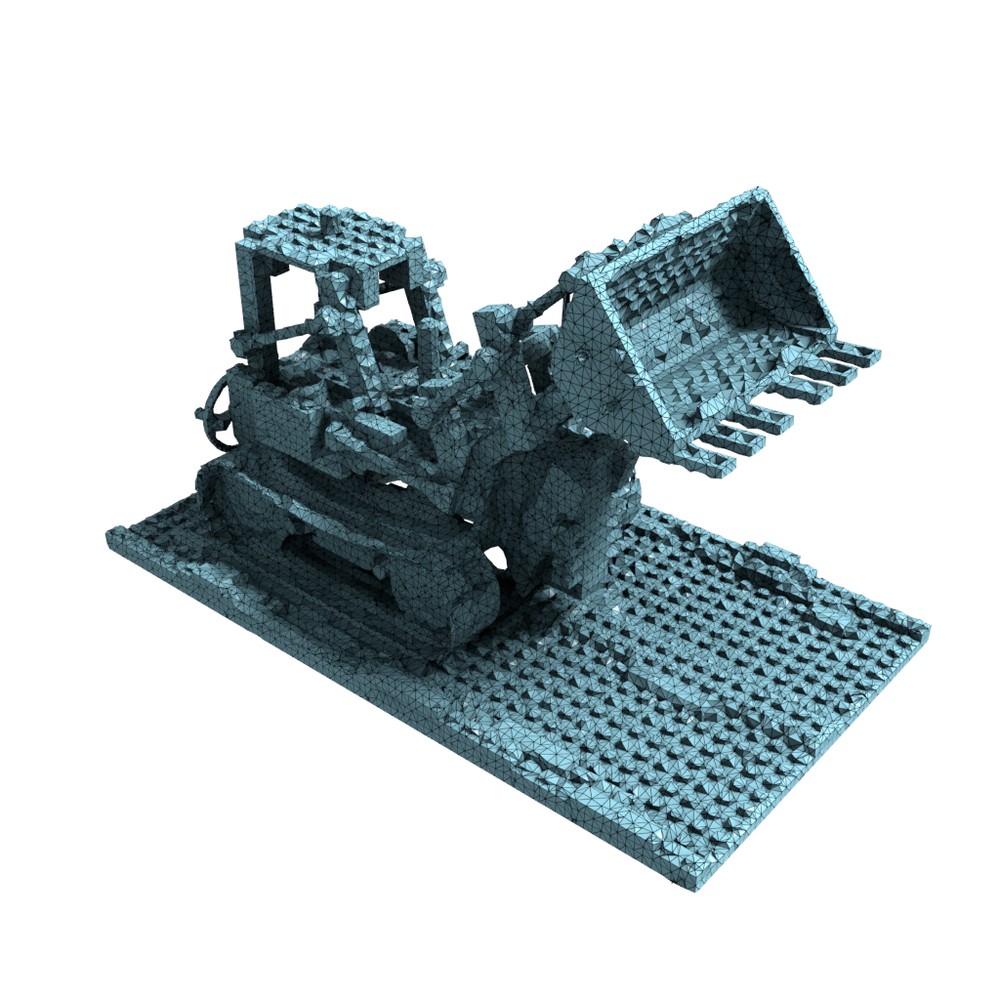}} &
			\raisebox{-0.5\height}{\includegraphics[width=0.24\columnwidth]{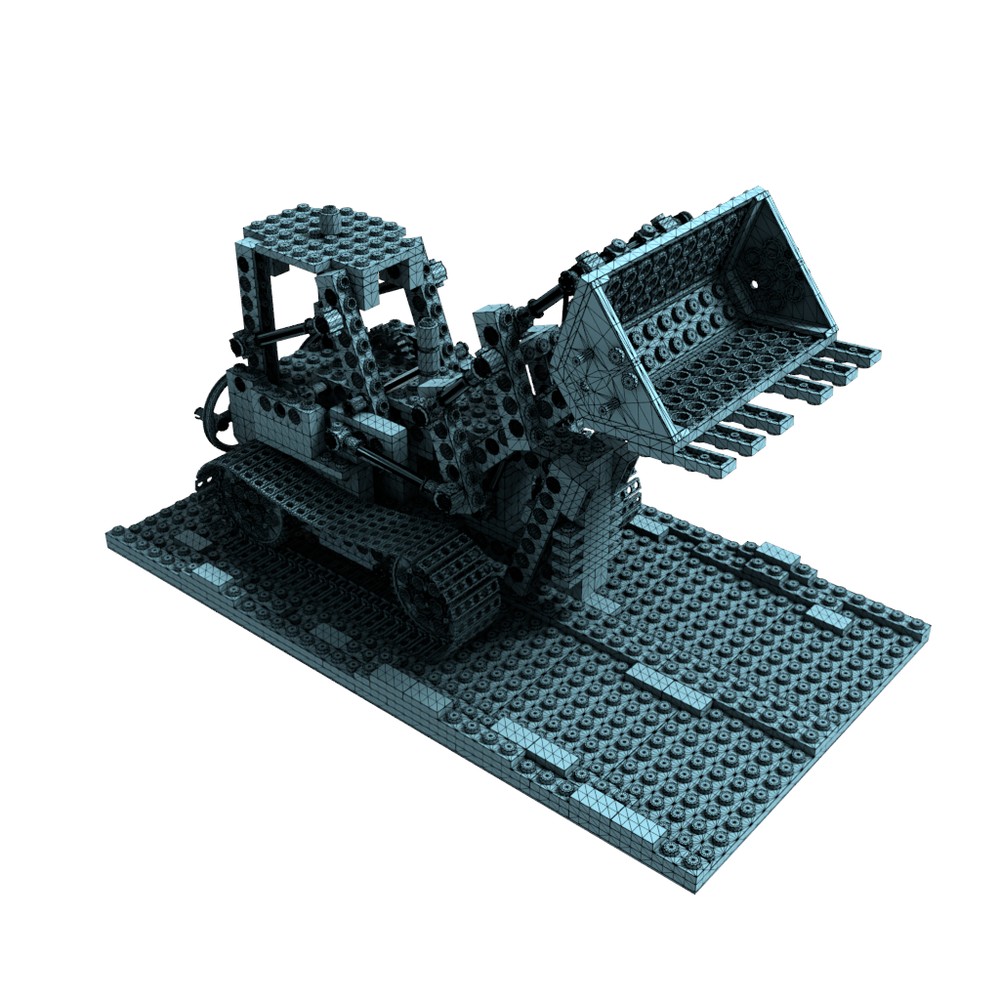}} &

			\rotatebox[origin=c]{90}{Materials} &
			\raisebox{-0.5\height}{\includegraphics[width=0.22\columnwidth]{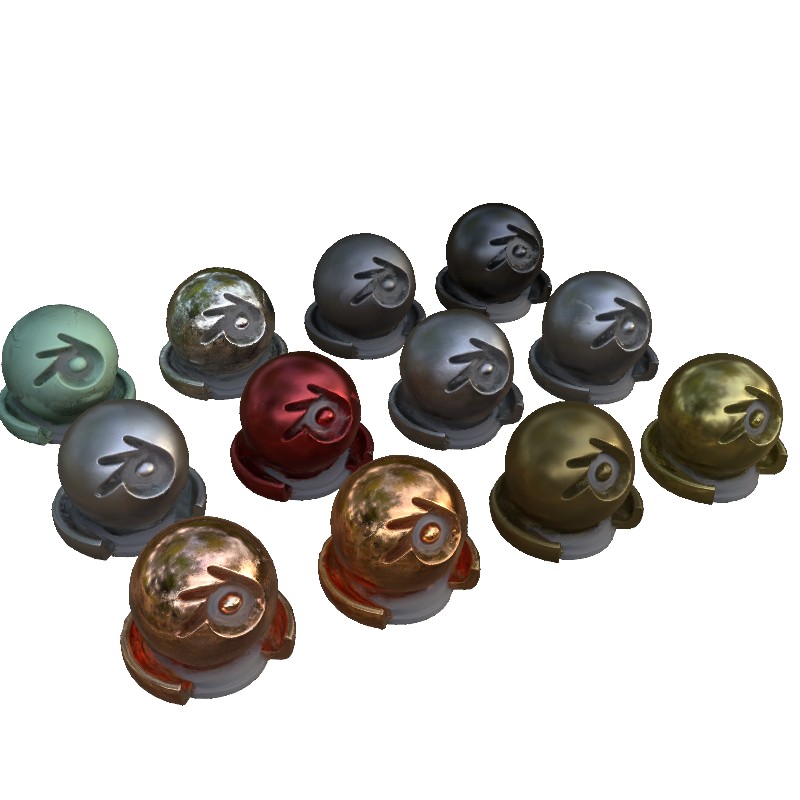}} &
			\raisebox{-0.5\height}{\includegraphics[width=0.24\columnwidth]{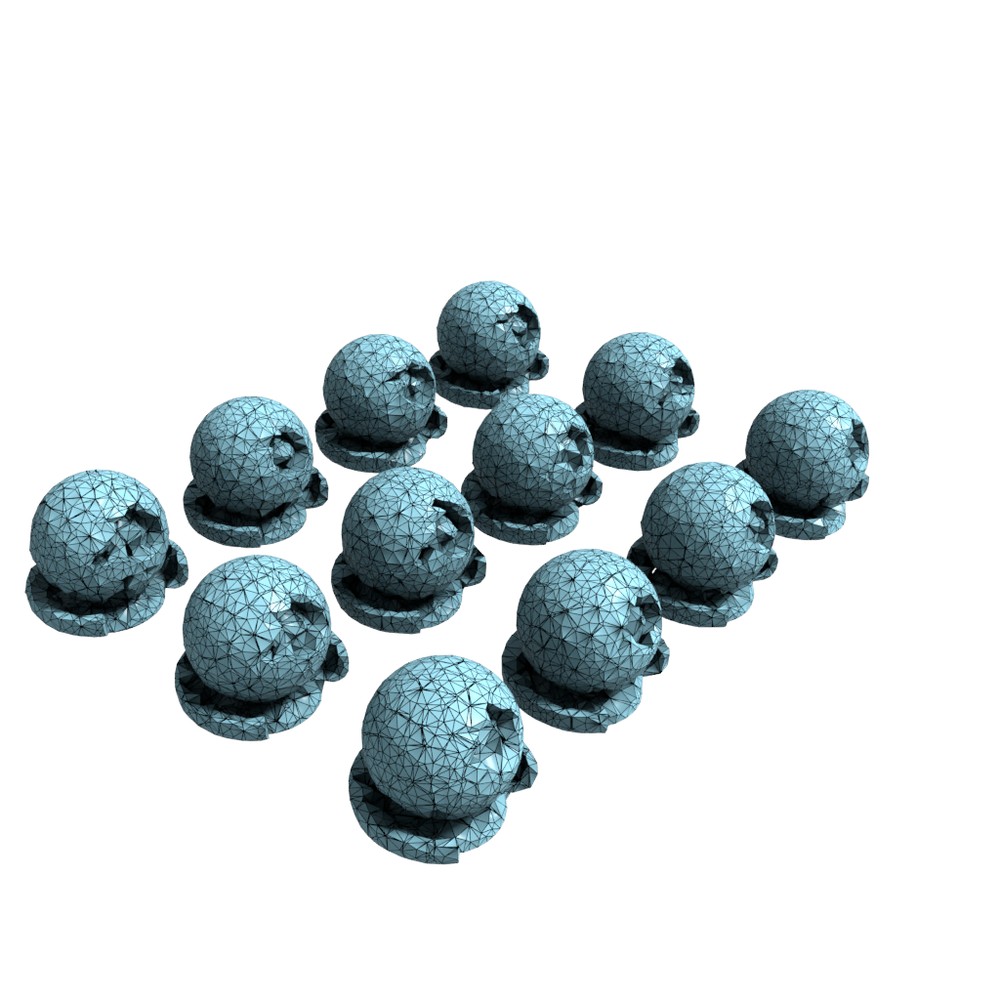}} &
			\raisebox{-0.5\height}{\includegraphics[width=0.24\columnwidth]{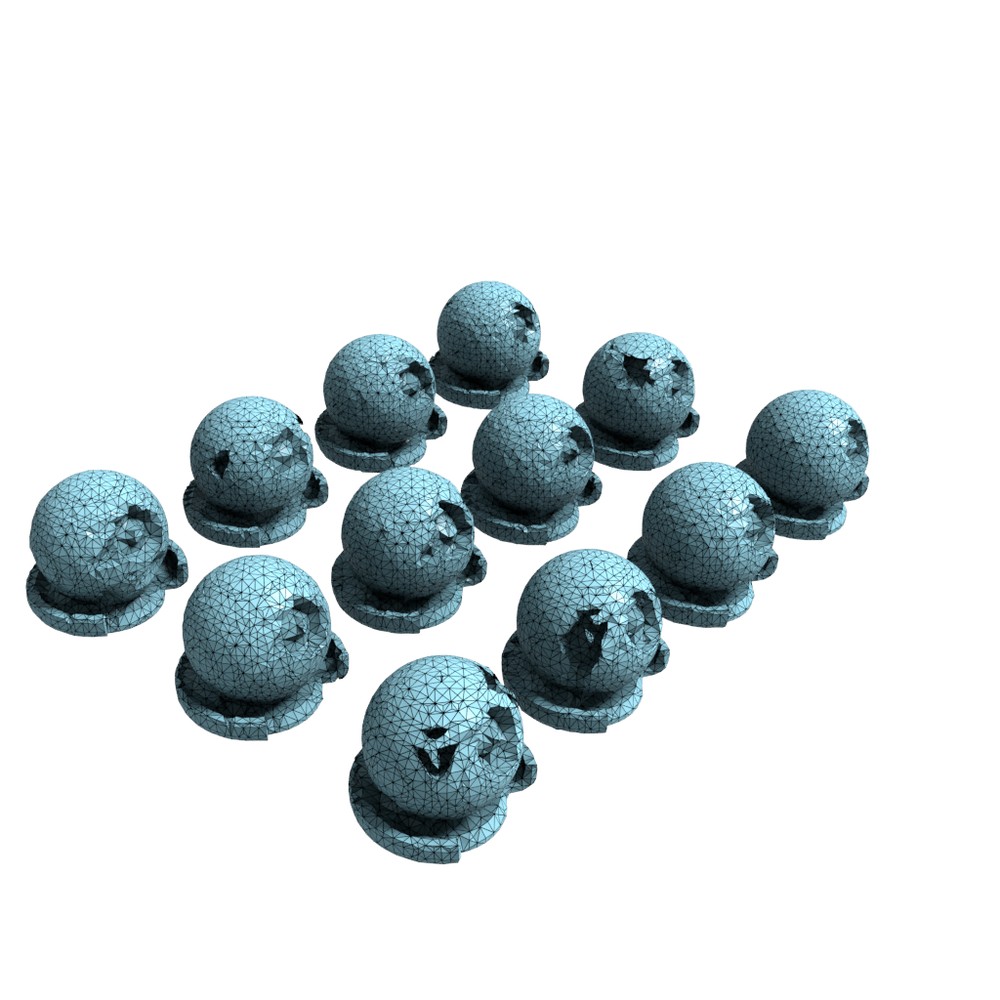}} &
			\raisebox{-0.5\height}{\includegraphics[width=0.24\columnwidth]{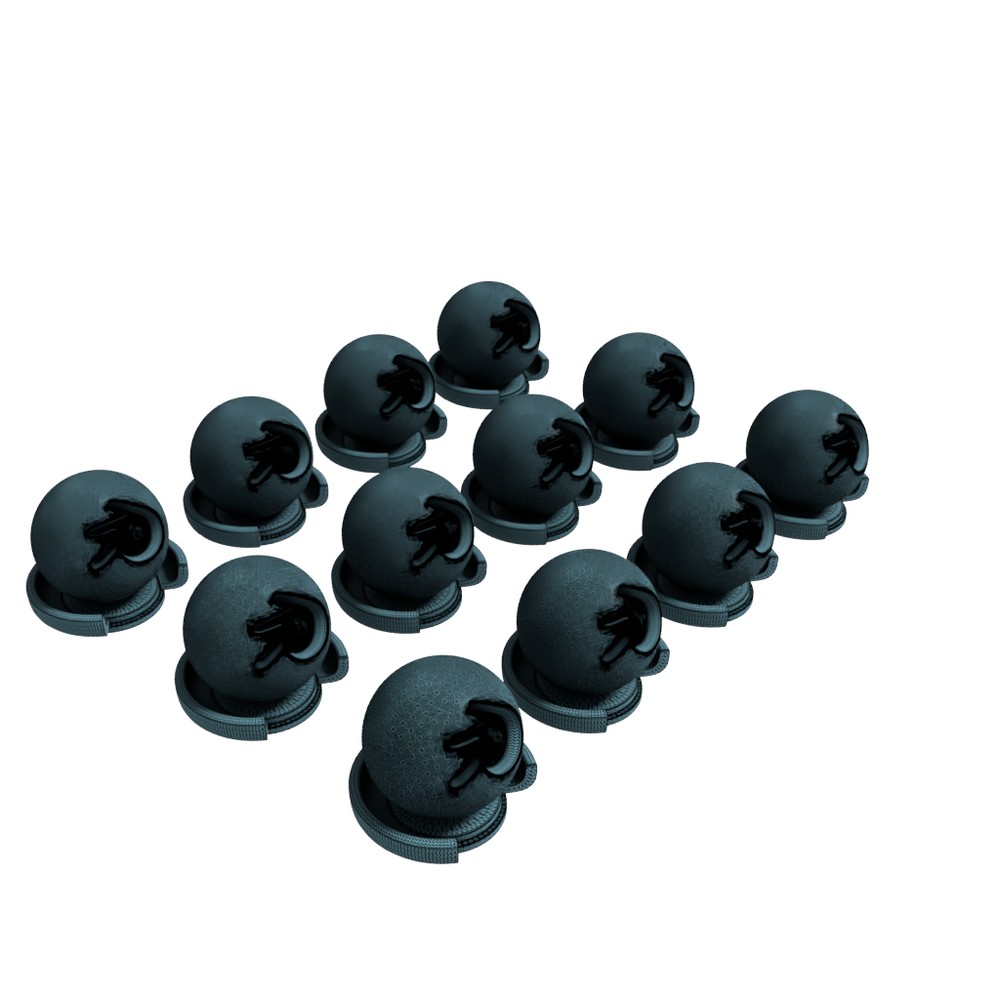}} \\

			\rotatebox[origin=c]{90}{Mic} &
			\raisebox{-0.5\height}{\includegraphics[width=0.24\columnwidth]{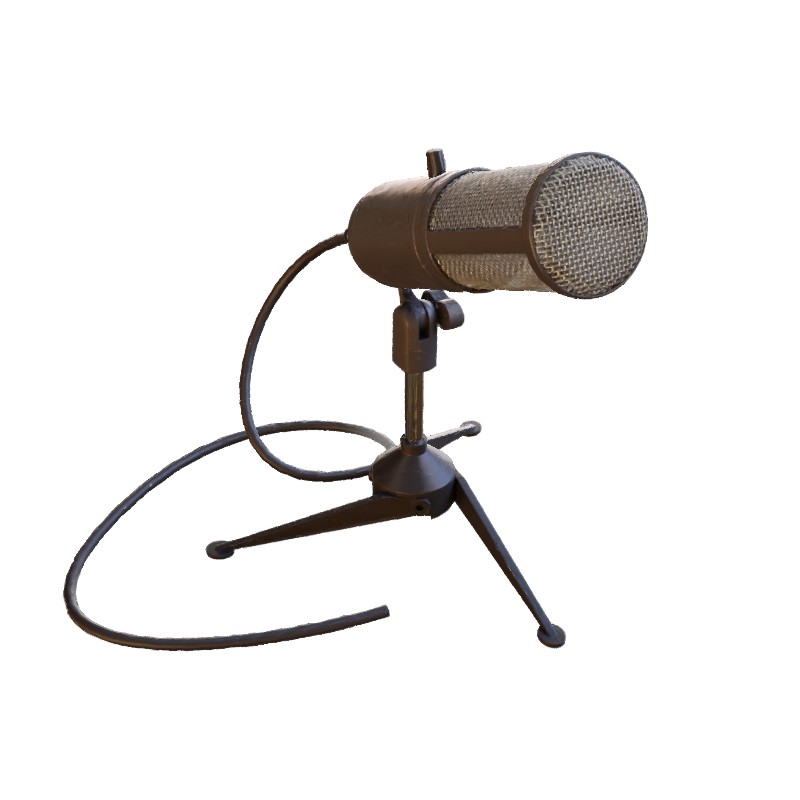}} &
			\raisebox{-0.5\height}{\includegraphics[width=0.24\columnwidth]{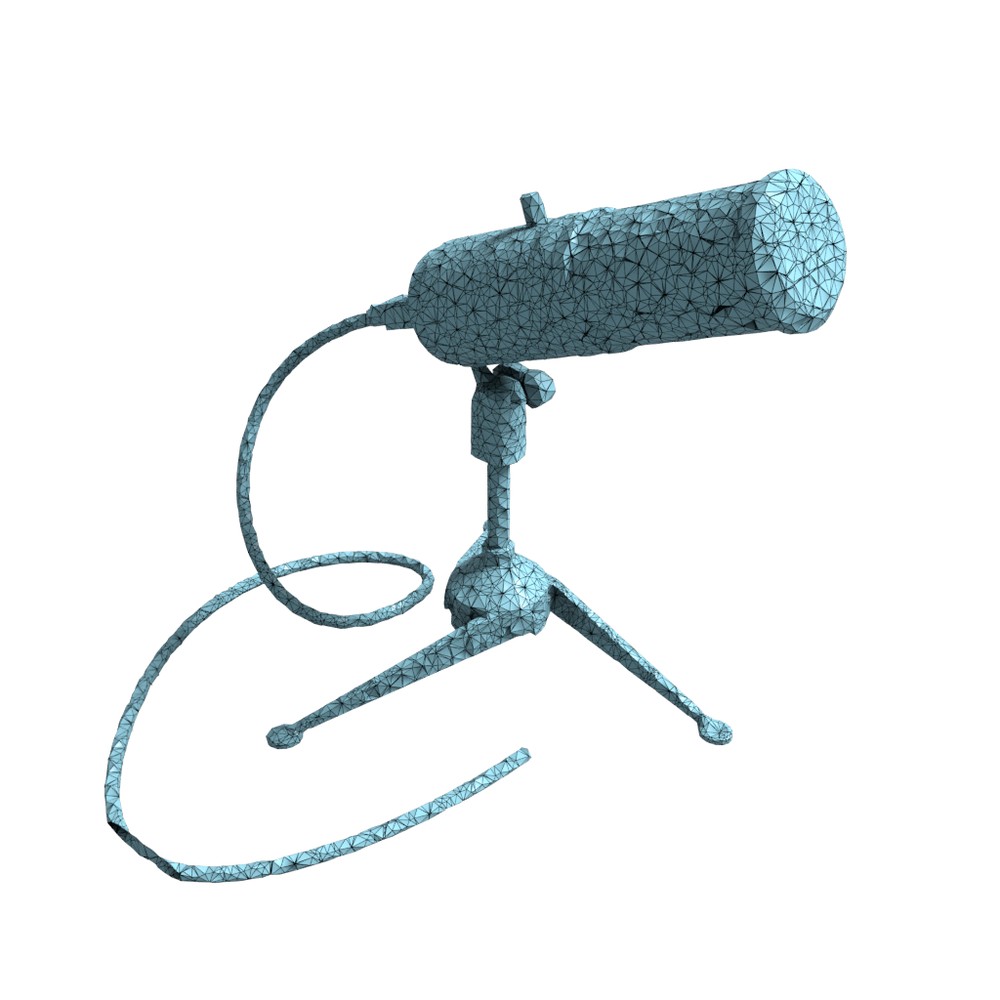}} &
			\raisebox{-0.5\height}{\includegraphics[width=0.24\columnwidth]{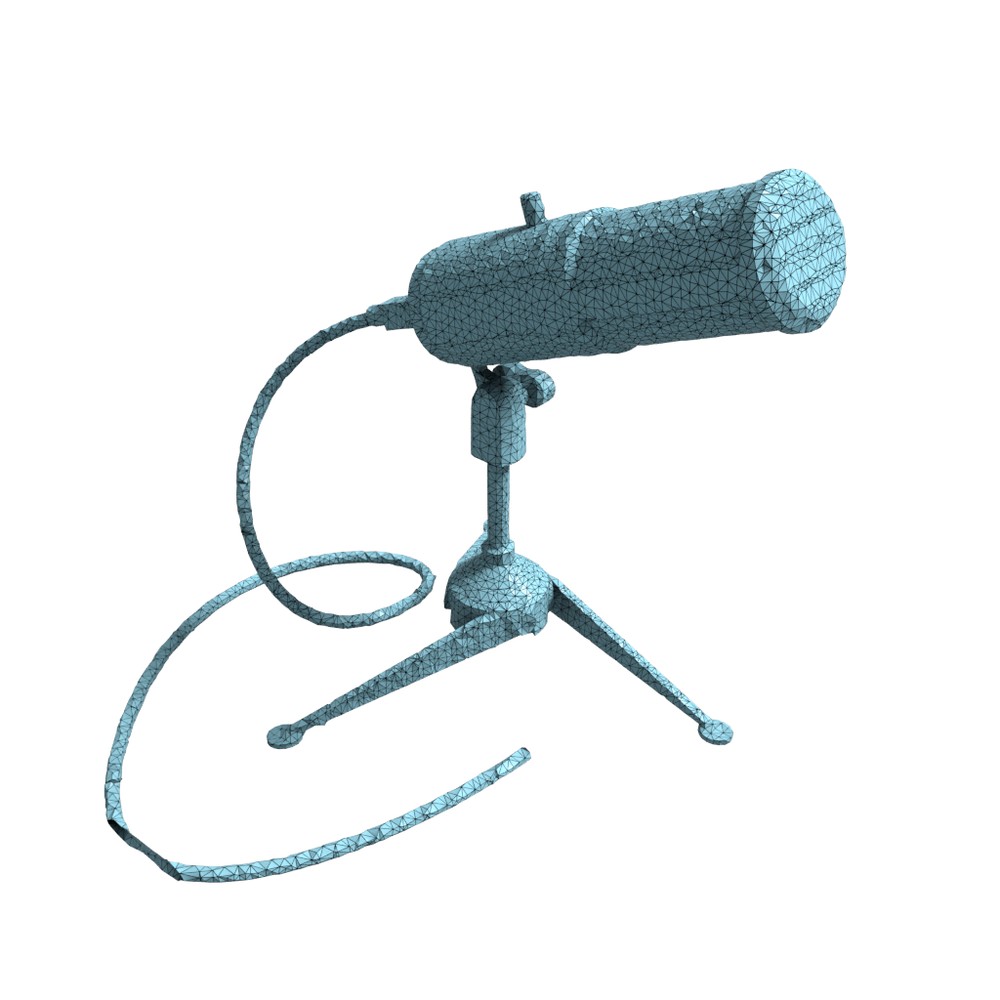}} &
			\raisebox{-0.5\height}{\includegraphics[width=0.24\columnwidth]{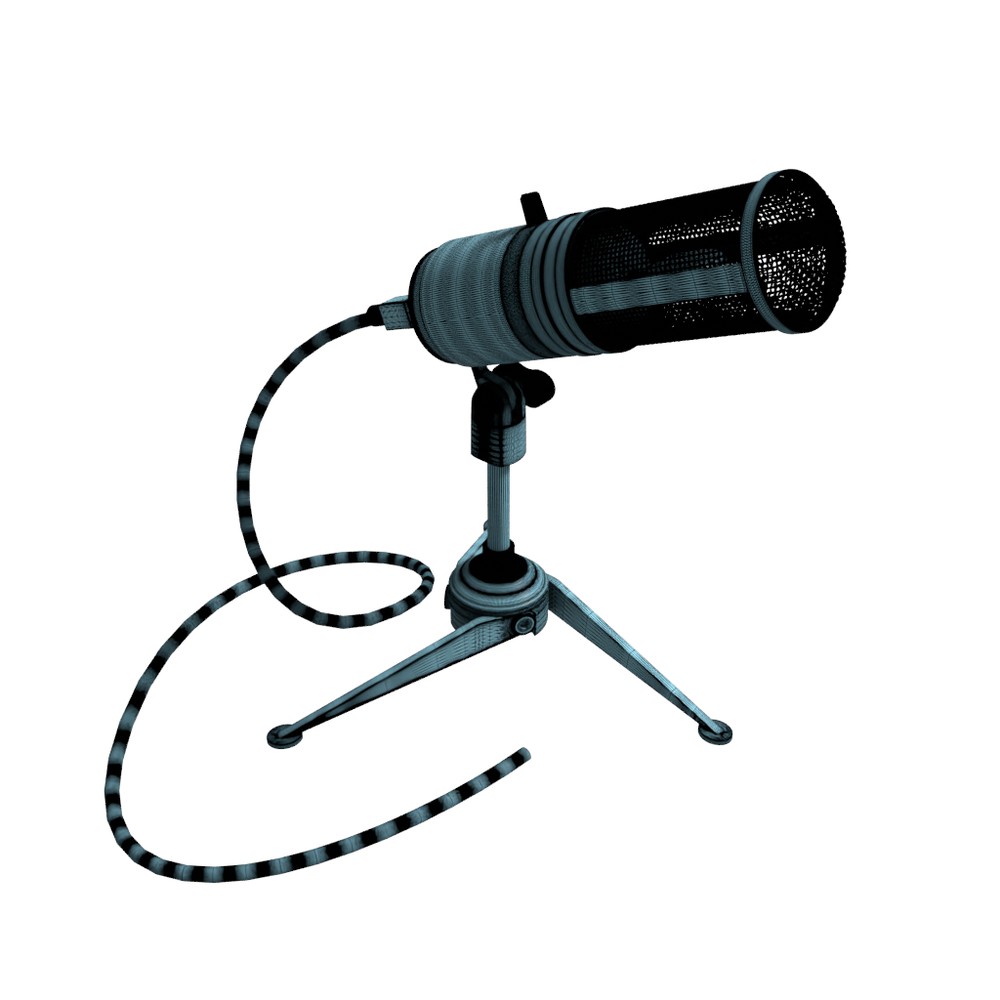}} &

			\rotatebox[origin=c]{90}{Ship} &
			\raisebox{-0.5\height}{\includegraphics[width=0.24\columnwidth]{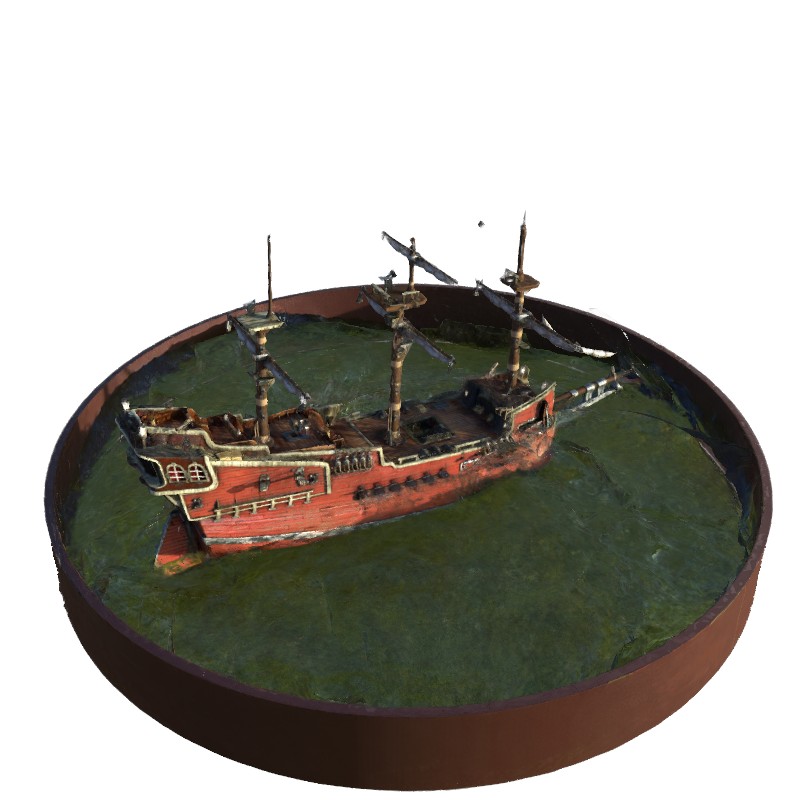}} &
			\raisebox{-0.5\height}{\includegraphics[width=0.24\columnwidth]{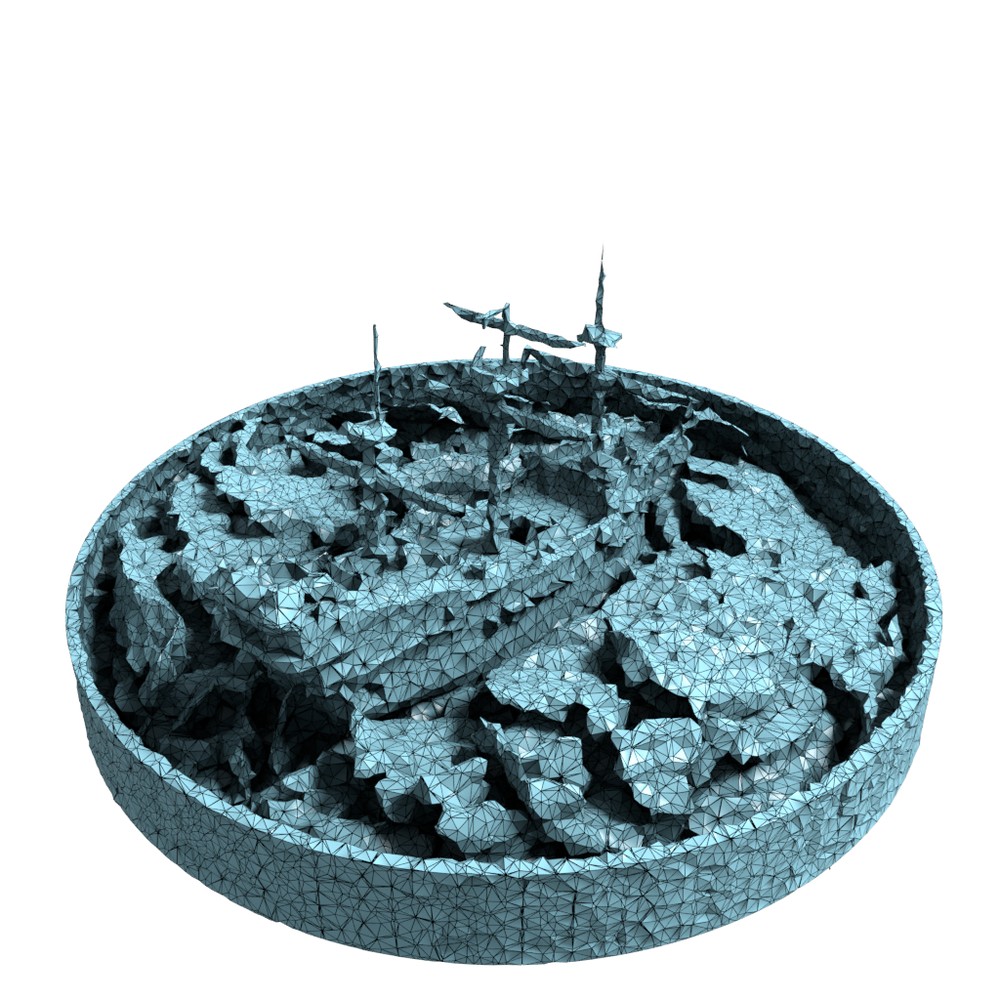}} &
			\raisebox{-0.5\height}{\includegraphics[width=0.24\columnwidth]{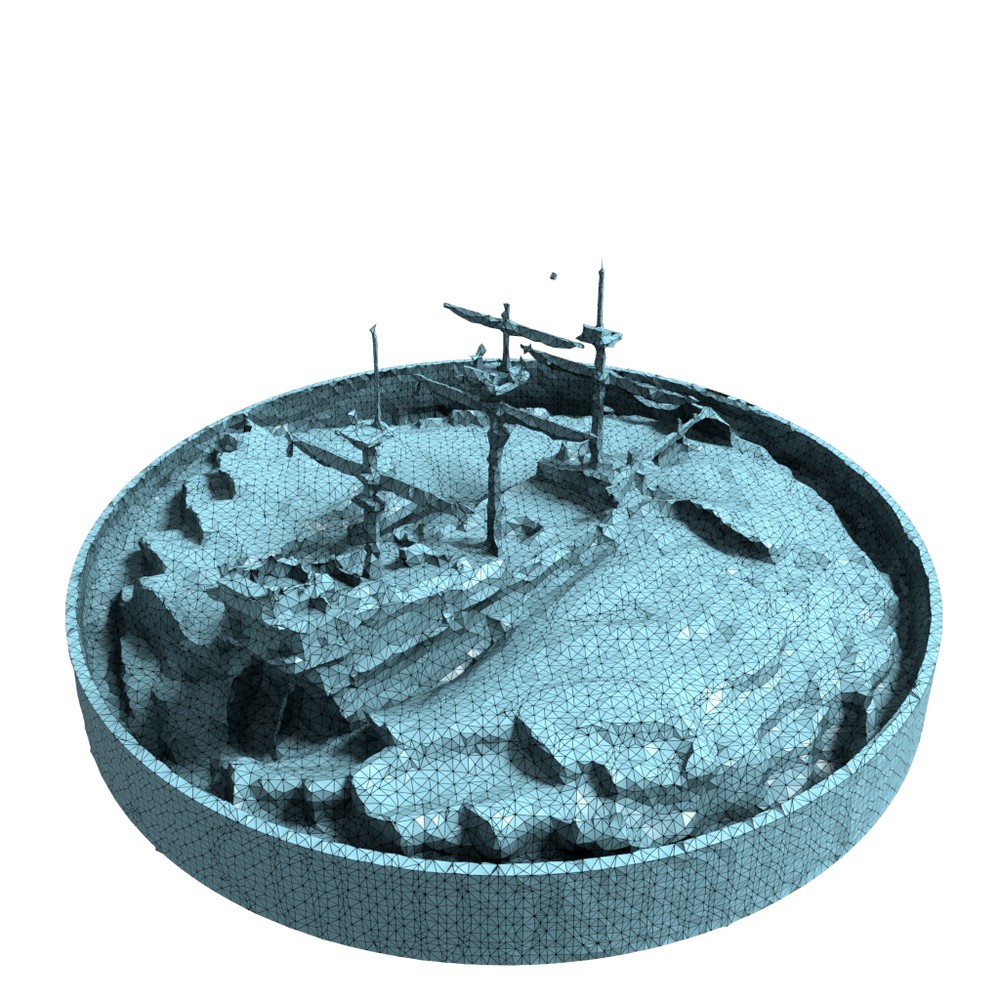}} &
			\raisebox{-0.5\height}{\includegraphics[width=0.24\columnwidth]{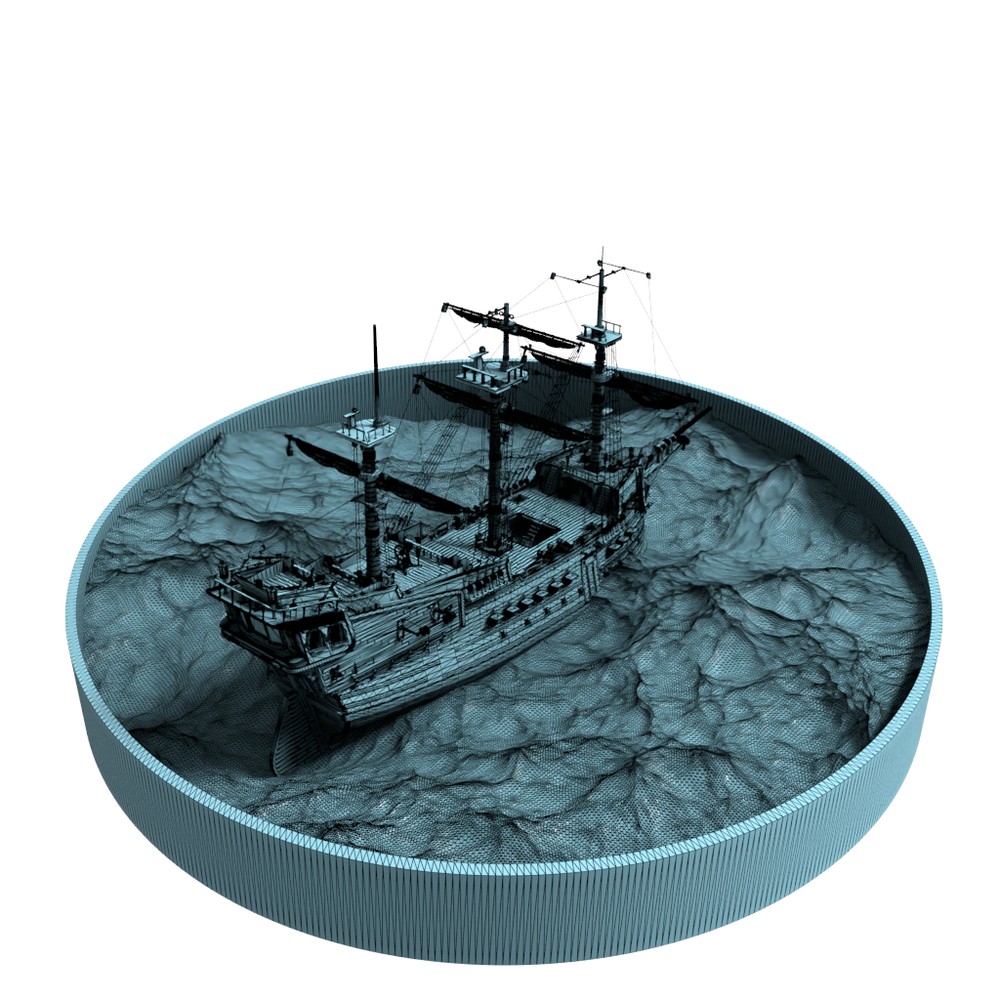}} \\
			
			& Shaded & \DMTet & \ours & Reference & & Shaded & \DMTet & \ours & Reference \\
		\end{tabular}
		\caption{Geometry breakdown for all scenes of the original NeRF dataset. We compare to the original \nvdiffrec{} implementation using \DMTet{}. 
		Note that \ours{} offers more uniform tessellation, while being better capturing small details, as seen in the Lego scene. }
		\label{fig:nvdiffrec_nerf_suppl}
	\end{figure*}
}

\newcommand{\figNVDIFFRECPhotoSupplemental}{
	\begin{figure*}
		\centering
		\setlength{\tabcolsep}{0mm}
		\begin{tabular}{cccccc}
			\rotatebox[origin=c]{90}{Family} &
			\raisebox{-0.5\height}{\includegraphics[width=0.195\textwidth]{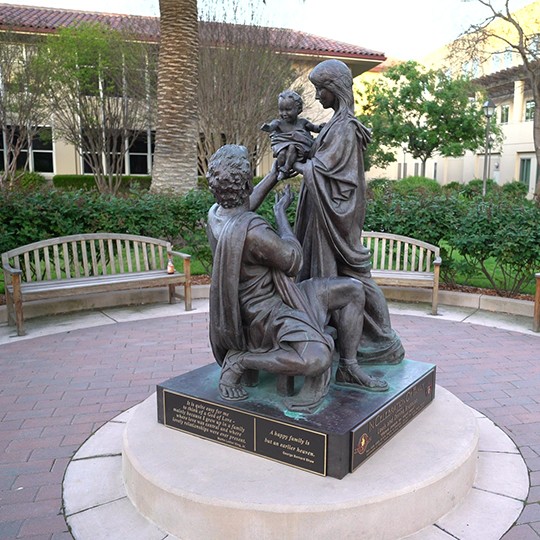}} &
			\raisebox{-0.5\height}{\includegraphics[width=0.195\textwidth]{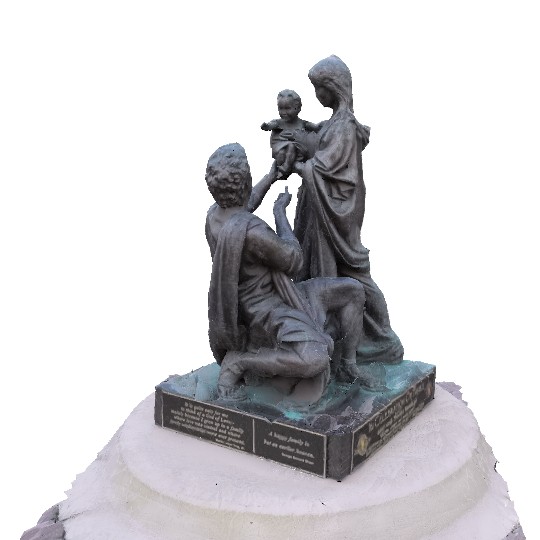}} &
			\raisebox{-0.5\height}{\includegraphics[width=0.195\textwidth]{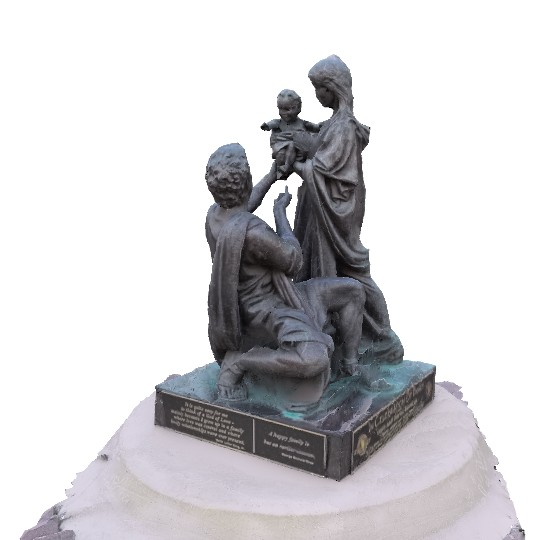}} &
			\raisebox{-0.5\height}{\includegraphics[width=0.195\textwidth]{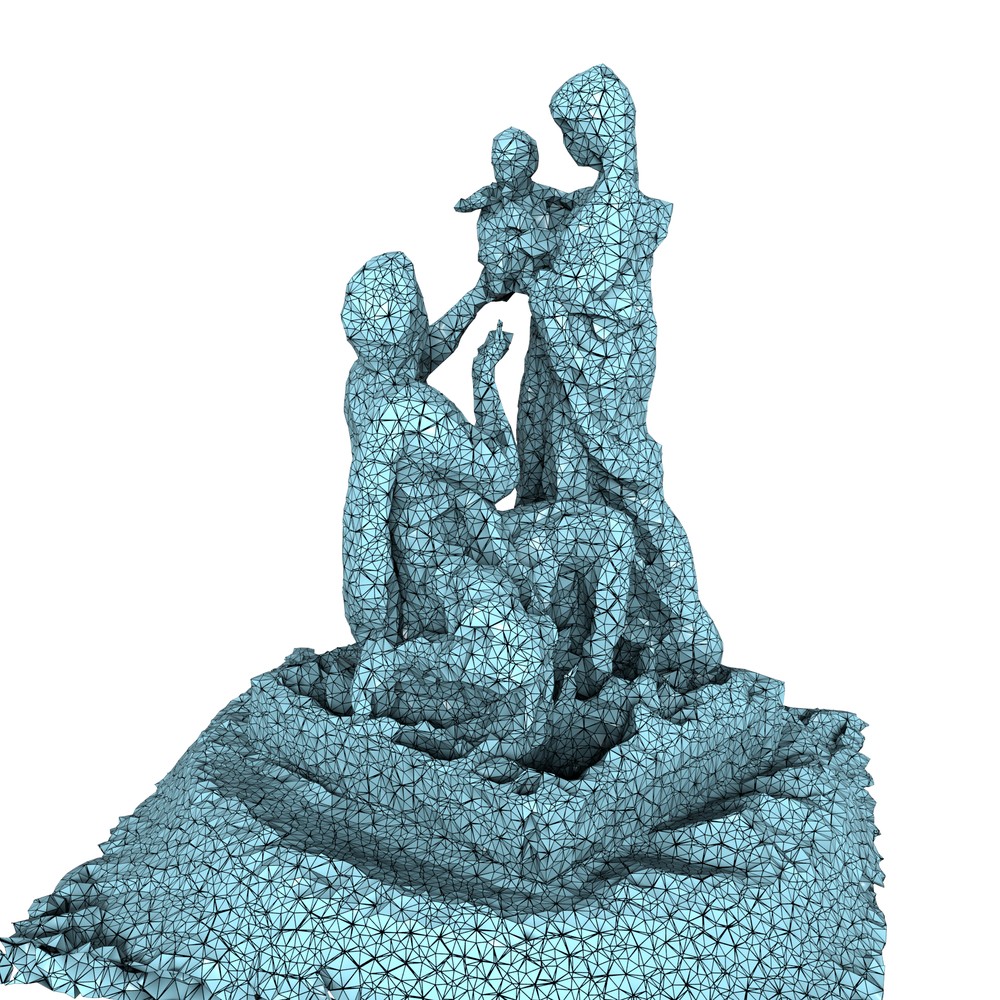}} &
			\raisebox{-0.5\height}{\includegraphics[width=0.195\textwidth]{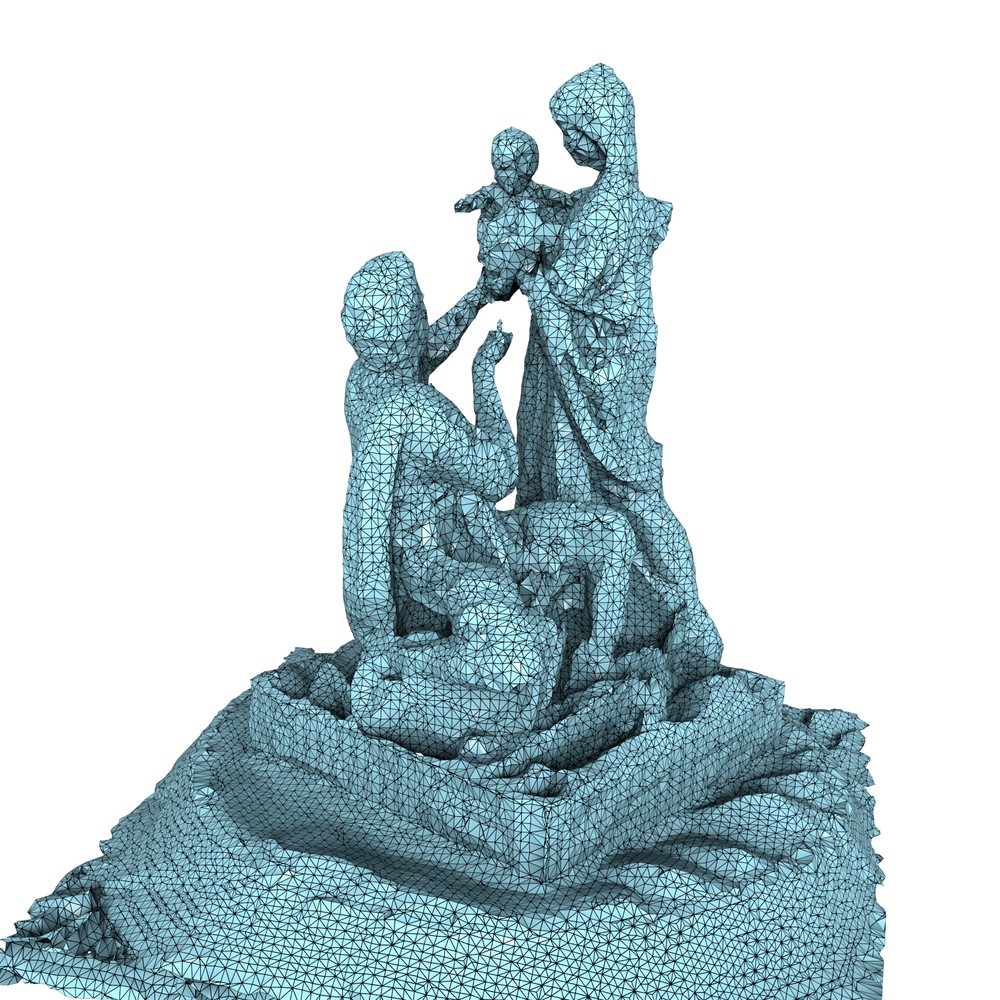}} \\
                & Reference & \DMTet{} & \ours{} & \DMTet{} (96k tris) & \ours{} (79k tris)\\
			\rotatebox[origin=c]{90}{GoldCape} &
			\raisebox{-0.5\height}{\includegraphics[width=0.195\textwidth]{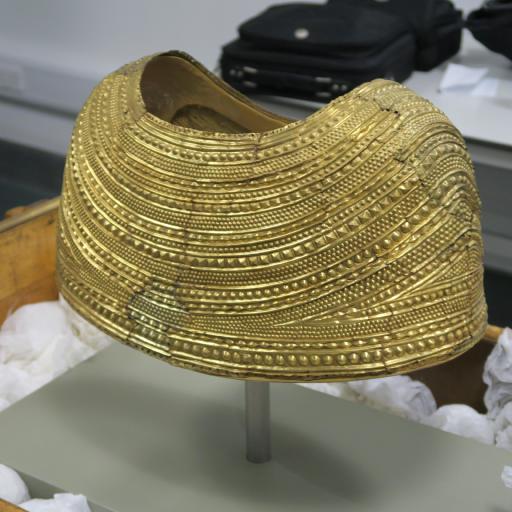}} &
			\raisebox{-0.5\height}{\includegraphics[width=0.195\textwidth]{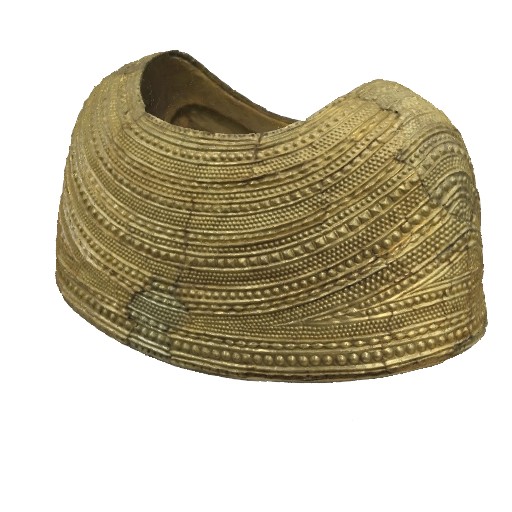}} &
			\raisebox{-0.5\height}{\includegraphics[width=0.195\textwidth]{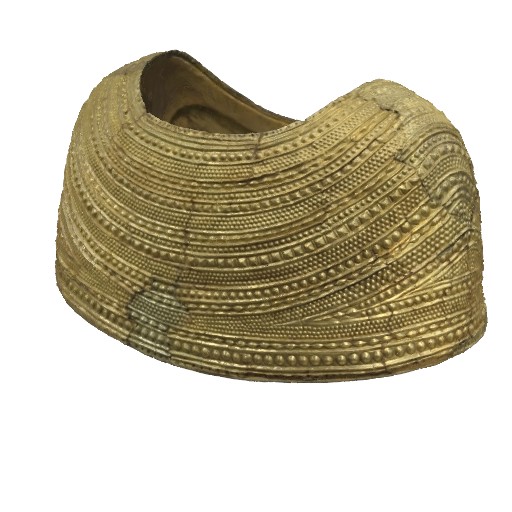}} &
			\raisebox{-0.5\height}{\includegraphics[width=0.195\textwidth]{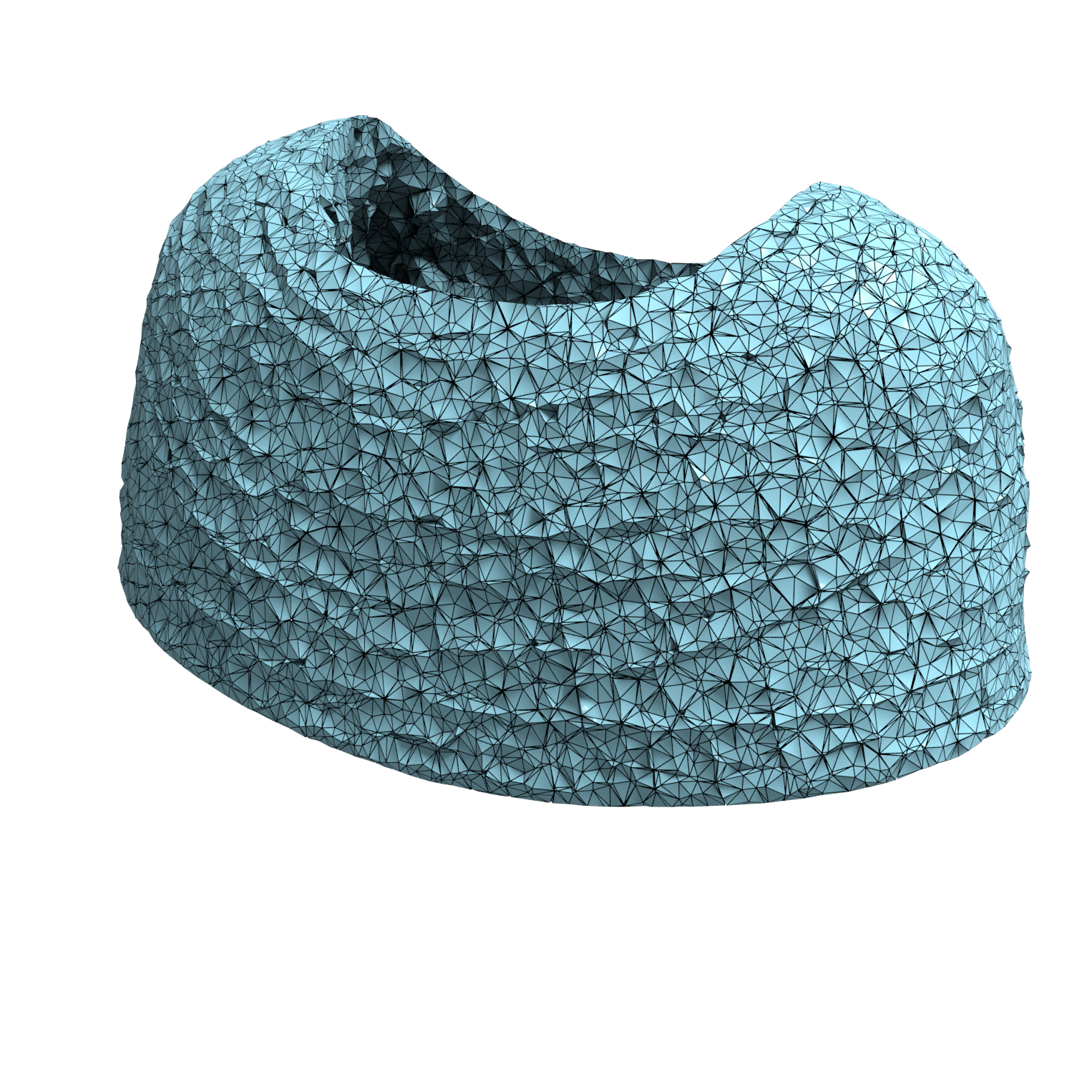}} &
			\raisebox{-0.5\height}{\includegraphics[width=0.195\textwidth]{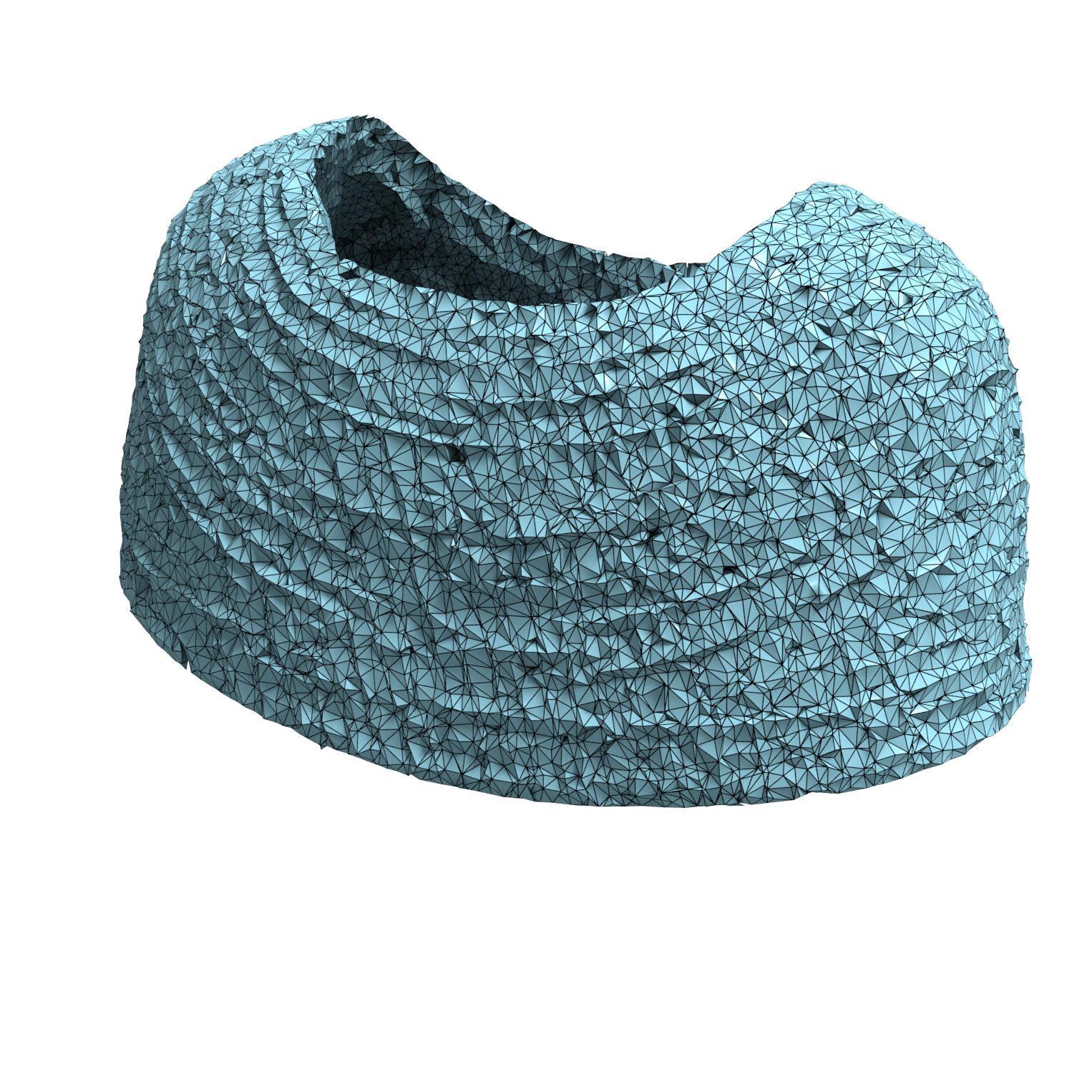}} \\
			& Reference & \DMTet{} & \ours{} & \DMTet{} (55k tris) & \ours{} (38k tris) \\
		\end{tabular}
		\caption{
            \updates{We compare \ours{} and \DMTet{} on real world photographic datasets using \nvdiffrecmc{} for the Family dataset from Tanks\&Temples~\cite{Knapitsch2017} and \nvdiffrec{} for the GoldCape dataset~\cite{Boss2021}.
            \ours{} offers more uniform tessellation and more faithfully captures small geometric details (e.g. the grooves in the GoldCape scene). PSNR view interpolation validation scores are 28.49 / 28.47 dB (\DMTet{} / \ours{}) for the Family scene and 24.44 / 24.56 dB for the GoldCape scene.}
        }
		\label{fig:nvdiffrec_photo_suppl}
	\end{figure*}
}

%% file: sources_suppl/supp_figs.tex
\newcommand{\figphysicssupp}{
	\begin{figure*}
		\centering
		\setlength{\tabcolsep}{0mm}
		\begin{tabular}{ccccccccc}
			& && \multicolumn{2}{c}{$t=0.0s$} & \multicolumn{2}{c}{$t=0.9s$} & \multicolumn{2}{c}{$t=1.5s$}  \\
			\rotatebox{90}{ \quad  Init.} & &
			\rotatebox{90}{ \quad  $D= 1.500$}
			&
			\includegraphics[width=0.30\columnwidth,trim={0cm 0cm 0cm 0cm},clip]{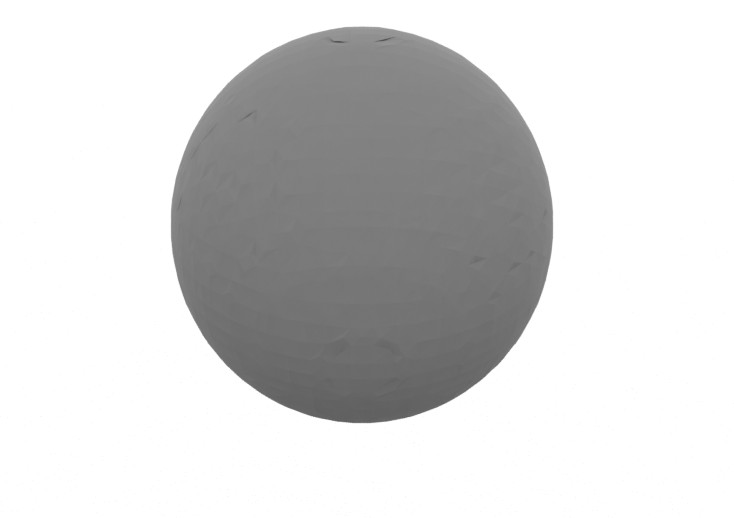} &
			\includegraphics[width=0.30\columnwidth,trim={0cm 0cm 0cm 0cm},clip]{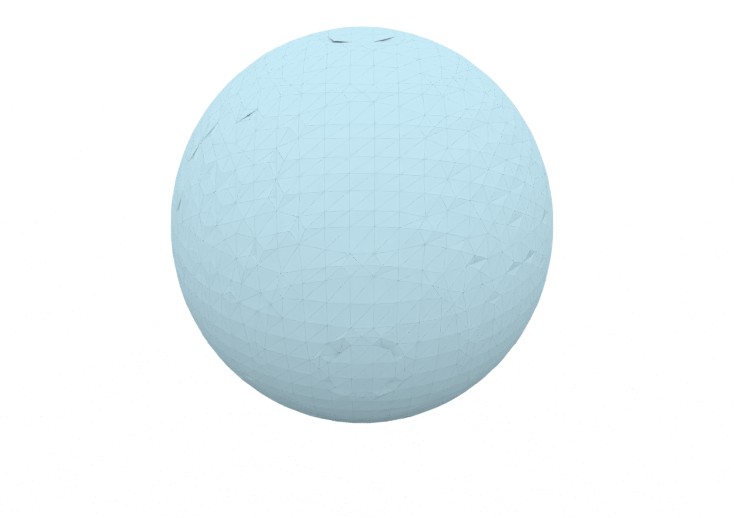} &
			\includegraphics[width=0.30\columnwidth,trim={0cm 0cm 0cm 0cm},clip]{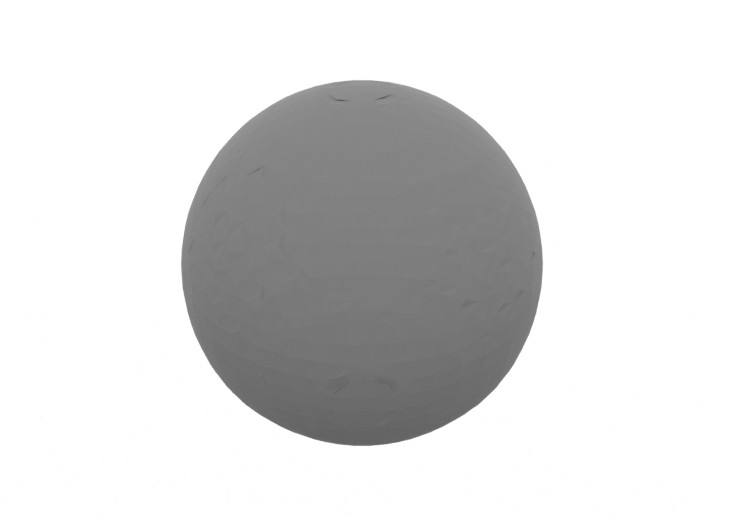} &
			\includegraphics[width=0.30\columnwidth,trim={0cm 0cm 0cm 0cm},clip]{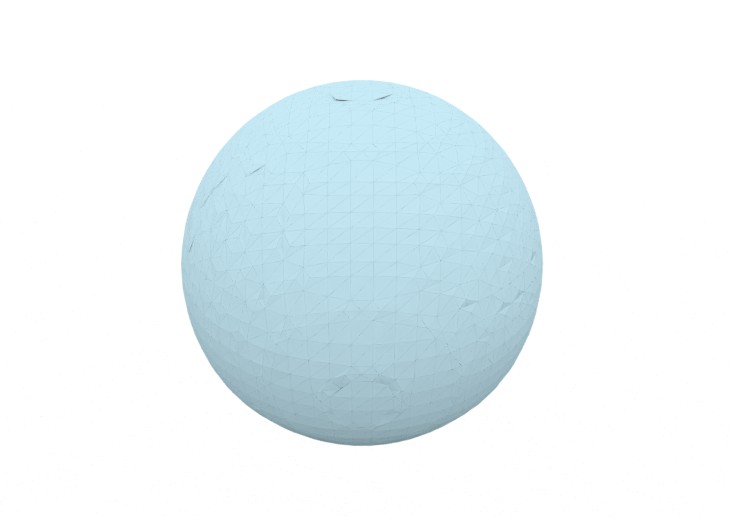} &
			\includegraphics[width=0.30\columnwidth,trim={0cm 0cm 0cm 0cm},clip]{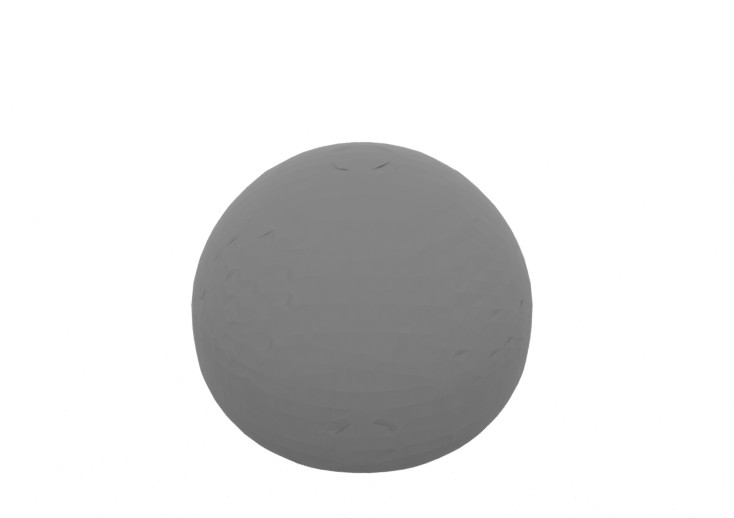}&
			\includegraphics[width=0.30\columnwidth,trim={0cm 0cm 0cm 0cm},clip]{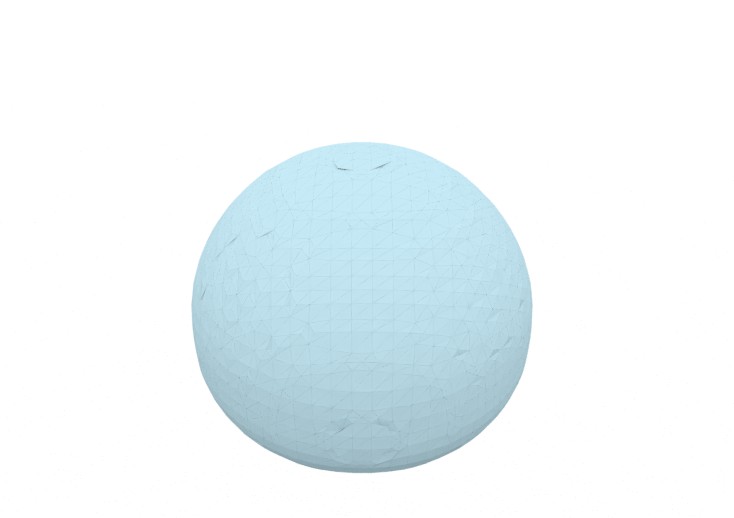} \\
			\rotatebox{90}{ \quad  Opt.} & &
			\rotatebox{90}{ \quad  $D= 0.511$}
			&
			\includegraphics[width=0.30\columnwidth,trim={0cm 0cm 0cm 1cm},clip]{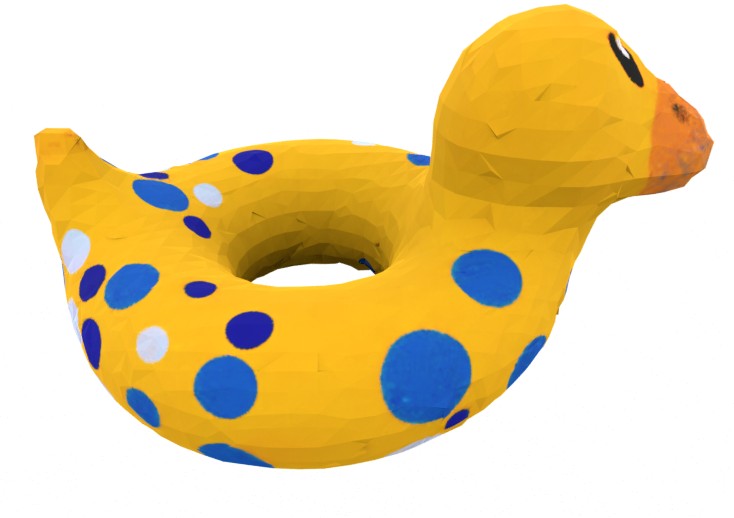} &
			\includegraphics[width=0.30\columnwidth,trim={0cm 0cm 0cm 1cm},clip]{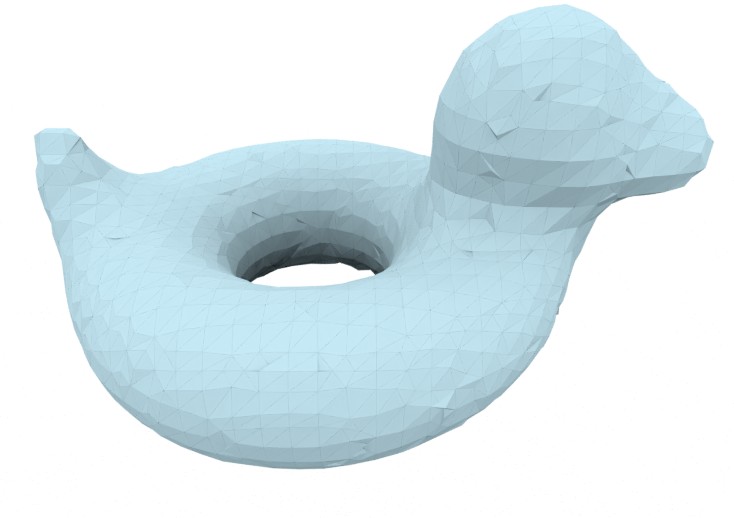} &
			\includegraphics[width=0.30\columnwidth,trim={0cm 0cm 0cm 1cm},clip]{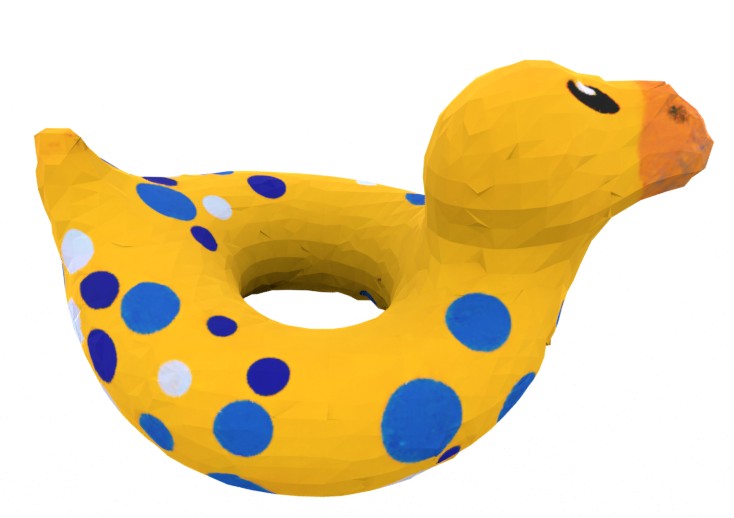} &
			\includegraphics[width=0.30\columnwidth,trim={0cm 0cm 0cm 1cm},clip]{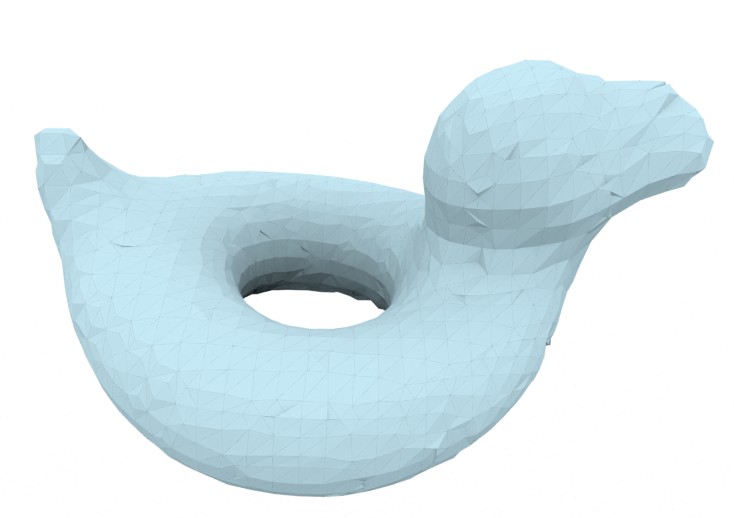} &
			\includegraphics[width=0.30\columnwidth,trim={0cm 0cm 0cm 1cm},clip]{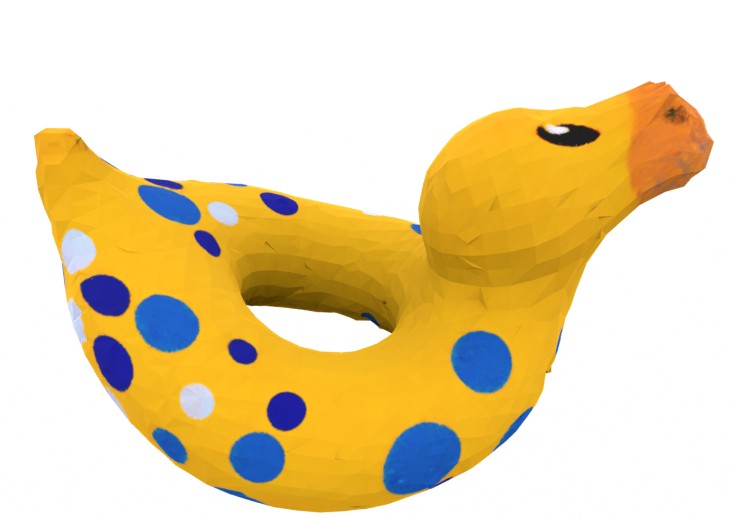} &
			\includegraphics[width=0.30\columnwidth,trim={0cm 0cm 0cm 1cm},clip]{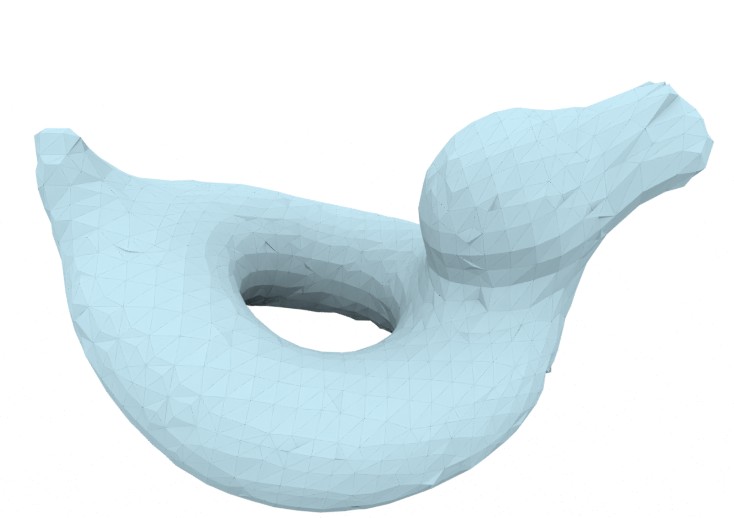}\\
			\rotatebox{90}{\quad   Ref.} & &
			\rotatebox{90}{\quad   $D= 0.500$}
			&
			\includegraphics[width=0.30\columnwidth,trim={0cm 0cm 0cm 1cm},clip]{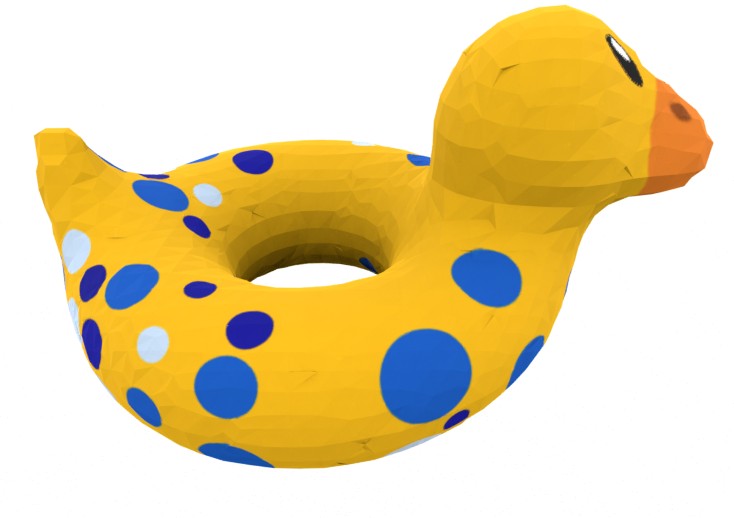} &
				\includegraphics[width=0.30\columnwidth,trim={0cm 0cm 0cm 1cm},clip]{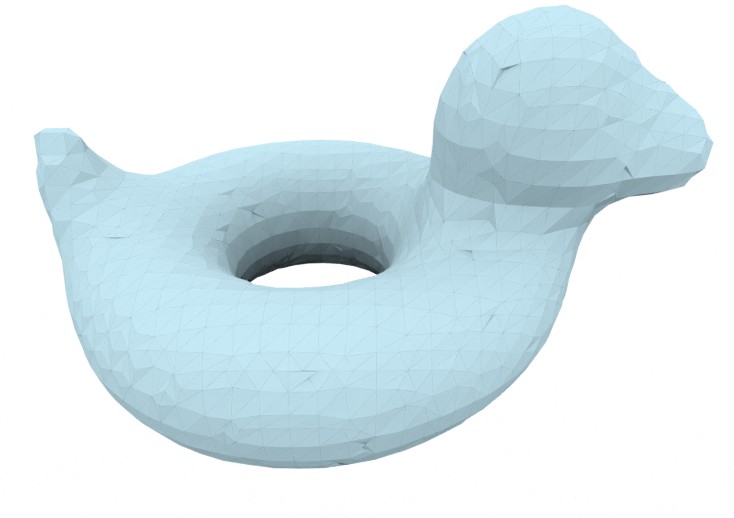} &
			\includegraphics[width=0.30\columnwidth,trim={0cm 0cm 0cm 1cm},clip]{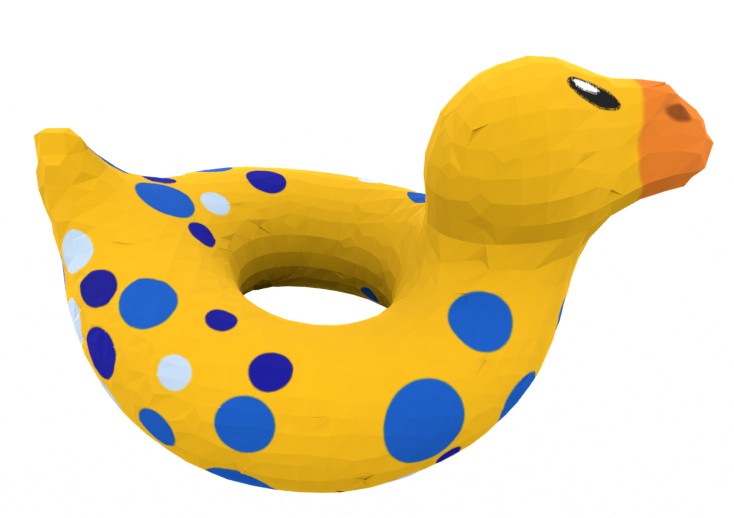} &
				\includegraphics[width=0.30\columnwidth,trim={0cm 0cm 0cm 1cm},clip]{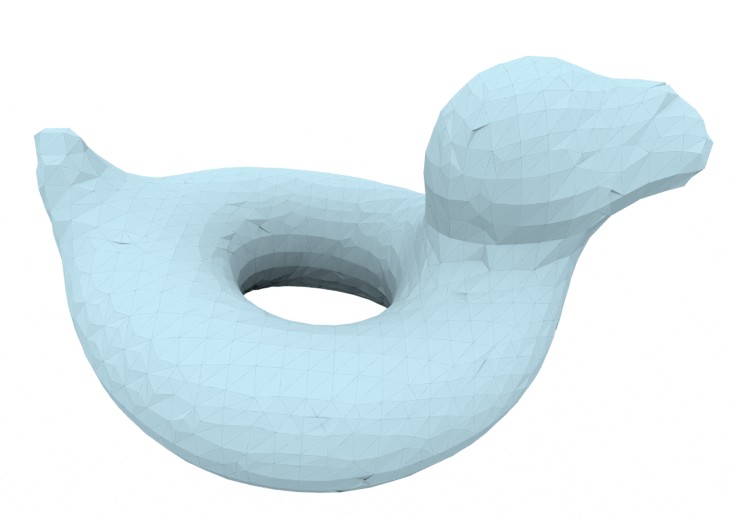} &
			\includegraphics[width=0.30\columnwidth,trim={0cm 0cm 0cm 1cm},clip]{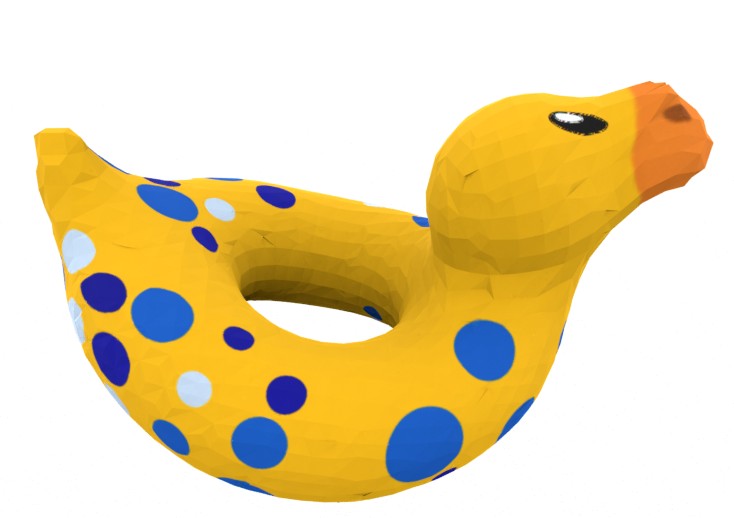}  &
			\includegraphics[width=0.30\columnwidth,trim={0cm 0cm 0cm 1cm},clip]{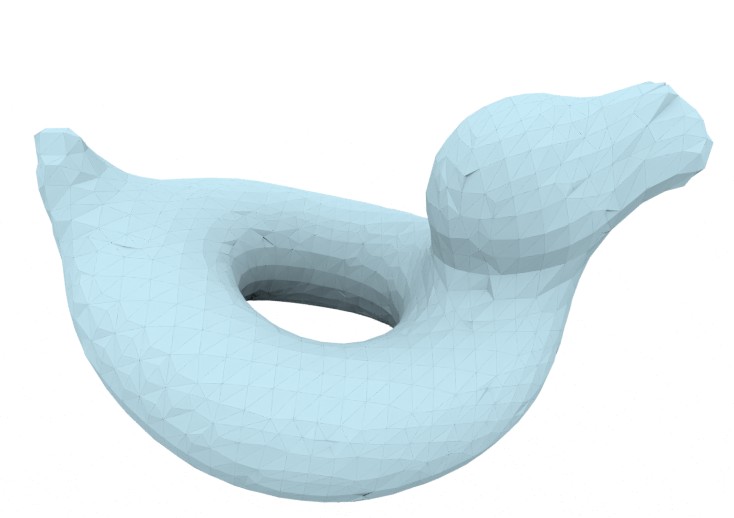}\\
			\hline
			
			& && \multicolumn{2}{c}{$t=0.0s$} & \multicolumn{2}{c}{$t=1.0s$} & \multicolumn{2}{c}{$t=1.8s$}  \\
			\rotatebox{90}{ \quad  Init.} & &
			\rotatebox{90}{ \quad  $D= 1.500$}
			&
			\includegraphics[width=0.30\columnwidth,trim={0cm 0cm 0cm 0cm},clip]{figures/physics/pred_init-frame0.png} &
			\includegraphics[width=0.30\columnwidth,trim={0cm 0cm 0cm 0cm},clip]{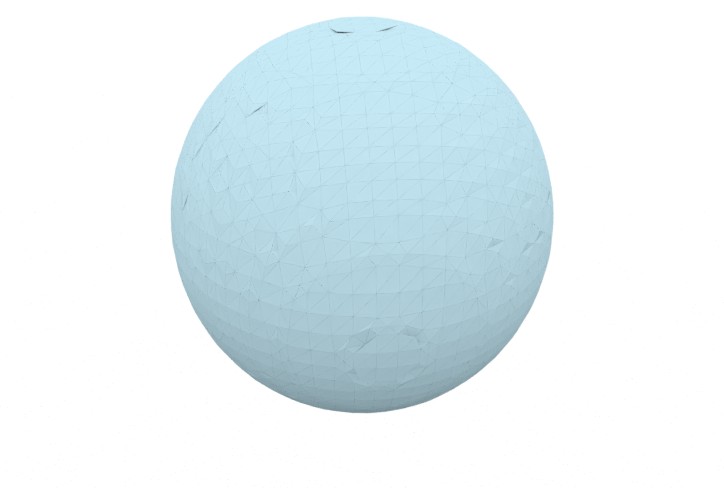} &
			\includegraphics[width=0.30\columnwidth,trim={0cm 0cm 0cm 0cm},clip]{figures/physics/pred_init-frame65.png} &
			\includegraphics[width=0.30\columnwidth,trim={0cm 0cm 0cm 0cm},clip]{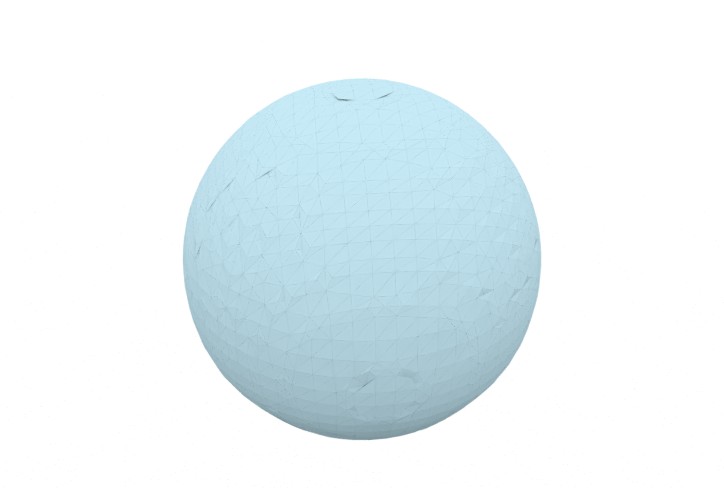} &
			\includegraphics[width=0.30\columnwidth,trim={0cm 0cm 0cm 0cm},clip]{figures/physics/pred_init-frame120.png}&
			\includegraphics[width=0.30\columnwidth,trim={0cm 0cm 0cm 0cm},clip]{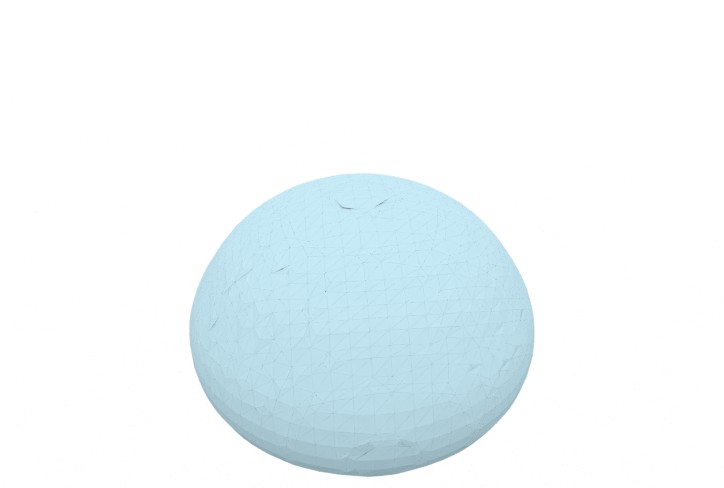} \\
			\rotatebox{90}{ \quad  Opt.} & &
			\rotatebox{90}{ \quad  $D= 0.311$}
			&
			\includegraphics[width=0.30\columnwidth,trim={0cm 0cm 0cm 1cm},clip]{figures/physics/pred_converage-frame0.png} &
			\includegraphics[width=0.30\columnwidth,trim={0cm 0cm 0cm 1cm},clip]{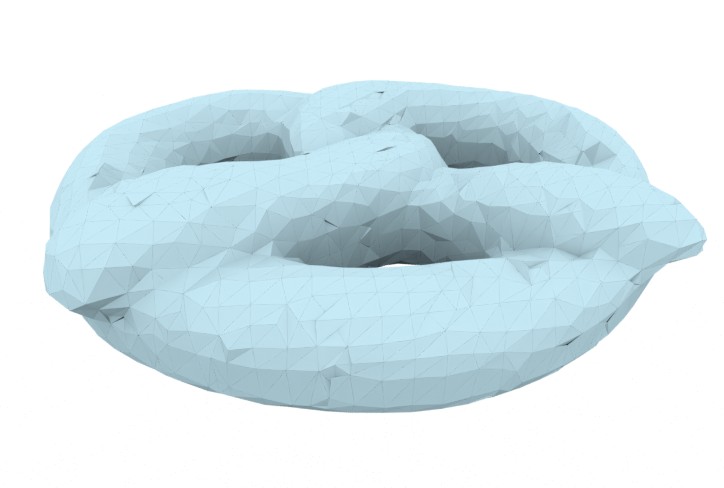} &
			\includegraphics[width=0.30\columnwidth,trim={0cm 0cm 0cm 1cm},clip]{figures/physics/pred_converage-frame65.png} &
			\includegraphics[width=0.30\columnwidth,trim={0cm 0cm 0cm 1cm},clip]{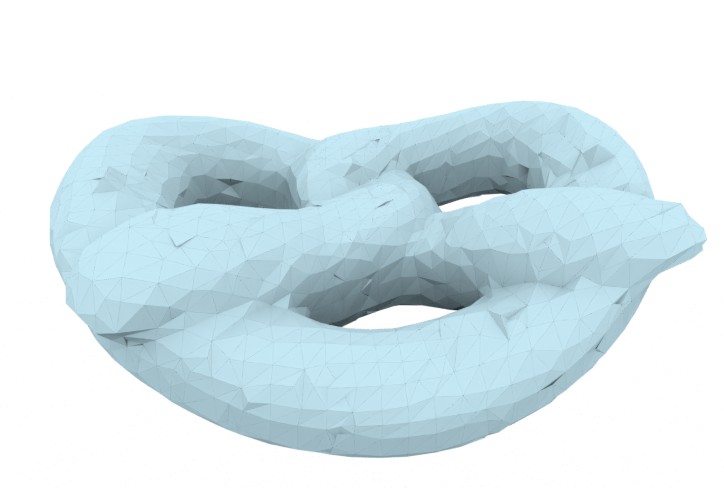} &
			\includegraphics[width=0.30\columnwidth,trim={0cm 0cm 0cm 1cm},clip]{figures/physics/pred_converage-frame120.png} &
			\includegraphics[width=0.30\columnwidth,trim={0cm 0cm 0cm 1cm},clip]{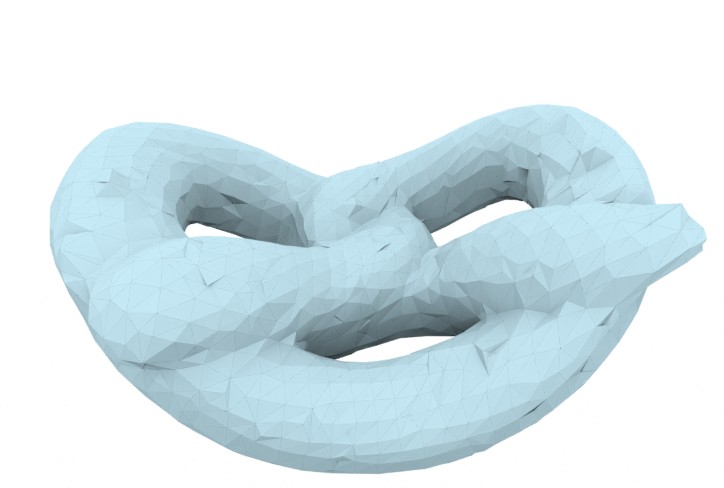}\\
			\rotatebox{90}{\quad   Ref.} & &
			\rotatebox{90}{\quad   $D= 0.300$}
			&
			\includegraphics[width=0.30\columnwidth,trim={0cm 0cm 0cm 1cm},clip]{figures/physics/gt-frame0.png} &
			\includegraphics[width=0.30\columnwidth,trim={0cm 0cm 0cm 1cm},clip]{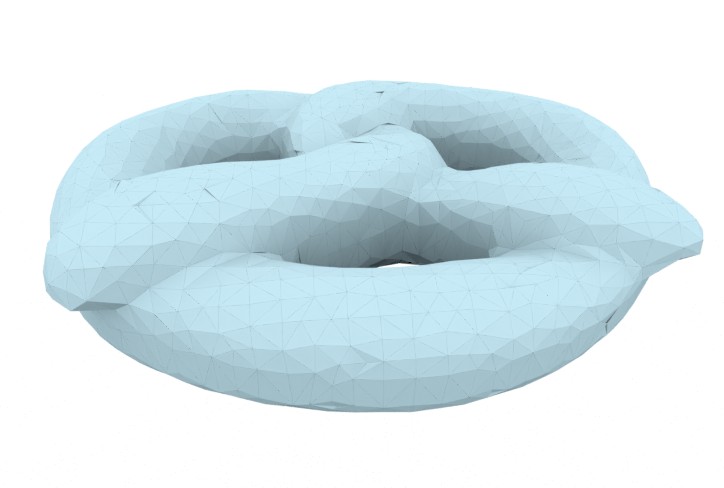} &
			\includegraphics[width=0.30\columnwidth,trim={0cm 0cm 0cm 1cm},clip]{figures/physics/gt-frame65.png} &
			\includegraphics[width=0.30\columnwidth,trim={0cm 0cm 0cm 1cm},clip]{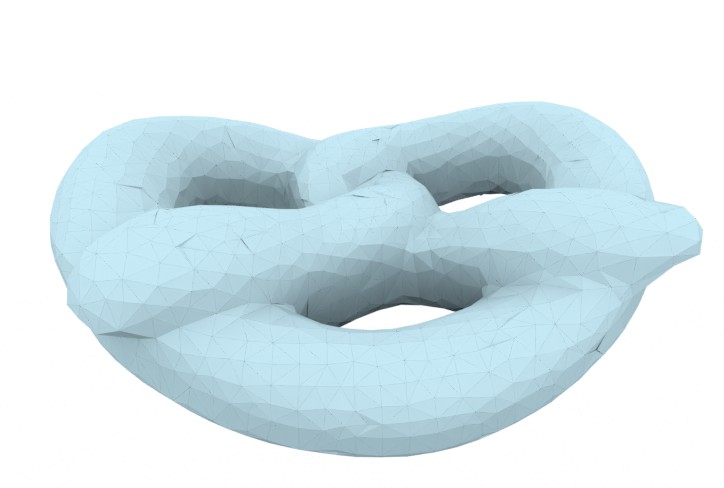} &
			\includegraphics[width=0.30\columnwidth,trim={0cm 0cm 0cm 1cm},clip]{figures/physics/gt-frame120.png}  &
			\includegraphics[width=0.30\columnwidth,trim={0cm 0cm 0cm 1cm},clip]{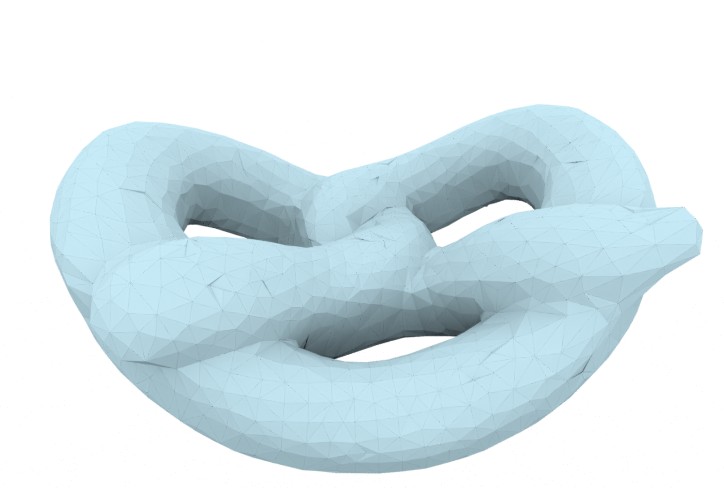}\\
			
		\end{tabular}
		\caption{Geometry details of the tetrahedral meshes in the physical simulation experiment. \ours{} generates tetrahedral meshes with well-defined topology which can be directly utilized in physical simulation. 
			We further bridge it with differentiable physical simulation and differentiable rendering frameworks to optimize shape, texture, and physical parameters from multi-view videos.
			In the two examples, we optimize the mass density $D$ of the object and get very close parameters after converging.
			We also show the wireframe of the deformed objects.
			To apply the extracted tetrahedral meshes in physical simulation, we delete  tetrahedrons with tiny  volume.  This results in some spiky parts of the shape, \emph{e.g.}, the shape has some black slits. 
			We find without deleting small tetrahedrons leads to unstable simulation results. This problem can be potentially addressed by designing new regularization terms, which we leave for future work. 
		}
		\label{fig:physicus2}
	\end{figure*}
}

%% file: sources/tables.tex
\definecolor{goodcolor}{rgb}{0.1, 0.7, 0.0}
\definecolor{badcolor}{rgb}{0.8, 0.0, 0.0}
\newcommand{\goodmark}{\ding{52}}%
\newcommand{\badmark}{\ding{54}}%
\newcommand{\cmark}{\color{goodcolor}{\goodmark}}
\newcommand{\xmark}{\color{badcolor}{\badmark}}

\newcommand{\tabTaxonomySimplified}{
    \begin{table}[b!]
    \vspace{-2mm}
    \centering
    \setlength{\tabcolsep}{8pt}
    \renewcommand{\arraystretch}{1.1}
    \caption{Taxonomy of isosurfacing methods. \textsc{Grad} means gradient-based based optimization is effective in practice, and \textsc{Uniform} means the resulting tessellations are generally uniform \updates{without sliver triangles.}}
    \resizebox{.46\textwidth}{!}{
    \begin{tabular}{l|ccccc}
    \hline
        Method & Grad. & Sharp Features & Uniform   & Intersection-Free & 2-Manifold \\ \hline
        MC~\cite{Lorensen1997} & \cmark& \xmark& \xmark& \cmark& \xmark   \\
        DC~\cite{Ju2002} & \xmark& \cmark& \xmark& \xmark& \xmark   \\
        NDC~\cite{ChenNDC2022} & \xmark& \cmark& \cmark& \cmark& \xmark   \\
        DMC~\cite{nielson2004dual} Centroid & \cmark& \xmark& \cmark&  \cmark& \cmark \\
        DMC~\cite{schaefer2007manifold} QEF & \xmark& \cmark & \cmark&  \cmark& \cmark \\
        Template Mesh~\cite{softras} & \xmark& \xmark& \cmark&  \xmark& \cmark  \\
        DMTet~\cite{Shen2021} & \cmark& \cmark& \xmark&  \cmark& \cmark   \\
        \textbf{\ours{}} & \cmark& \cmark& \cmark&  \xmark& \cmark   \\
        \hline
    \end{tabular}
    
    }
    \label{tbl:taxonomy}
    \end{table}
}

\newcommand{\tabMainOne}{
    \begin{table}[]
    \caption{Quantitative results on Mesh Reconstruction. CD: Chamfer distance (1e-5), F1: F1 score. ECD, edge chamfer distance (1e-2). EF1: edge F1. NV: Non-mainfold vertices, NE: Non-manifold edges, SI: self-intersection. IN>$5^{\circ}$: normal angle difference > $5^{\circ}$. SA: small angle. \# V: number of vertices. \#F: number of triangles.}
    \resizebox{\columnwidth}{!}{
    \begin{tabular}{lccccccccccc}
    \hline
    $32^3$ & CD$\downarrow$ & F1 $\uparrow$ & ECD $\downarrow$ & EF1 $\uparrow$& \#V & \#T & NV(\%) $\downarrow$& NE(\%) $\downarrow$& SI(\%)  $\downarrow$& IN>$5^{\circ}$(\%)  $\downarrow$& SA<$10^{\circ}$(\%)  $\downarrow$\\ \hline
     $MC_{SDF}$ & 22.65 & 0.28 &  5.56 & 0.08 & 2387 & 4771 & 0.0 & 0.0 & 0.0 & 85.6 & 24.7\\
     $DC_{hermite}$ & 17.15 & 0.38 &  4.82 & 0.11 & 2360 & 4775 & 0.131 & 0.349 & 1.882 & 74.43 & 9.59 \\
     $NDC_{SDF}$ & 17.61 & 0.42 & 3.55 & 0.13 & 1877 & 3801 & 0.155 & 0.378 & 0.232 & 72.60 & 0.77 \\
    \hline
    MC & 9.11 & 0.54 & 2.60 & 0.13 & 2573 & 5146 & 0.0 & 0.0 &  0.0 & 66.61 & 11.74 \\
    DMTet(32) & 11.56 & 0.52 & 3.64 & 0.17 & 1691 & 3387 & 0.0 & 0.0 &  0.0 & 66.22 & 12.32 \\
    DMTet(40) & 8.35 & 0.58 &  3.64 & 0.20 & 2626& 5259& 0.0& 0.0& 0.0 & 61.21 & 12.72\\
     \ours{} & 7.01 & 0.64 & 2.11 & 0.26 & 2400 & 4800 & 0.0 & 0.0 &  0.715  &  50.52  & 2.99 \\ \hline
    \end{tabular}
    
    }
    \resizebox{\columnwidth}{!}{
    \begin{tabular}{lccccccccccc}
    \hline
    $64^3$ & CD$\downarrow$ & F1 $\uparrow$ & ECD $\downarrow$ & EF1 $\uparrow$& \#V & \#T & NV(\%) $\downarrow$& NE(\%) $\downarrow$& SI(\%)  $\downarrow$& IN>$5^{\circ}$(\%)  $\downarrow$& SA<$10^{\circ}$(\%)  $\downarrow$\\ \hline
     $MC_{SDF}$ & 6.84 & 0.55 &  2.55 & 0.14 & 9881 & 19783 & 0.0 & 0.0 & 0.0 & 80.67 & 24.32\\
     $DC_{hermite}$ &5.90 & 0.61 &3.80 & 0.23 & 9828 & 19769 &  0.043 & 0.139  & 1.483 & 63.34 &8.67  \\
     $NDC_{SDF}$ & 6.16 & 0.57 &1.22 & 0.26 & 9828 & 19769 &  0.043 &  0.140  & 0.130 & 55.22 & 0.48 \\
    \hline
    MC & 6.33 & 0.66 &  1.25 & 0.25 & 10459 & 20801 &  0.0 & 0.0 &  0.0 & 52.37 & 11.90 \\
    DMTet(64) & 7.50 & 0.66 & 3.77 & 0.28 & 6783 & 13566 & 0.0 & 0.0 &  0.0 & 50.20 & 14.41 \\
    DMTet(80) & 5.17&0.66 &3.59 & 0.29& 10385& 20784&0.0 &0.0 &0.0 &48.66 & 17.88 \\
     \ours{} & 4.87 & 0.70 & 0.71 & 0.43 & 9916 & 19843 & 0.0 & 0.0 &  0.103 & 34.87 & 1.97\\ \hline
    \end{tabular}
    }
    \resizebox{\columnwidth}{!}{
    \begin{tabular}{lccccccccccc}
    \hline
    $128^3$ & CD$\downarrow$ & F1 $\uparrow$ & ECD $\downarrow$ & EF1 $\uparrow$& \#V & \#T & NV(\%) $\downarrow$& NE(\%) $\downarrow$& SI(\%)  $\downarrow$& IN>$5^{\circ}$(\%)  $\downarrow$& SA<$10^{\circ}$(\%)  $\downarrow$\\ \hline
     $MC_{SDF}$ & 4.72 & 0.68 &  1.13 & 0.33 & 40164 & 80374 & 0.0 & 0.0 & 0.0 & 77.23 & 24.25\\
     $DC_{hermite}$ & 4.59 & 0.69 &  3.82 & 0.40 & 40128 & 80360 & 0.017 & 0.069 & 1.280 & 53.98 & 7.95 \\
     $NDC_{SDF}$ & 5.04 & 0.65 &  0.79 & 0.43 & 40129 & 80361& 0.017 & 0.036 & 0.084 & 43.2 & 0.37  \\
    \hline
    MC & 4.51 & 0.72 &  1.32 & 0.44 & 38645 & 77212 & 0.0 & 0.0 &0.0 & 42.56 & 12.46  \\
    DMTet(128) & 4.98 & 0.74 &  1.50 & 0.39 & 23535 & 47001 & 0.0 & 0.0 &0.0 & 48.86 & 23.52 \\
     \ours{} & 4.31 & 0.71 & 0.42 & 0.51 & 38923 & 77845 & 0.0 & 0.0 &0.017 & 30.57 & 0.79  \\ \hline
    \end{tabular}
    }
    \label{tbl:main64large}
    \end{table}
}

\newcommand{\tabMainOneSimplified}{
    \begin{table}[]
    \caption{Quantitative results on Mesh Reconstruction. We report the following metrics: IN> $5^{\circ}$: normal angle difference > $5^{\circ}$, CD: Chamfer Distance, F1: F1 score, ECD: Edge Chamfer Distance, EF1: Edge F1 Score. \#V: number of vertices, \#F: number of faces.}
    \resizebox{\columnwidth}{!}{
    \begin{tabular}{lccccccc}
    \hline
    $32^3$ & IN>$5^{\circ}$(\%)$\downarrow$  & CD$(10^{-5})\downarrow$ & F1 $\uparrow$ & ECD $(10^{-2})\downarrow$ & EF1 $\uparrow$& \#V & \#T \\  \hline
     $MC_{SDF}$ & 85.60 & 22.65 & 0.28 & 5.56 & 0.08 & 2387  & 4771   \\
     $DC_{hermite} $ & 74.43 & 17.15 & 0.38 & 4.82 & 0.11 & 2360  & 4775  \\
     $NDC_{SDF}$ & 72.60 & 17.61 & 0.42 & 3.55 & 0.13 & 1877  & 3801  \\
    \hline
    MC & 66.61 & 9.11  & 0.54 & 2.60 & 0.13 & 2573  & 5146 \\
    DMTet(32) & 66.22 & 11.56 & 0.52 & 3.64 & 0.17 & 1691  & 3387  \\
    DMTet(40) & 61.21 & 8.35  & 0.58 & 3.64 & 0.20 & 2626  & 5259  \\
     \ours{} & \textbf{50.52} &\textbf{ 7.01}  & \textbf{0.64 }& \textbf{2.11} & \textbf{0.26} & 2400  & 4800  \\ \hline
    \end{tabular}
    
    }
    \resizebox{\columnwidth}{!}{
    \begin{tabular}{lccccccc}
    \hline
    $64^3$ & IN>$5^{\circ}$(\%)$\downarrow$  & CD$(10^{-5})\downarrow$ & F1 $\uparrow$ & ECD $(10^{-2})\downarrow$ & EF1 $\uparrow$& \#V & \#T \\ \hline
     $MC_{SDF}$ & 80.67 & 6.84  & 0.55 & 2.55 & 0.14 & 9881  & 19783 \\
     $DC_{hermite}$ & 63.34 & 5.90  & 0.61 & 3.80 & 0.23 & 9828  & 19769 \\
     $NDC_{SDF}$ & 55.22 & 6.16  & 0.57 & 1.22 & 0.26 & 9828  & 19769 \\
    \hline
    MC & 52.37 & 6.33  & 0.66 & 1.25 & 0.25 & 10459 & 20801 \\
    DMTet(64) & 50.20 & 7.50  & 0.66 & 3.77 & 0.28 & 6783  & 13566 \\
    DMTet(80) & 48.66 & 5.17  & 0.66 & 3.59 & 0.29 & 10385 & 20784  \\
     \ours{} & \textbf{34.87} & \textbf{4.87}  & \textbf{0.70} & \textbf{0.71} & \textbf{0.43} & 9916  & 19843 \\ \hline
    \end{tabular}
    }
    \resizebox{\columnwidth}{!}{
    \begin{tabular}{lccccccc}
    \hline
    $128^3$  & IN>$5^{\circ}$(\%)$\downarrow$  & CD$(10^{-5})\downarrow$ & F1 $\uparrow$ & ECD $(10^{-2})\downarrow$ & EF1 $\uparrow$& \#V & \#T \\ \hline
     $MC_{SDF}$ & 77.23 & 4.72  & 0.68 & 1.13 & 0.33 & 40164 & 80374 \\
     $DC_{hermite}$ & 53.98 & 4.59  & 0.69 & 3.82 & 0.40 & 40128 & 80360 \\
     $NDC_{SDF}$ & 43.20 & 5.04  & 0.65 & 0.79 & 0.43 & 40129 & 80361  \\
    \hline
    MC & 42.56 & 4.51  & 0.72 & 1.32 & 0.44 & 38645 & 77212  \\
    DMTet(128) & 48.86 & 4.98  & 0.74 & 1.50 & 0.39 & 23535 & 47001 \\
     \ours{} & \textbf{30.57} &\textbf{ 4.31}  & \textbf{0.71} & \textbf{0.42 }& \textbf{0.51} & 38923 & 77845  \\ \hline
    \end{tabular}
    }
    \label{tbl:main64}
    \end{table}
}

\newcommand{\tabEquiEdgeSimpNew}{
    \begin{table}[]
    \caption{Quantitative results on mesh reconstruction with equilateral triangle regularizer. Adding regularizer for DMTet and MC significantly impacts geometric metrics (IN>$5^{\circ}$(\%), CD), while \ours{} only sacrifices a bit.}
    \resizebox{\columnwidth}{!}{
    \begin{tabular} {lccccc}
    \hline
    $64^3$ & IN>$5^{\circ}$(\%)   $\downarrow$ & CD$(10^{-5})\downarrow$ & Aspect Ratio > 4 (\%)  $\downarrow$ & Radius Ratio > 4 (\%)  $\downarrow$ & Min Angle < 10 (\%)     $\downarrow$\\ \hline
    MC                & 52.37 & 6.33 &  11.71  & 11.71 & 11.84  \\
    DMTet(80)         & 48.66 & 5.17 &  17.31 & 16.68 & 17.83   \\
    \ours{}           & 34.87 & 4.87 &  2.93  & 4.49  &  2.04  \\ \hline
    MC + Reg.         & 50.16 & 8.56 &  11.46 & 11.43 & 11.62 \\
    DMTet(80) + Reg.  & 67.65 & 6.69 &  0.29  & 0.46  & 0.26 \\
    \ours{} + Reg.    & 41.05 & 5.46 &  0.39  & 0.69  & 0.24  \\

    \hline
    \end{tabular}
    }
    \label{tbl:equiedge_simp}
    \end{table}

}

\newcommand{\tabAblationSimplified}{
    \begin{table}[]
    \caption{Quantitative results on shape reconstruction ablating different formulations of the dual vertex.}
    \resizebox{\columnwidth}{!}{
    \begin{tabular}{lccccccc}
    \hline
    $64^3$ & IN>$5^{\circ}$(\%)  $\downarrow$& CD$(10^{-5})\downarrow$ & F1 $\uparrow$ & ECD $(10^{-2})\downarrow$ & EF1 $\uparrow$& \#V & \#T \\  \hline
     $DMC_{centroid}$\cite{nielson2004dual} & 53.02& 5.85 & 0.65 & 2.60 & 0.19 & 10240 & 20488\\
     + flex vertex (Section~\ref{sec:FlexibleDualVertexPositioning}) &  40.88 & 5.34 & 0.68  & 0.99 & 0.37 & 10022 & 20055 \\
     + grid deform (Section~\ref{sec:grid_deformation}) & 39.46& 5.01 & 0.69 & 0.98 & 0.41 & 9879 & 19766 \\
     + flex quad split (Section~\ref{sec:quad_spliting}) & \textbf{34.87}  &\textbf{4.87} & \textbf{0.70}& \textbf{0.71 }& \textbf{0.43} & 9916 & 19843 \\
      \hline
    \end{tabular}
    }
    \label{tbl:ablation}
    \end{table}
}

\newcommand{\tabNVDIFFRECA}{
    \begin{table}[tb]
    \caption{View interpolation results (PSNR) for \nvdiffrec{} reconstructions of the NeRF synthetic dataset, using either \textsc{\DMTet} or \ours{} for the topology step. The image metric scores are arithmetic means over all test images.
    We also include Chamfer distances (CD) computed on visible triangles (the set of triangles visible in at least one test view) using 2.5~M point. Lower scores indicate better geometric fidelity.}
    \setlength{\tabcolsep}{1.2mm}	
    \begin{tabular}{lcccccccc}
        \toprule
        \textbf{PSNR}~(dB) $\uparrow$ & \small{Chair} & \small{Drums} & \small{Ficus} & \small{Hotdog} & \small{Lego} & \small{Mats} & \small{Mic} & \small{Ship} \\
        \midrule
        \small{\textsc{DMTet}} & 31.8 & 24.6 & 30.9 & 33.2 & 29.0 & 27.0 & 30.7 & 26.0 \\
        \small{\ours{}} & 31.8 & 24.7 & 30.9 & 33.4 & 28.8 & 26.7 & 30.8 & 25.9 \\ %
        \midrule
        \textbf{CD} ($10^{-2}$) $\downarrow$ & \small{Chair} & \small{Drums} & \small{Ficus} & \small{Hotdog} & \small{Lego} & \small{Mats} & \small{Mic} & \small{Ship} \\
        \midrule
        \small{\textsc{DMTet}} & 4.51 & 3.98 & 0.30 & 2.67 & 2.41 & 0.41 & 1.20 & 55.8 \\
        \small{\ours{}}  & 0.45 & 2.27 & 0.37 & 1.44 & 1.60 & 0.53 & 1.51 & 10.5 \\
        \bottomrule
        \end{tabular}        
    \label{tbl:nvdiffreca}
    \end{table}
}

\newcommand{\tabApplicationGenerativeModels}{
    \begin{table}[th]
    \caption{Quantitative FID scores for a 3D generative modeling application. \ours{} can be applied as a differentiable mesh extraction module to GET3D~\cite{gao2022get3d}, producing significantly improved synthesis quality. 
    }
    \begin{tabular}{|l|ccc|}
    \hline
    Isosurfacing Method & Motorbike & Chair & Car \\ 
    \hline
    DMTet~\cite{Shen2021}   & $48.90$ & $22.41$ & 10.60 \\
    \ours{}                   & $\mathbf{44.87}$ & $\mathbf{17.51}$ & $\mathbf{9.55}$ \\
    \hline
    \end{tabular}
    \label{tbl:ApplicationGenerativeModels}
    \end{table}
}

\newcommand{\tabPerformanceIsosurface}{
{

    \begin{table}[th]
    \caption{\updates{Quantitative comparison of the performance of isosurfacing operations. \ours{} may significantly increase time and memory costs compared to simpler extractors, however these  costs are still generally small in the context of downstream applications (see Table~\ref{tbl:PerformanceApplicaitons}).}
    }
    \resizebox{\columnwidth}{!}{
    {
    \begin{tabular}{l|ccc}
    \hline
    $64^3$ & Forward Time (ms) & Backward Time (ms) & Memory (MB) \\ 
    \hline
MC   & $2.28$ & $0.43$ & 12.05 \\
    DMTet   & $2.33$ & $1.38$ & 22.44 \\
$DMC_{centroid}$ ~\cite{nielson2004dual}   & $4.97$ & $1.69$ & 25.08 \\
    \ours{}                   & $8.93 $ & $7.32$ & $116.56$ \\
    \hline
    \end{tabular}
    }
}
    \resizebox{\columnwidth}{!}{
     {
    \begin{tabular}{l|ccc}
    \hline
    $128^3$ & Forward Time (ms) & Backward Time (ms) & Memory (MB) \\ 
    \hline
MC   & $5.08$ & $0.58$ & 72.85 \\
    DMTet   & $6.94$ & $1.39$ & 168.27 \\
$DMC_{centroid}$ ~\cite{nielson2004dual}   & $7.34$ & $1.74$ & 150.75 \\
    \ours{}                   & $14.06$ & $9.53$ & $816.17$ \\
    \hline
    \end{tabular}
    }
    }

    \label{tbl:performanceIsosurface}
    \end{table}
} 
}

\newcommand{\tabPerformanceApplications}{

    \begin{table}[th]
    \centering
    \caption{\updates{Quantitative comparison of the performance of various applications using two isosurfacing methods: DMTet and \ours{}. Note that \textsc{nvdiffrecmc} stores the per-vertex parameters in memory, whereas GET3D uses MLPs to predict these parameters. As a result, the introduction of more parameters leads to a larger increase in memory usage for \textsc{nvdiffrecmc}. \ours{} may even lower the memory requirements of the overall application, because fewer triangles are needed to represent the same geometry.
    \label{tbl:PerformanceApplicaitons}}}
    \resizebox{\columnwidth}{!}{
        {
        
    \begin{tabular}{l|ll|ll}
    
    \hline
    Applications ($96^3$) & \multicolumn{2}{l|}{\textsc{nvdiffrecmc}} & \multicolumn{2}{l}{GET3D}        \\ \hline

    Isosurface     & DMTet               & \ours{}             & DMTet  & \ours{}  \\
    Time per iter. (ms)   & $307$               & $315$                                & $510$  & $610$                    \\
    Memory(GiB)           & $13.1$              & $15.3$                               & $11.6$ & $11.1$                   \\ \hline
    \end{tabular}
    }
    }
    \end{table}
}

%% file: sources/01_intro.tex
\section{Introduction}
\label{sec:Introduction}
Surface meshes serve a ubiquitous role in the representation, transmission, and generation of 3D geometry across fields ranging from computer graphics to robotics. Among many other benefits, surface meshes offer concise yet accurate encodings of arbitrary surfaces, benefit from efficient hardware accelerated rendering, and support solving equations in physical simulation and geometry processing.

However, not all meshes are created equal---the properties above are often realized only on a \emph{high quality} mesh. In fact, meshes which have an excessive number of elements, suffer from self-intersections and sliver elements, or poorly capture the underlying geometry, may be entirely unsuitable for downstream tasks. Generating a high-quality mesh of a particular shape is therefore very important, but far from trivial and often requires significant manual effort.  

The recent explosion of algorithmic content creation and generative 3D modeling tools has led to increased demand for automatic mesh generation. Indeed, the task of producing a high-quality mesh, traditionally the domain of skilled technical artists and modelers, is increasingly tackled via automatic algorithmic pipelines. These are often based on differentiable mesh generation, i.e. parameterizing a space of 3D surface meshes and enabling their optimization for various objectives via gradient-based techniques. For example, applications such as inverse rendering~\cite{munkberg2021nvdiffrec, hasselgren2022nvdiffrecmc}, structural optimization~\cite{subedi2020review}, and generative 3D modeling~\cite{gao2022get3d, lin2022magic3d} all leverage this basic building block. In a perfect world, such applications would simply perform na{\"\i}ve gradient descent with respect to some mesh representation to optimize their desired objectives. However, many obstacles have stood in the way of such a workflow, from the basic question of how to optimize over meshes of varying topology, to the lack of stability and robustness in existing formulations which lead to irreparably low-quality mesh outputs. In this work, we propose a new formulation that brings us closer towards this goal, significantly improving the ease and quality of differentiable mesh generation in a variety of downstream tasks.

\tabTaxonomySimplified

Directly optimizing the vertex positions of a mesh easily falls victim to degeneracy and local minima unless very careful initialization, remeshing, and regularization are used~\cite{softras,Nicolet2021Large,wang2018pixel2mesh}. As such, a common paradigm is to define and optimize a scalar field or a signed distance function (SDF) in space and then extract a triangle mesh approximating the level set of that function. The choice of scalar function representation and mesh extraction scheme greatly affects the performance of an overall optimization pipeline. A subtle but significant challenge of extracting a mesh from a scalar field is that the space of possible generated meshes may be restricted. As we will show later, the choice of the specific algorithm used to extract the triangle mesh directly dictates the properties of the generated shape. 

To capture these concerns, we identify two key properties that a mesh generation procedure should offer to enable easy, efficient, and high-quality optimization for downstream tasks:

\begin{enumerate}
    
    \item \textbf{\textsc{Grad.}} Differentiation with respect to the mesh is well-defined, and gradient-based optimization converges effectively in practice.

    \item \textbf{\textsc{Flexible.}} Mesh vertices can be individually and locally adjusted to fit surface features and find a high-quality mesh with a small number of elements.
\end{enumerate}

However, these two properties are inherently in conflict. Increased flexibility provides more capacity to represent degenerate geometry and self-intersections, which hinder convergence in gradient-based optimization.  As a result, existing techniques~\cite{Shen2021,Lorensen1997,remelli2020meshsdf} usually neglect one of the two properties (Table~\ref{tbl:taxonomy}). For example, the widely-used Marching Cubes procedure~\cite{Lorensen1997} is not \textsc{Flexible}, because the vertices always lie along a fixed lattice and hence generated meshes can never align with non-axis-aligned sharp features (Figure~\ref{fig:teaser}).
Generalized marching techniques can deform the underlying grid~\cite{gao2020deftet, Shen2021}, but still do not allow the adjustment of individual vertices, leading to sliver elements and imperfect fits. On the other hand, Dual Contouring~\cite{Ju2002} is popular for its ability to capture sharp features, but lacks \textsc{Grad.}; the linear system used to position vertices leads to unstable and ineffective optimization. Section~\ref{sec:prev_work} and Table~\ref{tbl:taxonomy} categorize past work in detail.

In this work, we present a new technique called \ours{}, which satisfies both desired properties. Our insight is to adapt a particular Dual Marching Cubes formulation and introduce additional degrees of freedom to \emph{flexibly} position each extracted vertex within its dual cell. We carefully constrain the formulation such that it still produces manifold and watertight meshes that are intersection-free in the vast majority of cases, enabling well-behaved differentiation (\textsc{Grad.}) with respect to the underlying mesh.

The most important property of this formulation is that gradient-based optimization of meshes succeeds consistently in practice. To assess this inherently empirical concern, we devote a significant part of this work to an extensive evaluation of \ours{} on several downstream tasks.  Specifically, we demonstrate that our formulation offers significant benefits for various mesh generation applications, including inverse rendering, optimizing physical and geometric energies, and generative 3D modeling. The resulting meshes concisely capture the desired geometry at low element counts and are easily optimized via gradient descent. 
Moreover, we also propose extensions of \ours{} such as adaptively adjusting the resolution of the mesh via hierarchical refinement, and automatically tetrahedralizing the interior of the domain. 
Benchmarks and experiments show the value of this technique compared to past approaches, which we believe will serve as a valuable tool for high-quality mesh generation in many application areas.

%% file: sources/02_related.tex
\section{Related Work}
\label{sec:prev_work}

In this section, we first provide a broad outline of related work before continuing with an in-depth analysis of the most relevant techniques in Section~\ref{sec:BackgroundAndMotivation}.

\subsection{Isosurface Extraction}
Traditional isosurfacing methods extract a polygonal mesh representing the level set of a scalar function, a problem that has been studied extensively across several fields. Here, we review particularly relevant work and refer the reader to the excellent survey of \citet{de2015survey} for a thorough overview. Following~\citet{de2015survey} we divide isosurfacing methods into three categories and taxonomize the most commonly used ones in Table~\ref{tbl:taxonomy}. 

\paragraph{Spatial Decomposition} Methods in the first category obtain the isosurface through spatial decomposition, which divides the space into cells like cubes or tetrahedrons and creates polygons within the cells that contain the surface~\cite{bloomenthal1988polygonization,bloomenthal1997introduction}. Marching Cubes (MC)~\cite{Lorensen1997} is the most representative method in this category.
As originally presented, Marching Cubes suffers from topological ambiguities and struggles to represent sharp features. 
Subsequent work improves the look-up table which assigns polygon types to cubes~\cite{nielson2003marching,montani1994discretized,scopigno1994modified,hege1997generalized,chernyaev1995marching,lewiner2003efficient} or divides cubes into tetrahedra~\cite{bloomenthal1994implicit} and uses the similar Marching Tetrahedra~\cite{marchingtet} to extract the isosurface.  
To better capture sharp features, Dual Contouring (DC)~\cite{Ju2002} moved to a \emph{dual} representation where mesh vertices are extracted per-cell, and proposed to estimate vertex position according to the local isosurface details. 
Dual Contouring was extended to adaptive meshing~\cite{azernikov2005anisotropic} and can output tetrahedral meshes.
Another improved approach is Dual Marching Cubes (DMC)\updates{~\cite{nielson2004dual}}, which leverages the benefits from both Marching Cubes and Dual Contouring. 
Recently, Neural Marching Cubes~\cite{ChenNMC2021} and Neural Dual Contouring (NDC)~\cite{ChenNDC2022} propose a data-driven approach to position the extracted mesh as a function of input field.
Despite much progress in extraction from known scalar fields, applying isosurfacing methods to gradient-based mesh optimization remains challenging.

\paragraph{Surface Tracking}
Methods in the second category utilize surface tracking and exploit the neighboring information between surface samples to extract the isosurface. Marching Triangles~\cite{hilton1996marching,hilton1997marching}, one of the first representative methods, iteratively triangulates the surface from an initial point under a Delaunay constraint. Following works aim to incorporate adaptivity~\cite{akkouche2001adaptive,karkanis2001curvature} or alignment to sharp features~\cite{mccormick2002edge}. However, gradient-based mesh optimization in the framework of surface tracking would require differentiating through the discrete, iterative update process, which is a non-trivial endeavor. 

\paragraph{Shrink Wrapping}
The methods from the third category rely on shrinking a spherical mesh~\cite{van2004shrinkwrap}, or inflating critical points~\cite{stander1995interactive} to match the isosurface. By default, these methods apply only in limited topological cases and require manual selection of critical points~\cite{bottino1996shrinkwrap} to support arbitrary topology. Moreover, the differentiation through the shrinking process is also not straightforward and hence these methods are not well suited for gradient-based optimization.

\subsection{Gradient-Based Mesh Optimization in ML}
With recent advances in machine learning (ML), several works explore generating 3D meshes with neural networks, whose parameters are optimized via gradient-based optimization under some loss function. 
Early approaches seek to predefine the topology of the generated shape, such as a sphere~\cite{wang2018pixel2mesh,dibr,hanocka2020point2mesh,n3mr}, a union of primitives~\cite{Paschalidou2021CVPR,abstractionTulsiani17} or a set of segmented parts~\cite{coalesce, zhu_siga18,ComplementMe}. However, they are limited in their ability to generalize to objects with complex topologies. To remedy this issue, AtlasNet~\cite{Groueix2018} represents a 3D shape as a collection of parametric surface elements, though it does not encode a coherent surface. Mesh R-CNN~\cite{Gkioxari2019} first predicts a coarse structure which is then refined to a surface mesh. Such a two-stage approach can generate meshes with different topologies, but since the second stage still relies on mesh deformation, topological errors from the first stage can not be rectified. PolyGen~\cite{PolyGen} autogressively generates mesh vertices and edges, but they are limited in requiring 3D ground truth data. CvxNet~\cite{convexnet} and BSPNet~\cite{bspnet} seek to use convex decomposition of the shape or binary planes for space partitioning, however extending them for various objectives defined on the meshes is non-trivial.

More recently, many works explore differentiable mesh reconstruction schemes, which extract an isosurface from an implicit function, often encoded via convolutional networks or implicit neural fields. Deep Marching Cubes~\cite{Liao2018DMC} computes the expectation over possible topologies within a cube, which scales poorly with increasing grid resolution.
MeshSDF~\cite{remelli2020meshsdf} proposes a specialized scheme for sampling gradients through mesh extraction, while \citet{Mehta2022} carefully formulates level set evolution in the neural context. DefTet~\cite{gao2020deftet} predicts a deformable tetrahedral grid to represent 3D objects. Most similar to our method is DMTet~\cite{Shen2021}, which utilizes a differentiable Marching Tetrahedra layer to extract the mesh. An in-depth analysis of DMTet is provided in Section~\ref{sec:BackgroundAndMotivation}.

%% file: sources/03_method.tex
\section{Background and Motivation}
\label{sec:BackgroundAndMotivation}

Here, we first discuss common existing isosurface extraction schemes, to understand their shortcomings and motivate our proposed approach in Section~\ref{sec:Method}.

\paragraph{Problem Statement}
As outlined in Section~\ref{sec:Introduction}, we seek a representation for differentiable mesh optimization, where the basic pipeline is to: i) define a scalar signed-distance function in space, ii) extract its $0$-isosurface as a triangle mesh, iii) evaluate objective functions on that mesh, and iv) back-propagate gradients to the underlying scalar function.
Several popular algorithms in widespread use for isosurface extraction still have significant issues in this differentiable setting.
The main challenge is that the effectiveness of gradient-based optimization depends dramatically on the particular mechanism for isosurface extraction: restrictive parameterizations, numerically unstable expressions, and topological obstructions all lead to failures and artifacts when used in gradient-based optimization.

We emphasize that our \ours{} representation is \emph{not} intended for isosurface extraction from fixed, known scalar fields, the primary case considered in past work.
Instead, we particularly consider differentiable mesh optimization, where the underlying scalar field is an unknown and extraction is performed many times during gradient-based optimization. This setting offers new challenges and motivates a specialized approach.

\paragraph{Notation} All methods we consider extract an isosurface from a scalar function $s : \mathbb{R}^3 \to \mathbb{R}$, sampled at the vertices of a regular grid and interpolated within each cell.
The function $s$ may be discretized directly as values at grid vertices, or evaluated from an underlying neural network, \etc{}, the exact parameterization of $s$ makes no difference for isosurface extraction.
For clarity, the set $X$ denotes the vertices of the grid with cells $C$, while $M = (V,F)$ denotes the resulting extracted mesh with vertices $V$ and faces $F$.
We implicitly overload $v \in V$  or $x \in X$ to refer to either a logical vertex, or that vertex's position in space \eg{} $x \in \mathbb{R}^3$.

\subsection{Marching Cubes \& Tetrahedra}

The most direct approach is to extract a mesh with vertices on the grid lattice, and one or more mesh faces within each grid cell, as in Marching Cubes~\cite{Lorensen1997}, Marching Tetrahedra~\cite{marchingtet}, and many generalizations.
Mesh vertices are extracted along grid edges where the linearly-interpolated scalar function changes sign
\begin{equation}
    u_e = \frac{x_a \cdot s(x_b) - x_b \cdot s(x_a)}{s(x_b) - s(x_a)}.
    \label{eq:linear_interpolated_sf}
\end{equation}
\citet{remelli2020meshsdf,Liao2018DMC} observe that this expression contains a singularity when $s(v_a) = s(v_b)$, which might obstruct differential optimization, although \citet{Shen2021} note that Equation~\ref{eq:linear_interpolated_sf} is never evaluated under the singular condition during extraction. 
The resulting mesh is always self-intersection-free and manifold.

However, the mesh vertices resulting from marching extraction can only lie along a sparse lattice of grid edges, by construction.
This prevents the mesh from fitting to sharp features, and unavoidably creates poor-quality sliver triangles when the isosurface passes near a vertex.
Recent methods propose schemes beyond naive auto-differentiation to compute improved gradients on the underlying scalar field \cite{remelli2020meshsdf,Mehta2022}, but this does not address the restricted output space for the mesh. 

A promising remedy is to allow the underlying grid vertices to deform~\cite{Shen2021,gao2020deftet}. 
Although this generalization significantly improves performance, the extracted mesh vertices are still not able to move independently, leading to star-shaped skinny triangle artifacts as mesh vertices cluster around a degree of freedom on the grid.
Our method takes inspiration from~\citet{Shen2021} and also leverages grid deformation, but augments the representation with additional degrees of freedom to allow independent repositioning of the vertices, as shown in Figure~\ref{fig:differentiability}.

\figQEF

\figDMCCompare

\figDifferentiability

\subsection{Dual Contouring}

As the name suggests, Dual Contouring (DC)~\cite{Ju2002} moves to a \emph{dual} representation, extracting mesh vertices that can be generally positioned within grid cells to better capture sharp geometric features.
The position of each mesh vertex is computed by minimizing a local quadratic error function~(QEF) depending on the local values and spatial gradients of the scalar function $s$
\begin{equation}
  v_d = \operatorname*{argmin}_{v_d} \sum_{u_e \in \mathcal{Z}_e} \nabla s(u_e) \cdot (v_d - u_e).
\label{eq:qef}
\end{equation}
where $u_e \in \mathcal{Z}_e$ are the zero-crossings of the linearly-interpolated scalar function along the cell edges.

Dual Contouring excels at fitting sharp features when extracting a single mesh from a fixed scalar function, but several properties impede its use in differential optimization.
Most importantly, Equation~\ref{eq:qef} does not guarantee that the extracted vertex lies inside the grid cell. In fact, co-planar gradient vectors $\nabla s(u_e)$ create degenerate configurations in which the vertex explodes to a distant location, leading to self-intersections and numerically unstable optimization when differentiating through the formulation.
Explicitly constraining the vertex to lie in the cell zeros out the gradient, and regularizing Equation~\ref{eq:qef} enough to resolve the issue removes the ability to fit sharp features (Figure~\ref{fig:qef} \& \ref{fig:differentiability}).
Additionally, the resulting mesh connectivity may be nonmanifold, and the output mesh contains non-planar quadrilaterals which introduce error as they are split in to triangles (Figure~\ref{fig:dmc_compare}).

Recent generalizations \cite{ChenNDC2022} of Dual Contouring replace Equation~\ref{eq:qef} with a learned neural network, improving extraction quality from imperfect but \emph{fixed} scalar functions.
However, when optimizing with respect to the underlying function, differentiating through an additional neural network further complicates the optimization landscape and impedes convergence (Figure~\ref{fig:differentiability}). 

Our approach takes inspiration from these methods and the importance of positioning each vertex freely within a cell.
However, rather than explicitly positioning the extracted vertex as a function solely of a scalar field, we introduce additional carefully-chosen degrees of freedom which are optimized to locally adjust the vertex position.
We are able to resolve manifoldness by instead basing our scheme on the similar but lesser-known \emph{Dual Marching Cubes}.

\subsection{Dual Marching Cubes}
\label{sec:dmc}
Much like Dual Contouring, Dual Marching Cubes\updates{~\cite{nielson2004dual}} extracts vertices positioned within grid cells.
However, rather than extracting a mesh along the dual connectivity of the \emph{grid}, it extracts a mesh along the dual connectivity of the mesh that \emph{would be extracted by Marching Cubes}.
This allows for manifold mesh outputs for all configurations, by emitting multiple mesh vertices within a single grid cell when needed.
The extracted vertex locations are defined either as the minimizer of a QEF akin to Dual Contouring\updates{~\cite{schaefer2007manifold}}, or as a geometric function of the primal mesh geometry~\cite{nielson2004dual}, such as the face centroid.

In general, Dual Marching Cubes improves the connectivity of the extracted mesh \vs{} Dual Contouring, but if a QEF is used for vertex positioning, it suffers from many of the same drawbacks as Dual Contouring.
If vertices are positioned at the centroids of the primal mesh, then the formulation lacks the freedom to fit individual sharp features.
In the subsequent text, whenever we refer to Dual Marching Cubes we mean the centroid approach, unless otherwise clarified.

Our approach builds on Dual Marching Cube extraction, but we introduce additional parameters for positioning vertices which generalize the centroid approach.
Basing our method off a scheme which can emit correct topology even in difficult configurations is one key to our success.

\section{Method}
\label{sec:Method}

We propose the \ours{} representation for differentiable mesh optimization.
The core of the method is a scalar function on a grid, from which we extract a triangle mesh via Dual Marching Cubes.
Our main contribution is to introduce three additional sets of parameters, carefully chosen to add flexibility to the mesh representation while retaining robustness and ease of optimization:
\begin{itemize}[leftmargin=2em,itemsep=0.4em]
  \item \textbf{Interpolation weights} $\alpha\in \mathbb{R}_{> 0}^{8},\beta\in \mathbb{R}_{> 0}^{12}$ per grid cell, to position dual vertices in space (Section~\ref{sec:FlexibleDualVertexPositioning}).
  \item \textbf{Splitting weights} $\gamma\in \mathbb{R}_{> 0}$ per grid cell, to control how quadrilaterals are split into triangles (Section~\ref{sec:quad_spliting}).
  \item \textbf{Deformation vectors} $\delta \in \mathbb{R}^{3}$ per vertex of the underlying grid for spatial alignment, as in \citet{Shen2021} (Section~\ref{sec:grid_deformation}).
\end{itemize}
These parameters are optimized along with the scalar function $s$ via auto-differentiation to fit a mesh to the desired objective.
We also present extensions of the \ours{} representation to extract a tetrahedral mesh of the volume (Section~\ref{sec:tetmesh}) and represent hierarchical meshes with adaptive resolution (Section~\ref{sec:adaptive}).

\subsection{Dual Marching Cubes Mesh Extraction}
\label{sec:DualMarchingCubesMeshExtraction}

We begin by extracting the connectivity of the Dual Marching Cubes mesh based on the value of the scalar function $s(x)$ at each grid vertex $x$, just as in \updates{~\citet{nielson2004dual,schaefer2007manifold}}.
The signs of $s(x)$ at cube corners determine the connectivity and adjacency relationships (Figure~\ref{fig:dmc_configs}).
Unlike ordinary Marching Cubes, which extracts vertices along grid edges, Dual Marching Cubes extracts a vertex for each primal face in the cell; typically a single vertex, but possibly up to four (Figure~\ref{fig:dmc_configs}, case C13).
Extracted vertices in adjacent cells are linked by edges to form the dual mesh, composed of quadrilateral faces (Figure~\ref{fig:dmc}).
\updates{The resulting mesh is guaranteed to be manifold, although due to the additional degrees of freedom described below, it may rarely contain self-intersections; see Section~\ref{sec:Limitations}.}

\subsection{Flexible Dual Vertex Positioning}
\label{sec:FlexibleDualVertexPositioning}

Our method generalizes ordinary Dual Marching Cubes in how the extracted mesh vertex locations are computed.
Recall that Marching Cubes primal vertices are located at scalar zero-crossings along grid cell edges 
\begin{equation}
    u_e = \frac{x_a \cdot s(x_b) - x_b \cdot s(x_a)}{s(x_b) - s(x_a)},
    \label{eq:linear_interpolation}
\end{equation}
and ordinary Dual Marching Cubes then defines the location of each extracted vertex to be the centroid of its primal face
\begin{equation}
v_d = \frac{1}{|V_E|} \sum_{u_e\in V_E} u_e,
\label{eq:v_d}
\end{equation}
where $V_E$ is the set of crossings which are the primal face vertices.

To introduce additional flexibility into this representation, we first define a set of weights in each grid cell $\alpha \in \mathbb{R}_{> 0}^8$ associating a positive scalar with each cube corner. 
These weights adjust the location of the crossing point $c_e$ 
along each edge, and Equation~\ref{eq:linear_interpolation} then becomes
\begin{equation}
    u_e = \frac{s(x_i) \alpha_i x_j - s(x_j) \alpha_j x_i}{s(x_i)\alpha_i-s(x_j)\alpha_j}.
    \label{eq:flexible_edge_interpolation}
\end{equation}
\updates{In our implementation, we apply a $\textrm{tanh}(\cdot)+1$ activation function to restrict $\alpha \in [0, 2]$, and do not observe any convergence problems due to degeneracy.}

Likewise, rather than naively positioning the dual vertex at the centroid of the primal face, we introduce a set of weights in each cell $\beta \in \mathbb{R}_{> 0}^{12}$, associating a positive scalar with each cube edge.
These weights adjust the location of the dual vertex inside each face, and Equation~\ref{eq:v_d} then becomes
\begin{equation}
    v_d = \frac{1}{\sum_{u_e \in V_E} \beta_e} \sum_{u_e \in V_E} \beta_e u_e.
    \label{eq:flexible_face_interpolation}
\end{equation}
\updates{In practice, we again apply a $\textrm{tanh}(\cdot)+1$ activation to restrict the range of $\beta$, similar to $\alpha$.}

Together these weights $\alpha \in \mathbb{R}_{> 0}^8$,$\beta \in \mathbb{R}_{> 0}^{12}$ amount to $20$ scalars per grid cell.
In both cases, weights are defined independently per cell, \emph{not} shared at adjacent corners or edges; independent weights offer more flexibility, and there is no continuity condition to maintain at adjacent elements in our dual setting.

Notice that both Equation~\ref{eq:flexible_edge_interpolation} \& \ref{eq:flexible_face_interpolation} are intentionally parameterized as convex combinations, and thus the resulting extracted vertex position is necessarily within the convex hull of its grid cell vertices.
Furthermore, when a convex cell emits multiple dual vertices (Figure~\ref{fig:dmc_configs}), the corresponding primal faces in which the dual vertices are positioned are non-intersecting, which prevents nearly all self-intersections in the resulting mesh (see Section~\ref{sec:discussion} and Supplement).

\figDMC

\figDualVertex

\figDMCConfigs

\subsection{Flexible Quad Splitting}
\label{sec:quad_spliting}

\figSpliting

Dual Marching Cubes, and thus also \ours{}, extracts pure quadrilateral meshes with non-planar faces, which are typically split to triangles for processing in downstream applications.
Simply splitting along an arbitrary diagonal can lead to significant artifacts in curved regions (Figure~\ref{fig:spliting}), and there is in general no single ideal policy to split non-planar quads to represent unknown geometry.
Our next parameter is introduced to make the choice of split flexible, and optimize it as a continuous degree of freedom.

We define a weight $\gamma \in \mathbb{R}_{>0}$ in each grid cell, which is propagated to the emitted vertices in the extracted mesh.
At optimization-time \emph{only}, each quadrilateral mesh face is split into 4 triangles by inserting a midpoint vertex $\overline{v_d}$ (Figure~\ref{fig:spliting}).
The location of this midpoint is computed as
\begin{equation}
  \overline{v_d} = 
      \frac{\gamma_{c_1} \gamma_{c_3} (v_d^{c_1}+v_d^{c_3})/2 + 
            \gamma_{c_2} \gamma_{c_4} (v_d^{c_2}+v_d^{c_4})/2  }
           {\gamma_{c_1} \gamma_{c_3}+\gamma_{c_2} \gamma_{c_4}}
           \label{eq:spliting}
\end{equation}
with notation is as in Figure~\ref{fig:spliting}.
This is a weighted combination of the midpoints of the two possible diagonals of the face, where the weights come from the $\gamma$ parameters on the corresponding vertices.
Intuitively, adjusting the $\gamma$ weights smoothly interpolates between the geometries resulting from the two possible splits.
Optimizing $\gamma$ encourages the choice of split which fits the objective of interest. 
For final extraction when optimization is complete, we do \emph{not} insert the midpoint vertex $\overline{v_d}$, but simply split each quadrilateral along whichever diagonal has larger product of $\gamma$ values.

\figAblQual

\subsection{Flexible Grid Deformation}
\label{sec:grid_deformation}

Inspired by DefTet~\cite{gao2020deftet} and DMTet~\cite{Shen2021}, we furthermore allow the vertices of the underlying grid to deform according to displacements $\delta \in \mathbb{R}^3$ at each grid vertex.
These deformations allow the grid to locally align with thin features, and give additional flexibility in positioning vertices.
We limit the deformation to at most half of the grid spacing to ensure that grid cells never invert.

\subsection{Tetrahedral Mesh Extraction}
\label{sec:tetmesh}
\figTetmesh

Many applications such as physical simulation and character animation require a tetrahedralization of the shape volume.
We augment \ours{} to additionally output a tetrahedral mesh when desired, which exactly conforms to the boundary of the extracted surface and supports automatic differentiation in the same sense as our surface extraction.

Our approach adapts the strategy proposed by~\citet{liang2014octree} for Dual Contouring.
The vertex set for the tetrahedral mesh is the union of the grid vertices, our extracted mesh vertices in cells, and the midpoint of any cell for which no surface vertex was extracted.
We then emit tetrahedra as shown in Figure~\ref{fig:tetmesh}, \emph{left}.
For each grid edge connecting two grid vertices with the same sign, four tetrahedra are generated, each formed by the two grid vertices and two vertices in consecutive adjacent cells.
For each grid edge connecting two grid vertices with different signs, two four-sided pyramids are generated, each formed by one grid vertex and a vertex from each adjacent cell. 
These pyramids are then split at the base as in Section~\ref{sec:quad_spliting} to yield two tetrahedra each.
When working with Dual Marching Cubes connectivity, there is an additional complexity that a cell may contain multiple extracted mesh vertices, and we must choose the correct vertex when forming tetrahedra.
In most cases, this choice can be read-off unambiguously from Figure~\ref{fig:dmc_configs}; although rare difficult deformed configurations lead to small mesh defects--we detail these in the Supplement, and find that they do not obstruct downstream applications.
The resulting meshes are visualized in Figure~\ref{fig:tetmesh}, \emph{right}, and Figure~\ref{fig:physicus} demonstrates an application of differentiable physical simulation.

\subsection{Adaptive Mesh Resolution}
\label{sec:adaptive}
\figOctree
\figOctreeFix

We also augment \ours{} to leverage adapative hierarchical grids, and represent meshes which variably increased spatial resolution in areas of high geometric detail.
The policy of where to refine the octree grid representation is application-specific, \eg{} thresholds on local curvature in geometric fitting or visual error in inverse rendering; our representation is responsible for extracting hierarchically adaptive meshes while maintaining the key properties of flexibility and effective gradient-based optimization.
Here we again mimic approaches designed for Dual Contouring~\cite{Ju2002,schaefer2007manifold}, adapting them to our \ours{} extension of Dual Marching Cubes.

The approach is to locally refine our background grid into a hierarchical octree with varying resolution.
Most of our algorithm applies unchanged on an octree, except for the challenge of connecting adjacent dual vertices to form quadrilateral meshes faces when they span different levels of the octree.
On a general octree there may not exist any dual face connectivity which yields a closed manifold mesh (Figure ~\ref{fig:octree}); existing methods mitigate this problem by constraining the topology of the octree or signs of the implicit function~\cite{Ju2002,schaefer2007manifold}.
However, these rules are not applicable in an optimization setting, where the topology is unknown and constantly changing. 
We adopt the approach shown in Figure~\ref{fig:octree_fix}; refined octree grid vertices adjacent to coarser cells always take their value as the interpolated value from coarse face vertices, guaranteeing consistency of signs.
This projection yields nearly watertight adaptive meshes in our experiments.
Here again the combination of all possible configurations from Figure~\ref{fig:dmc_configs} at adjacent octree nodes of different hierarchies leaves a small number of cases where the extracted mesh contains a hole. 
Nonetheless, the adaptively refined mesh yields significant improvements, as shown in Figure~\ref{fig:adaptive_qual}.

\subsection{Regularizers}

Our method is a general-purpose tool, which can be optimized according to application-specific objectives and regularizers, including geometric depth and SDF losses, image-space rendering losses, and mesh-quality regularizers.
In the next section, we will detail several examples utilizing such terms.
Here, we first propose two regularization terms which are specific to the internals of our representation.

\updates{
Our over-parameterization of the location of each vertex, described in Section ~\ref{sec:Method},  is intentional and beneficial, allowing for properties such as the convex weighting in Section~\ref{sec:FlexibleDualVertexPositioning}, and the bounded grid deformation in Section~\ref{sec:grid_deformation}, as well as easing stochastic optimization. As such, we introduce two terms to regularize the internal representation, and encourage non-degenerate parameters which can easily “flex” to accommodate any local vertex movement. These regularizers are used for all examples shown in this work.
}

The first term penalizes the deviation of the distances between each dual vertex and the edge crossings which compose the face in which it sits
\begin{equation}
\mathcal{L}_{\textrm{dev}} := \sum_{v \in V} \textrm{MAD} \big[ \{ |v - u_e|_2: u_e \in \mathcal{N}_v \}  \big],
\end{equation}
where $|\cdot|_2$ is Euclidean distance, $\textrm{MAD}$ denotes the \emph{mean absolute deviation} ${\textrm{MAD}(Y) = \frac{1}{|Y|} \sum_{y \in Y} |y - \textrm{mean}(Y)|}$, and $u_e \in \mathcal{N}_v$ are the edge crossings which bound the primal face for dual vertex $v$.
This term regularizes the extracted connectivity, and encourages vertices to lie near the center of their cell so they have a margin in which to flex and adapt.

The second term discourages spurious geometry in regions of the shape which receive no supervision in the application objective, such as internal cavities.
We follow~\citet{munkberg2021nvdiffrec} and penalize sign changes of the implicit function on all grid edges.
First, we let $\vec{\mathcal{E}}_g$ be the set of all pairs of scalar function values $(s_a,s_b)$ at grid vertices $(a,b)$ connected by an edge and with $sign(s_a) \neq sign(s_b)$.
Then the loss is given by
\begin{equation}
\mathcal{L}_{\textrm{sign}} := \sum_{(s_a,s_b)\in \vec{\mathcal{E}}_g} H\big(\sigma(s_a), \textrm{sign}(s_b))\big),
\end{equation}
where $H$, $\sigma$ are cross-entropy and sigmoid functions respectively.

%% file: sources/04_experiments.tex
\section{Experiments}

\figMainQual

In this section, we evaluate \ours{} in various mesh optimization tasks. First, we analyze the capacity of \ours{} in reconstructing 3D geometry under perfect 3D supervision defined on the surface and compare with other iso-surfacing techniques in Section~\ref{sec:shape_reconstruction}. Next, we show that benefiting from differentiably extracting an explicit mesh, \ours{} can further optimize for various mesh-based regularization losses to improve the mesh quality for downstream applications.

\tabMainOneSimplified
    
\subsection{Mesh Reconstruction} 
\label{sec:shape_reconstruction}
\paragraph{Motivation and Experimental settings}
To evaluate the performance of optimizing 3D meshes using isosurfacing methods and avoid the inefficiency that could be introduced by imperfect objective functions, we experiment in an ideal setting where we define the objective functions directly on the geometric difference between the extracted mesh and a ground truth mesh. More specifically, in each iteration we reconstruct a mesh, render depth and silhouette images from a randomly sampled camera pose and compute the differences with a ground truth depth and silhouette images. We also compute the SDF loss, where we randomly sample 1000 points and evaluate their SDF values w.r.t the ground truth mesh as well as the extracted mesh, and minimize the differences between two SDF values. Please refer to the Supplement for details of the objective functions and their weighting factors.

\paragraph{Dataset} We use the dataset collected by~\citet{MPZ14}, which contains 3D shapes from the AIM@Shape database and popular assets from other community repositories. This shape collection has a great diversity in geometric features and topology complexities, ranging from noisy scanned surfaces to highly-detailed CAD models.  Following~\citet{ChenNMC2021}, we remove the non-watertight and very skinny (e.g. wires) shapes, which are not suitable for isosurfacing methods to reconstruct. In total, we use 79 different shapes in our evaluation.

\paragraph{ Baselines}
As shown in Figure~\ref{fig:main_qual}, we compare \ours{} with different methods split into two categories. \ours{} is grouped with the \emph{differentiable} isosurfacing algorithms, MC and \textsc{DMTet}, which provides the most direct comparisons. We reconstruct meshes through optimization with objective functions mentioned above. Note that the resolution of tetrahedral grid used by DMTet is not directly comparable with voxel grids used by our method, as the number of vertices are different under the same resolution. Thus, we additionally report DMTet at different resolution to match the triangle counts. The resolution of DMTet is specified in the brackets.

In the other category we group the \emph{non-differentiable} isosurfacing methods. To ensure a fair comparison, we use the \textit{ground truth} SDF field and extract the mesh using vanilla MC ($MC_{SDF}$), DC ($DC_{hermite}$) and NDC ($NDC_{SDF}$).
For DC we complement the ground truth SDF with normal vectors computed using finite differences, 
and for NDC, we use a pretrained model provided by the authors\footnote{\url{https://github.com/czq142857/NDC}}.

\paragraph{Evaluation Metrics} We evaluate the reconstructed meshes in terms of reconstruction accuracy and intrinsic quality of the reconstructed mesh. For the former, we follow NDC~\cite{ChenNDC2022} and compute Chamfer Distance (CD), F-Score (F1), Edge Chamfer Distance (EDC), Edge F-score (EF1), and the percentage of Inaccurate Normals (IN> $5^{\circ}$) w.r.t to the ground truth mesh. For the latter, we compute triangle aspect ratios, radius ratios, and min and max angles. A detailed description of the evaluation metrics is provided in the Supplement.

\figAdaptiveQual
\tabAblationSimplified
\figTriQualityHist
\paragraph{Results} 

The quantitative results of reconstruction quality are provided in Table~\ref{tbl:main64} with qualitative examples depicted in Figure~\ref{fig:main_qual}. Figure~\ref{fig:tri_quality_hist} shows the quantitative results of intrinsic mesh quality. Methods that extract the mesh as a post-processing step fail to achieve competitive performance in terms of reconstruction quality, highlighting the importance of end-to-end optimization that mitigates the discretization errors introduced in post-processing. When compared with other methods that use differentiable iso-surfacing for mesh reconstruction (MC and DMTet), \ours{} extracts meshes that align significantly better with ground truth geometry, while maintaining superior mesh quality which is on par with the best performing NDC method.

We further ablate each component we introduced in \ours{}, and provide quantitative results in Table~\ref{tbl:ablation} with qualitative examples in Figure~\ref{fig:ablqual}. 
In the Supplement we also include reconstructions of the same object under different rotations. 

\subsection{Mesh Optimization with Regularizations}

Our \ours{} representation is flexible enough that objectives and regularizers which depend on the extracted mesh itself can be directly evaluated with automatic differentiation and incorporated into gradient-based optimization.
Some surface-based regularizers such as surface area may be easily expressed directly as functions of the underlying scalar field, while others, especially those which depend on the mesh discretization itself, have no direct equivalent.
This same simple strategy does not succeed with more rigid representations like Marching Cubes, because the extracted mesh does not have the degrees of freedom to adapt to arbitrary objectives.
We provide two examples of mesh regularizations below.

\figEdgeReg
\tabEquiEdgeSimpNew

\paragraph{Equilateral Edge Length}
In many applications, such as physics simulation, generating equilateral triangles is preferable over thin triangles. We penalize the variance of the edge  lengths on the extracted mesh in this regularization. In particular, we compute the average edge length $\bar{e} = \frac{1}{|\mathcal{E}|}\sum_{e\in\mathcal{E}} |e|_2$, where $\mathcal{E}$ denotes the set of all the edges in the extracted mesh. The regularization is computed as:  $ R_{\text{edge}} = \frac{1}{|T|} \sum_{e \in t,  t \in T } (|e - \bar{e}|^2)$, where $T$ is the set of extracted triangles. 
We combine this regularization with the reconstruction loss mentioned in Section~\ref{sec:shape_reconstruction}. We first run the optimization to reconstruct an input mesh without the regularization term for 1000 steps, then we further run  300 steps using both the reconstruction loss and the regularization loss, with the regularization weight progressively increasing from 0 to 100.
Adding equilateral triangle regularization allows \ours{} to generate more uniform triangles with a slight degradation in the reconstruction quality.
We compare \ours{} with MC and DMTet, and provide qualitative results in Figure~\ref{fig:EdgeReg}. 
The quantitative comparison in Table~\ref{tbl:equiedge_simp} shows that both our method and DMTet can gain a significant improvement in triangle quality after adding the regularization, as measured by percentages of triangles having Aspect Ratio > 4, Radius Ratio > 4, or Min Angle < 10.
However, our method has a significantly smaller drop in geometric quality, as measured by the first two metrics in Table~\ref{tbl:equiedge_simp}, thanks to the flexibility of our surface extraction formulation.

\figdevelopable

\paragraph{Developability}

As a more complex mesh-based term, we consider the \emph{developability} energy of \citet[Equation 4]{Stein:2018:DSF}, which amounts to penalizing the smallest eigenvalue of the covariance matrix of face normals about each vertex.
Developability is a geometric measure penalizes stretching of the surface relative to a flat sheet, but does not penalize bending in a single direction; it has applications to manufacturing from sheets of material like sheet metal or plywood.
Although developability could in-principle be quantified directly on an implicit function, it has a significant relationship to discrete mesh connectivity, as discussed by \citet{Stein:2018:DSF}.
Figure~\ref{fig:developable} shows the result of incorporating this term into a synthetic reconstruction problem.
Attempting to do the same with Marching Cubes is much less successful, failing to preserve shape features and achieve the desired style.

%% file: sources/05_applications.tex
\section{Applications}
\subsection{Photogrammetry Through Differentiable Rendering}

\figNerfVisu
\figNVDIFFRECPhotoSupplemental

The differentiable isosurfacing technique DMTet~\shortcite{Shen2021} is at the core of the recent
work, \nvdiffrec{}, which jointly optimizes shape, materials, and lighting from 
images~\cite{munkberg2021nvdiffrec,hasselgren2022nvdiffrecmc}. By simply replacing DMTet with \ours{} in the topology optimization step, leaving the remainder 
of the pipeline unmodified, we observe improved geometry reconstructions at equal triangle count, which is illustrated in Figure~\ref{fig:roller}. We also 
report \nvdiffrec{} result with DMTet vs. \ours{} on the NeRF synthetic dataset~\cite{Mildenhall2020}. View interpolation scores and Chamfer distances are shown in Table~\ref{tbl:nvdiffreca}. \updates{We show additional results on datasets of real-world photographs in Figure~\ref{fig:nvdiffrec_photo_suppl}.}
In general, \ours{} produces fewer sliver triangles as can be observed in the visual examples (Figure~\ref{fig:nerf_visu}) and the min angle histogram.
Additionally, the nicer triangulation of \ours{} leads to easier creation of unique texture coordinates (UV unwrapping) and improved UV layouts when running the extracted meshes through an off-the-shelf unwrapping tool~\cite{xatlas}. Figure~\ref{fig:uv} illustrates this property. 

\tabNVDIFFRECA

\figRoller

\figUV

\subsection{Mesh Simplification of Animated Objects}

We show the benefit of mesh optimization using the explicit mesh representation from \ours{} in an animated meshing task, illustrated in Figure~\ref{fig:skinning}. Given a known animated skeleton and images of the target animated object, we leverage \nvdiffrec{} to extract a concise mesh which accurately represents the object throughout the animation. Here, we use an animation sequence from 
RenderPeople~\shortcite{RenderPeople2020}.

Rather than fitting a single mesh in a reference pose, \ours{} allows us to differentiably skin and deform the mesh via off-the-shelf skinning tools, and simultaneously optimize with respect to the entire animated sequence. This is in contrast to neural volumetric~\cite{Mildenhall2020} or implicit surface representations~\cite{Wang2021neus}, where a geometry deformation system either needs to be redesigned for the specific neural representation, e.g., D-NeRF~\cite{pumarola2020d}, or where skinning is applied only \emph{after} mesh optimization, without end-to-end mesh optimization and gradient flow through the deformation.

As a baseline, we use \ours{} and optimize using images from randomized cameras viewing only the mesh in its T-pose. The mesh is then re-skinned in a post-processing step. To illustrate the benefit of optimizing over the animation, we combine \ours{} with the differentiable mesh skinning approach of Hasselgren~et~al.~\shortcite{hasselgren2021nvdiffmodeling}, re-skin the mesh in each training iteration, and optimize for image loss using images rendered from randomized cameras and animation frames. 
As shown in the bottom part of Figure~\ref{fig:skinning}, optimizing for the appearance over the entire animation helps re-distribute triangle density to avoid mesh stretching. %
Note that the differentiable skinning approach from Hasselgren~et~al. deforms a template mesh with fixed topology, while the \ours{} version presented here additionally optimizes topology, hence provides a more flexible approach to mesh simplification of animated assets.

\figSkining

\subsection{3D Mesh Generation}

Generation of 3D meshes, typically with the goal of facilitating 3D content creation, is an important task for computer graphics and vision, and benefits industries such as gaming and social platforms. 
Recent 3D generative models~\cite{gao2022get3d,Chan2022,Schwarz2022,zhou2021CIPS3D,gu2021stylenerf}  differentiably render a 3D representation into 2D images, and combine with a classic generative adversarial framework~\cite{karras2019style,Karras2019stylegan2} to synthesize 3D content using only 2D image supervision. The recent state-of-the-art GET3D~\cite{gao2022get3d} directly synthesizes high-quality textured 3D meshes, enabled by the differentiable iso-surfacing module DMTet~\cite{Shen2021}.

In this application, we demonstrate that \ours{} can serve as a plug-and-play differentiable mesh extraction module in a 3D generative model, and produce significantly improved mesh quality. 
Specifically, we use GET3D~\cite{gao2022get3d} and replace DMTet with \ours{} in the mesh extraction step. We only modify the last layer of the 3D generator in GET3D to additionally generate 21 weights for every cube in \ours{}. The training procedure, dataset (we use ShapeNet~\cite{shapenet}) and other hyperparameters of GET3D are kept unchanged. 

\tabApplicationGenerativeModels
\figApplicationMeshGeneration

Qualitative comparisons and quantitative results are provided in Figure~\ref{fig:ApplicationMeshGeneration} and Table~\ref{tbl:ApplicationGenerativeModels}, respectively.
\ours{} achieves better FID scores across all categories, demonstrating the higher capacity in generating 3D models. Qualitatively, the shapes generated using the \ours{} version of GET3D are of significantly higher quality, with more details and fewer sliver triangles.

\subsection{Differentiable Physics Simulation}

\figphysics

To leverage \ours{}'s ability to differentiably extract tetrahedral meshes, we combine it with differentiable physics simulation~\cite{gradsim} and a differentiable rendering pipeline \cite{Laine2020diffrast} to jointly recover 3D shapes and physical parameters from multi-view videos.
Given a video sequence of an object deforming, we aim to recover a tetrahedral mesh of the rest pose as well as material parameters which reproduce the motion under simulation.
In particular, we focus on FEM simulation with neo-Hookean elasticity to model elastic objects. After extracting the tetrahedral mesh from \ours{}, we feed it into GradSim~\cite{gradsim} to obtain deformed shapes at different time steps, these shapes are then differentiably rendered into multi-view images. We optimize both the 3D geometry and the physical density of the 3D shape in two-stage manner as in past work. See Figure~\ref{fig:physicus} and the Supplement for more details. The optimized physical parameters and 3D geometry with texture are close to the ground truth.

%% file: sources/06_conclusion.tex
\section{Discussion}
\label{sec:discussion}

\subsection{Performance}
Introducing additional degrees of freedom into the extraction representation incurs a moderate increase in runtime and memory usage. However, in many applications, the cost of mesh extraction is often small compared to the overall computation, and the ability to work with more concise extracted meshes may ultimately reduce the memory requirements of the overall pipeline. \updates{Concretely, we show a performance benchmark of different isosurfacing methods in Table~\ref{tbl:performanceIsosurface}. FlexiCubes is indeed slower and more memory-intensive than DMTet, and significantly more so than ordinary Marching Cubes, but all of these costs are small compared to the downstream task, which we benchmark in Table~\ref{tbl:PerformanceApplicaitons}. The maximum grid resolution is not constrained by isosurface extraction, but rather by other components of the applications, such as rendering or neural network evaluations. We consistently choose the highest resolution that can be supported by high-end GPUs for all of our applications.}

\tabPerformanceIsosurface
\tabPerformanceApplications

\subsection{Limitations}
\label{sec:Limitations}

\paragraph{\updates{Self-intersections}}
Although our approach generally produces high-quality meshes with improved element shapes in practice, \updates{and our core algorithm guarantees manifoldness, we do not guarantee non-self-intersecting output}.
\updates{
Intersections arise because our flexible dual representation (Section~\ref{sec:Method}) allows the extracted vertices to move into intersecting configurations; we found that strictly constraining the motions to non-intersecting configurations unacceptably worsened the expressivity and ease of optimization of the method.}
At a grid resolution of $64^3$, in our experiment with 79 3D shapes in Table~\ref{tbl:main64}, we observed self intersections on $0.10\%$ of the triangles, and we note that this is lower than Dual Contouring variants (DC: $1.48\%$, NDC: $0.13\%$). Our optional extensions to tetrahedral and hierarchical meshing have slightly weaker guarantees, occasionally containing small cavities or nonmanifold elements in ambiguous cases arising from the Dual Marching Cubes topology. In our evaluation, we find that these small imperfections are not detrimental for downstream applications, but note that additional consideration may be required if a watertight mesh is imperative for a given application.

\paragraph{\updates{Continuity}}
More fundamentally, although we consider differentiable mesh extraction, our method is actually not even globally continuous.
When the isosurface slips over a grid vertex, the mesh jumps discontinuously, a property we inherit from Dual Contouring and Dual Marching Cubes.
\updates{
Fortunately, because we apply our extraction in stochastic optimization settings, such as stochastic gradient descent with Adam, small local discontinuities do not obstruct optimization in practice.
For this reason, we focus on our analysis and experiments on the property of effective optimization in downstream applications  (Figure \ref{fig:differentiability}), rather than on formal notions of differentiability or smoothness.
}

\subsection{Future Work}

Looking forward, one opportunity to advance this approach is to integrate volumetric rendering with mesh-based representations for improved gradient approximation on visual tasks~\cite{chen2022mobilenerf}.
Furthermore, 4D spatiotemporal meshing has important applications in dynamic geometry representation and optimization~\cite{park2021hypernerf}.
Very directly, we also hope to integrate adaptive hierarchical mesh extraction (Section~\ref{sec:adaptive}) into generative modeling applications.
More broadly, in our experiments, we have found \ours{} to be a powerful tool for mesh optimization in visual computing, and we are eager to continue to build on top of it both in our own work and across the larger community.

%% file: sources_suppl/overview.tex
We start the supplement by providing more details on our method in Section~\ref{sec:suppl_method}, as well as further details of different isosurfacing methods in Section~\ref{sec:suppl_analysis}. Section~\ref{sec:suppl_exp_details} describes the experimental setting along with the baselines and depicts more qualitative results. Finally, in Section~\ref{sec:suppl_application}, we provide additional details and results for our applications.

%% file: sources_suppl/method_details.tex
\section{Details on \ours}
\label{sec:suppl_method}

\subsection{Tetrahedral Mesh Extraction}
In Section~4.5 of the main paper, we describe the ambiguity in connectivity when extending the tetrahedralization strategy proposed by~\citet{liang2014octree} to Dual Marching Cubes (DMC). The ambiguity arises when a cell contains multiple extracted mesh vertices. Connecting the incorrect vertices leads to intersections or holes in the extracted tetrahedral mesh. We propose two rules to address the ambiguity cases. Recall that the tetrahedra are formed in two cases: 
\begin{enumerate}
    \item For each grid edge connecting two grid vertices with \emph{different} signs, we first form a four-sided pyramid by connecting one of the grid vertices with four mesh vertices that correspond to the grid edge and then subdivide the pyramid into two tetrahedra. This case is uniquely determined in all DMC configurations (see bottom-left subfigure of Figure~10 in the main paper). %
    \item For each grid edge connecting two grid vertices with the \emph{same} sign, the tetrahedron is formed by the two grid vertices and two vertices in consecutive adjacent cells (referring to top-left subfigure in Figure~10 in the main paper). In this case, we first identify the face shared by the adjacent cells. We then identify an edge of the face with different signs and select the mesh vertex corresponding to the identified edge. Referring to Figure~7 in the main paper, in cases C6, C10, C12, and C15, it is obvious that the formed tetrahedra following this rule are always inside the primal faces. Note that the described rule can be implemented efficiently with a precomputed lookup table.
\end{enumerate}
 In most cases, these rules address the ambiguity and result in correct tetrahedralization of the interior volume. However, for case C18, the interior volume is not completely filled by the formed tetrahedra. Although this case rarely happens during optimization and does not obstruct the downstream application, it remains a limitation of our method. %

\subsection{Adaptive Mesh Resolution}
\label{sec:adaptive_meshing_suppl}
In Section~4.6 of the main paper, we describe a constraint applied to the SDF on octree vertices to avoid cracks or non-manifold surfaces being produced by DMC. Here we provide more details on how this constraint is enforced. As a preliminary, \citet{Ju2002} propose a method to generate adaptive contours from an octree representation. Their method identifies the \textit{minimal edges} of the octree, i.e., the edges which do not contain a finer edge of the adjacent cell, as well as four cells sharing each minimal edge with a recursive call. We refer readers to the original paper for more details about the algorithm. We repurpose this algorithm to identify the vertices whose SDF values we constrain. Specifically, for the four cells sharing a minimal edge, we check the four pairs of adjacent faces among these cells. If a finer face $F_{f}$ is adjacent to a coarser face $F_{c}$, for every vertex $v_f$ of $F_{f}$ that is not a vertex of $F_{c}$, we compute and store the bilinear weights of $v_f$ with respect to the vertices of $F_{c}$. During optimization, the SDF value of $v_f$ is not optimized. Instead, it is directly interpolated using precomputed bilinear weights applied to the SDF values on $F_{c}$. %
Note that the constrained vertices only need to be re-identified when there is an update to the octree structure.

\subsection{Addressing Ambiguity in DMC Configurations}
In rare cases, the original Dual Marching Cubes algorithm can produce non-manifold meshes. We follow the solution described in ~\citet{wenger2013isosurfaces} to address the ambiguity of cases C16 and C19 in the DMC configurations. With this modification applied, the extracted surface of ~\ours{} in the uniform grid setting is always 2-manifold.

%% file: sources_suppl/analysis_details.tex
\section{Analysis}
\label{sec:suppl_analysis}

This section provides further details of the experiment shown in Figure~4 in the main paper, 
where we show the optimization process for a collection of isosurfacing algorithms.

In the analysis, we follow a similar experimental setup as Section~5.1 in the main paper. Specifically, we start by initializing the SDF to represent a sphere for all methods. In each iteration, we then extract the surface mesh from the SDF (defined on a grid). Finally, we render the reconstructed mesh from randomly sampled camera views (same for all methods) and compute the differences with the ground truth depth and silhouette image. We
also compute the SDF loss, where we randomly sample 1000 points and evaluate their SDF values w.r.t the ground truth mesh, as well as the extracted mesh, and minimize the differences between two SDF
values. We use L1 loss for the silhouette image, L2 norm for the depth image, and MSE loss for the SDF. 
The combinedd loss is back-propagated to the SDF through the differentiable isosurfacing layers, which we detail in the next paragraph. We use the same optimizer and learning rate for all methods. For \ours{}, 
we leverage the regularizers described in Section 4.7 of the main paper. We also leverage  $\mathcal{L}_{\textrm{sign}}$ for all methods to remove floater and internal geometry. %
In this analysis, the grid resolution is set to 64 for the tetrahedral grid used by DMTet, and 48 for the voxel grid used by the other methods to roughly match the number of triangles in the output mesh. 

\subsection{Baselines}
We use the official implementation of \DMTet{} from the \nvdiffrec{} codebase\footnote{\url{https://github.com/NVlabs/nvdiffrec}}. For Dual Contouring (DC) and Marching Cubes (MC), there are, to the best of our knowledge, no \emph{differentiable} implementations available, so we implemented these methods in PyTorch to leverage the autograd functionalities. Specifically, we adapted the MC variant from~\citet{Lorensen1997}, following the implementation of the Marching Tetrahedra function in DMTet, such that the zero-crossings on grid edges are computed in a differentiable manner. For DC, we utilize  PyTorch's linear solver, \texttt{lstsq}\footnote{\url{https://pytorch.org/docs/stable/generated/torch.linalg.lstsq.html\#torch.linalg.lstsq}}, to solve the quadratic error function (QEF) given in Equation~2 (main paper). The gradient direction at any point, $\nabla s(u_e)$, is approximated by local differentiation over the SDF computed via trilinear interpolation. %
To avoid the solution exploding to a distant location when $\nabla s(v_e)$ are nearly coplanar, we followed the DC implementation in the NDC source code\footnote{\url{https://github.com/czq142857/NDC}} and add a regularization term which biases the solution toward the centroid of associated zero-crossings $V_E$. Specifically, Equation~2 is regularized as:
\begin{equation}
  v_d = \operatorname*{argmin}_{v_d} \sum_{u_e \in \mathcal{Z}_e} \nabla s(u_e) \cdot (v_d - u_e) + \lambda | v_d - \overline{u_e }|,
\label{eq:qef_reg}
\end{equation}
where $\overline{u_e }= \frac{1}{|V_E|} \sum_{u_e\in V_E} u_e$ is the centroid of the zero-crossing points and $\lambda $ is the scalar weight that controls the strength of the regularizer. We ablate two DC versions with $\lambda=1$ and $\lambda=0.01$, denoted as $DC_{reg1}$ and $DC_{reg001}$ respectively. In addition, we compare with a DC variant which directly takes the centroid as the mesh vertex, denoted as $DC_{centroid}$.
Please refer to Figure~2 in the main paper for illustrations of these regularized versions of DC.
For NDC, we use a pretrained neural network provided by the authors to extract the isosurface. During optimization, we freeze the network parameters and only optimize the SDF values. 

%% file: sources_suppl/main_experiments_details.tex
\section{Experimental Details}
\label{sec:suppl_exp_details}

This section provides additional details for the experiment settings in Section~5 in the main paper.

\subsection{Baselines}
The implementation of the baseline methods used in this experiment is described in Section~5.1 of the main paper. The inputs to $MC_{SDF}$ and $NDC_{SDF}$ are the ground truth SDF values evaluated at grid vertices. Since the pretrained neural network in $NDC_{SDF}$ is sensitive to the scale of the SDF inputs, we use the function\footnote{\url{https://github.com/czq142857/NDC/blob/9054e20f55013d031af9e3a2c91f5cab75837bc4/data_preprocessing/get_groundtruth_NDC/SDFGen}} provided by the authors to compute the SDF. For $DC_{hermite}$, which requires gradients as input, we additionally compute the gradient of the SDF at zero-crossings using finite differences. The regularizer weight $\lambda$ in Eqn.~\ref{eq:qef_reg} is determined independently for each cube in an adaptive manner, following the DC implementation by~\citet{ChenNDC2022}. Specifically, we begin by solving Eqn.~\ref{eq:qef_reg} with a small $\lambda$, and iteratively double the value of $\lambda$ until the solution with the updated QEF falls inside the cube or $\lambda$ reaches the limit. The initial $\lambda$ is set to 0.01 and the limit is $10^6$ in our experiment. This approach is very time-consuming to evaluate and hard to leverage in a general gradient-based mesh optimization framework. Therefore, we only adopt it in this main experiment (Section~5 in the main paper) but used fixed values of $\lambda$ in the optimization process discussed in Section~\ref{sec:suppl_analysis} (Figure~4 in the main paper) for completeness.

For the mesh reconstruction pipeline, we leverage the codebase of \nvdiffrec{}\footnote{\url{https://github.com/NVlabs/nvdiffrec}}, and replace the image loss with depth and SDF losses described in the main paper. We also leverage the mask loss in \nvdiffrec{}, and the two regularization losses $\mathcal{L}_{\textrm{sign}}$ and $\mathcal{L}_{\textrm{dev}}$ in Eqnuation~8 and Equation~9 from the main paper. We use L1 loss for mask, L2 norm for depth and MSE loss for SDF. We scale the mask loss, depth loss, SDF loss and $\mathcal{L}_{\textrm{dev}}$ by 1, 10, 2000, and 1 respectively. The scale of $\mathcal{L}_{\textrm{sign}}$ decays from 0.2 to 0.01 linearly during training. We optimize each shape for 1000 iterations with a learning rate of 0.01.

For all isosurfacing methods, we use uniform grids in $[-1,1]^3$. We center each object around the origin and scale it such that the longest side of its bounding-box equals 1.8. For NDC, the effective resolution of the grid is reduced by the padding of the network. Therefore, we increase the grid resolution for NDC by the padding size for a fair comparison with other methods.

\subsection{Evaluation Metrics}
We provide details on all the evaluation metrics we used in the main paper. 

\paragraph{Chamfer Distance (CD)} This metric measures the distance between two point clouds by nearest neighbor search.  To measure the CD between meshes, we sample each mesh to get a point cloud of size 100,000. Note that for the \nvdiffrec{} NeRF synthetic dataset evaluation (Table~5 in the main paper), we use a different version of Chamfer Distance, computed only on visible triangles, using 2.5M points on meshes with another scale than in the main experiment, so the CD numbers from Table~5 cannot be directly compared against the CD scores reported in Section~5.

\paragraph{F1-score} The harmonic mean of precision and recall. To compute precision and recall, we sample each mesh into a point cloud of the same size as CD and search for the nearest neighbor points. When computing the precision, if the distance from a point on the predicted mesh to the GT point cloud is small enough (threshold = 0.003), we count it as a \emph{true positive} point. Otherwise, it is counted as a \emph{false positive} point. When computing the recall, if the distance from a point on the GT mesh to the predicted mesh is small enough (threshold = 0.003), we count it as a true positive point. Otherwise, we count it as a false negative point.  

\paragraph{Edge Chamfer Distance (ECD) and Edge F-score (EF1)} These metrics are used in prior works~\cite{ChenNMC2021,ChenNDC2022} for evaluating the reconstruction of sharp features (edge points). First, for each point in the sampled point cloud, we look at the dot products between its normal and the normal of its neighbor points. If the mean dot product is smaller than a threshold (0.2), the point is treated as an edge point.  ECD and EF1 measure the Chamfer Distance and F1-score between edge point sets.

\paragraph{The percentage of inaccurate normals (IN> $5^{\circ}$)} For each point in the sampled point cloud, we store the normal of the face that the point was generated from. Given a predicted mesh and a GT mesh, we search for nearest point pairs from one to another. We compute the angles between the normals stored on two points, and report the percentage of pairs having angles larger than 5 degrees.

\paragraph{Aspect Ratio (AR) and Radius Ratio (RR)} AR and RR are different measures of triangle regularity, with a smaller value indicates better triangle quality. While there exist different definitions of AR and RR in the literature, we follow the definition in the PyVista\footnote{\url{https://docs.pyvista.org/api/core/_autosummary/pyvista.DataSet.compute_cell_quality.html}} codebase in our evaluations.

\paragraph{Min and max angles} Given a triangle on the extracted mesh, we compute its three angles in degrees and select the max and min angles.

\paragraph{NV(\%)} The average percentage of non-manifold vertices.

\paragraph{NE(\%)} The average percentage of non-manifold edges.

\paragraph{SI(\%)} The average percentage of self-intersecting triangles.

\paragraph{SA<$10^{\circ}$} The average percentage of triangles with the smallest angle <$10^{\circ}$.

\subsection{Adaptive Meshing}

We provide experimental details for results demonstrated in Figure~14 in the main paper. Our goal is to reconstruct the target object with adaptive mesh resolution achieved by jointly optimizing mesh topology and the octree structure. We begin the optimization with a low-resolution, uniform voxel grid. During optimization, we keep a running average of the objective loss for each cell, computed by averaging the loss from all mesh vertices extracted from it. After the shape converges, we subdivide the cells in the octree with objective loss larger than a preset threshold of 0.04. Iterating this process, we obtain the adaptive mesh shown in Figure~14 without precomputed octree structure on GT geometry. Note that we apply the constraint described in Section~\ref{sec:adaptive_meshing_suppl} during optimization.

\subsection{Additional Results}
We include more visual examples in Figure~\ref{fig:sup_main_qual},
comparing all methods. Please zoom in the pdf to see the differences. Additional statistic are included in Table~\ref{tbl:main64large}.

\tabMainOne

\figRobustness

In Figure~\ref{fig:robustness} we show how \ours{} robustly reconstructs the same object
under different rotations. Note that the MC
reconstruction quality deteriorates when features are not axis-aligned. We show the influence of the regularizer $\mathcal{L}_{\textrm{dev}}$ (Equation~8 in the main paper) in Figure~\ref{fig:reg_abl}. 
\figRegAbl

\figSupplementalMainQual

%% file: sources_suppl/application_details.tex
\section{Applications}
\label{sec:suppl_application}
In this section, we provide a detailed description of the experimental setting and additional results for each application we did in the main paper.

\subsection{Photogrammetry Through Differentiable Rendering}
\label{sec:photogrammetry}
\figNVDIFFRECNerfSupplemental

Our photogrammetry application is based on \nvdiffrec{}, which jointly optimizes shape, materials, and lighting from image supervision~\cite{munkberg2021nvdiffrec,hasselgren2022nvdiffrecmc}. We followed the official codebase closely with minimal changes and only replaced the \textsc{DMTet}~\shortcite{Shen2021} geometry
extraction stage with \ours{}. We leverage the regularizer term, $\mathcal{L}_{\textrm{dev}}$, of Equation 8 from the main paper with a factor $\lambda_{dev} = 0.25$ in all experiments. The regularizer, $\mathcal{L}_{\textrm{sign}}$, (Equation 9, main paper) is already present with \textsc{DMTet} in \nvdiffrec{}. Additionally, we scale the silhouette mask loss of \nvdiffrec{}, with a factor, $\lambda_{mask} = 5.0$, to further emphasize geometry.
The combined objective function is:
\begin{equation}
\mathcal{L}_{\textrm{total}} = \lambda_{dev} \mathcal{L}_{\textrm{dev}} + \mathcal{L}_{\textrm{sign}} + \lambda_{mask} \mathcal{L}_{\textrm{mask}}.
\end{equation}
We leave the second pass of \nvdiffrec{} (which performs light, material, and shape optimization with fixed topology) unmodified.

We show shaded results and mesh illustrations in Figure~\ref{fig:nvdiffrec_nerf_suppl} for the entire NeRF dataset. 

\subsection{Mesh Simplification of Animated Objects}

We extend our photogrammetry application (from Section~\ref{sec:photogrammetry}) by adding support for mesh based, synthetic, datasets with skinned animations. In practice, we use the skinning provided by Universal Scene Description (USD)~\shortcite{USD2016} and source our animated meshes from RenderPeople~\shortcite{RenderPeople2020}. We achieve mesh simplification by using a coarse \ours{} grid ($32^3$) and optimize geometry using image supervision. In this application, we assume that the skeleton animation and 
reference skinning weights are known, and the task at hand is to learn a simplified mesh, and retarget the animation using the same skeleton, without manual adjustments. 

Mesh simplification using only the T-pose is straightforward and can be done in \nvdiffrec{} out of the box, with either \textsc{DMTet} or \ours{} for the topology extraction step. In each iteration, we pick a random viewpoint and optimize the parameters using photometric loss. We then re-skin the reconstructed, simplified mesh in a post-processing step as follows: for each vertex in the simplified mesh, we find the k-nearest neighbors ($k=10$) in the reference mesh and use inverse distance weighting to compute skinning weights for the simplified vertex. More formally, given a vertex, $\mathbf{v}_{\mathrm{lod}}$, in the simplified mesh and the k-nearest neighbors, $\mathbf{v}_\mathrm{ref}^i$, and their skinning weights, $w_\mathrm{ref}^i$, from the reference mesh, the skinning weight, $w_\mathrm{lod}$,  for the simplified mesh is computed as:
\begin{eqnarray}
	d\left(\mathbf{v}_\mathrm{ref}^i, \mathbf{v}_{\mathrm{lod}}\right) &=& \frac{1}{\max(10^{-3}, \lVert \mathbf{v}_\mathrm{ref}^i - \mathbf{v}_{\mathrm{lod}} \rVert_2)} \nonumber \\
	w_\mathrm{lod} &=& \frac{
		\sum_i d\left(\mathbf{v}_\mathrm{ref}^i, \mathbf{v}_{\mathrm{lod}}\right)w_\mathrm{ref}^i
	}{
		\sum_i d\left(\mathbf{v}_\mathrm{ref}^i, \mathbf{v}_{\mathrm{lod}}\right)
	}.
\label{eq:guess_skinning}
\end{eqnarray}

Performing end-to-end mesh simplification over the animation is more challenging. We first optimize for 300 iterations using the T-pose, as described above, to get a reasonable initial guess for geometry. We then enable the animation and optimize for random viewpoints and random animation frames. For this to work, our pipeline must support animation of a mesh with changing topology in a consistent way with gradients propagating back to the topology representation.
In each iteration we re-skin the \ours{} mesh using a differentiable version of the same k-nearest neighbor method outlined in Equation~\ref{eq:guess_skinning}, and animate the re-skinned mesh using the differentiable skinning approach of Hasselgren~et~al.~\shortcite{hasselgren2021nvdiffmodeling}. 
Note that, while we do not explicitly optimize skinning weights, each operation must be differentiable to enable end-to-end training. An interesting avenue for future work
is to also include optimization of the skinning weights, but that would require a consistent parametrization, as vertices can be added or removed during optimization as the topology evolve. This is similar to the texture parameterization problem in \nvdiffrec~\shortcite{munkberg2021nvdiffrec} work, which is solved by using 3D texturing 
(encoded in a MLP) during the topology optimization phase.

\subsection{3D Mesh Generation}

In the application of mesh generation, we adopt the recent state-of-the-art 3D generative model GET3D~\cite{gao2022get3d}, and show that \ours{} as a plug-and-play differentiable mesh extraction module can produce significantly improved mesh quality. 

GET3D~\cite{gao2022get3d} is a learning-based model trained on 2D images, and can directly synthesize high-quality textured 3D meshes at inference time. The framework combines classic generative adversarial networks~\cite{karras2019style,Karras2019stylegan2}, differentiable iso-surfacing~\cite{Shen2021} and differentiable rasterization-based rendering~\cite{Laine2020diffrast}. 
Given a sampled noise vector from a Gaussian distribution, the generator of GET3D predicts a signed distance field. Then, a mesh is extracted by DMTet~\cite{Shen2021}, and a differentiable renderer renders one RGB image and one 2D silhouette, which are fed into 2D discriminators to classify whether they are real or fake. The differentiable iso-surfacing module enables the ability to directly generate meshes, and also largely affects the quality of the produced meshes. 

In this application, we replace the DMTet module with \ours{}. In particular, we modify the last layer of the 3D generator in GET3D to output the SDF and deformation for each vertex, and additionally generate 21 weights for every cube defined in \ours{}, including the 8 vertex weights, 12 edge weights, and the remaining 1 parameter for quad splitting. \ours{} is adopted to differentiably extract meshes, and the remaining architecture, training procedure and other hyperparameters of GET3D are kept unchanged following the official released implementation\footnote{\url{https://github.com/nv-tlabs/GET3D}}. We follow GET3D and train the 3D generative model on ShapeNet~\cite{shapenet} Car, Chair and Motorbike categories using the same dataset rendering pipeline and train/validation/test split released by the official implementation.
Note that the architecture changes happen only at the last layer of the generator while the rest of the backbone remains the same. Thus, the computational overhead of the modification is relatively negligible. 
We include more qualitative visualization in Figure~\ref{fig:ApplicationMeshGenerationSupp}. 

\figApplicationMeshGenerationSupp

\input{sources_suppl/supp_physics.tex}

%% file: sources_suppl/supp_physics.tex
\subsection{Physics Simulation}

\figphysicssupp

\ours{} enables extracting tetrahedral meshes with well-defined topology, which can be directly utilized in physical simulation.  
It can be further combined with differentiable physical simulation frameworks~\cite{gradsim,hu2019difftaichi} and differentiable rendering pipelines~\cite{chen2019dibrender,Laine2020diffrast,munkberg2021nvdiffrec} to optimize shape, material and physical parameters from multi-view videos. We have shown two examples in teaser and Figure~\ref{fig:physicus} in the main paper. In this subsection, we talk about the details of our physical simulation experiments. 

\paragraph{Ground Truth Generation}

We first prepare multi-view ground truth videos with \ours{}. 
Given a surface mesh, we apply \ours{} to learn the shape with 3D supervision, as well as employing texture field~\cite{mueller2022instant,munkberg2021nvdiffrec} to learn the texture map from multi-view 2D images (Section~6.1 in the main paper). 
\ours{} supports directly exporting a tetrahedral mesh. We then send it to the physical simulation framework. Here we adopt a low-res grid ($32^3$ resolution) to extract the tetrahedral mesh such that the physical simulation can be more efficient.
We choose to use GradSim~\cite{gradsim}, a differentiable physics simulation framework which has shown both forward simulation and derivatives w.r.t. the physical parameters in the backward pass.
We focus on FEM simulation and use neo-Hookean elasticity to model elastic objects during the simulation.
During the forward simulation, we fix the two ending points of an object, letting it drop and deform under gravity. 
We choose time step as $\frac{1}{8192}s$ and set the mass density $D=0.5$,  For the Lamé parameters, $\lambda, \mu$, which control the element’s resistance to shearing and volumetric strains, we set  $\mu=1000$,  $\lambda=1000$,  and a damping coefficients $d=1.5$ for the example in the teaser.  For Figure~\ref{fig:physicus} in the main paper, we choose  $D=0.3$, $\mu=1000$,  $\lambda=1000, d=1.5$.
With the deformed meshes, we then employ the differentiable rendering pipeline~\cite{Laine2020diffrast} to render them into images and composite into videos.
We generate videos with 512 views, where we randomly set circular camera positions around the object.
Note here we do not render images at all the time steps, but instead, we choose video fps as 64.

\paragraph{Training}

Given multi-view video sequences, we then optimize shape, texture, and physical materials from the video input only. As a challenging task, it is extremely hard to jointly optimize geometry and physical parameters together.  In practice, we find FEM simulation is quite unstable, {\emph{e.g.},} joint optimization always has NaN values and crashed in the training. Therefore, following recent work~\cite{pacnerf} (one anonymous ICLR submission 2023 at the time of our submission), we optimize the shape, texture, and physical parameters in a two-stage manner.

First, we apply the beginning frame of the video to optimize the shape and texture only. We assume the object doesn't deform in the first frame. Therefore, it is equivalent to a rigid-body mesh reconstruction (Section 6.1 in the main paper). We use the same losses but slightly tune the weights (we set $\lambda_{dev} = 1.0$ and $\lambda_{mask} = 10.0$).
The first-stage optimization allows us to start from an descent shape to make the physical parameter optimization easier.  

In the second stage, we start from the initial guess of the mass density (we initialize it as $D=1.5$) and apply GradSim~\cite{gradsim} to compute the gradients of the video loss w.r.t. to the mass density. 
At each iteration, we send the tetrahedral mesh to GradSim and execute forward simulation to deform the shape. Then, we render the deformed shape into images and compare them with the ground truth images.
We use the same loss as in the first stage and backpropagate it to the mass density. We use a different learning rate schedule here. At the beginning, we use the learning rate 0.1 and decay it 10 times smaller every 50 iterations).
The whole optimization converges in 200 iterations.  

It is worth mentioning that when extracting the tetrahedral mesh, some tetrahedra can have tiny volume, which could lead to unstable physical simulation due to numerical issues.  Therefore, we conduct a tetrahedra filtering process after tetrahedral mesh extraction. 
Specifically, we check the volume of each tetrahedron and remove the one whose volume is less than a threshold ($2e^{-7}$ for a shape normalized in $(-0.45, 0.45)$). We find this step significantly improves the stability of the physics simulation, though at the cost of introducing some slits of the shape, as shown in Figure~\ref{fig:physicus2}. We hope this can be further addressed by designing new regularization terms, which we leave for future work. 
In Figure~\ref{fig:physicus2}, we provide more details of the physics simulation examples in the teaser and Figure~\ref{fig:physicus}.